\documentclass[11pt,a4paper,twoside,openright]{report}

\let\hbarorig\hbar

\usepackage{framed}
\usepackage[all]{xy}

\usepackage{amsmath}
\usepackage{xcolor}
\usepackage{mathtools}
\usepackage{amssymb}
\usepackage{braket}
\usepackage[bottom]{footmisc}
\usepackage{cancel}
\usepackage[margin=2.5cm, includehead, includefoot]{geometry}  
\usepackage{graphicx}
\usepackage[Bjornstrup]{fncychap}
\graphicspath{{Images}}
\usepackage[english]{babel}
\usepackage[utf8x]{inputenc}
\usepackage[T1]{fontenc}
\usepackage{cancel}
\usepackage{amsmath}
\usepackage{comment}
\usepackage{bbm}
\usepackage{amssymb}
\usepackage{mathtools}
\usepackage{centernot}
\usepackage{braket}
\usepackage{cite}
\usepackage{mathrsfs}
\usepackage{cases}
\usepackage[colorinlistoftodos]{todonotes}
\usepackage[colorlinks=true, allcolors=blue, linktocpage=true]{hyperref}
\usepackage{fancyhdr}
\usepackage{enumerate}

\newcommand{\bn}{{\boldsymbol{n}}}
\newcommand{\bm}{{\boldsymbol{m}}}
\newcommand{\br}{{\boldsymbol{r}}}
\newcommand{\bx}{{\boldsymbol{x}}}
\newcommand{\bv}{{\boldsymbol{v}}}
\newcommand{\bu}{{\boldsymbol{u}}}
\newcommand{\bw}{{\boldsymbol{w}}}
\newcommand{\by}{{\boldsymbol{y}}}
\newcommand{\bs}{{\boldsymbol{s}}}
\newcommand{\bq}{{\boldsymbol{q}}}
\newcommand{\bp}{{\boldsymbol{p}}}

\newcommand{\bA}{{\boldsymbol{A}}}
\newcommand{\bB}{{\boldsymbol{B}}}
\newcommand{\bE}{{\boldsymbol{E}}}
\newcommand{\bJ}{{\boldsymbol{J}}}
\newcommand{\bR}{{\boldsymbol{R}}}
\newcommand{\bS}{{\boldsymbol{S}}}

\newcommand{\bomega}{{\boldsymbol{\omega}}}
\newcommand{\bsigma}{{\boldsymbol{\sigma}}}
\newcommand{\boldeta}{{\boldsymbol{\eta}}}
\newcommand{\bmu}{{\boldsymbol{\mu}}}
\newcommand{\bnu}{{\boldsymbol{\nu}}}
\newcommand{\bbeta}{{\boldsymbol{\beta}}}
\newcommand{\bkappa}{{\boldsymbol{\kappa}}}
\newcommand{\bupsilon}{{\boldsymbol{\upsilon}}}
\newcommand{\bgamma}{{\boldsymbol{\gamma}}}
\newcommand{\bxi}{{\boldsymbol{\xi}}}

\newcommand{\bLambda}{{\boldsymbol{\Lambda}}}
\newcommand{\bOmega}{{\boldsymbol{\Omega}}}
\newcommand{\bUpsilon}{{\boldsymbol{\Upsilon}}}

\newcommand{\de}{{\rm d}}

\newcommand{\clearemptydoublepage}{\newpage\phantom{}\thispagestyle{empty}\newpage}

\makeatletter
\def\cleardoublepage{\clearpage\if@twoside \ifodd\c@page\else
\hbox{}
\vspace*{\fill}
\begin{center}
\end{center}
\vspace{\fill}
\thispagestyle{empty}
\newpage
\if@twocolumn\hbox{}\newpage\fi\fi\fi}
\makeatother

\let\fn\footnote
\renewcommand{\footnote}[1]{\linespread{1.1}\fn{#1}\linespread{1.29}}

\newtheorem{theorem}{Theorem}[chapter]
\newtheorem{remark}[theorem]{Remark}
\newtheorem{definition}[theorem]{Definition}
\newtheorem{corollary}[theorem]{Corollary}
\newtheorem{proposition}[theorem]{Proposition}
\newtheorem{lemma}[theorem]{Lemma}

\let\hbar\hbarorig

\begin{document}
  
\pagenumbering{gobble}
\definecolor{surreygold}{HTML}{907c09}
\definecolor{surreyblue}{HTML}{001d4d}

 \begin{titlepage}
   \centering
   
   \begin{figure}[htbp]
\centering
\includegraphics[scale=0.2]{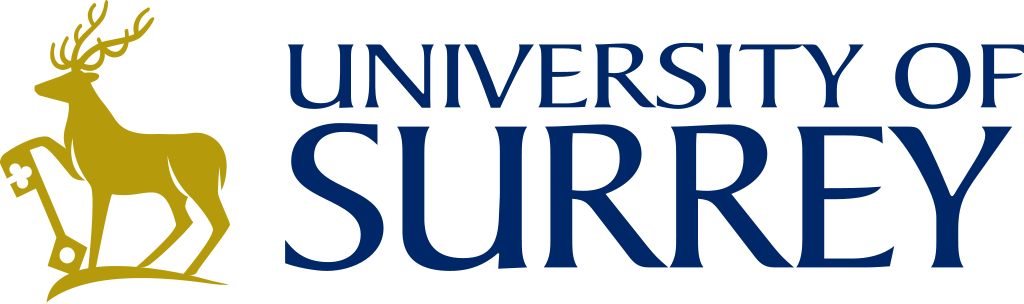}
\end{figure}
\vspace{1cm}
   
{\bfseries \huge \color{surreyblue} Geometry of quantum hydrodynamics in theoretical chemistry \par}
\vspace{1cm}
{\bfseries\huge \color{surreygold} Michael S. Foskett\par}
\vspace{1.5cm}
{\normalsize Thesis submitted to the University of Surrey\\ for the degree of Doctor of Philosophy \par}
\vspace{1.5cm}
{\large\it Department of Mathematics}\\
{\large\it University of Surrey }\\
{\large\it Guildford GU2 7XH, United Kingdom }\\
\vspace{1cm}

\begin{figure}[htbp]
\centering
\includegraphics[scale=0.1]{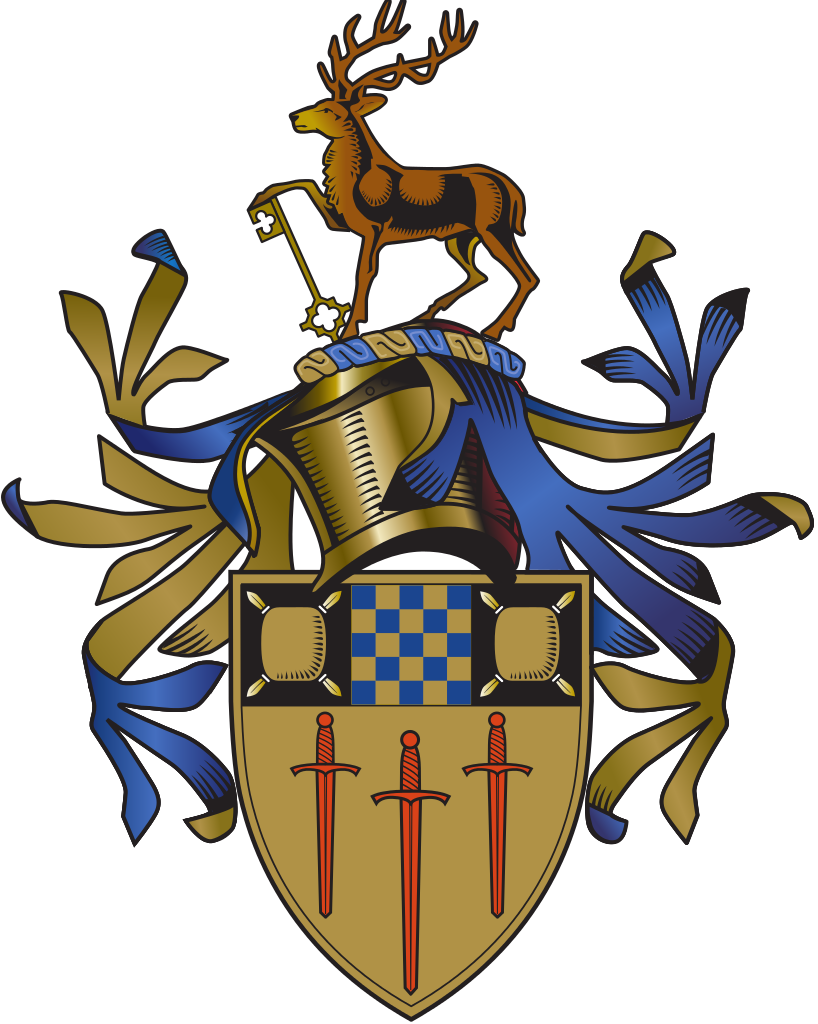}
\end{figure}

\vspace{1cm}
{ \normalsize Copyright © 2020 by Michael Foskett. All rights reserved. \par}
{ \normalsize E-mail address: \href{mailto:m.foskett@surrey.ac.uk}{\texttt{m.foskett@surrey.ac.uk}} \par}
\end{titlepage}
\null\newpage
\pagenumbering{roman}
\section*{Scientific abstract}

This thesis investigates geometric approaches to quantum 
hydrodynamics (QHD) in order to develop applications in theoretical quantum chemistry. 

Based upon the momentum map geometric structure of QHD and the associated 
 Lie-Poisson and  Euler-Poincar{\'e} equations, alternative geometric approaches to the classical limit in QHD are 
presented. Firstly, a new regularised Lagrangian is introduced, allowing for singular solutions called `Bohmions' for which the associated trajectory equations are finite-dimensional and depend on a smoothened quantum potential. 
Secondly, the classical limit is considered for quantum mixed states. By applying a cold fluid closure to the density matrix the quantum potential term is eliminated from the Hamiltonian entirely. 

The momentum map approach to QHD is then applied to the nuclear dynamics in a chemistry model known as exact factorization. 
A variational derivation of the coupled electron-nuclear dynamics is presented, comprising an Euler-Poincar{\'e} structure for the nuclear motion.
 The QHD equations for the nuclei possess a Kelvin-Noether circulation theorem which returns a new equation for the evolution of the electronic Berry phase. 
 The geometric treatment is then extended to include unitary electronic 
 evolution in the frame of the nuclear flow, with the resulting dynamics carrying both 
Euler-Poincar{\'e} and Lie-Poisson structures. A new mixed quantum-classical model is then derived by applying both the QHD regularisation and cold 
fluid closure to a generalised factorisation ansatz at the level of the 
molecular density matrix.

A new alternative geometric formulation of QHD is then constructed. Introducing a $\mathfrak{u}(1)$ connection 
 as the new fundamental variable provides a new method for incorporating holonomy in QHD,  which follows from its constant non-zero 
 curvature. The associated fluid flow is no longer constrained to be irrotational, thus possessing a non-trivial circulation theorem 
and allows for vortex filament solutions.  This approach is naturally extended to include the coupling of vortex filament dynamics to the Schr\"odinger equation.
This formulation of QHD is then applied to Born-Oppenheimer molecular dynamics suggesting new insights into the role of Berry phases in adiabatic 
phenomena.

Finally, non-Abelian connections are then considered in quantum mechanics. The dynamics of 
the spin vector in the Pauli equation allows for the introduction of an $\mathfrak{so}(3)$ 
connection whilst a more general $\mathfrak{u}(\mathscr{H})$ connection can be 
introduced from the unitary evolution of a quantum system. This is used to 
provide a new picture for the Berry connection and quantum geometric tensor 
and well as derive more general systems of equations which feature explicit dependence on the curvature
of the connection. Relevant applications to quantum chemistry are then considered.

\vfill
\noindent
Keywords: Geometric mechanics, Euler-Poincar\'e, 
Lie-Poisson, Quantum hydrodynamics, Geometric phase, Mathematical physics, Chemical physics, Quantum 
physics.
\vspace{1cm}

\clearemptydoublepage

\section*{Lay summary}

The advantages of applying geometric approaches to mechanical systems are both varied and numerous. 
One of the most powerful features of geometric mechanics is that it provides a unified framework
under which all manner of physical systems can be described. In addition, 
through its abstract mathematical formalism, often seemingly unrelated physics 
can be shown to possess fundamentally equivalent geometric structures, often 
allowing previous knowledge to provide new insights into other fields.
In this thesis we apply the tools of geometric mechanics to study the hydrodynamic picture of quantum mechanics to develop applications in theoretical 
quantum chemistry. 

The main problem in quantum chemistry is the computational difficulty in solving the equations governing molecular 
processes involving a large number of constituent particles when written in terms of their true quantum mechanical description.
As a result, much of the effort in the field of quantum chemistry is focused
 on creating approximate models to describe such systems. A standard approach is to separate the system in two parts, treating one part in terms of classical mechanics whilst retaining the crucial quantum description in the other.
A convenient first step for performing this classical limit in one sector is to 
use the hydrodynamic description of quantum mechanics.

In this thesis, the geometric approach is used to study a wide class of models in quantum chemistry
based around the so-called Born-Oppenheimer factorisation and its recent extension, exact factorization.
 In particular, the geometry of these systems is investigated and for the latter, shown to have the same 
 structure as the previous geometric approaches in the study of liquid crystal 
 flows. In addition, new methods of invoking a classical description are used to 
 construct an entirely new model which extends the exact factorization approach currently in use.
 Finally, inspired by the analogy with liquid crystal flows, we transfer some of 
 the existing geometric tools in this field into quantum chemistry. The 
 result is a new alternative geometric formulation for quantum hydrodynamics 
as well as a new geometric object measuring the interaction between electrons and nuclei in quantum chemistry.
In both cases, this leads to interesting new perspectives which are hoped to result in physically 
relevant applications.

\clearemptydoublepage

\section*{Acknowledgements}

I would like to thank my PhD supervisor, Cesare Tronci, for his continued support and guidance
during the course of this research as well as giving me the freedom and belief to pursue my own interests.
It has been a great pleasure to explore the quantum world together through the illuminating lens of geometric mechanics.

A very special thanks also to Darryl Holm for sharing his expertise and enthusiam in my research.
Your invaluable input and advice is so much appreciated. I am also very grateful 
to Darryl for giving me the opportunity to share my research and participate in his group meetings at Imperial College, London.

A further thank you must go to the Applied Geometric Mechanics Network for inviting me to present my research on two occasions and
to participate in many enlightening meetings.

Thank you also to everyone I have had the pleasure of interacting with during my PhD from the University of Surrey. 
Thank you to Eran Ginossar from the physics department for acting as my secondary 
supervisor. A particular mention must go to Alessandro Torrielli, whose infectious enthusiam for quantum 
mechanics captured my interest in way that has continued to inspire my academic path over the last five 
years.

I feel very privileged to have had the opportunity to spend time among world leading mathematicians in the field of Hamiltonian systems 
as a program associate at the Mathematical Sciences Research Institute in Berkeley, CA, during the fall semester of 2018.
In addition, it has been a great pleasure to interact with cutting edge research chemists. Thank you to CECAM for a fascinating school on nonadiabatic dynamics
 at EPFL in 2018, as well as to the CHAMPS program for inviting me to your Bristol meeting in April 2019. 

Finally, the greatest thank you of all is to my family. To my parents, for 
encouraging my inquisitive nature from a young age and for supporting me in 
every possible way since then. Thank you to my wife, Nika, for all your support
and encouragement. You have always believed in my abilities and given me the confidence to aim
for the highest goals. Without you, none of this would have been possible.

\paragraph{Funding:}
I also kindly acknowledge the joint studentship funding from EPSRC Grant No. EP/M508160/1 and the University of Surrey.
Thank you also to the University of Surrey FEPS Faculty Research Support Fund, the Institute of Mathematics and its Applications 
Small Grant Scheme 2018 and the MSRI for assisting with my overseas trips during my PhD. 



%


\tableofcontents
\null\newpage
\thispagestyle{empty}

\pagenumbering{gobble}


%
%

\clearemptydoublepage

\pagestyle{fancy}
\renewcommand{\sectionmark}[1]{\markboth{\thesection~#1}{}} 
\fancyhead[RO,LE]{\thepage}
\fancyhead[CO,CE]{}
\fancyhead[LO,RE]{\nouppercase{\leftmark}}
\fancyfoot{}

\pagenumbering{arabic}
\chapter{Introduction and outline}\label{Chap1}
The purpose of this introductory chapter is to familiarise the reader with the 
 material contained within this thesis.  The first two sections provide a brief overview of the background physics and chemistry and are focused mostly on setting the scene, whilst a more in-depth mathematical introduction is given in Chapter \ref{Chap2}.
 
Section \ref{Graham} reviews the fundamental principles of quantum mechanics as well as outlining two additional formulations that will be also be discussed in this chapter.

Section \ref{QuantChem} introduces the field of quantum chemistry, specifically the Born-Oppenheimer approximation of molecular dynamics, which serves as a seed 
from which all of the key topics that informed the development of the author's work can be established. 

The remainder of this chapter outlines the motivation for this research and open questions
that this thesis aims to answer (Section \ref{Sec:Motivation}) as well as presenting a summary of the key results (Section 
\ref{Sec:Outline}).

\section{Preliminary quantum mechanics}\label{Graham}
At any time, the state of a quantum system is defined by a normalised vector $\psi(t)$ 
belonging to some Hilbert space $\mathscr{H}$. Observables (corresponding to measurable quantities) 
are described by self-adjoint operators on $\mathscr{H}$. In particular, the 
total energy of the quantum system is associated with the Hamiltonian operator 
$\widehat{H}$ which, analogous to the role of the Hamiltonian in classical mechanics, governs the 
time evolution of the system. In quantum mechanics, the dynamics of $\psi$ are determined by the celebrated Schr\"odinger equation 
\cite{Schrodinger1926}
\begin{align*}
  i\hbar\dot{\psi} = \widehat{H}\psi\,,
\end{align*} 
which enjoys both variational and Hamiltonian structures as described in Section 
\ref{Sec:1-GeomQM}.

When a quantum system has a classical analog, the state of the system is given by a 
wavefunction $\psi(\bx,t)$ belonging to the Hilbert space $\mathscr{H}=L^2(\mathbb{R}^3)$ 
of square-integrable functions, so that normalisation of the wavefunction is written as $\int |\psi(\bx,t)|^2\,\de^3x=1$, where
$|\psi|^2:=\psi^*\psi$ is the probability density. For such systems the 
Hamiltonian is of the type $\widehat{H}=\widehat{H}(\widehat{Q},\widehat{P})$,
 where $(\widehat{Q},\widehat{P})$ are the position and momentum operators satisfying 
$[\widehat{Q},\widehat{P}]=i\hbar\mathbbm{1}$. Usually, the Hamiltonian is expressed in the form $\widehat{H}=\widehat{P}^2/2m+{V}(\widehat{Q})$ (kinetic + potential) 
so that the Schr\"odinger equation reads
\begin{align*}
i\hbar\partial_t\psi(\bx,t)=-\frac{\hbar^2}{2m}\Delta\psi(\bx,t) + 
V(\bx)\psi(\bx,t)\,.
\end{align*}
There are two additional formulations of quantum theory 
which it is helpful to introduce to complete this first section.
\subsubsection{Mixed states and density matrices}
Whilst the above formulation of quantum mechanics is perhaps the most well-known, it in 
fact only accounts for a particular class of quantum systems known as pure 
states. More generally, one considers mixed quantum states described by a density matrix $\rho$, itself a special type of operator on the Hilbert space $\mathscr{H}$.  
Without loss of generality this can be expressed in terms of a collection of pure states $\{\psi_a\}$, given by
\begin{align*}
  \rho = \sum_a w_a \psi_a \psi_a^{\dagger}\,,
\end{align*}
where $\text{Tr}(\rho)=\sum_{a} w_a=1$ ensures the normalisation and 
pure states are recovered by the projection operator $\rho=\psi\psi^{\dagger}$.

The dynamics of such a density matrix can be computed by 
combining this expression with the Schr\"odinger equation for each 
pure state $\psi_a$ to yield
\begin{align*}
  i\hbar\dot{\rho}=[\widehat{H}, \rho]\,,
\end{align*}
which is called the {\it von Neumann equation} or sometimes the {\it Liouville-von Neumann equation}, due to the natural analog with classical mixed states \cite{Tronci2018}.
Density matrices representing mixed states reflect systems of a classical statistical nature in which the precise preparation is not 
known and are required to describe any system which is not isolated, instead interacting with an external environment through a process known as {\it decoherence}. 

We remark that mixed states are also described in the phase space picture of 
quantum mechanics \cite{ZaFaCu2005}, in which the state of the system is given 
by a {\it Wigner function} satisfying the {\it Wigner-Moyal equation}. This 
formulation makes only a brief appearance in this thesis, however an overview of the general theory
is provided in Appendix \ref{App:Wigner}.

\subsubsection{Hydrodynamic picture}
That quantum mechanics admits a hydrodynamic formulation was recognised in the earliest days of quantum theory by Madelung 
\cite{Madelung1926,Madelung1927}. This equivalent formulation of quantum theory emerges by rewriting the wavefunction in its polar form ({\it Madelung transform})
\begin{align*}
\psi(\bx,t)=\sqrt{D(\bx,t)}e^{{i}{\hbar}^{-1} S(\bx,t)}\,,
\end{align*}
thus taking the Schr\"odinger equation into the coupled system
\begin{align*}
\partial_t D + \text{div}\left(\frac{D\nabla S}{m}\right)&=0\,,\\
   \partial_t S+\frac{|\nabla S|^2}{2m}-\frac{\hbar^2}{2m}\frac{\Delta\sqrt{D}}{\sqrt{D}}+V  
   &=0\,.
   \end{align*}
The first equation is clearly the continuity equation for the probability density $D=|\psi|^2$, whilst the second equation resembles the Hamilton-Jacobi equation of classical mechanics, albeit with an additional 
term, referred to as the {\it quantum potential}
\begin{align*}
  V_Q := -\frac{\hbar^2}{2m}\frac{\Delta\sqrt{D}}{\sqrt{D}} \,. 
\end{align*}
Madelung's insight was to recognise that, upon defining a fluid velocity field $\bu:=m^{-1}\nabla 
S$, these equations can be rewritten in terms of the new variables $(D,\bu)$ so that 
one obtains the hydrodynamical system
\begin{align*}
\partial_t D + \text{div}\left(D\bu\right)&=0 \,,\\
 m (\partial_t+\bu\cdot\nabla)\bu&= -\nabla(V+V_Q)\,,
   \end{align*}
   corresponding to a quantum version of the Euler equations of fluid mechanics.
   These quantum hydrodynamic (QHD) equations were the starting point for 
   several further interpretations of quantum mechanics, most notably 
   Bohmian mechanics \cite{Bohm1952}, inspired by the earlier works of de Broglie \cite{deBroglie1927}, in which one considers a physical particle that is guided by the quantum 
   fluid. The interpretative nature of such theories is not considered in this thesis.
   
Having previously explained that pure states only comprise a subset of all possible quantum states,
 we remark that one can also formulate a hydrodynamic picture for quantum mixed states.
 In the standard approach, hydrodynamic variables are defined by a set of projections of the density matrix  
known as {\it moments}. As explained in greater detail in Section \ref{Sec:1-QHD}, 
these hydrodynamic variables form a countably infinite hierarchy of equations 
which, in the special case of pure states, close at first order resulting in the QHD 
equations above. As the equations do not close in general, there is much practical and theoretical interest in 
finding suitable truncations and closures of the system. 
     
Overall, the hydrodynamic picture of quantum mechanics will be central to the work in this thesis and hence will be considered in further detail in Section \ref{Sec:1-QHD}.

\section{The seed of quantum chemistry}\label{QuantChem}
Ever since Bohr’s model of the atom in 1913 \cite{Bohr1913}, it has been known that one needs to employ a quantum mechanical description in order to accurately describe structures on an atomic scale.
Further developments in quantum theory in the 1920s, including the discovery of the Schr\"odinger equation \cite{Schrodinger1926}, provided a theoretical description of all atomic and molecular processes, giving a rigorous mathematical foundation for all of the chemistry that preceeded it.
Indeed, the quantum description of the simplest atom, hydrogen, is celebrated as one of the greatest scientific achievements of the 20th century.

For our purposes, a molecule can be considered as a collection of $N_n$ nuclei with Cartesian coordinates $\{\br_k\}$, each having mass $M_k$ and charge $Z_k \,q_e$, and $N_e$ electrons 
with Cartesian coordinates $\{\bx_{\ell}\}$, each with mass $m_e$ and charge $-q_e$, where $q_e \approx 1.6\times 10^{-19}$C is the elementary charge, $m_e\approx 9.1\times 10^{-31}$kg is the rest mass of the electron and $Z_k$ is the atomic number of each nuclei.
The complete quantum mechanical description of the molecule is given by a complex wavefunction 
$\Psi(\{\br_k\},\{\bx_{\ell}\},t)$, whose time evolution is  governed by the {\it molecular Schr{\"o}dinger 
equation} 
\begin{equation}
i\hbar \frac{\partial}{\partial t}\Psi=\left(\widehat{T}_n + \widehat{T}_e + \widehat{V}_{n} + \widehat{V}_{e} + 
\widehat{V}_{I}\right)\Psi := 
\widehat{H}\Psi \label{molecularSchrodinger}\,,
\end{equation}
in which $\widehat{T}$ and $\widehat{V}$ refer to the kinetic energy and potential energy operators, while the subscripts $n$ and $e$ denote nuclear and electronic energies, respectively and the subscript $I$ refers to the interaction potential between nuclei and electrons.
The Hamiltonian operator $\widehat{H}$ is referred to as the {\it molecular Hamiltonian} and is more explicitly 
written as
\begin{multline}
\widehat{H}= -\sum^{N_n}_{k=1}\frac{\hbar ^2}{2M_k}\Delta _{{\br}_k}-\sum^{N_e}_{\ell=1}\frac{\hbar ^2}{2m_e}\Delta_{{\bx}_{\ell}} \\
+  \sum_{j}^{N_n}\sum_{k>j} \frac{Z_j \, Z_k \,q_e^2}{|{\br}_j - {\br}_k|} +\sum_{m}^{N_e}\sum_{\ell>m} \frac{q_e^2}{|{\bx}_{m}- {\bx}_{\ell}|}-\sum_{\ell = 1}^{N_e}\sum_{k = 1}^{N_n} \frac{Z_k \,q_e^2}{|{\bx}_{\ell} - {\br}_k|}\,, \label{molecularHamiltonian}
\end{multline}
in which we have set the physical constant $(4\pi \epsilon_0)^{-1}$ equal to unity in each of the 
potential terms.

Despite the conceptual success of applying quantum mechanics to molecular dynamics, solving \eqref{molecularSchrodinger} for a molecule with a large number of constituent particles
 is computationally very costly, with numerical simulations scaling exponentially with respect to the number of particles \cite{Lubich2008}.
Hence, there currently exists a rich field of study in finding reduced models and algorithms to simplify computation, serving as one of the main goals of quantum chemistry. 

To ease the notational and computational complexity of the material, for the remainder of this thesis we will consider the simplest case of two particles, one nucleus and one electron, 
with the knowledge that it is straightforward to generalise to the full multi-particle dynamics. 

Traditionally, the starting point for such approximations is due to Born and Oppenheimer in 1927 \cite{BornOppenheimer1927}, in which the nuclear motion is separated from the electronic 
motion, motivated by the fact that the proton is approximately 1836 times more massive than the electron. This electron-nuclear separation is first performed in the molecular Hamiltonian in which \eqref{molecularHamiltonian} 
is written as
\begin{align}
  \widehat{H} = \widehat{T}_n + \widehat{H}_e \label{BOHamiltonianSep}\,,
\end{align}
where $\widehat{H}_e$ is called the {\it electronic Hamiltonian}. Since $\widehat{H}_e= \widehat{T}_e+\widehat{V}_I(\bx,\br)$, one may write $\widehat{H}_e=\widehat{H}_e(\br)$, understood as an operator on the electronic Hilbert space $\mathscr{H}_e$,
usually identified with the space of $L^2-$functions of the electronic coordinate, $\boldsymbol{x}$, which depend parametrically on the nuclear coordinate, $\br$. 
It therefore posesses the eigenvalue equation
\begin{align}
\widehat{H}_e(\br)\phi_n(\boldsymbol{x};\br)=E_n(\br)\phi_n(\boldsymbol{x};\br)\label{stationaryelectronicSchrodinger}
\,,
\end{align}
known as the time-independent electronic Schr\"odinger equation. At every nuclear configuration $\br$, the electronic eigenfunctions $\phi_n(\boldsymbol{x};\br)$ provide an orthonormal frame in $\mathscr{H}_e$
 and the level sets of the eigenvalues $E_n(\br)$ comprise hypersurfaces in the nuclear coordinate space, referred to as {\it potential energy surfaces} (PES) in the quantum chemistry literature \cite{MarxHutter2009,Wyatt2006}.
 These surfaces are used in the chemistry field as key interpretative tools in understanding molecular 
processes, for example with stable nuclear geometries corresponding to minima of the PES and 
transition states identified by saddle points. 

Having performed the separation in the Hamiltonian, the standard approach also applies to the molecular wavefunction itself. Specifically, motivated by the electron-nuclear mass ratio, the nuclei are thought of as moving more slowly than the 
electrons, so that the electrons respond almost instantaneously to the changes in the nuclear positions, remaning in a given eigenstate 
of $\widehat{H}_e$. This is exactly the situation described by the celebrated quantum adiabatic theorem \cite{BornFock1928}, and for this reason the Born-Oppenheimer approach is also referred to as the {\it adiabatic 
approximation}.
 Then, despite not appearing in their original work \cite{BornOppenheimer1927}, which instead considers perturbations around a stationary state, the modern interpretation of the Born-Oppenheimer (BO) approximation is given by the following factorisation of the molecular 
wavefunction,
\begin{align}
  \Psi(\br,\bx,t)=\Omega(\br,t)\phi(\bx;\br)\label{BOAnsatz}\,,
\end{align}
in which $\phi$ is a specific eigenfunction of \eqref{stationaryelectronicSchrodinger}. This BO factorisation ansatz 
has survived the test of time in the chemistry community and is still widely used due to its intrinsic 
simplicity and appealing interpretative power. Indeed, in addition to the imagery of a PES, the dynamics of the 
nuclear wavefunction, {following from the substitution of \eqref{BOAnsatz} in the molecular Schr\"odinger equation \eqref{molecularSchrodinger},} 
can be written in the suggestive form
\begin{align}
  i\hbar\partial_t\Omega = \frac{(-i\hbar\nabla_{\br}+{\boldsymbol{\cal 
  A}})^2}{2M}\Omega + \epsilon\,\Omega\,,\label{BOnuclearEOMMinCoup2}
\end{align}
analogous to the dynamics of a particle in a magnetic field, in which $\boldsymbol{\cal A}$ is the {\it Berry connection}, playing the role of a vector potential and $\epsilon$ is an 
effective electronic potential. 

The BO approximation as presented above is the first port-of-call for 
much of quantum chemistry and, following this brief introduction, the remainder of this section outlines additional directions that naturally arise.
In particular, we consider three further avenues: nuclear hydrodynamics, Berry's phase and nonadiabatic models, 
each serving as a specific example of the three, more general, topics which form cornerstones of this thesis: QHD, gauge connections and applications in nonadiabatic dynamics.

\subsubsection{Nuclear QHD and classical trajectories}
Whilst the BO approximation reduces the full molecular system to considering only nuclear motion, the dynamical 
equation \eqref{BOnuclearEOMMinCoup2} still requires solving a (usually high-dimensional) 
quantum Schr\"odinger equation. Hence, for further computational simplicity it is common 
practice to take a classical limit and model the nuclei as point 
particle trajectories $\bq(t)$ by first considering a 
QHD description for the nucleus. As usual, one writes the nuclear wavefunction in polar form
\begin{align*}
  \Omega(\br,t)=\sqrt{D(\br,t)}e^{i\hbar^{-1}S(\br,t)}\,,
\end{align*}
taking the nuclear equation \eqref{BOnuclearEOMMinCoup2} into a 
pair of coupled PDEs for the density $D$ and phase $S$. Then, upon defining the shifted fluid velocity $\bu:=M^{-1}(\nabla S+\boldsymbol{\cal A})$, 
these coupled equations become 
\begin{align*}
     \partial_tD  +\text{div}(D\bu)&=0\,,\\
     M(\partial_t+\bu\cdot\nabla)\bu &=-\bu\times\boldsymbol{\cal B}-\nabla\left(\epsilon + 
  V_Q\right)\,,
\end{align*}
where $\boldsymbol{\cal B}:= \nabla\times \boldsymbol{\cal 
A}$ is the {\it Berry curvature} and $V_Q$ is the {quantum potential} for the nucleus. Compared with the standard QHD equations of quantum mechanics,
 one clearly recognises the presence of an additional Lorentz force term 
generated by the electron, in which the Berry curvature acts as a magnetic 
field.

 At this point, the standard practice is to make further {approximations}, including neglecting the quantum potential, so that 
 eventually the only remaining contribution is the force from the gradient of the potential energy 
 $E(\br)$. Selecting a single-particle solution of the form $\bu(\br,t)=\delta(\br-\bq(t))$,
 the fluid equation ultimately reduces to Newton's equation $\ddot{\bq}=-\nabla 
E$ \cite{deCarvalhoEtAl2014,Tully1998}. This process motivates chemists to picture classical nuclear dynamics evolving on the electronic PES $E(\br)$, providing an intuitive, although extremely simplified, picture for BO nuclear dynamics.

\subsubsection{Berry phase and conical intersections}
The tell-tale sign of the breakdown of the BO approximation corresponds to 
degeneracies in the electronic Hamiltonian $\widehat{H}_e$. Specifically, 
it is possible that two (or more) separate energy surfaces intersect for a given nuclear configuration $\br_0$, that is $E_i(\br_0) = E_j(\br_0), i\neq j$, 
and it is well-known that locally such intersections often form the shape of a double cone and are therefore referred to as {\it conical intersections} in the quantum chemistry literature \cite{Yarkony1996}.
Whilst near to these conical intersections, electron-nuclear coupling terms tend 
to infinity, thus signalling the need for a nonadiabatic description, the topological 
nature of the these singularities lead to interesting effects even within the BO 
regime. 

The origin of such phenomena dates back to the celebrated Jahn-Teller effect \cite{JahnTeller1937}, and how,
 as was first noticed by Longuet-Higgins et. al. \cite{LonguetHigginsEtAl1958} in the context of the BO factorisation, the electronic 
factor undergoes a sign change when transported around a conical intersection. 
Since the seminal work of Berry \cite{Berry1984}, in which it was discovered that quantum systems  
undergoing an adiabatic cyclic evolution attain an additional, purely geometric, phase 
factor, such effects have been found to occur frequently in the BO approach and are understood as examples of this so-called {\it Berry phase}. In the general setting the electronic wavefunction 
undergoes the transformation $\phi \mapsto e^{i\hbar^{-1}\alpha}\phi$, where $\alpha$ is the Berry phase defined by
\begin{align}
  \alpha(c_0):=\oint_{c_0}\boldsymbol{\cal A}(\br)\cdot\de\br\,,
\end{align}
for a fixed loop $c_0$. The Berry phase is non-trivial whenever $c_0$ encloses a conical 
intersection, for example equal to $\pi$ for the Jahn-Teller system, and also depends on the winding number of the loop. 
Over the years the Berry phase effect in BO systems has attracted much attention, see for example \cite{HerzbergLonguetHiggins1963,LonguetHiggins1975,Mead1992, Kendrick2003, Baer2006, IzFr2016, RyabinkinIzmaylov2013, RyabinkinEtAl2014, RyabinkinEtAl2017, BohmKendrickLoewe1992, Mead1980} as well as many others,
and in Chapter \ref{Chap2} an overview of the typical approach to dealing with the double-valued nature of the electronic wavefunction, known as the {\it Mead-Truhlar method}, will be presented.   

Berry phase's is the original and architypal example of a quantum geometric phase and was given a geometric interpretation as the holonomy associated to a connection on a principal 
bundle by Simon \cite{Simon1983}.  Since then, a wide variety of physical phenomena have been unified by the concept of geometric phase, both in the classical and quantum domains.
A key example of each is the Pancharatnam phase in classical optics, \cite{Pancharatnam1956}, which has been experimentally verified using experiments involving laser interferometry \cite{BhandariSamuel1988} and the celebrated Aharonov-Bohm effect of quantum mechanics, discovered in 1959 \cite{AharonovBohm1959} and experimentally verified in the late 80s \cite{TonomuraEtAl1986}, which was given a geometric phase interpretation in \cite{Berry1984}.

\subsubsection{Nonadiabatic molecular dynamics}

Despite the several successes of the BO approximation and its enduring practical and research interests, 
there are many situations in which the adiabatic separation of nuclear and electronic motion in the BO approximation does not accurately describe molecular 
dynamics. Such situations are thus described as nonadiabatic processes and occur whenever one considers excited electronic dynamics or electronic transitions, for example in photochemistry, which underpins the understanding of crucial physical phenomena including photosynthesis and vision (see, for example, \cite{AgostiniEtAl2015,AgostiniEtAl2016,CurchodAgostini2017,AgostiniCurchod2018} and references therein). 

The natural bridge from the BO ansatz \eqref{BOAnsatz} to nonadiabatic dynamics is via the {\it Born-Huang (BH) expansion} \cite{BornHuang1954}
\begin{align}
\Psi(\br,\boldsymbol{x},t)=\sum_{n=0}^\infty 
\Omega_n(\br,t)\phi_n(\boldsymbol{x};\br)\label{BornHuang}\,,
\end{align}
in which one writes the molecular wavefunction on the basis of the electronic 
eigenstates \eqref{stationaryelectronicSchrodinger}. This exactly solves the full 
molecular Schr\"odinger equation \eqref{molecularSchrodinger} and provides equations for the coefficients $\Omega_n(\br,t)$, identified with nuclear wavefunctions in the chemistry community.  
From this one clearly sees that the ground-state BO approximation can be identified as the lowest-order truncation of the 
BH expansion \eqref{BornHuang}. This representation of the molecular 
wavefunction is {particularly} appealing for chemists as it retains the picture of 
potential energy surfaces defined by \eqref{stationaryelectronicSchrodinger} so 
that nonadiabatic dynamics is often thought of as nuclear dynamics evolving on multiple PESs with the ability to
transfer between them \cite{deCarvalhoEtAl2014}.

Presently, the search for mathematical models that go beyond the BO regime is an active area of research and dates back many decades, with different approaches capturing 
nonadiabatic effects to varying extents. Today, some of the most popular techniques are known as {\it surface hopping models}. These models comprise mixed quantum-classical methods that consider dynamics on multiple PESs by approximating the 
  nuclear dynamics as a swarm of classical trajectories, each evolving according to the force generated by the specific PES it evolves on, whilst 
  simultaneously running a stochastic algorithm that assigns a probability of 
  `hopping' between these electronic energy surfaces. Perhaps the most 
  well-known such algorithm is due to Tully \cite{Tully1990}. Whilst surface hopping has, in some situations, allowed more accurate numerical modelling 
compared with mean-field models (below), it can not be derived from consistent approximations of the molecular Schr{\"o}dinger equation\footnote{There have been many attempts to reconcile this issue, see for example \cite{deCarvalhoEtAl2014, SubotnikEtAl2013} and references therein.} and hence surface hopping approaches do not shed light on fundamental processes or the theory underpinning the phenomena they are designed to model.  

Another main branch of nonadiabatic models consist of factorisations of the molecular 
wavefunction. The most simple ansatz in this setting is the {\it mean-field model} 
\begin{align*}
  \Psi(\br,\bx,t)=\Omega(\br,t)\psi(\bx,t)\,,
\end{align*}
where $\Omega$ and $\psi$ are thought of as nuclear and electronic wavefunctions, respectively \cite{deCarvalhoEtAl2014}, both containing time-dependence. 
Up to a global phase factor, this model is also known as the {\it time-dependent self-consistent field} or TDSCF 
approximation in the chemistry literature\cite{Tully1998, MarxHutter2009}. Upon 
computing the equations of motion {for the nuclear and electronic wavefunctions}, one 
sees that the dynamics of the nuclei are governed by an average potential generated by 
the electronics and vice versa. At this stage, one again typically models the nucleus as a classical particle,
leading to a mixed quantum-classical system known as the Ehrenfest method \cite{deCarvalhoEtAl2014, 
Tully1998}. Whilst mean-field models capture some degree of nonadiabatic 
effects, they neglect particle correlations between nuclei and electrons, which in the BH picture,
results in nuclei evolving on an unphysical averaged PES and hence have limited applicability \cite{Tully1998, MarxHutter2009}.

In recent years, another product ansatz has been suggested for the molecular 
wavefunction \cite{AbediEtAl2010,AbediEtAl2012}. Originating in the earlier works \cite{Hunter1975, Cederbaum2008}, with the concept 
even dating back to von Neumann \cite{vonNeumann2018}, the  {\it exact factorization} (EF)
of the molecular wavefunction is given by
 \begin{align*}
  \Psi(\br,\bx,t)=\Omega(\br,t)\psi(\bx,t;\br)\,,
\end{align*}
where $\Omega$ and $\psi$ are again interpreted as nuclear and electronic 
wavefunctions, respectively. Immediately one can recognise this factorisation as 
a time-dependent generalisation of the BO ansatz where, in addition to the 
parametric dependence on the nuclear coordinate, $\psi$ also carries its own 
time dependence. Whilst such an ansatz does not simplify the computational 
difficulty of solving the full molecular Schr\"odinger equation, the EF approach to nonadiabatic dynamics has generated much interest in the 
    chemistry community \cite{AbediEtAl2010,AbediEtAl2012,abedi2013response, alonso2013comment, deCarvalhoEtAl2014, SuzukiEtAl2015, RequistEtAl2016, AgostiniEtAl2015, AgostiniEtAl2016, CurchodAgostini2017, AgostiniCurchod2018, MinEtAl2014, MinEtAl2015,SuzukiWatanabe2016, MinEtAl2017},
    serving as a key interpretive tool for nonadiabatic dynamics, as it generalises all of the intuitive concepts associated to the BO picture.
    In particular, one obtains coupled nuclear and electronic equations (see Section \ref{NonadiabaticSection}), with the nuclear dynamics in an effective time-dependent magnetic field associated to a
    generalised Berry connection $\boldsymbol{\cal A}(\br,t)$, also governed by the force due to a time-dependent generalisation of the electronic potential energy surface.

\section{Motivation and open questions}\label{Sec:Motivation}
Whilst the previous sections have presented the background material in quantum mechanics and its application to chemistry,
this section outlines the motivation for the work in this thesis and covers some of the open questions that it aims to answer.   
Initially motivated by the general idea of applying geometric mechanics to quantum chemistry,  as demonstrated above, investigating
the Born-Oppenheimer approximation naturally leads into multiple fields of study and hence
unfurled a wide variety of research opportunities. We present some of the corresponding thoughts and questions organised 
under three headings, although naturally there is some degree of overlap.

\subsubsection{Quantum hydrodynamics}
Having seen that QHD is commonly used in the field of quantum chemistry, it is 
natural to further investigate this direction. The first questions regard the quantum potential and 
 the common practice of taking a classical limit to allow point particle trajectories. Specifically:
\begin{itemize}
  \item Whilst the structure of $V_Q$ is a barrier to single particle solutions, the standard $\hbar^2 \to 0$ limit also neglects crucial electron-nuclear coupling effects in chemistry.
  Are there alternative methods that allow for {\bf singular solutions} without taking the $\hbar^2 \to 0$ limit?  
\item The quantum potential is well-known as the source of all non-local effects in quantum mechanics. Is it possible to {\bf smoothen the quantum potential}, thus allowing for particle trajectory solutions, whilst 
retaining its potentially long-range effects?
\end{itemize}
Geometric approaches to QHD have already been pursued, for example the geometry of the Madelung transform as in \cite{KhesinEtAl2018}. 
From such works it is known that pure state QHD possesses a Hamiltonian formulation in terms of collective momentum map variables.
In line with the spirit of this work, we ask:
\begin{itemize}
  \item Will this geometric formulation shed light on the use of QHD for 
  {\bf modelling nuclear dynamics}?
  \item There has been much study on the geometry of {\bf singular solutions} in 
continuum theories, specifically that the `peakon' solutions of the Camassa-Holm 
equation are momentum maps \cite{HoMa2005, HoTr2009}. Do the singular solutions used in QHD fit 
into this geometric picture? Could they be helpful for modelling purposes in 
quantum chemistry?
\item Can we generalise these ideas to a {\bf density matrix description}? Is there a geometric approach to mixed state
 QHD in terms of momentum maps? 
\end{itemize}

\subsubsection{Quantum chemistry}
As well as questions regarding the use of QHD in modelling nuclear dynamics, having seen the prevelance of the Born-Oppenheimer approximation in all facets 
of quantum chemistry, one is immediately led to ask: what is the geometric nature of the BO approach and nonadiabatic models?
In particular:

\begin{itemize}
  \item Much of the existing literature on the BO factorisation uses a variational approach to derive the dynamics. In the spirit of geometric mechanics,
are there {\bf symmetries} of the Lagrangian? Can we insert Lie groups and formulate 
an Euler-Poincar{\'e} theory? 
  \item Current expositions of the exact factorization model present cumbersome equations of motion. Are there better ways to 
  express the dynamics? Perhaps the structure of the equations become clearer for an arbitrary energy functional?
  \item Can we relate the BO and EF models to other areas of physics previously studied  
 from a geometric mechanics perspective? What could we take or learn from them? 
 \item EF is claimed to capture decoherence \cite{MinEtAl2015, MinEtAl2017}. Thus, what are the dynamics for density matrices? Are there 
 {\bf generalisations of EF} at the level of density matrices which may capture more general phenomena?
\end{itemize}

\subsubsection{Connections and geometric phase}
As we have seen, another area of particular interest is the appearance of 
 Berry's phase in BO dynamics. As EF can be considered as the natural time-dependent 
 generalisation of BO, the question arises: what is the nature of Berry phase in nonadiabatic dynamics? 
This last question has indeed started to be asked in the chemistry community 
\cite{MinEtAl2014,RequistEtAl2016}, in which it is suggested that the topological nature of such effects are simply `artifacts' of 
the adiabatic separation of nuclei and electrons in the BO factorisation. In 
tackling such ideas one inherently encounters the concept of connections in 
principal bundles, which have applications in a multitude of physical systems. This
leads to many additional questions relating to the scope of this research, 
including
\begin{itemize}
   \item Can the tools of geometric mechanics shed light on the {\bf nature of the 
   Berry phase} in both nonadiabatic and BO dynamics?
   \item A dynamical connection has been introduced in the geometric approach to liquid crystal dynamics \cite{GBRaTr2012,GBRaTr2013} and more generally in the study of complex fluids \cite{GayBalmazRatiu2009, Holm2002, Tronci2012}
   which led to new interpretations of fundamental objects arising in the theory.   
   Are there analogous objects in the BO and/or EF settings? Could we apply some technology transfer between these fields?
     \item Recent work considers the QHD fluid velocity as a $\mathcal{U}(1)$ connection \cite{Spera2016}. In this picture, geometric phase again arises via
     a {\bf quantised circulation} theorem. Can we insert Lie group actions to QHD and again construct a connection-based dynamical theory analogous to those in complex fluids?
   \end{itemize}

\section{Outline and main results}\label{Sec:Outline}
The material in this thesis is based largely on the work of the two papers published 
by the author \cite{FoHoTr2019, FoTr2020} during the course of this research. 
As a general rule of thumb, Chapters \ref{Chap:QHD}, \ref{Chap:EFOLD} and \ref{Chap:EFNEW} correspond to \cite{FoHoTr2019}, Chapter \ref{Chap:Holonomy} corresponds to \cite{FoTr2020} 
whilst Chapter \ref{Chap:Gamma} contains as yet largely unpublished material.
We proceed by presenting a summary of the subsequent chapters outlining the main results of this work, noting that Chapter \ref{Chap2} contains no new results
 and is solely devoted to further developing the background material employed in 
this thesis, including the previous state-of-the-art research in both 
quantum geometric mechanics and nonadiabatic quantum chemistry.

\subsubsection{Chapter \ref{Chap:QHD}: New geometric closures of QHD}
After reviewing the geometric approach to QHD in terms of momentum map 
collective variables and the corresponding Euler-Poincar{\'e} and Lie-Poisson structures, 
new regularisation techniques and the 
geometric nature of singular solutions are presented. Specifically,
\begin{itemize}
  \item Smoothened fluid variables are defined that circumvent the troublesome 
  nature of the quantum potential allowing for singular solutions named {\bf`Bohmions'} that result in 
  finite-dimensional dynamical equations that depend on a {regularised quantum 
  potential}.
  \end{itemize}
Then, a geometric approach to mixed state QHD is considered in terms of density matrices: 
\begin{itemize}
  \item The generalisation of the standard QHD {\bf momentum map} is found for mixed 
  states, resulting in a fluid velocity that no longer corresponds to a potential flow and hence possesses non-zero vorticity.
  \item A {\bf cold fluid closure} is applied to the density matrix, eliminating the contribution of the quantum potential 
  to the total energy, without taking any $\hbar$ limit.
\end{itemize}

\subsubsection{Chapter \ref{Chap:EFOLD}: Geometry of wavefunction factorisations in quantum chemistry}
This chapter takes a geometric approach to factorisations of the molecular wavefunction used in the study of nonadiabatic quantum chemistry.
The mean-field model is considered first, acting as an example demonstrating the application of the geometric approach to QHD to model nuclear dynamics.
From there, the remainder of the chapter considers in detail the geometric viewpoint of the existing approach to exact factorization dynamics. 
The key results of this chapter are as follows:
\begin{itemize}
  \item A {\bf generalised set of Euler-Lagrange equations} are derived for an arbitrary total 
  energy, leading to a compatibility condition for permissable Hamiltonian 
  functionals.
  \item The momentum map QHD approach is applied to the 
nuclear dynamics allowing for an equivalent {\bf Euler-Poincar{\'e} formulation} of the equations 
of motion.
\item The {\bf circulation theorem} for the nuclear fluid is found to comprise the 
dynamics of the {\bf electronic Berry phase}, which can be equivalently formulated in 
terms of the {\bf quantum geometric tensor} \cite{PrVa1980}.
\end{itemize}
In addition further relevant comments about the prominant features of the EF theory are discussed, including 
 the analogy with electromagnetism and the Newtonian limit. 

\subsubsection{Chapter \ref{Chap:EFNEW}: New geometric approaches to exact factorization dynamics}
Whilst the previous chapter focused on using geometry to better understand the existing approaches to EF dynamics, this chapter uses the tools of geometric mechanics to go beyond the scope of existing literature:
\begin{itemize}
  \item Electronic motion is assumed to evolve unitarily and is written in the 
  frame of the nuclear hydrodynamic flow. This key step
  allows the dynamics to be expressed exactly in terms of a pure state {\bf electronic density matrix}.
  \item The corresponding geometric structure is identical to that previously studied in complex fluids, on the Hamiltonian side corresponding to a {\bf Lie-Poisson bracket} on a semidirect product Lie algebra.
\end{itemize}
 From here, the electronic dynamics are specialised to two-level systems, capturing spin-boson dynamics as well as the hydrodynamic form of the Pauli 
  equation, considered in depth in Chapter \ref{Chap:Gamma}. The final section builds on the results of Chapter \ref{Chap:QHD}, 
 constructing a generalised EF model for the molecular density matrix, allowing for a more general description
 of molecular dynamics in terms of {\bf mixed nuclear states}. 
 \begin{itemize}
   \item Upon taking the cold 
 fluid closure for the nuclear density matrix, one obtains a {\bf generalised EF 
model} analogous to the pure state case, without the quantum potential term and 
with non-zero vorticity dynamics.
\item The regularisation procedure from Chapter \ref{Chap:QHD} is applied using {\bf Bohmion solutions} 
resulting in a new mixed quantum-classical nonadiabatic model comprising a set of finite-dimensional coupled equations for the nuclear trajectories and electronic pure state density matrix. 
 \end{itemize}

\subsubsection{Chapter \ref{Chap:Holonomy}: Holonomy in quantum hydrodynamics}
Motivated by the appearance of gauge connections throughout this research, this 
chapter considers a new formulation of QHD by expressing the Madelung transform in terms of Lie group elements
as opposed to the usual exponential form. This leads to interesting new 
features and potential applications:
\begin{itemize}
  \item An {\bf alternative Euler-Poincar{\'e} formulation} of QHD is derived in which
  the fundamental variable is a Lie algebra-valued connection 1-form derived from the wavefunction's phase factor.
  \item This formulation provides a new method for incorporating {\bf holonomy in 
  QHD}, endowing the fluid flow with constant non-zero vorticity and appearing 
  via minimal coupling in the reconstructed Schr\"odinger equation.
  \item Upon specialising the curvature of the connection to capture {\bf hydrodynamic vortex filaments}, 
a new method for coupling the Schr\"odinger equation to dynamical quantum vortices 
  \cite{RasettiRegge1975} emerges.
  \item Finally, the concepts of this chapter are considered in the setting of 
  BO dynamics, resulting in a trajectory description for the nucleus in which 
  terms that are known to diverge near conical intersections have been smoothened.
  \end{itemize} 

\subsubsection{Chapter \ref{Chap:Gamma}: Non-Abelian gauge connections}
The penultimate chapter of this thesis considers the more general setting of non-Abelian gauge connections in mechanical systems.
 As a first port-of-call, the hydrodynamic interpretation of the Pauli equation 
 is examined. In particular, the original approaches \cite{BohmSchillerTiomnoA1955,BohmSchillerB1955, Takabayasi1955} turn out to be analogous to the two-level EF system from Section \ref{Sec:2Level} and the subsequent system of equations are derived using
 the momentum map QHD formulation of Chapter \ref{Chap:QHD}. 
From there, a new non-Abelian connection is introduced in the setting of general two-level EF systems and is derived from the 
evolution of the spin vector under the rotation group. This leads to:
\begin{itemize}
  \item A new reduced system of equations for two-level EF systems, derived from 
  Euler-Poincar{\'e} reduction, in which the {\bf non-Abelian connection} appears as an additional dynamical 
  variable. These equations also feature explicit dependence on the curvature of 
  the connection.
  \item The {\bf Kelvin-Noether circulation theorem} again 
  describes the evolution of the Berry phase, and an equation for the {\bf circulation of the 
  connection} is derived.
  \item  This general system is specialised to describe Pauli QHD leading to a new set of equations in which the quantum geometric tensor is
  expressed in terms of the connection.
\end{itemize}
Motivated by the successful connection-based theory of QHD from Chapter \ref{Chap:Holonomy} and the derivation of a non-Abelian connection associated to the spin vector in two-level systems, a general 
  geometric formulation for introducing such connections in a general setting is presented. 
\begin{itemize}
  \item 
  Whilst intially a zero curvature relation is used to construct a Lagrangian for a given system, 
 the resulting {\bf Euler-Poincar{\'e} equations} then allow for {\bf non-trivial curvature}. 
\end{itemize}
Applying this theory to a broad class of quantum systems, a new non-Abelian $\mathfrak{u}(\mathscr{H})$ connection is derived from the unitary evolution of quantum states. 
This allows the reformulation of many key geometric objects resulting in stimulating new 
interpretations. In particular
\begin{itemize}
  \item The Berry connection is shown to be the $\mathcal{U}(1)$ projection of 
  the larger non-Abelian connection.  
  \item The quantum geometric tensor is identified as the {\bf covariance} between 
  components of the connection. This allows the derivation of {\bf novel uncertainty 
  relations}, imposing constraints on the components of the QGT.
\end{itemize}
Finally, this non-Abelian connection is introduced for the EF system in 
  quantum chemistry, derived from the nonadiabatic coupling between electrons and nuclei. 
  This connection acts a type of {\bf correlation 
  observable}, measuring the strength of electron-nuclear coupling and leading to 
  new interpretations of familiar objects in quantum chemistry.

\chapter{Mathematical background and prequisite material}\label{Chap2}
 
Having presented the introduction and outline for this thesis in Chapter \ref{Chap1}, we now
further develop the necessary mathematical formalism and consider in more detail the preqrequisite material
 that will be encountered time and again throughout this work.
 
  In particular, Section \ref{QMGeometry} will present a more complete approach to
 the principles of quantum mechanics as well as spending some additional time developing other useful aspects of quantum theory 
 that go beyond a basic exposition. 
 
 Then, in Section \ref{Sec:Chem}, the material covers a more in-depth exploration of the
 relevant concepts in quantum chemistry, providing the context for many of the
applications of the results that follow in later chapters.

\section{Quantum mechanics}\label{QMGeometry}
 
 In this section we will present an overview of the necessary mathematical and physics principles of quantum mechanics, with particular emphasis on the underlying geometric structures.
Specifically, Section \ref{Sec:1-Postulates} introduces the general principles of quantum mechanics as well as covering the additional key concepts that will be used later in this thesis, namely the celebrated uncertainty principle, the density 
matrix description of quantum mechanics and finally spin-½ systems and the Pauli equation. 

From there, Section \ref{Sec:1-GeomQM} considers the geometric principles behind such ideas, beginnning 
with the Dirac-Frenkel variational principle, which gives rise to the Schr\"odinger 
equation as the corresponding Euler-Lagrange equation. We then switch to an 
alternative viewpoint in which the Hamiltonian structure of 
quantum mechanics is presented, arising from the symplectic structure on the 
Hilbert space and giving rise to a non-canonical Poisson bracket. Then, 
a summary of the recent geometric approach to quantum mechanics \cite{BonetLuzTronci2015} is presented, in which Euler-Poincar{\'e} 
 reduction is used to provide a variational approach for both pure and mixed state dynamics as well as showing the corresponding Lie-Poisson 
 structure for density matrices. 
 The general definitions and constructions of geometric mechanics can be found 
 in Appendix \ref{App:GMTheory}.
 
 Finally Section \ref{Sec:1-QHD} concludes by considering in further detail the hydrodynamic picture of quantum 
 mechanics as it arises from the Madelung transform of the wavefunction. Using the previously introduced variational 
 principle we rederive the Hamilton-Jacobi type 
 equation (augmented by the quantum potential) and continuity equation 
 for the probability density. We then move to the hydrodynamic picture,
leading to discussions of quantised circulation and the nature of classical trajectory 
solutions.
 
\subsection{Fundamental principles}\label{Sec:1-Postulates}

It is important to note that throughout this thesis, we shall not dwell upon various complications that may emerge in infinite-dimensional Hilbert spaces
 and we assume convergence where necessary. When convenient, we shall consider dynamics on finite-dimensional spaces and rely on the possibility of extending the results to the infinite-dimensional case.
With this in mind, in this section we present an abridged version of the principles of quantum 
mechanics (see, for example, \cite{Shankar2012} for a full treatment), paying closer attention to the ideas relevant to the subsequent material in this thesis.

\subsubsection{I: States}
A quantum state is described by a normalised vector $\ket{\psi}$ in a Hilbert space $\mathscr{H}$. 
That $\ket{\psi}$ is normalised is to say that $\|\psi\|^2:=\braket{\psi|\psi}=1$, in 
which the angled bracket denotes the inner product on $\mathscr{H}$. For example if $\mathscr{H}=\mathbb{C}^n$, then the inner product is the mapping $\mathbb{C}^n\times\mathbb{C}^n\to \mathbb{C}$, defined by 
\begin{align}
  \braket{\alpha|\beta} = \alpha^{\dagger}\beta\,,\qquad \forall 
  \alpha,\beta\in\mathbb{C}^n\label{CnInnerProduct}\,,
\end{align}
where $^{\dagger}$ denotes the conjugate transpose operation. As demonstrated by the last equality, 
throughout this thesis we will employ Dirac's `bra-ket' notation as 
appropriate, and if not, the meaning of a state vector should be clear from context. 

\subsubsection{II: Observables}
 Physical observables, i.e. measurable quantities, are described by Hermitian operators on 
$\mathscr{H}$. An operator $\widehat{A}:\mathscr{H}\to\mathscr{H}$ is Hermitian (or self-adjoint) if $\widehat{A}^{\dagger}=\widehat{A}$. 
Considering the inner product for an arbitrary $\mathscr{H}$, one has that
\begin{align}
  \braket{\phi|\widehat{A}\psi} = \braket{\widehat{A}^{\dagger}\phi|\psi}=\braket{\widehat{A}^{}\phi|\psi}, \qquad \forall \ket{\phi},\ket{\psi} \in 
  \mathscr{H}\,.
\end{align}

Without commenting on the delicate act of measurement in quantum mechanics, we define the expectation value of the observable $\widehat{A}$ in the state $\psi$ as
\begin{align}
  \braket{\widehat{A}}_{\psi}:= \braket{\psi|\widehat{A}\psi}\label{expectationdef}\,.
\end{align}
This value corresponds to the average measured value of $\widehat{A}$ in this state, which, as $\widehat{A}$ is Hermitian, is guaranteed to be real.

For systems with a classical analog one promotes the canonical phase space coordinates $(\bq,\bp)$ 
to their quantum counterparts $(\widehat{Q},\widehat{P})$ which, as the notation 
suggests, are now Hermitian operators on $\mathscr{H}$. If one utilises a 
position space representation, in which one has the wavefunction $\psi(\bx):= \braket{\bx|\psi}\in L^2(\mathbb{R}^3)$, these so-called {\it canonical observables} are defined 
as follows 
\begin{align}
  \braket{\bx|\widehat{Q}|\psi} = \bx \psi(\bx)\,,\qquad \braket{\bx|\widehat{P}|\psi} = -i\hbar\nabla 
  \psi(\bx)\,,\label{canonicaloperatorsdef}
\end{align}
and satisfy the canonical commutation relation
 \begin{align}
 [\widehat{Q},\widehat{P}]:=\widehat{Q}\widehat{P}-\widehat{P}\widehat{Q}=i\hbar\label{CCR}\,.
  \end{align}
As much of the emphasis in this thesis is on quantum chemistry and quantum hydrodynamics,
we are often in the case when the system in consideration has a classical counterpart 
and hence frequently use the position representation of the wavefunction.

\subsubsection{III: Dynamics}
The time evolution of $\psi$ is governed by the {Schr\"odinger equation} \cite{Schrodinger1926}
\begin{align}
  i\hbar\ket{\dot{\psi}}&= \widehat{H}\ket{\psi}\label{BasicSchrodinger}\,,
\end{align}
where $\widehat{H}:\mathscr{H}\rightarrow\mathscr{H}$ denotes the Hamiltonian operator 
on the Hilbert space. Analogous to classical mechanics, the Hamiltonian operator corresponds to the total energy of the system, which, in the quantum realm, is given by the expectation value $\braket{\psi|\widehat{H}\psi} $. 
Then, in the case that the quantum system has a classical analog one constructs the Hamiltonian operator from the canonical 
observables as $\widehat{H}=\widehat{H}(\widehat{Q},\widehat{P})$. Hence, for 
classical systems of the type $H = |\bp|^2/2m + V(\bx)$, in the position representation the Schr\"odinger equation 
reads
\begin{align}
  i\hbar\partial_t\psi(\bx,t)&= -\frac{\hbar^2}{2m}\Delta\psi(\bx,t)+ 
  V(\bx)\psi(\bx,t)\,.\label{PositionSchrodinger}
\end{align}
\hfill \\ \newline
Having outlined the fundamental aspects, we 
now turn our attention to three other important features of quantum mechanics that will be considered in this thesis:
 uncertainty relations, the density matrix formalism and two-level systems including the Pauli equation.  

\subsubsection{Uncertainty relations}

Firstly, we consider the celebrated Heisenberg uncertainty relation \cite{Heisenberg1927} between the canonical observables. In fact, more generally, there exists an uncertainty relation between any two non-commuting observables. 
To write these relations, we first consider an arbitrary observable $\widehat{A}$ and state $\ket{\psi}$ so that we have the expectation $\braket{\widehat{A}}_{\psi}$ defined by \eqref{expectationdef}. We then introduce the Hermitian operator $\Delta\widehat{A}:= \widehat{A}-\braket{\widehat{A}}_{\psi}\mathbbm{1}$, (noting that $\Delta$ is purely notational) where $\mathbbm{1}$ denotes the identity operator on $\mathscr{H}$, so that
we have the following definitions.
\begin{definition}[Variance]\label{Def:Variance}
  The {\bf variance} of a Hermitian operator $\widehat{A}$ is defined as
  \begin{align}
  \sigma_{{A}}^2 := \braket{(\Delta \widehat{A})^2}_{\psi}=
  \braket{\widehat{A}^2}_{\psi}-\braket{\widehat{A}}^2_{\psi}\,.\label{Goo}
\end{align}
\end{definition}
\begin{definition}[Standard Deviation]
The {\bf standard deviation} of $\widehat{A}$ is simply the square root of \eqref{Goo}, denoted by 
$\sigma_A$.
\end{definition}
Having defined these quantities, we can now write the following theorem, for which we adopt the notation $\braket{\,\cdot\,}_{\psi}\equiv\braket{\,\cdot\,}$ 
and use $\{\,\cdot\,,\,\cdot\,\}$ and $[\,\cdot\,,\,\cdot\,]$ to denote the
anticommutator and commutator of operators respectively.

\begin{theorem}[Schr\"odinger uncertainty relation \cite{Schrodinger1930}]\label{Theorem:SchrUR}
\hfill \\ 
For any two arbitrary Hermitian operators $\widehat{A}$ and $\widehat{B}$, the product of their 
variances $\sigma_A^2$ and $\sigma_B^2$ satisfies
\begin{align}
  \sigma_{A}^2\sigma_{B}^2 \geq \left|{\frac {1}{2}}\langle \{{ {\widehat{A}}},{ {\widehat{B}}}\}\rangle -\langle { {\widehat{A}}}\rangle \langle { {\widehat{B}}}\rangle \right|^{2}+\left|{\frac {1}{2i}}\langle [{ {\widehat{A}}},{ {\widehat{B}}}]\rangle 
  \right|^{2}\label{SchrodingerUR}\,.
\end{align}
\end{theorem}
\paragraph{Proof:} For ease of notation, consider the new state $\ket{f}:= \Delta \widehat{A}\ket{\psi}$, so 
that $\ket{f}=0$ implies that $\ket{\psi}$ is an eigenvector of $\widehat{A}$. Then upon introducing a second observable $\widehat{B}$ and the corresponding state $\ket{g}:= \Delta 
\widehat{B}\ket{\psi}$, the proof follows from simple algebraic manipulations of the Cauchy-Schwarz 
inequality $\braket{f|f}\braket{g|g}\geq|\braket{f|g}|^2$.
\hfill$\square$\\

\noindent From this theorem one can immediately derive the weaker result:
\begin{corollary}[Robertson uncertainty relation \cite{Robertson1929}]\label{Cor:RobertsonUR}
\hfill \\ 
For any two arbitrary Hermitian operators $\widehat{A}$ and $\widehat{B}$, the product of their standard deviations
$\sigma_A$ and $\sigma_B$ satisfies
\begin{align}
  \sigma_{A}\sigma_{B} \geq \frac{1}{2}\left|\braket{[\widehat{A},\widehat{B}]}\right| 
  \label{RobertsonUR}\,.
\end{align} 
\end{corollary}
Then the celebrated Heisenberg uncertainty principle \cite{Heisenberg1927} arises as the specialisation to 
the canonical observables $(\widehat{Q},\widehat{P})$. That is, upon using the canonical commutation relation \eqref{CCR}, one has $\sigma_{Q}\sigma_{P} \geq 
\hbar/2$. Together these uncertainty relations show that, one cannot know the exact values of two observables 
simultaneously, if their corresponding operators do not commute.

\subsubsection{Density matrix formalism}

Having briefly introduced the concepts of density matrices and mixed states in Chapter \ref{Chap1}, we now reiterate and expand on this topic.
Whilst in the typical presentation of quantum mechanics states are described by normalised vectors in a Hilbert space, an alternative description 
can be given by a class of operators known as density operators. 
\begin{definition}[Density operator]\label{Def:DensityOp}
  A {\bf density 
operator} is an operator $\widehat\rho\,:\mathscr{H}\to\mathscr{H}$ such that
\begin{enumerate}[(i)]
  \item ${\rm Tr}(\widehat\rho\,) =1\,,\qquad \qquad \qquad\quad\text{(unit trace)}$,
  \item $\braket{\phi|\widehat\rho\,|\phi}\geq 0\,,\quad \forall \phi \in \mathscr{H}\,,\qquad \text{(positive 
  semi-definite)}$\,,
\end{enumerate}
where the second condition ensures that $\widehat\rho\,$ is Hermitian\footnote{This can be seen by using that $\braket{\phi|\widehat\rho\,|\phi}$ is real. Then, $\braket{\phi|\widehat\rho\,\phi}=\braket{\widehat\rho\,^{\dagger}\phi|\phi} = \braket{\phi|\widehat\rho\,^{\dagger}\phi}^* = \braket{\phi|\widehat\rho\,^{\dagger}\phi}$. One can then write that $\braket{\phi|(\widehat\rho -\widehat\rho\,^{\dagger} )\,\phi}=0$ which implies that $\widehat\rho =\widehat\rho\,^{\dagger}$ as $\phi$ is arbitrary.}, that is 
$\widehat\rho\,^{\dagger}=\widehat\rho\,$. 
\end{definition}
Such operators not only capture states given by vectors in a Hilbert space, in which case $\widehat\rho=\psi\psi^{\dagger}$ so that $\widehat\rho\,^2 = \widehat\rho$ and are known as {\it pure states}, but also can describe more general quantum states known as {\it mixed states}. 
In general, these mixed states refer to an ensemble of pure states $\{\psi_a\}$ 
given by
\begin{align}
  \widehat\rho\, = \sum_a w_a \psi_a\psi_a^{\dagger}\label{rhomixture}\,,
\end{align} 
with $\sum_{a} w_a=1$, and are required to describe a variety of systems, e.g. those of uncertain preparation or entangled with another system. {We remark that  \eqref{rhomixture}} is the equivariant momentum map  for unitary transformations {of the $\{\psi_a\}$} recently studied in \cite{Tronci2018}.
 A measure of the degree of mixture is given by the {\it purity}, defined as $\text{Tr}(\widehat\rho\,^2)\leq 1$, which in 
the case of a pure state is given by $\text{Tr}(\widehat\rho\,^2)=\text{Tr}(\widehat\rho\,)=1$. Hence, loss of purity over time  
describes the process of a pure state evolving into a mixed state and can therefore be understood as a sign of {\it decoherence}.  

In terms of density matrices, the expectation value of an observable $\widehat{A}$ \eqref{expectationdef} generalises to 
\begin{align}
  \braket{\widehat{A}}:= \text{Tr}(\widehat\rho\,\widehat{A})\,,\label{DenExp}
\end{align}
whilst the time evolution is given by the {\it von Neumann} or {\it Liouville-von Neumann equation}
\begin{align}
  i\hbar\frac{\de }{\de 
  t}{\widehat\rho\,}=[\widehat{H},\widehat\rho\,]\label{LvN}\,,
\end{align} 
as can be seen from using the mixture \eqref{rhomixture} in the Schr\"odinger equation 
\eqref{BasicSchrodinger}.
One can also consider the position representation of a density matrix (or indeed any other operator on 
$\mathscr{H}$), given by `sandwiching' the operator with Dirac's `bra-ket' 
\begin{align}
  \rho(\bx,\bx'):= \braket{\bx|\widehat\rho\,|\bx'}\,,
\end{align}
also referred to as expressing $\widehat\rho\,$ in terms of its {\it matrix elements} in the physics literature.
Mathematically, this corresponds to the writing the kernel of the integral form of density 
operator, for which the transformation $\phi\mapsto \widehat{\rho}\phi$ is written 
as
\begin{align}
\phi(\bx) \mapsto  (\widehat{\rho}\phi)(\bx) = \int 
\rho(\bx,\bx')\phi(\bx')\,\de^3x'\,.
\end{align}
 In this matrix element notation, the above equations \eqref{rhomixture} - \eqref{LvN} are rewritten as follows
 \begin{align*}
   \rho(\bx,\bx')&=\sum_a w_a \psi_a(\bx)\psi_a^*(\bx')\,,\\
   \braket{\widehat{A}} &= \iint \rho(\bx',\bx)A(\bx,\bx')\,\de^3x\,\de^3x'\,,\\
   i\hbar\partial_t\rho(\bx,\bx')&= \int \Big[H(\bx,\by)\rho(\by,\bx')-\rho(\bx,\by)H(\by,\bx')\Big]\,\de^3y \,.
 \end{align*}
 We complete this discussion by giving the matrix elements of the canonical observables
\begin{align}
\widehat{Q}(\bx,\bx')=\bx\delta(\bx-\bx')
\,,\qquad
\widehat{{P}}(\bx,\bx')=-i\hbar\nabla_{\bx}\delta(\bx-\bx')
\,,
\end{align}
in agreement with their earlier defintion in terms of their action on states \eqref{canonicaloperatorsdef}.

\subsubsection{Two-level systems and the Pauli equation}
With the content of later sections in mind, here we introduce the formalism for 
treating the dynamics of a two-level system, for example a spin-½ particle interacting with an external magnetic field. 
Firstly, the state of such a particle (assuming charge $q=1$ and mass $m$) is 
given by a two-component wavefunction or {\it Pauli spinor}
\begin{align}\label{PauliSpinor}
  \Psi(\bx) = \begin{pmatrix}
  \Psi_1(\bx)
 \\
 \Psi_2(\bx)
  \end{pmatrix}\,,
\end{align}
where now $\Psi \in L^2(\mathbb{R}^3)\otimes\mathbb{C}^2$ and is normalised such that $\int \Psi^{\dagger}\Psi \,\text{d}^3x=1$. Then, the dynamics are 
given by the Pauli equation \cite{Pauli1927} which amounts to the Schr{\"o}dinger equation with specific Hamiltonian operator 
\begin{align}
  i\hbar\partial_t\Psi = \widehat{H}\Psi\,,\qquad \widehat{H}=\frac{(-i\hbar\nabla-\bA)^2}{2m} - 
  \frac{\hbar}{2m}\bB\cdot\widehat{\boldsymbol{\sigma}} + V\,.\label{PauliEquation}
\end{align}
Here, $\bA(\bx)$ is the constant magnetic potential, $\bB:=\nabla\times\bA$ is the magnetic field and $V(\bx)$ is a scalar potential. 
In writing the Pauli equation, we have also introduced the {\it Pauli matrices}, expressed in vector form as $\widehat{\boldsymbol{\sigma}}:=(\sigma_1, \sigma_2, \sigma_3)^T$ 
where 
\begin{align}
  \sigma_1 =
    \begin{pmatrix}
      0&1\\
      1&0
    \end{pmatrix}\,,\qquad
  \sigma_2 =
    \begin{pmatrix}
      0&-i\\
      i&0
    \end{pmatrix} \,,\qquad
  \sigma_3 =
    \begin{pmatrix}
      1&0\\
      0&-1
    \end{pmatrix} \,,
\end{align}
each of which being Hermitian $(\sigma_i^{\dagger}=\sigma_i)$ and unitary $(\sigma_i^{\dagger}=\sigma_i^{-1})$.
In this context, the identity matrix is sometimes denoted as $\sigma_0=\mathbbm{1}$.
 We also comment that the set $\{i\sigma_1,i\sigma_2,i\sigma_3\}$ forms a basis for the Lie algebra $\mathfrak{su}(2)$ of traceless skew-Hermitian matrices with commutation relations
\begin{align}
  [\sigma_i,\sigma_j]=2i\,\epsilon_{ijk}\sigma_k\,,\qquad i,j,k\in \{1,2,3\}\,,
\end{align}
where $\epsilon_{ijk}$ is the Levi-Civita symbol. Interestingly, there exists a 
Lie algebra isomorphism from  $(\mathbb{R}^3, \,\cdot\,\times\,\cdot\,)$ to $(\mathfrak{su}(2),[\,\cdot\,,\,\cdot\,])$  \cite{MaRa2013}, 
where $\times$ denotes the cross product of $\mathbb{R}^3$ vectors, so that for $\bxi \in 
\mathbb{R}^3$, one has
\begin{align}
\bxi \mapsto \widetilde{\xi} := -\frac{i}{2}\boldsymbol{\xi}\cdot\widehat{\boldsymbol\sigma} \in 
\mathfrak{su}(2)\,,\label{TILDEMAP}
\end{align}
which is also known as the {\it tilde map} \cite{DarrylBook2}. Using \eqref{TILDEMAP} is straightforward to prove that 
$\widetilde{\bxi\times\bUpsilon}=[\widetilde{\xi},\widetilde{\Upsilon}]$.

We complete this section by considering the density matrix for a two-level system, which as it is Hermitian, 
can be written on the basis of Pauli matrices (including the identity) using the 
well-known relation
\begin{align}\label{rhoblochvector}
  \rho = \frac{1}{2}(\mathbbm{1}+\bn\cdot\widehat{\boldsymbol{\sigma}})\,,
\end{align}
where $\bn$ is called the {\it Bloch vector} \cite{NielsenChuang}. Whilst we will not go into further detail on the geometry of
the Bloch sphere, we comment that $\bn$ itself is intimately connected to the Pauli matrices via the relation $\bn=\braket{\widehat{\bsigma}} $, where 
 $\braket{\widehat{\bsigma}}=\text{Tr}(\rho\widehat{\bsigma})$ is the expectation 
of the Pauli vector. This relation can be verified by a direct computation and 
leads naturally to the definition of the {\it spin vector} $\bs:= \hbar\bn/2$, 
corresponding to the expectation of the spin observable $\widehat{\bS}:=\hbar 
\widehat{\bsigma}/2$. For normalised pure states, one can verify that $|\bn|^2 = 1$ or equivalently $|{\boldsymbol{s}}|^2=\hbar^2 /4$.

\subsection{Geometry of quantum mechanics}\label{Sec:1-GeomQM}
Geometric approaches to quantum mechanics have become increasingly popular ever 
since Kibble's seminal work of 1979 \cite{Kibble1979} investigating the geometry of quantum state spaces.
Furthermore, with the fruitful field of geometric mechanics emerging from the early 
1960s, it is only natural to focus the attention of these tools on quantum mechanics. Among the many rationales for the application of geometry to quantum 
mechanics, one of particular practical interest is the understanding the transition between quantum and classical mechanics and in 
particular for chemistry purposes, developing quantum-classical hybrid models in which one divides the total system in two, allowing one subsystem to be modelled classically whilst another retains crucial quantum features. 
Therefore, in this section the focus is on the geometric approaches to quantum mechanics, familiarising the reader with concepts that will be used repeatedly throughout this thesis.
The relevant general theory can be found in Appendix \ref{App:GMTheory}.

\subsubsection{Variational and Hamiltonian structures}
The Schr\"odinger equation \eqref{BasicSchrodinger} possesses a variational 
structure corresponding to applying Hamilton's principle $0=\delta \int_{t_1}^{t_2} L \,\de t$ 
to the Lagrangian $L:T\mathscr{H}\to\mathbb{R}$, given by
\begin{align}
  L(\psi,\dot{\psi})=\braket{\psi,i\hbar\dot\psi - 
  \widehat{H}\psi}\label{DFStateLagrangian}\,,
\end{align}
where the bracket $\braket{\,\cdot\,,\,\cdot\,} := \text{Re}\braket{\,\cdot\,|\,\cdot\,}$ denotes the real-valued pairing on $\mathscr{H}$. Then, \eqref{BasicSchrodinger} simply 
corresponds to the Euler-Lagrange equation for the arbitrary quantity $\delta 
\psi$. This variational approach was first observed by Dirac \cite{DiracApprox}, inspriring the later developments by Frenkel \cite{Frenkel1934} and is therefore 
referred to as the {\it Dirac-Frenkel (DF) variational principle} with \eqref{DFStateLagrangian} 
known as the DF Lagrangian.

At this point, we notice that the functional $h(\psi)=\langle\psi,\widehat{H}\psi\rangle$ corresponds to the total energy and is hence the Hamiltonian functional of the system. By considering an arbitrary $h(\psi)$, sometimes called {\it Dirac Hamiltonian} to distinguish it from the Hamiltonian operator, $\widehat{H}$, the DF Lagrangian corresponds to a whole class of Schr\"odinger equations.

Then, in the general setting of arbitrary $h(\psi)$, the normalisation condition $\|\psi\|^2=1$ must be incorporated as a constraint, \cite{Ohta2000}, 
by considering the augmented Lagrangian
\begin{align}
L(\psi,\dot{\psi}, \lambda)=\braket{\psi,i\hbar\dot\psi}-h(\psi)+\lambda(\|\psi\|^2-1)
\,,\label{ConstrainedDFLagrangian}
\end{align}
where $\lambda(t)$ is a real-valued Lagrange multiplier. Then, the Euler-Lagrange equation for $\psi$ reads
\begin{align}
  i\hbar\dot{\psi}-\frac{1}{2}\frac{\delta h}{\delta \psi} = -\lambda\psi\,,
\end{align}
where we have defined the variational derivative by the real-valued pairing
\begin{align*}
  \delta h(\psi)=\left\langle \frac{\delta h}{\delta 
  \psi},\delta\psi\right\rangle\,,
\end{align*}
 so that upon taking the inner product with $\bra{\psi}$, separating real and imaginary parts yields the relations
\begin{align}
  \lambda =\left\langle\psi, \frac{1}{2}\frac{\delta h}{\delta \psi} -i\hbar\dot\psi
  \right\rangle\,,\qquad
  0 = \text{Im}\left\langle\psi\Bigg|\frac{\delta h}{\delta \psi} 
  \right\rangle\,.\label{CompCond1}
\end{align}
The first of these, specifying $\lambda$, implies the {\it projective Schr\"odinger equation} \cite{Kibble1979}
\begin{align}
(\mathbbm{1}-\psi\psi^\dagger)\bigg(i\hbar\dot\psi-\frac12\frac{\delta 
h}{\delta\psi}\bigg)=0
\,,\label{ProjectiveSchrodinger}
\end{align}
whilst the second \eqref{CompCond1} can be thought of as a compatibilty condition which ensures that $h(\psi)$ is $\mathcal{U}(1)-$invariant, since for a phase shift $\psi\mapsto e^{i\alpha} \psi$, $\delta h \mapsto \delta h + \delta\alpha \langle i\psi ,{\delta h}/{\delta\psi}\rangle$.
Analogously, Noether's theorem for the $\mathcal{U}(1)$ symmetry of the constrained Lagrangian \eqref{ConstrainedDFLagrangian} again implies conservation of $\|\psi\|^2=\langle\psi,\psi\rangle$, since the Lagrangian is invariant under infinitesimal phase shifts.
 Consequently, the constraint $\|\psi\|^2-1=0$ is satisfied and we may write, simply,
\begin{align}
i\hbar\dot\psi=\frac12\frac{\delta h}{\delta \psi}
\,,
\label{GenSchr}
\end{align}
which is the Hamiltonian form of the class of Schr\"odinger equations. 

The Hamiltonian structure for such Schr\"odinger equations is encoded in the following symplectic form on $\mathscr{H}$:
\begin{align}
\omega(\psi_1,\psi_2)=2\hbar\text{Im}\langle\psi_1|\psi_2\rangle=2\hbar\langle i\psi_1, 
\psi_2\rangle\label{QMsympform}\,,
\end{align}
which, when combined with the Hamiltonian vector field $X_h(\psi)$, defined by 
the rearrangement
\begin{align}
\dot\psi=-\frac{i}{2\hbar}\frac{\delta h}{\delta \psi}=:X_h(\psi)
\,,
\end{align}
results in the Poisson bracket given by
\begin{align}
\{f,g\}(\psi)=\frac1{2\hbar}\text{Im}\left\langle\frac{\delta f}{\delta \psi}\bigg|\frac{\delta g}{\delta \psi}\right\rangle 
= \left\langle \frac{\delta f}{\delta \psi},-\frac{i}{2\hbar}\frac{\delta g}{\delta \psi} \right\rangle 
\,, \label{SchrPB}
\end{align}
having used standard relation $\omega(X_f,X_g)=\{f,g\}$ on a symplectic manifold \cite{MaRa2013}. This Poisson bracket then takes us full circle, yielding the corresponding Hamiltonian equation \eqref{GenSchr} via the relation $\dot{f}=\{f,h\}$ and using that $\dot{f}=\braket{\delta f/\delta \psi, \dot {\psi}}$.

\subsubsection{Euler-Poincar{\'e}  variational principle}
In this section, we introduce the Euler-Poincar{\'e} approach to quantum 
dynamics, one of the key results of 
\cite{BonetLuzTronci2015}, which will be used and developed throughout this thesis. The general approach can be found in Appendix \ref{App:EPReduction}. 
The fundamental ingredient in this construction is allowing
 the quantum state $\psi\in \mathscr{H}$ to evolve unitarily, that is, the time 
 evolution of the state is given by the action
\begin{align}
  \psi(t)=U(t)\psi_0\,,\label{UnitaryEvo}
\end{align}
where $U(t)$ corresponds to a curve in $\mathcal{U}(\mathscr{H})$, the group of unitary 
operators on the Hilbert space and $\psi_0$ is an initial state such that $\|\psi_0\|^2=1$. 
Combined with the unitary evolution \eqref{UnitaryEvo}, this last condition ensures that 
the state is normalised at all times, $\|\psi(t)\|^2=1$. Then, the time 
derivative of the state is written
\begin{align}
  \dot{\psi} = \xi\psi\,,\label{Debens}
\end{align}
having defined the skew-Hermitian ($\xi^{\dagger}=-\xi$) operator $\xi:=\dot{U}U^{-1}$ belonging to the 
Lie algebra $\mathfrak{u}(\mathscr{H})$. {Identifying $\psi$ as a symmetry breaking parameter in the general Euler-Poincar{\'e} 
theory \cite{HolmEtAl1998}, equation \eqref{Debens}} transforms a Lagrangian of the type $L(\psi,\dot{\psi})$, 
defined on the tangent bundle $T\mathscr{H}$, to a reduced Lagrangian $\ell(\xi,\psi)$ 
defined on $\mathfrak{u}(\mathscr{H})\times\mathscr{H}$. Then, upon computing 
the variations 
\begin{align}
  \delta\psi = \eta\psi\,,\qquad \delta\xi = \dot\eta + [\eta,\xi]\label{QMEPVariations}\,,
\end{align}
 where $\eta:=(\delta U)U^{-1}\in \mathfrak{u}(\mathscr{H})$ is an arbitrary quantity, Hamilton's 
principle $0=\delta \int_{t_1}^{t_2} \ell(\xi,\psi)\,\de t$ yields the quantum 
mechanical Euler-Poincar{\'e} equations
\begin{align}
  \frac{\de}{\de t}\left(\frac{\delta \ell}{\delta \xi}\right) - \left[\xi,\frac{\delta \ell}{\delta \xi}\right] 
  = \frac{1}{2}\left(\frac{\delta \ell}{\delta \psi}\psi^{\dagger}-\psi\frac{\delta \ell}{\delta 
  \psi}^{\dagger}\right)\,,\\
  \dot{\psi}=\xi\psi \label{DotPsiisXiPsi}\,,
\end{align}
corresponding to the general equations \eqref{EPGenRight}.
Applied to the DF Lagrangian \eqref{DFStateLagrangian}, one has that $\ell(\xi,\psi)=\braket{\psi\psi^{\dagger}, i\hbar\xi - \widehat{H}}$ 
specialising the $\xi$ equation to read
\begin{align}
  [i\hbar\xi - \widehat{H},\psi\psi^{\dagger}]=0\,.
\end{align}
Whilst the general solution to this equation is presented in 
\cite{BonetLuzTronci2015}, here we simply notice that upon acting with this equation on $\psi$ 
and using \eqref{DotPsiisXiPsi}, one regains the projective Schr\"odinger equation 
\eqref{ProjectiveSchrodinger}. This demonstrates how the correct geometric interpretation, invoking the unitary 
evolution of quantum states (which automatically enforces the normalisation for all time), means that the Euler-Poincar{\'e} approach returns the dynamics on the correct space, in this case the unit sphere on the Hilbert space, $S(\mathscr{H})$.  

A further upshot of this approach is that upon recalling the definition of the 
density matrix for pure states $\rho:=\psi\psi^{\dagger}\in\mathbb{P}\mathscr{H}$ (now omitting the hat notation for $\rho$ and denoting the projective Hilbert space by $\mathbb{P}\mathscr{H}$\footnote{The projective Hilbert space $\mathbb{P}\mathscr{H}$ can be defined as the set of equivalence classes $\mathbb{P}\mathscr{H}:=\{[\psi]:\psi\sim e^{i\alpha}\psi\,,\, \alpha\in\mathbb{R}\}$. Hence, states which differ by a $\mathcal{U}(1)$ phase factor are identified in $\mathbb{P}\mathscr{H}$ so that each element $[\psi]$ uniquely corresponds to a pure state density matrix  $[\psi] = \psi\psi^{\dagger}$. The geometry of the projective Hilbert space naturally invokes a principal fibre bundle description (Hopf fibration) in which $\mathbb{P}\mathscr{H}=S(\mathscr{H})/\mathcal{U}(1)$ with projection operator $\pi:\psi\mapsto\psi\psi^{\dagger}$. For further information on the geometry of principal fibre bundles, see Appendix \ref{App:Principalbundles}.}), the Euler-Poincar{\'e} construction 
above can be rewritten in terms of $\rho$, corresponding to the reduced 
Lagrangian $\ell:\mathfrak{u}(\mathscr{H})\times\mathbb{P}\mathscr{H} \to \mathbb{R}$ given by
\begin{align}
  \ell(\xi,\rho)=\braket{\rho, i\hbar\xi-\widehat{H}}\,,\label{DFLagrDensityMatrix}
\end{align}
with variations $\delta \rho = [\eta, \rho]$ and $\delta\xi$ as given by the second of \eqref{QMEPVariations}. 
The corresponding equations now read
\begin{align}
  \frac{\de}{\de t}\left(\frac{\delta \ell}{\delta \xi}\right) - \left[\xi,\frac{\delta \ell}{\delta \xi}\right] 
  =\left[\frac{\delta \ell}{\delta \rho},\rho\right]\,,\label{EPdensitymatrixequation}\\
  \dot{\rho}=[\xi,\rho] \,,
\end{align}
and, importantly, {the entire construction is valid for the more general case of mixed states (i.e. $\rho\notin \mathbb{P}\mathscr{H}$)}. In the mixed state formulation
{$\rho \in {\rm Her}(\mathscr{H})$} and the unitary evolution is written as $\rho(t)=U(t)\rho_0 U^{\dagger}(t)$, 
allowing us to postulate that the Lagrangian $\ell(\xi,\rho)=\braket{\rho, i\hbar\xi - \widehat{H}}$ is valid for mixed state density matrices, from  
which we immediately see that the equation for $\xi$ returns the von Neumann 
equation \eqref{LvN}. Interestingly, upon considering the Wigner-Moyal picture of quantum dynamics (Appendix \ref{App:Wigner}), 
 this construction also results in a new variational principle  
for the Moyal equation on quantum phase space \cite{BonetLuzTronci2015}.

\subsubsection{Lie-Poisson structure}

Having seen how the von Neumann equation follows from the Euler-Poincar{\'e} variational principle, 
here we demonstrate how mixed state dynamics also possess a non-canonical 
Hamiltonian structure in the form of a Lie-Poisson bracket. To see this, we 
start with the Poisson bracket for pure states \eqref{SchrPB} and then consider that the Hamiltonian functional can be expressed solely in terms of the density matrix, i.e. that 
$h(\psi)=h(\rho)$. Notice this is certainly true for the physical Hamiltonian whereby $h(\psi)=\braket{\psi|\widehat{H}\psi}=\braket{\rho|\widehat{H}}=h(\rho)$.
Then, using that $\rho=\psi\psi^{\dagger}$ we use the chain 
rule to compute
\begin{align*}
  \frac{\delta h}{\delta \psi} = 2\frac{\delta h}{\delta \rho}\psi\,,
\end{align*}
which upon substitution into the bracket \eqref{SchrPB} returns
\begin{align}
  \{f,g\} = \left\langle \frac{i}{\hbar}\rho, \left[\frac{\delta f}{\delta \rho},\frac{\delta g}{\delta \rho}\right] \right\rangle 
  \label{LPBDens}
  \,.
\end{align}
This bracket is Lie-Poisson on $\mathfrak{u}(\mathscr{H})^*$ where the pairing is defined on $\mathfrak{u}(\mathscr{H})^*\times\mathfrak{u}(\mathscr{H})$ {and $[\,\cdot\, \,,\,\cdot\,]$ denotes the commutator, being the Lie bracket on the Lie algebra $\mathfrak{u}(\mathscr{H})$ of skew-Hermitian operators. } 
The bracket \eqref{LPBDens} is minus the general form of the Lie-Poisson bracket \eqref{LPBDEF} after recognising 
the change of variable $\rho \mapsto i\hbar\rho \in 
\mathfrak{u}(\mathscr{H})^*$ and defining $\mu:=i\hbar\rho$. As usual, the dynamics for $\rho$ can be obtained using the relation $\dot{f}(\rho)=\{f,h\}(\rho)$ returning
\begin{align}
  i\hbar\dot{\rho}=\left[\frac{\delta h}{\delta \rho},\rho\right]\,,
\end{align} 
which specialises to the von Neumann equation \eqref{LvN} for the physical 
Hamiltonian functional $h(\rho)=\braket{\rho|\widehat{H}}$.
 At this point, we also comment that $-i\hbar\psi\psi^{\dagger}$ 
is in fact an example of a momentum map associated to the action of the unitary group on 
the Hilbert space, a topic which we now consider in more detail.

\subsubsection{Momentum Maps}
With the general theory discussed in detail in Appendix \ref{App:Momaps}, here we specifically consider relevant examples of momentum maps in quantum mechanics,
 which will play a key role throughout this thesis, particularly in Chapters \ref{Chap:QHD}, \ref{Chap:EFOLD} and \ref{Chap:EFNEW}. 

As we have seen, generic quantum pure states belong to a Hilbert space 
$\mathscr{H}$, which is a symplectic vector space endowed with a symplectic form given by 
\eqref{QMsympform}. {Hence, for an arbitrary Hamiltonian (left) group action (see definitions in Appendix \ref{App:GMTheory})} of $G$ on $\mathscr{H}$, momentum maps will be of the form $J:\mathscr{H}\to\mathfrak{g}^*$, 
thus specialising the general result \eqref{VSMomapFormula} from Appendix \ref{App:Momaps} 
to give
\begin{align}
  \braket{J(\psi), \xi} = \hbar\braket{i\xi_{\mathscr{H}}(\psi),\psi}\,,\qquad \forall \psi\in\mathscr{H}\,,\qquad \forall \xi \in \mathfrak{g}\,,\label{QMVSmomap}
\end{align}
where $\xi_{\mathscr{H}}$ denotes the infinitesimal generator associated to the 
action of $G$ on $\mathscr{H}$. 

We now consider two typical examples in quantum mechanics, noting that we will 
consider in detail momentum maps in the hydrodynamic picture in Chapter 
\ref{Chap:QHD}. First, let us consider the action $\Phi_U:\mathscr{H}\to \mathscr{H}$ corresponding to the unitary group on the 
Hilbert space $\mathcal{U}(\mathscr{H})$, given by
\begin{align}
  \Phi_U(\psi) = U\psi\,.
\end{align}
Then, upon considering a curve $U(t)$ in $\mathcal{U}(\mathscr{H})$ such that $U(0)=\mathbbm{1}$ and $\dot{U}(0)=\xi\in \mathfrak{u}(\mathscr{H})$,
we compute the infinitesimal generator according to Definition \ref{Def:InfGen} as follows 
\begin{align}
  \begin{split}
    \xi_{\mathscr{H}}(\psi)&:= \frac{\de}{\de t}\bigg|_{t=0}\Phi_{U(t)}(\psi)\\
    &= \frac{\de}{\de t}\Big[U(t)\psi\Big]_{t=0}\\
    &= \dot{U}(0)\psi = \xi\psi\,.
  \end{split}
\end{align}
At this point, simply applying the momentum map formula \eqref{QMVSmomap} and 
rearranging using the properties of the inner product on $\mathscr{H}$ results 
in the momentum map
\begin{align}
  J(\psi) = -i\hbar\psi\psi^{\dagger}\,,
\end{align}
which can be identified with the density matrix for pure states \cite{CGMa2008}. 

The other key momentum map we point out here is the total probability 
$\|\psi\|^2$, which we have already seen must be conserved according to Noether's theorem.
 Indeed, upon considering the action of phase transformations $G=\mathcal{U}(1)\cong S^{1}$ 
on $\mathscr{H}$ given by $\Phi_{\varphi}(\psi)=e^{-i\hbar^{-1}\varphi}\psi$, a 
similar calculation for the infinitesimal generator returns 
$\xi_{\mathscr{H}(\psi)}=-i\hbar\xi\psi$\,, where $\xi \in \mathfrak{g}=\mathfrak{u}(1)\cong 
i\mathbb{R}$. Again, simply applying the momentum map formula yields 
$J(\psi)=\|\psi\|^2$ as required. Furthermore, upon considering the Hilbert space of wavefunctions, that is $\mathscr{H}=L^2(\mathbb{R}^3)$, simply extending the 
group to local phase transformations so that $\psi(\bx) \mapsto e^{-i\hbar^{-1}\varphi(\bx)}\psi(\bx)$, results in the momentum map identifying the probability density $J(\psi) = |\psi(\bx)|^2=:D(\bx)$. 
This already gives a flavour for the momentum map approach to quantum 
hydrodynamics given in Section \ref{Chap:QHD}.

We close this section by noting that, whilst we will consider momentum maps for density matrices in the hydrodynamic picture in Section \ref{Sec:2-ColdFluid}, 
 further details on the more general momentum map aspects of quantum mixed states can be found in \cite{Montgomery1991,Tronci2018}.

\subsection{Quantum hydrodynamics}\label{Sec:1-QHD}
In this section, we reintroduce the hydrodynamic picture of quantum mechanics, which proves to be a unifying 
 theme in this thesis, providing a common framework across the various topics. We begin by rederiving the hydrodynamic equations 
from a variational approach before discussing some of the more detailed aspects 
of the theory that are relevant to the subsequent material.

As we have seen, in the standard approach to quantum hydrodynamics, one writes the 
wavefunction $\psi(\bx,t)$ in the polar form
\begin{align}
  \psi(\bx,t)=\sqrt{D(\bx,t)}e^{i\hbar^{-1}S(\bx,t)}\label{StandardMadelung}\,,
\end{align}
where both $D$ and $S$ are real functions of $\bx$. Equation \eqref{StandardMadelung} is known as the {\it Madelung transform} after his celebrated works of 1926 and 1927 \cite{Madelung1926,Madelung1927}, the first of which was published even before Schr{\"o}dinger's seminal paper 
\cite{Schrodinger1926}!
Upon identifying the probability density $D=|\psi|^2$, the normalisation of the wavefunction reads $\int D(\bx,t) \,\de^3x =1$. 
Next, as we have already introduced the variational structure underlying quantum 
dynamics, we substitute this expression into the DF Lagrangian \eqref{DFStateLagrangian} for the physical Hamiltonian operator $\widehat{H}=-\hbar^2 \Delta/2m+V(\boldsymbol{x})$, to 
obtain
\begin{align}
  L(D,S,\partial_t S)=\text{Re}\int\!D\left(\partial_tS + \frac{|\nabla S|^2}{2m} +\frac{\hbar^2}{8m}\frac{|\nabla D|^2}{D^2} 
  +V\right)\,\de^3x\label{DSLagrangian}\,,
\end{align}
(having made an overall change of sign) thereby leading to
\begin{align}
\partial_t D + \text{div}\left(\frac{D\nabla S}{m}\right)&=0\label{DSEqn1} \,,\\
   \partial_t S+\frac{|\nabla S|^2}{2m}-\frac{\hbar^2}{2m}\frac{\Delta\sqrt{D}}{\sqrt{D}}+V  
   &=0\label{DSEqn2}\,.
   \end{align}
As we saw in Section \ref{Graham}, these are understood as a continuity equation for the probability density and a Hamilton-Jacobi type equation of classical mechanics $\partial_t S + H =0$, augmented by the {\it quantum potential}
\begin{align}\label{BohmPot}
  V_Q := -\frac{\hbar^2}{2m}\frac{\Delta\sqrt{D}}{\sqrt{D}} \,.
\end{align}
{Using this variational approach\footnote{We remark that equations \eqref{DSEqn1} and \eqref{DSEqn2} possess an additional variational structure corresponding to a Lagrangian $L(D,\partial_tD)$ on $T{\rm Den}(\mathbb{R}^3)$ in which the kinetic energy corresponds to the Wasserstein-Otto metric \cite{KhesinEtAl2018}.}}, we now see how $V_Q$ arises due to the third term in \eqref{DSLagrangian}, which can  
be equivalently written in the following ways
\begin{align}
\frac{\hbar^2}{8m}\int \frac{|\nabla D|^2}{D}\,\de^3x=\frac{\hbar^2}{2m}\int |\nabla\sqrt{D}|^2\,\de^3x 
= \int D \,V_Q\,\de^3x\,,
\end{align}
where in the final equality we have integrated by parts.
\begin{remark}[Effects of the quantum potential]
\hfill \\ 
The appearance of the amplitude of the wavefunction in the denominator of the quantum potential in \eqref{BohmPot} implies that its effects do not necessarily fall off with distance. That is, the effects of the quantum potential need not decrease, as the amplitude of the wavefunction decreases. Moreover, the third term in \eqref{DSLagrangian} is known as the Fisher-Rao norm, which is well-known in information theory. For further discussion of the information geometry in quantum mechanics, see e.g. \cite{BrHu1998}.
\end{remark}

To obtain the final form of the hydrodynamic equations one notices that, written in terms of the variables $(D,S)$, the equation for conservation of probability takes the form of a fluid continuity equation for a fluid velocity field $\bu:=m^{-1}\nabla 
S$, consistent with the interpretation of $\nabla S$ as a momentum in the Hamilton-Jacobi type equation. Then, upon taking the gradient of the quantum Hamilton-Jacobi equation, both equations are expressed in terms of $(D,\bu)$ as 
\begin{align}
\partial_t D + \text{div}\left(D\bu\right)&=0 \label{QHD1}\,,\\
 m (\partial_t+\bu\cdot\nabla)\bu&= -\nabla(V+V_Q)\label{QHD2}\,.
   \end{align}
It is important to notice that as $\bu$ is a potential flow, the vorticity $\bomega := \nabla\times\bu$ of the flow is identically zero unless there are points where $S$ is multi-valued {(see equation \eqref{MultivaluedS})}. 
This consideration naturally leads to the following remark.

\begin{remark}[Quantised circulation condition]
\hfill \\ 
Whilst the hydrodynamic picture of quantum mechanics dates back to Madelung \cite{Madelung1926,Madelung1927}, it was Takabayasi in 1952 \cite{Takabayasi1952} who first realised the requirement of the condition 
\begin{align}
  \oint_{c_0} \bu \cdot \de\bx = \frac{2\pi\hbar}{m}n\,,\label{QHDQuantCirc}
\end{align}
around any closed loop $c_0:[0,1]\to\mathbb{R}^3$ and for $n \in \mathbb{Z}$. This condition, asserting that the circulation $\Gamma:=\oint\bu\cdot\de\bx$ of the fluid velocity must be quantised, 
is non-zero only if the loop $c_0$ encloses a region where the phase $S$ is multi-valued, which 
itself occurs only at points where the wavefunction vanishes (nodes) \cite{GriffinKan1976,HirschfelderEtAl1974}. This can easily be seen by inverting the Madelung transform so that
\begin{align}
  S = -i\hbar\ln\left(\frac{\psi}{\sqrt{D}}\right)  = \hbar\left(\arctan\left(\frac{{\rm Re}(\psi)}{{\rm Im}(\psi)}\right)+ 
  n\pi\right)\,,\label{MultivaluedS}
\end{align} 
where $n\in \mathbb{Z}$ arises from the multi-valued nature of the inverse 
tangent function. Then, the quantisation condition \eqref{QHDQuantCirc} ensures that the total wavefunction is single-valued and hence that the QHD 
equations \eqref{QHD1}, \eqref{QHD2} are equivalent to the Schr\"odinger equation 
\eqref{PositionSchrodinger}, a fact which was later rediscovered by Wallstrom 
\cite{Wallstrom1994} in 1994.
\end{remark}

We continue the discussion by demonstrating that in the classical limit the QHD equations possess single particle solutions. To do so, we perform the $\hbar^2\to 0$ limit thereby neglecting the quantum potential term so that we have the classical hydrodynamic 
equations
\begin{align}
m(\partial_t+\bu\cdot\nabla)\bu&= -\nabla V\,,\qquad\partial_tD+ 
\text{div}(D\bu)=0\,,\label{QHDClassicalLimit}
\end{align}
and then write the (weak) solution for the density as $D(\bx,t) = \delta(\bx − \bq(t))$ where $\bq(t)$ corresponds to a classical trajectory.
Then, upon integrating the classical-limit QHD equations \eqref{QHDClassicalLimit} over a test function, 
the continuity equation tells us that $\dot{\bq}=\bu(\bq)$ whilst the Euler 
equation returns
\begin{align}
  m\ddot\bq = -\nabla_{\bq}V(\bq)\label{Restricted-qHcH-eqns}\,,
\end{align}
which we recognise as Newton's second law for a point particle in a conservative force field. 

\begin{remark}[Problems with the $\hbar^2\to0$ classical limit]\label{ultra}
\hfill \\ 
This Newtonian limit of QHD, obtained by simply neglecting the contribution of the quantum potential \eqref{BohmPot}, turns out to be problematic.
 Firstly, as described in detail in \cite{FoHoTr2019}, the Newtonian limit system \eqref{QHDClassicalLimit} is not strictly 
 hyperbolic, meaning that the solution behaviour can become singular.
  Secondly, as we will see in later sections, the typical QHD 
 approach to modelling molecular dynamics results in a nuclear fluid equation with additional terms containing $\hbar^2$ coefficients. 
 Hence, the standard classical limit $\hbar^2\to0$ allowing single particle 
 solutions also has the effect of removing these important coupling terms, thus 
 proving ineffective for deriving quantum-classical models in much of quantum chemistry. 
 Alternative solutions to this problem are given in Chapter \ref{Chap:QHD} in which
 new regularisation techniques and classical closures in terms of density matrices are presented. 
\end{remark}

We conclude this section by giving a brief overview for the QHD approach to mixed quantum states.
Whilst the standard approach invokes the Wigner phase space formulation of quantum mechanics, 
here the hydrodynamic variables for mixed quantum states are introduced as {\it moments} 
of the density matrix $\rho(\bx,\bx')$:
\begin{align*}
  \bar{p}_n(\bx,t):= \int \left[\frac{\hbar}{2i}\Big(\nabla_{\bx}-\nabla_{\bx'}\Big)\right]^n \rho(\bx,\bx') \delta(\bx-\bx')\,\de^3x'\,.
\end{align*}
 These quantities, essentially projections onto coordinate space, transform the von Neumann equation \eqref{LvN} into a countably infinite hierarchy
 of moment equations. For pure states $\rho(\bx,\bx')=\psi(\bx)\psi^*(\bx')$, the 0th and 1st order moments are given by
 \begin{align*}
    \bar{p}_0(\bx,t)&= \int \psi(\bx)\psi^*(\bx') \delta(\bx-\bx')\,\de^3x' = 
    |\psi(\bx)|^2 =: D(\bx)\,,\\
    \begin{split}
      \bar{p}_1(\bx,t)&= \int \frac{\hbar}{2i}\Big(\nabla_{\bx}\psi(\bx)\,\psi^*(\bx')- \psi(\bx)\,\nabla_{\bx'}\psi^*(\bx') \Big)\delta(\bx-\bx')\,\de^3x' 
      \\
      &\qquad\qquad\qquad\qquad\qquad\qquad= \hbar\text{Im}(\psi^*(\bx)\nabla\psi(\bx)) =: \bmu(\bx)\,,
  \end{split}
 \end{align*}
 and the hierarchy closes at first order, reproducing the QHD equations \eqref{QHD1} and \eqref{QHD2} upon identifying $\bmu = mD\bu$. 
 As we shall see in Chapter \ref{Chap:QHD}, both $D$ and $\bmu$ are collective momentum 
 maps and indicate the rich geometry underlying the moment hierarchy. In fact, an analogous picture 
 occurs in kinetic theory for the Vlasov equation, whereby so-called kinetic moments are taken of the particle distribution function, a topic which has been extensively studied from a geometric viewpoint, for example in \cite{GiHoTr08FOCK,GiHoTr08SING, HolmTronci2009}.
 As in the general case of mixed quantum states the hierarchy is infinite, one often invokes a {\it closure} of the density matrix, expressing 
 the dynamics in terms of a finite number of moments. These differ from 
 truncation schemes which instead simply neglect equations of higher order, 
 breaking the geometric structure.

As mentioned in the opening chapter, the striking similarity between quantum mechanics and classical fluid dynamics
has resulted in fundamental debates over the interpretation of quantum theory which persist to this day.
 Whilst we will not consider the implications of such theories in this work, we remark that 
   the use of (Bohmian-like) trajectory based descriptions of quantum mechanics are used to simulate 
      a wide variety of physical processes \cite{Wyatt2006}, particularly in the field of quantum chemistry \cite{deCarvalhoEtAl2014},
       and that QHD approaches continue to be at the forefront of a wide variety of academic research, for example in the context of quantum gravity \cite{Volovik2006} and more recently in a geometric approach to hybrid quantum-classical dynamics \cite{GayBalmazTronci2019}.

\section{Quantum chemistry}\label{Sec:Chem}

The work in this thesis was originally motivated by the idea of applying geometric mechanics to some of the open questions in the field of quantum chemistry, namely the development of nonadiabatic models going beyond the Born-Oppenheimer approximation as well as mixed quantum-classical models which aim to couple classical and quantum subsystems in a consistent manner.
The process of applying geometry to these ideas paved the road to understanding and developing geometric techniques in quantum hydrodynamics which now form a central unifying role in this work, although throughout we go back and consider the application of these results to quantum chemistry. 
Hence, it is the goal of this section to provide a more in-depth look at the relevant aspects of quantum chemistry first introduced in Chapter \ref{Chap1}, as they will feature throughout this thesis.

\subsection{Born-Oppenheimer approximation}\label{SectionBOApprox}
Recall that we will consider the simplest non-trivial example of an atom comprising a single electron and nucleus (of mass $M$), 
so that the molecular wavefunction is given by $\Psi(\br,\bx,t)\in {L}^2(\mathbb{R}^3\times 
\mathbb{R}^3)$. Then, the molecular Schr\"odinger equation \eqref{molecularSchrodinger} now reads
\begin{align}
  i\hbar\partial_t\Psi= \left(-\frac{\hbar^2}{2M}\Delta_{\br}+ \widehat{H}_e(\br)\right)\Psi\,,\label{MolSchr2Part}
\end{align}
where the electronic Hamiltonian is given by $\widehat{H}_e= \widehat{T}_e + 
\widehat{V}_I$. At this point, we consider some further details of the BO approach, adopting Dirac's notation so that $\ket{\phi(\br)}:=\phi(\bx;\br)$ and henceforth drop the label on nuclear derivatives so that $\nabla$ is identified with $\nabla_{\br}$.
Then, recall that the BO factorisation ansatz \eqref{BOAnsatz} is given by
\begin{align*}
  \Psi(t)=\Omega(\br,t)\ket{\phi(\br)}\,,
\end{align*}
where $\phi$ is a given eigenstate of $\widehat{H}_e$ with eigenvalue $E$. Hence, the electronic factor satisfies a partial normalisation condition 
(PNC)
\begin{align}
\int |\phi(\bx;\br)|^2\,\de^3 x = 1\label{BOPNC}\,,
\end{align}
which, due to the normalisation of the total wavefunction, also ensures the nuclear normalisation $\int|\Omega(\br,t)|^2\,\de^3r=1$ at all times.
\subsubsection{Nuclear dynamics and electromagnetic analogy}
In order to compute the nuclear dynamics explicitly, we substitute the BO ansatz directly into the molecular Schr\"odinger equation \eqref{molecularSchrodinger} 
and take the inner product with $\bra{\phi}$ to obtain
\begin{align}
  i\hbar\partial_t\Omega = -\frac{\hbar^2}{2M}\Delta\Omega - 
  \frac{\hbar^2}{M}\braket{\phi|\nabla\phi}\cdot\nabla\Omega 
  -\frac{\hbar^2}{2M}\braket{\phi|\Delta\phi}\Omega + E\,\Omega\label{BOnuclearEOM}\,,
\end{align}
having used the normalisation of $\phi$ and that 
$\braket{\phi|\widehat{H}_e\phi}=E$. At this point, we comment that the nuclear 
dynamics correspond to the standard Schr{\"o}dinger equation \eqref{PositionSchrodinger} 
(where $E$ plays the role of the external potential) augmented by two additional 
terms, both involving derivatives of the electronic state with respect to the 
nuclear parameter. These additional terms $\braket{\phi|\nabla\phi}$ 
and $\braket{\phi|\Delta\phi}$ are known as the {\it first and second order 
electron-nuclear couplings}\cite{deCarvalhoEtAl2014}.

However, as we have seen, the nuclear equation \eqref{BOnuclearEOM}
can be more elegantly packaged by expressing it in the following form
\begin{align}
  i\hbar\partial_t\Omega = \frac{(-i\hbar\nabla+{\boldsymbol{\cal 
  A}})^2}{2M}\Omega + \epsilon\,\Omega \label{BOnuclearEOMMinCoup}\,,
\end{align}
in which we have defined the electronic {\it Berry connection} \cite{Berry1984}
\begin{align}
  {\boldsymbol{\cal 
  A}}(\br):=\braket{\phi|-i\hbar\nabla\phi}\label{BOBerryConnection}\,,
\end{align}
 as well as the {\it effective electronic potential}
 \begin{align}
   \epsilon(\br):= E + \frac{\hbar^2}{2M}\|\nabla\phi\|^2-\frac{|{\boldsymbol{\cal 
  A}}|^2}{2M}\,,\label{effectiveelecpot}
 \end{align}
which can also be thought of as a functional 
$\epsilon=\epsilon(\phi,\nabla\phi)$ and in which the last two terms correspond to the trace of the {\it quantum geometric tensor} (QGT) \cite{PrVa1980}. 
This point will be considered further in the context of exact factorization in Chapter \ref{Chap:EFOLD}, whilst new perspectives on the interpretation of the QGT will arise in Chapter 
\ref{Chap:Gamma}.

There are several important comments to 
make here. Firstly, the appearance of a Berry connection is well-known in the 
BO approach \cite{BohmEtAl2003, MeadTruhlar1979,Mead1992,Kendrick2003} and arises due to the parametric dependence of the electronic factor 
on the nuclear coordinate. Indeed, in the BO approximation, many molecular systems exhibit non-trivial 
Berry phases, given by the loop integral of \eqref{BOBerryConnection}. We will 
expand on this topic shortly. We also notice the formal equivalence between the nuclear equation \eqref{BOnuclearEOMMinCoup} and that for a particle in the presence of a constant electromagnetic field.
Here, the Berry connection acts as a magnetic vector 
potential, appearing via minimal coupling in the kinetic energy, whilst $\epsilon$ 
plays the role of the scalar potential. This analogy is maintained in the time-dependent setting, when in a later section we consider exact factorization dynamics. 
The Berry connection is an example of a connection
in a $\mathcal{U}(1)$ gauge theory, for which the geometric setting is a principal fibre bundle (see Appendix \ref{App:Principalbundles}). This concept will form a key part of this thesis in 
Chapters \ref{Chap:Holonomy} and \ref{Chap:Gamma}.
\begin{remark}[Variational structure of nuclear dynamics]\label{BODFVP}
  \hfill \\ 
  The nuclear equation \eqref{BOnuclearEOMMinCoup} can also be derived as the Euler-Lagrange equation associated to the DF 
  Lagrangian \eqref{DFStateLagrangian}, which for the BO factorisation ansatz 
  \eqref{BOAnsatz} can be written in the form $L={\rm Re}\int i\hbar\Omega^*\partial_t\Omega\,\de^3r - 
  h(\Omega)$, in which the total energy is given by
  \begin{align}
    \begin{split}
    h &= {\rm Re}\int \braket{\Psi|\widehat{H}\Psi}\,\de^3r\\
&= \int \left[\Omega^*\frac{(-i\hbar\nabla+ \boldsymbol{\cal A})^2}{2M}\Omega + 
|\Omega|^2\epsilon(\phi,\nabla\phi)\right]\,\de^3r\,.
  \end{split}
  \end{align}
\end{remark}

\subsubsection{Hydrodynamic picture and classical nuclear trajectories}\label{Sec:BOClassicalNuclear}
The nonadiabatic coupling terms in the nuclear equation \eqref{BOnuclearEOM} are well-known to diverge near specific nuclear geometries corresponding to degeneracies in the electronic Hamiltonian \cite{Baer2006}.
To alleviate this computational difficulty, it is standard practice in many areas of quantum chemistry to simply neglect both of these terms so that 
the nuclear equation becomes
\begin{align}
  i\hbar\partial_t\Omega = -\frac{\hbar^2}{2M}\Delta\Omega + E\,\Omega\,,\label{adiabaticBOapprox}
\end{align}
in what is referred to as the {\it adiabatic BO approximation} 
\cite{deCarvalhoEtAl2014}. Here, we again utilise a hydrodynamic picture for the nucleus 
to consider more carefully how such approximations must be made.
To do so, we follow the procedure given in Section \ref{Sec:1-QHD},  
inserting the Madelung transform $\Omega(\br,t) = \sqrt{D(\br,t)}e^{i\hbar^{-1}S(\br,t)}$ into the 
Lagrangian from Remark \ref{BODFVP} yielding
\begin{align}
\label{BO-Lagr}
L(D,S,\partial_t S)   &= \int D\left(\partial_t S + \frac{|\nabla S+\boldsymbol{\cal A}|^2}{2M} + \frac{\hbar^2}{8M}\frac{|\nabla D|^2}{D^2} + 
\epsilon (\phi,\nabla\phi)\right)\,\text{d}^3r\,.
\end{align}
We proceed by applying Hamilton's principle $\delta\int_{t_1}^{t_2} \!L\,\de t=0$ for arbitrary variations $\delta D$ and $\delta S$, which returns the Euler-Lagrange 
equations
\begin{align}\label{BOMadelung}
  \frac{\partial D}{\partial t}+\text{div}\left(D\,\frac{\nabla S 
  +\boldsymbol{\cal A}}{M}\,\right)&= 0
  \,,\qquad\qquad\ 
    \frac{\partial S}{\partial t} + \frac{|\nabla S+\boldsymbol{\cal A}|^2}{2M} + V_Q + \epsilon = 0\,,
\end{align}
as usual understood as a continuity equation for the nuclear density $|\Omega|^2 = D$ and a quantum Hamilton-Jacobi equation for the nuclear phase. 
Finally, we write the Madelung equations in hydrodynamic form in terms of the shifted velocity $\bu:=M^{-1}(\nabla S +\boldsymbol{\cal A})$, 
given by
\begin{align}
     \partial_tD  +\text{div}(D\bu)&=0\,,\\
     M(\partial_t+\bu\cdot\nabla)\bu &=-\bu\times\boldsymbol{\cal B}-\nabla\left(\epsilon + 
  V_Q\right)\,,\label{BOnuclearFLUID}
\end{align}
where $\boldsymbol{\cal B}:=\nabla\times\boldsymbol{\cal A}$ is the {\it Berry curvature} \cite{BohmKendrickLoewe1992}. 

 We now proceed to summarise the possible simplifications by considering their subsequent effects on the nuclear fluid equation 
 \eqref{BOnuclearFLUID}.

\begin{enumerate}

 \item{\bf Second order coupling:} In the 
 quantum chemistry literature it is often the case that the second order  coupling term, namely $\braket{\phi|\Delta\phi}$ is neglected on the grounds that it has a negligible effect on the nuclear dynamics \cite{Tully1998,Martinez1997, MeadTruhlar1979}. In particular, as described in \cite{Tully1998}, if $L$ describes a typical length scale for the system (e.g. bond length) then the full second order coupling term is of order $\hbar^2/ML^2$ and is usually negligibly small. As can be verified directly upon expanding the real part one has that $\|\nabla\phi\|^2= -\text{Re}\braket{\phi|\Delta\phi}$, and hence such an approximation transforms the effective electronic potential \eqref{effectiveelecpot} to
\begin{align}
  \epsilon(\phi,\nabla\phi):=E
  - \frac{|\boldsymbol{\cal A}|^2}{2M}\,.
\end{align}
 At this stage, one is left with the both the Lorentz force acting on the nuclei as well as the potential given by the sum of the 
 new effective electronic energy and nuclear quantum potential.

 \item{\bf Real electronic eigenstate:} Next, we consider when the electronic eigenstate $\phi(\br)$ is 
 real-valued, which is valid when the electronic Hamiltonian is non-degenerate \cite{Tully1998,BredtmannEtAl2015}. The immediate consequence of the reality of $\phi$ is that the Berry connection $\boldsymbol{\cal A}:=\braket{\phi|-i\hbar\nabla\phi}$ 
 vanishes since the electronic phase is spatially constant. 
 In this case the nuclear fluid equation becomes
 \begin{align}
   (\partial_t+\bu\cdot\nabla)\bu &=- M^{-1}\nabla\left(E + 
  V_Q\right)\,.
 \end{align}
Clearly we still have  the nuclear quantum potential as well as the potential energy surface capturing electron-nuclear coupling. 
This is the QHD equation for the adiabatic BO approximation described by 
\eqref{adiabaticBOapprox}.

 \item{\bf Quantum potential:} As detailed in \cite{FoHoTr2019}, the quantum potential can also cause difficulties in numerical simulations. If we also consider
 neglecting the quantum potential term $V_Q$, the nuclear hydrodynamic equation can be written in its simplest form, 
 given by
 \begin{align}
    (\partial_t+\bu\cdot\nabla)\bu &=- M^{-1}\nabla E\,.\label{BONuclearClassicalFluid}
 \end{align}
 \end{enumerate}

 It is only after this extreme level of approximation, neglecting all quantum terms (involving $\hbar$), that one can then consider a classical equation of motion for the nuclei, in which one may describe the picture of a nucleus evolving on a single 
 potential energy surface $E(\br)$ \cite{deCarvalhoEtAl2014,Tully1998}.
  To make this point clear we again follow the approach given in Section \ref{Sec:1-QHD} and write the nuclear density as $D(\br,t)=\delta (\br-\bq(t))$, so that $\bu(\bq)=\dot\bq$, thus modelling the nucleus as a classical particle with a trajectory $\bq(t)$.
Hence, \eqref{BONuclearClassicalFluid} becomes Newton's law $\ddot{\bq}=-\nabla E$ in which the classical nuclear trajectory evolves in the effective potential given by the BO electronic eigenstate. 

Whilst in this section we have considered adiabatic dynamics in the hydrodynamic picture via the Madelung 
transform, one can also proceed with an alternative approach in which the nuclear wavefunction is modelled by a frozen Gaussian wavepacket \cite{Heller1976,Littlejohn1986}.
As previously mentioned, this idea is briefly considered in 
the appendix of the author's work \cite{FoTr2020}, employing Gaussian coherent states within the 
variational principle \eqref{BO-Lagr}, thus providing an alternative approach to regularising the singularities that are known to arise in BO systems.

\subsubsection{Berry phase and Mead-Truhlar method}\label{SECBOBerryPhase}
Recall from the discussion in Chapter \ref{Chap1} that the electronic 
Hamiltonian can be degenerate for specific nuclear geometries, corresponding to 
the intersection of PESs, known as conical intersections. It is near these singular points that the electron-nuclear coupling terms play a 
significant role and suggests that nonadiabatic models may be better suited to 
describing molecular processes. In addition we saw that, in the BO approximation, the electronic wavefunction changes sign around 
such a conical intersection, due to the presence of a non-trivial Berry phase. 

Here, we demonstrate how the Berry phase appears in the BO approach, following the derivation found in \cite{Berry1984, Baer2006}.
To do so, let us now consider the time-dependent wavefunction $\ket{\psi(\br,t)}$ which 
solves the electronic time-dependent Schr\"odinger equation
\begin{align}
  i\hbar\partial_t\ket{\psi(\br,t)} &= \widehat{H}_e\ket{\psi(\br,t)}\label{timedepelectronicSchr}\,.
\end{align}
In accordance with the adiabatic nature of the BO approximation we 
assert that this time-dependent electronic wavefunction must remain in an 
eigenstate of $\widehat{H}_e$, up to a phase factor. Hence, if we select the 
initial condition that $\ket{\psi}$ is in the ground state, that is 
$\ket{\psi(\br,0)}=\ket{\phi_0(\br)}$, then we restrict to consider solutions such that
\begin{align}
 \ket{\psi(\br,t)}=e^{i\hbar^{-1}\alpha(t)}\ket{\phi_0(\br)}\,,
\end{align}
for all future times.

Substituting this into the electronic Schr\"odinger equation \eqref{timedepelectronicSchr}, we obtain an evolution equation for 
the phase $\alpha$, given by
\begin{align*}
  \dot{\alpha} = E_0 - \braket{\phi_0|i\hbar\dot{\phi}_0}\,.
\end{align*}
Now, if we use the chain rule to evaluate $\ket{\dot{\phi}_0}=\dot{\br}\cdot\nabla\ket{\phi_0}$ and consider the nuclear coordinates evolving round a fixed loop $c_0$ such that $\br(0) = \br(T)$, 
we can take the loop integral of $\text{d}\alpha = E_0\, \text{d}t - \braket{\phi_0|i\hbar\,\text{d}{\phi}_0}$ 
to obtain
\begin{align}
  \alpha = \int_0^T E_0(t)\,\text{d}t + \oint_{c_0} 
  \braket{\phi_0(\br)|-i\hbar\nabla\phi_0(\br)}\cdot\text{d}\br\,,\label{Dynamic+Geometric}
\end{align}
from which we see that $\alpha$ is composed of two parts. The first is dynamical in nature depending on $E_0(t)$, whilst the second is purely geometric depending only on the fixed loop $c_0$ and not on the rate of its traversal.
One clearly recognises that the second term is simply the loop integral of the 
ground state Berry connection ${\boldsymbol{\cal A}_0}$ \eqref{BOBerryConnection} and this term is similarly named 
{\it Berry's phase} \cite{Berry1984}. The Berry phase is the 
original and architypal example of a quantum geometric phase, a topic which has a rich underlying geometric picture in terms of the holonomy associated to a connection on a principal bundle, which we will consider 
in more detail in Chapters \ref{Chap:Holonomy} and \ref{Chap:Gamma}. For now, we are interested in 
the role of the Berry phase in the BO approach. 

The standard approach for dealing with the multi-valued electronic factor appearing in BO dynamics is known as the {\it Mead-Truhlar method} 
\cite{MeadTruhlar1979} and utilises the invariance of the electronic eigenvalue problem under the transformation 
\begin{align}
\ket{\phi(\br)}\mapsto 
\ket{\phi'(\br)}=e^{i\alpha(\br)/\hbar}\ket{\phi(\br)}\label{MTalphaequation}
\end{align}
 which in turn implies the transformation of the Berry connection
 \begin{align*}
 \boldsymbol{\cal A} \mapsto  \boldsymbol{\cal A}'=\boldsymbol{\cal A} +\nabla 
 \alpha\,.
 \end{align*}
 Such a transformation leaves the effective electronic potential \eqref{effectiveelecpot} 
 invariant and hence transforms the nuclear equation \eqref{BOnuclearEOMMinCoup} 
 into 
 \begin{align}
  i\hbar\partial_t\Omega = \frac{(-i\hbar\nabla+\nabla\alpha+{\boldsymbol{\cal 
  A}})^2}{2M}\Omega + \epsilon\,\Omega \label{BOnuclearEOMMinCoupMTM}\,,
\end{align}
and one can equivalently consider this as the shift in nuclear momentum operator 
$-i\hbar\nabla\mapsto -i\hbar\nabla + \nabla\alpha$.

At this point, let us consider the case when $\ket{\phi}$ is real, which as we have seen implies that $\boldsymbol{\cal A}=0$. Then, it is possible to select $\alpha$ such that the phase $e^{i\alpha/\hbar}$ exactly compensates the double-valuedness of $\ket{\phi}$ arising in Berry phase phenomena
resulting in the new (complex) electronic state $\ket{\phi'(\br)}$ being single-valued \cite{Kendrick2003,Mead1992}. However, such a transformation has the cost that the corresponding transformed vector potential $\boldsymbol{\cal A}' = \nabla\alpha$ is singular at the point of the conical intersection, 
as $\alpha$ itself is double-valued. This means that the corresponding {\it Berry curvature}, defined by $\boldsymbol{\cal B}:=\nabla\times\boldsymbol{\cal A}$ and effectively a magnetic field, is singular at this point \cite{Kendrick2003}. 
Hence, in this case the Berry phase is expressed as
\begin{align}
  \oint _{c_0}
 \boldsymbol{\cal A}'\cdot\text{d}\br = \oint_{c_0} \nabla \alpha \cdot \de\br = 
 \int_{S_0} \nabla\times\nabla\alpha\cdot\de\bS\,,
\end{align}
where $S_0$ is a surface such that its boundary defines the loop $\partial S_0 = c_0$, and is a topological quantity due to the nature of the singularity. For further 
information regarding the use of multi-valued functions in physics, the reader 
is referred to \cite{Kleinert}. Due to the obvious similarities, this effect is also known as the molecular Aharonov-Bohm effect \cite{Mead1980}. 
Thus the Mead-Truhlar method trades the double-valuedness of the 
electronic wavefunction at the conical intersection for a singular connection at that same point.
 As we shall see in Chapter \ref{Chap:Holonomy}, one of the main results of this thesis is a new approach
 to introducing connections in quantum mechanical systems, which will give rise to a purely geometric phase without the
  need for multi-valued functions or singular connections.

\subsection{Nonadiabatic models}\label{NonadiabaticSection}
Having outlined the requirement of nonadiabatic models in the first chapter, here we 
will further consider the implications of two of the most popular factorisations of the molecular wavefunction, mean-field models and exact factorization. 

\subsubsection{Mean-field model}
Among the most commonly used models for capturing nonadiabatic effects are 
mean-field models of the form
\begin{align}
  \Psi(\br,\bx,t)=\Omega(\br,t)\psi(\bx,t)\,,\label{GeneralMeanField}
\end{align}
where $\Omega$ and $\psi$ are identified with the nuclear and electronic wavefunctions, respectively. 
Once again adopting Dirac's notation so that the electronic state is written 
$\ket{\psi(t)}:=\psi(\bx,t)$, replacing the mean-field ansatz \eqref{GeneralMeanField}  
into the DF Lagrangian for the molecular wavefunction results in
\begin{align}
  \begin{split}
    L &= \text{Re}\int 
    \braket{\Psi|i\hbar\partial_t\Psi-\widehat{H}\Psi}\,\de^3r\\
    &= \braket{\psi,i\hbar\dot{\psi}} + \text{Re}\int \Omega^*\left(i\hbar\partial_t\Omega 
    + \frac{\hbar^2}{2M}\Delta\Omega\right) 
    -|\Omega|^2\braket{\psi|\widehat{H}_e(\br)\psi}\,\de^3r\,.
  \end{split}
\end{align}
At this stage, we again choose to express the nuclear wavefunction in polar form $\Omega(\br,t)=\sqrt{D(\br,t)}e^{i\hbar^{-1}S(\br,t)}$ thus taking the above Lagrangian into 
the form
\begin{align}
  L(D, S, \partial_tS, \psi, \dot{\psi}) = \int D\left(\partial_t S + \frac{|\nabla S|^2}{2M}+V_Q + 
  \braket{\psi|\widehat{H}_e(\br)\psi}\right)\,\de^3r - 
  \braket{\psi,i\hbar\dot{\psi}}\,,
\end{align} 
where $V_Q$ denotes the nuclear quantum potential and we have made an irrelevant overall 
sign change. Computing the Euler-Lagrange equations results in the equations of 
motion
\begin{align}
  \partial_tD + \text{div}\left(\frac{D\nabla S}{M}\right)&= 0\,,\\
  \partial_t S +  \frac{|\nabla S|^2}{2M}+V_Q + 
  \braket{\psi|\widehat{H}_e(\br)\psi} &= 0\,,\label{Jonas}\\
  i\hbar\dot\psi - \int D\widehat{H}_e(\br)\psi\,\de^3r&=0\,,
\end{align} 
in which we recognise the now familiar continity and quantum Hamilton-Jacobi 
equations for the nucleus along with an electronic Schr\"odinger equation. 
Performing the usual procedure, taking the gradient of \eqref{Jonas} and identifying the nuclear fluid velocity $\bu:= M^{-1}\nabla S$, 
these equations can be rewritten as
  \begin{align}
  \partial_tD + \text{div}\left({D\bu}\right)&= 0\,,\\
  M(\partial_t +\bu\cdot\nabla)\bu&= -\nabla V_Q - 
  \braket{\psi|(\nabla{V}_I)\psi}\,,\label{Jonas2}\\
  i\hbar\dot\psi &= \left(\widehat{T}_e +  \int D{V}_I\,\de^3r\right)\psi\,,
\end{align} 
also having expanded the electronic Hamiltonian as $\widehat{H}_e = \widehat{T}_e + 
V_I(\br,\bx)$.

These coupled equations demonstrate how the electronic motion is governed by the average field generated by the nuclear wavefunction, whilst the nuclear motion is governed by an average field provided by the electrons and hence warrant the name mean-field equations.

At this point, the nuclear subsystem is often modelled using classical 
trajectories via the hydrodynamic limiting process described in Section 
\ref{Sec:1-QHD}, in which one drops the quantum potential term and writes $D(\br,t) = \delta(\br - \bq(t))$, so that one obtains a mixed quantum-classical system
\begin{align}
  M\ddot{\bq} &= -\nabla_{\bq}\braket{\psi|{V}_I(\bq, \bx)\psi}\,,\\
  i\hbar\dot\psi &=\Big( \widehat{T}_e + V_I(\bq, \bx)\Big)\psi\,.
\end{align}
These equations are known as the quantum-classical mean-field model or {\it Ehrenfest method} in the chemistry literature \cite{deCarvalhoEtAl2014, Tully1998} and represent the standard mean-field model as it is usually implemented in molecular dynamics simulations \cite{MarxHutter2009} (although here we have focused on the simplest case of one nucleus and one electron). 

Whilst the geometric structure of these equations has been studied in \cite{BonetLuzTronci2015} and, in possessing a variational formulation, conserve energy, 
there are many limitations to such approaches. Firstly, we once again encounter the problems with performing the classical limit $\hbar^2\to 0$ 
as descibed in Remark \ref{ultra}.
More importantly, whilst indeed capturing nonadiabatic effects to some extent and even proving advantageous for modelling certain systems, it is clear the only quantum-classical coupling occurs solely through the interaction potential $V_I$, and this averaged nature of the mean-field model does not adequately reproduce the correlations between nuclei and electrons, \cite{MarxHutter2009}.
Hence mean-field models are apparently too simple to apply accurately to realistic situations for many nonadiabatic phenomena.

\subsubsection{Exact factorization}
The final method discussed in this section and a topic which we will investigate geometrically in depth in Chapter \ref{Chap:EFOLD} is the {\it exact factorization} model \cite{AbediEtAl2010,AbediEtAl2012}
\begin{align}
  \Psi(\br,\bx,t)=\Omega(\br,t)\psi(\bx,t;\br)\,,\label{ExactFactAnsatz}
\end{align} 
subject to the partial normalisation condition
\begin{align}
  \int |\psi(\bx,t;\br)|^2\,\de^3x = 1\,,
\end{align}
at all nuclear configurations $\br$ and times $t$. As described in \cite{AbediEtAl2012} equation \eqref{ExactFactAnsatz} is `exact' in the sense that one can always make the
 factorisation provided the nuclear factor $\Omega$ has no nodes, i.e. is non-zero for all nuclear geometries $\br$ and at all times $t$. 
  Following the earlier discussion of Chapter \ref{Chap1}, one immediately notices that this method serves as a time-dependent generalisation of the BO ansatz 
\eqref{BOAnsatz}, having now promoted the electronic factor to be 
time-dependent. Correspondingly, upon direct substitution into the molecular Schr\"odinger equation \eqref{MolSchr2Part}, one obtains equations of motion of the form
\begin{align}
  i\hbar\partial_t\Omega &= - \frac{\hbar^2}{2M}\Delta\Omega -\frac{\hbar^2}{2M}\braket{\psi|\Delta\psi}\Omega - \frac{\hbar^2}{M}\braket{\psi|\nabla\psi}\cdot\nabla\Omega + \braket{\psi|\widehat{H}_e\psi}\Omega  -\braket{\psi|i\hbar\partial_t\psi} \Omega
 \,, \label{EFNuclearEqnSEC1}\\
  0&= (\mathbbm{1}-\psi\psi^{\dagger})\left(i\hbar|\Omega|^2\partial_t\psi - |\Omega|^2\widehat{H}_e\psi + \frac{\hbar^2}{2M}|\Omega|^2\Delta\psi + 
  \frac{\hbar^2}{M}\Omega^*\nabla\Omega\cdot\nabla\psi\right)\label{EFElectronicEqnSEC1}\,.
\end{align}
\begin{remark}[Normalised electronic dynamics]
\hfill \\
  Notice here that a projective electronic Schr\"odinger equation has been obtained simply by substituting the EF ansatz into the molecular Schr\"odinger equation.
  As we have seen in Section \ref{Sec:1-GeomQM} such an equation is usually derived upon enforcing 
  the normalisation as a Lagrange multiplier at the level of the Lagrangian and ensures that the 
electronic state evolves in such a way that it remains normalised at all times. 
In fact this is exactly the approach we will apply for arbitrary Hamiltonian functionals in Chapter 
\ref{Chap:EFOLD},
which, for the physical total energy, formed much of the subsequent discussion 
 in \cite{alonso2013comment,abedi2013response} in relation to the original work \cite{AbediEtAl2012}.
\end{remark}

At this stage, one recognises the form of the nuclear equation with that from the 
BO case \eqref{BOnuclearEOM} and analogously one can define time-dependent 
generalisations of the Berry connection and effective electronic potential
\begin{align}
  {\boldsymbol{\cal 
  A}}(\br,t)&:=\braket{\psi|-i\hbar\nabla\psi}\label{GenBOBerryConnection}\,,\\
   \epsilon(\br,t)&:= \braket{\psi|\widehat{H}_e\psi} + \frac{\hbar^2}{2M}\|\nabla\psi\|^2-\frac{|{\boldsymbol{\cal 
  A}}|^2}{2M}\,,\label{GenBOEffectivePotential}
 \end{align}
 where the latter is referred to as the gauge invariant part of the {\it time-dependent potential energy surface (TDPES)} in the literature. Hence, the nuclear equation can be rewritten as
 \begin{align}
     i\hbar\partial_t\Omega = \frac{(-i\hbar\nabla+{\boldsymbol{\cal 
  A}})^2}{2M}\Omega +\Big( \epsilon-\braket{\psi|i\hbar\partial_t\psi}\Big)\,\Omega 
  \,,
   \end{align}
   analogous to \eqref{BOnuclearEOMMinCoup} up to the term 
   $\braket{\psi|i\hbar\partial_t\psi}$ which, as we will discuss further in Chapter \ref{Chap:EFOLD},
    can be made to vanish by selecting a gauge.
    
    Despite not simplifying the computational task of the molecular Schr\"odinger equation,
    the EF approach has generated significant interest due to its appealing correspondence 
    with the BO factorisation, in which all of the concepts and related objects 
    simply are promoted to their time-dependent counterparts. Of particular 
    interest is the role of geometric phases in the EF approach. As reported in 
    \cite{MinEtAl2014, RequistEtAl2016} the topological nature of the Berry 
    phase appears to be an artifact of the BO approximation, becoming a true (path-dependent) 
    geometric phase in the fully time-dependent setting. This can be seen in the TDPES which shows no sign of a singularity at the nuclear geometries which correspond to conical intersections in the BO approach. 
    Due to its intrinsic geometric features, both the EF and some new generalisations 
    are considered from a hydrodynamic geometric mechanics perspective in Chapters \ref{Chap:EFOLD} 
    and \ref{Chap:EFNEW} respectively, and form a key part of this thesis.

\chapter{Geometry of QHD and new closure schemes}\label{Chap:QHD}
In this chapter, we will investigate the hydrodynamic picture of quantum mechanics, as presented in 
Section \ref{Sec:1-QHD}, but now from the geometric perspective in terms of momentum maps. 

In Section \ref{Sec:2-HD} we shall see how the fundamental
hydrodynamical variables are themselves momentum maps 
under which the Hamiltonian collectivises \cite{Fusca2016,KhesinEtAl2018}. We then demonstrate how, via the reduced Legendre transform,
the equations of motion can be derived equivalently via Euler-Poincar{\'e} reduction or as Lie-Poisson equations corresponding to a Lie-Poisson bracket.

This chapter then presents some of the new results from the author's publication \cite{FoHoTr2019}.
In Section \ref{Sec:2-Bohmions} the geometric 
setting of QHD is used to provide a new regularisation method on the Lagrangian side
 allowing for particle-like singular solutions which we call Bohmions.
  This regularisation procedure results in finite-dimensional 
Newtonian equations in which Bohmion dynamics are governed by a smoothened 
version of the quantum potential, thus having the benefit of classical 
trajectories without resorting to fully classical dynamics.

Finally, in Section \ref{Sec:2-ColdFluid} we generalise the momentum map approach to include a density matrix 
description of QHD which allows for fluid velocities possessing non-trivial vorticity.  A classical closure is then presented using a technique well-known 
in fluids and plasma physics, resulting in a new method for eliminating the quantum potential without taking a formal $\hbar^2\to 0$ limit.

\section{Geometry of pure state QHD}\label{Sec:2-HD}

We commence this chapter by presenting the geometric structure of quantum 
hydrodynamics in terms of momentum maps \cite{Fusca2016, KhesinEtAl2018} under which the Hamiltonian functional collectivises \cite{FoHoTr2019}. 
From there, we derive the QHD dynamical equations using the associated Lie-Poisson structure before moving back to the Lagrangian side,
following the standard Euler-Poincar{\'e} reduction procedure for continuum theories \cite{HolmEtAl1998}.
This section also comments on the geometric interpretation of so-called Bohmian trajectories from the quantum physics literature.

\subsection{Momentum maps and collectivisation}
In order to unveil the geometry of QHD, we now consider an equivalent formalism whereby the wavefunction $\psi(\bx)\in L^2(\mathbb{R}^3)$ is considered as a 
half-density \cite{Fusca2016,BatesWeinstein1997}. In this sense we write $\psi \in \text{Den}^{1/2}(\mathbb{R}^3)$ such that $D(\bx):= |\psi(\bx)|^2 \in  
\text{Den}(\mathbb{R}^3)$. More generally, for $\psi_1, \psi_2 \in  
\text{Den}^{1/2}(\mathbb{R}^3)$, one has that $\text{Re}(\psi_1^*\psi_2) \in 
\text{Den}(\mathbb{R}^3)$. Having characterised the space of half-densities, the next step is to consider the natural (left) action
 of the diffeomorphism group $ \text{Diff}(\mathbb{R}^3)$ on the space of half-densities 
$ \text{Den}^{1/2}(\mathbb{R}^3)$, $\Phi:\text{Diff}(\mathbb{R}^3)\times \text{Den}^{1/2}(\mathbb{R}^3) \to \text{Den}^{1/2}(\mathbb{R}^3)$, given by
\begin{align}
  \Phi(\eta, \psi) =: \Phi_{\eta}(\psi) = 
  \frac{\psi\circ\eta^{-1}}{\sqrt{\text{det}\nabla\boldeta^T}}\label{half-dens-action}\,.
\end{align} 
Here, $\circ$ denotes composition of functions and ${\text{det}\nabla\boldeta^T}$ denotes the Jacobian of the smooth invertible map, $\eta$, acting on coordinates $\bx\in\mathbb{R}^3$ as $\eta:\bx\mapsto\boldeta(\bx)\in\mathbb{R}^3 $.
The notation in \eqref{half-dens-action} defines the mapping $\Phi_\eta:\text{Den}^{1/2}(\mathbb{R}^3)\to\text{Den}^{1/2}(\mathbb{R}^3)$ that is naturally induced by the group action {$\Phi(\eta,\psi)$ of the diffeomorphism $\eta\in\text{Diff}(\mathbb{R}^3)$ on the half-density $\psi\in\text{Den}^{1/2}(\mathbb{R}^3)$. Indeed, the left action in \eqref{half-dens-action} of diffeomorphisms on half-densities can be thought of as defining the push-forward of a half-density by a diffeomorphism.

As derived explicitly in Appendix \ref{App:HalfDen}, the infinitesimal generator 
corresponding to \eqref{half-dens-action} can be computed to obtain
\begin{align}
  u_{\,\text{Den}^{1/2}}(\psi)
= -\bu\cdot\nabla\psi-\frac12(\nabla\cdot\bu)\psi = 
{-\frac{i}{2\hbar}\{\widehat{u}^k,\widehat{P}_k\}_{+\,}\psi\,,}
\label{Diff-action}
\end{align}
 where $\bu(\bx)\in \mathfrak{X}(\mathbb{R}^3)$ is a smooth vector field on 
$\mathbb{R}^3$, $\widehat{P}_k:=-i\hbar\partial_k$ is the momentum operator, $\widehat{u}^{\,k\!}$ denotes the multiplicative operator associated to $u^k(\boldsymbol{x})$ and we adopt the anticommutator notation $\{A,B\}_+:=AB+BA$.
Following the general procedure outlined in Section \ref{Sec:1-GeomQM}, we now compute the equivariant momentum map $\bJ:\text{Den}^{1/2}(\mathbb{R}^3)\to\mathfrak{X}^*(\mathbb{R}^3)$ for the left action \eqref{Diff-action} from the standard formula \eqref{QMVSmomap}. 
Here we have identified the Hilbert space as $\mathscr{H}=L^2(\mathbb{R}^3)=\text{Den}^{1/2}(\mathbb{R}^3)$  
so that $\text{Den}^{1/2}(\mathbb{R}^3)$ inherits the standard symplectic form \eqref{QMsympform} on $L^2(\mathbb{R}^3)$ 
and the momentum map reads
\begin{align}
{\boldsymbol{J}}(\psi)=\hbar\text{Im}(\psi^*\nabla\psi) 
=D\nabla S\,.
\label{QHDmomap}
\end{align}
This is in agreement with \cite{Fusca2016,KhesinEtAl2018}, where in the last equality we have used the exponential form \eqref{StandardMadelung}.
Once again the reader is referred to Appendix \ref{App:HalfDen} for a full proof. 
Up to the mass factor, $m$, this momentum map $\bJ(\psi)$ agrees with the standard definition of the probability current in the physics literature on quantum mechanics.

At this point, we recall that $D:=|\psi|^2$ is also a momentum map relating 
the action of local phase transformations $\psi(\bx)\mapsto 
e^{-i\hbar^{-1}\varphi(\bx)}\psi(\bx)$. Hence, upon defining $\bmu:= \bJ(\psi)$ 
it can be shown that the total energy $h(\psi)=\braket{\psi|\widehat{H}\psi}$ can also be written as
\begin{align}
  h(\bmu, D) = \int\left[\frac{|\bmu|^2}{2mD} + \frac{\hbar^2}{8m}\frac{|\nabla D|^2}{D} + 
  DV\right]\,\de^3x\,,\label{collective1}
\end{align}
solely in terms of the momentum maps $(\bmu, D)$ and hence is a collective Hamiltonian \cite{GuilleminSternberg1980,GuilleminSternberg1990} as described in 
Appendix \ref{App:GMTheory}. We point out that both variables can be considered as components of the momentum map
to the dual of the semidirect product Lie algebra given by $(\bmu, D)\in 
(\mathfrak{X}(\mathbb{R}^3)\,\circledS\,\mathcal{F}(\mathbb{R}^3))^*$, 
corresponding to the 
semidirect product Lie group 
$\text{Diff}(\mathbb{R}^3)\,\circledS\,\mathcal{F}(\mathbb{R}^3,S^1)$, 
whose elements $(\eta,\varphi)$ act from the left on $\text{Den}^{1/2}(\mathbb{R}^3)$ 
by
\begin{align}
  \Phi_{(\eta, \varphi) }(\psi)=
  \frac{(e^{-i\hbar^{-1}\varphi}\psi)\circ\eta^{-1}}{\sqrt{\text{det}\nabla\boldeta^T}}\label{SEMIhalf-dens-action}\,,
\end{align} 
extending \eqref{half-dens-action} to include a local phase transformation \cite{Fusca2016} (For proof of the momentum map for the semidirect product action see Appendix \ref{App:HalfDen3}). 
The important feature here is that, under this collectivisation, the Hamiltonian $h(\bmu,D)$ given by \eqref{collective1} belongs to a widely studied class of Hamiltonians possessing the Lie-Poisson bracket structure 
\eqref{LPBDEF}, associated to the Euler-Poincar\'e formulation of ideal classical continuum dynamics with advected quantities in \cite{HolmEtAl1998}.

\subsection{Lie-Poisson structure}
The Lie-Poisson structure of QHD follows from the above collectivisation of the Hamiltonian \eqref{collective1}. 
Specifically, we specialise the general treatment of Lie-Poisson systems from Appendix \ref{App:GMTheory} to the case of continuum systems,
hence writing the QHD equations \eqref{QHD1}-\eqref{QHD2} in Hamiltonian form,
 with a Lie-Poisson bracket written symbolically as
 
\begin{equation} 
\frac{\partial}{\partial t}
    \begin{bmatrix}
    \boldsymbol{\mu} \\ D 
    \end{bmatrix}
=
\left\{
    \begin{bmatrix}
    \boldsymbol{\mu} \\ D 
    \end{bmatrix}
    \,,\,
    h(\boldsymbol{\mu},D)
\right\}
= -\, 
\begin{bmatrix}
   {\rm ad}^\ast_\Box\,\boldsymbol{\mu} &
   \Box\diamond D \\
   \pounds_\Box\,D & 0 
    \end{bmatrix}
   \begin{bmatrix}
   \delta h/\delta \boldsymbol{\mu} \\
   \delta h/\delta D    
    \end{bmatrix}    \,,
\label{LP-Ham-1}
\end{equation}
(see for example \cite{HolmEtAl1998}) in which each box $\Box$ in \eqref{LP-Ham-1} indicates where to substitute elements of the last column 
of variational derivatives of the Hamiltonian  \eqref{collective1} in the matrix multiplication. 
Here ${\rm ad}^\ast$ denotes the coadjoint action \eqref{Def:CoadjAlgAction} of the Lie algebra $\mathfrak{X}(\mathbb{R}^3)$ on its dual, the space of 1-form densities $\mathfrak{X}(\mathbb{R}^3)^*=\Lambda^1(\mathbb{R}^3)\otimes \text{Den}(\mathbb{R}^3)$, where  
 $\Lambda^1(\mathbb{R}^3)$ denotes the space of differential 1-forms on $\Bbb{R}^3$. 
The symbol $\pounds$ denotes the Lie derivative and is the infinitesimal generator for smooth flows, appearing in \eqref{LP-Ham-1} 
with respect to the vector field $\delta h/\delta \bmu$ and acting on the density $D\in 
\text{Den}(\mathbb{R}^3)$. Finally, the diamond operation $(\diamond)$ is generally defined 
by \eqref{DIAMOND}, in the case of continuum systems specialised to
\begin{align*}
\left\langle \frac{\delta h}{\delta a} \diamond a \,,\,\xi \right\rangle_{\mathfrak{g}^*\times \mathfrak{g}}
:=
\left\langle \frac{\delta h}{\delta a}\,,\, - \pounds_\xi a  \right\rangle_{V^*\times V}
\,,
\end{align*}
in the $L^2(\mathbb{R}^3)$ pairing $\langle \,\cdot\,,\,\cdot\,\rangle_{V^*\times V}: V^*\times V \to \mathbb{R}$ for elements of the tensor space $a\in V$ and on its dual $\delta h/\delta a\in V^*$. 
In particular, for \eqref{LP-Ham-1}, we have an advected density $D\in 
\text{Den}(\mathbb{R}^3)$.
The corresponding notation is defined explicitly for QHD {\em in components} by 
\begin{align}
\begin{split}
({\rm ad}^\ast_{\delta h/\delta \boldsymbol{\mu}}\,\boldsymbol{\mu})_i 
&= 
\partial_j \left(\mu_i \frac{\delta h}{\delta \mu_j}\right)+ \mu_j \partial_i\frac{\delta h}{\delta \mu_j} 
\in\Lambda^1(\mathbb{R}^3)\otimes \text{Den}(\mathbb{R}^3)\,,
\\
\pounds_{\delta h/\delta \boldsymbol{\mu}}\,D &:= {\rm div} \left( D \frac{\delta h}{\delta \boldsymbol{\mu}} \right)
\in \text{Den}(\mathbb{R}^3)\,,
\\
\frac{\delta h}{\delta D}\diamond D &:= D \nabla \frac{\delta h}{\delta D} 
\in \Lambda^1(\mathbb{R}^3)\otimes \text{Den}(\mathbb{R}^3)\,.
\end{split}
\label{LPB-notation}
\end{align}
Hence, the QHD equations can be expressed for an arbitrary Hamiltonian functional $h(\bmu, D)$ as 
\begin{align}
    \partial_t \bmu + \partial_j\left(\frac{\delta h}{\delta \mu_j}\bmu\right)+ 
  \mu_j\nabla\frac{\delta h}{\delta \mu_j} &= -D\nabla\frac{\delta h}{\delta D} \,,\\
  \partial_t D + \text{div}\left(D \frac{\delta h}{\delta \bmu} \right)&= 0\,,
\end{align}
and arise from the explicit form of the Lie-Poisson bracket 
\begin{align}
\begin{split}
\{f,g\}(\bmu,D) 
=  \int \bmu \cdot\bigg[ \left(\frac{\delta g}{\delta \boldsymbol{\mu}} \cdot \nabla \right)  \frac{\delta f}{\delta \bmu}
-  &\left(\frac{\delta f}{\delta \boldsymbol{\mu}} \cdot \nabla \right)  \frac{\delta g}{\delta \bmu} \bigg]\\
&+ D \left[  \left(\frac{\delta g}{\delta \boldsymbol{\mu}} \cdot \nabla \right) \frac{\delta f}{\delta D}
- \left(\frac{\delta f}{\delta \boldsymbol{\mu}} \cdot \nabla \right)  \frac{\delta g}{\delta D} 
\right]\,\text{d}^3x\,,
\label{LPB}
\end{split}
\end{align}
after using $\dot{f}(\bmu, D)=\{f,h\}(\bmu, D)$. This bracket is well-known as the {\it compressible fluid bracket} \cite{MaRa2013, Fusca2016} defined
on the dual of the semidirect product Lie algebra $\mathfrak{X}(\mathbb{R}^3)\,\circledS\, \mathcal{F}(\mathbb{R}^3)$, with dual coordinates $\boldsymbol{\mu}\in \mathfrak{X}^*(\mathbb{R}^3)=\Lambda^1(\mathbb{R}^3)\otimes {\rm Den}(\mathbb{R}^3)$ (1-form densities) and $D\in {\rm Den}(\mathbb{R}^3)$. For further discussions of {Lie-Poisson brackets,} see e.g. \cite{MaRa2013,HoScSt2009} and references therein.

We complete this discussion by recovering the QHD equations \eqref{QHD1} and \eqref{QHD2} explicitly. 
First, we compute the variational derivatives of the 
physical Hamiltonian \eqref{collective1}, obtaining 
\begin{align}
  \frac{\delta h}{\delta \bmu}=\frac{\bmu}{mD}\,,\qquad \frac{\delta h}{\delta 
  D}=-\frac{|\bmu|^2}{2mD^2}+ V + V_Q\,.
\end{align}
Then, upon recalling that $\bmu = D\nabla S$ and hence defining the fluid velocity $\bu:=m^{-1}\nabla S$, the 
general fluid momentum equations specialise to
\begin{align}
    \partial_t \bmu +(\text{div}\bu)\bmu+ 
  (\bu\cdot\nabla)\bmu &= -D\nabla(V+V_Q) \,,\\
  \partial_t D + \text{div}\left(D\bu \right)&= 0\,,
\end{align}
which indeed recover the QHD velocity equations in $(\bu, D)$ upon replacing $\bmu = mD\bu$ throughout.
 In fact, the relation $\delta h/\delta \bmu =\bu$ comes as no suprise as it has a geometric 
interpretation, defining the passage to the Euler-Poincar{\'e} formulation via 
the reduced Legendre transform, as we shall now see.

\subsection{Euler-Poincar{\'e} structure}
Having expressed the Lie-Poisson structure of the QHD equations, we now 
return to the Lagrangian setting. In particular, upon using reduced Legendre transform,
 \begin{align}
  \ell(\bu, D):= \int \bu 
  \cdot \bmu\,\de^3x - h(\bmu, D)\label{reducedLegendreQHD}\,,
 \end{align}
in which we define the velocity vector field $\bu:=\delta h/\delta \bmu$, the Euler-Poincar\'e Lagrangian $\ell:\mathfrak{X}(\mathbb{R}^3)\times {\rm Den}(\mathbb{R}^3)\to \mathbb{R}$ associated to the Hamiltonian \eqref{collective1} reads
\begin{align}
\ell(\bu,D)=\int\!\left[\frac{1}{2}mD|\bu|^2-\frac{\hbar^2}{8m}\frac{|\nabla D|^2}{D}-{D}V\right]\de^3x
\,.
\label{hydro-Lagr}
\end{align}  
In this new picture, following \cite{HolmEtAl1998}, the fluid velocity is the 
vector field defined by 
$\bu:=\dot{\eta}\circ\eta^{-1}\in\mathfrak{X}(\mathbb{R}^3)$, while the Eulerian density $D$ is defined 
by the {\it Lagrange-to-Euler map}
\begin{align}
 \label{defns}
 D(\bx,t)=\eta_*D_0:=\int \!D_0(\bx_0)\,\delta(\bx-{\boldeta}(\bx_0,t))\,\de^3x_0\in\text{Den}(\mathbb{R}^3)\,,
 \end{align}
for a reference density, $D_0=D_0(\bx)\,\de^3x$}. In the last definition, the symbol $\eta_*$ denotes the operation of {\em push-forward} by the map $\eta\in \text{Diff}(\mathbb{R}^3)$; so, $\eta_* D_0$ denotes the push-forward of the reference density, $D_0$ by the map $\eta$.
Push-forward by the smooth flow $\eta$ is called {\em advection} in hydrodynamics. In this context, the Lagrangian particle path of a fluid parcel is given
by the smooth, invertible, time-dependent map, $\eta_t:\mathbb{R}^3\to\mathbb{R}^3$, as follows,
$\bx(t)=\eta_{t}\boldsymbol{x}_0=\boldsymbol{\eta}(\boldsymbol{x}_0,t) \in\mathbb{R}^{3} $
known as the {\it back-to-labels map} {for initial reference position}  
$\eta_{0}\boldsymbol{x}_0=\boldsymbol{\eta}(\boldsymbol{x}_0,0)=\boldsymbol{x}_0$.
After this definition, there should be no confusion between $\eta_{t}\in {\rm Diff}(\mathbb{R}^3)$
and $\eta_{t}\boldsymbol{x}_0 = \boldsymbol{\eta}(\boldsymbol{x}_0,t) \in\mathbb{R}^{3}$. For brevity, the subscript $t$ will be omitted in most the subsequent material.
At this stage, we consider Hamilton's principle $\delta \int_{t_1}^{t_2} \ell(\bu,D)\,\text{d}t=0$ for an arbitrary reduced Lagrangian, with constrained variations from the Euler-Poincar\'e theory of ideal fluids with advected quantities, derived in \cite{HolmEtAl1998},
\begin{align}
\delta\bu=\delta(\dot\eta\circ\eta^{-1})=\partial_t\bw+(\bu\cdot\nabla)\bw-(\bw\cdot\nabla)\bu
\,,\qquad
\delta D=\delta (\eta_*D_0) =-\text{div}(D\bw)
\,.
\label{EPvar}
\end{align}
Here, the arbitrary vector field $\bw=\delta\eta\circ\eta^{-1}\in\mathfrak{X}(\mathbb{R}^3)$ vanishes at the endpoints in time.  
This results in the general equations of motion
\begin{align}
  (\partial_t + \pounds_{\bu})\frac{\delta \ell}{\delta \bu} &= D\nabla\frac{\delta \ell}{\delta 
  D}\,,\label{generalEPFluid1}\\
    (\partial_t + \pounds_{\bu})D &= 0\label{generalEPFluid2}\,,
\end{align}
where $\pounds_{\bu}$ denotes the Lie derivative with respect to the vector field $\bu=\dot{\eta}\circ\eta^{-1}\in\mathfrak{X}(\mathbb{R}^3)$. For example, the corresponding Lie derivative of the density {$D(\bx,t)\text{d}^3x$} is given by
\begin{align}
\pounds_\bu (D(\bx,t){\text{d}^3x}) = {\frac{\text{d}}{\text{d}s}}\bigg|_{s=0}\big(D(\boldeta(\bx,s),t)\,{\text{d}^3\eta(\bx,s)}\big)
=
{\rm div} \big( \bu D(\bx,t)\big){\text{d}^3 x}
\,,
\label{Lie-der}
\end{align}
for the Lagrangian path, $\boldeta(\bx,s)$ such that $\boldeta(\bx,0)=\bx$.
Hence, specialising to the Lagrangian \eqref{hydro-Lagr}, we obtain exactly the QHD 
equations 
\begin{align}
\partial_t D + \text{div}\left(D\bu\right)=0 \,, \qquad
(\partial_t+\bu\cdot\nabla)\bu= -\frac{1}{m}\nabla(V+V_Q)\,,
   \end{align}
in agreement with \eqref{QHD1} and \eqref{QHD2}.

\begin{remark}[Bohmian trajectories and Lagrangian paths]
  \hfill \\
  From the geometric formalism presented above, we see how, in QHD, the role of the Lagrangian path $\eta\in{\rm Diff}(\mathbb{R}^3)$ is of particular importance.
   Specifically, it plays the role of a {\it hidden variable} in the Bohmian interpretation of quantum dynamics \cite{Bohm1952}. Indeed, in the geometric framework the path $\eta$ is the fundamental dynamical variable,
    while the wavefunction is simply transported in time along the Lagrangian motion of $\boldeta(\bx_0,t)$, which in turn satisfies
\begin{align}
\dot{\boldeta}(\bx_0,t)=\bu(\boldeta(\bx_0,t),t)\,.
\label{EPvelocity}
\end{align}
However, in the {Bohmian mechanics} interpretation of quantum theory, the so-called \emph{Bohmian trajectory} defined by the relation \eqref{EPvelocity}
is in fact postulated as a physical particle trajectory $\bq(t)$ in addition to the quantum wavefunction.  
The geometric approach does no such thing, instead
given the Lagrangian path $\boldeta$,  the wavefunction $\psi$ is advanced in time simply according to its evolution as a half-density, that is
\begin{align}
\psi(\bx,t)=\Phi_{\eta(t)}(\psi(\bx_0,0))=\frac{\psi_0(\boldeta^{-1}(\bx,t))}{\sqrt{{\rm det}\nabla\boldeta(\bx,t)^T}}
\,.
\end{align}
Furthermore, it is important to emphasise that the (infinite-dimensional) Lagrangian paths \\
$\boldeta(\bx_0,t)$ are completely different from the point particle (Bohmian) trajectories $\bq(t)$ (finite-\\
dimensional), which can be only be consistently derived when the quantum potential is neglected, a point which we clarify below.
\end{remark}
Rewriting the Lagrangian \eqref{hydro-Lagr} in terms of $\boldeta$ by using \eqref{defns} and \eqref{EPvelocity}
takes it into the form $L:T{\rm Diff}(\mathbb{R}^3)\to \mathbb{R}$ given by
 \begin{align}
L(\boldeta,\dot\boldeta)=\int\!\left[\frac{mD_0}{2}{|\dot\boldeta|^2}-D_0(\bx_0)\Big(V_Q(\boldeta(\bx_0,t),t)+V(\boldeta(\bx_0,t))\Big)\right]\de^3x_0
\,,
\label{hydro-Lagr1}
\end{align}
where the quantum potential is  written in terms of $\boldeta$ as
\begin{align}
V_Q(\bx,t)=-\,\frac{\hbar^2}{2m}\sqrt{
\frac{
{\rm det}\nabla\boldeta(\bx,t)^T
}
{D_0(\boldeta^{-1}(\bx,t))}}
\,
\Delta\, \sqrt{\frac{D_0(\boldeta^{-1}(\bx,t))}{{\rm det}\nabla\boldeta(\bx,t)^T}}
\,.
\end{align}
The corresponding dynamics of the Lagrangian path $\eta$ is given by the Euler-Lagrange equation on the Lie group  
\cite{Wyatt2006} 
\begin{align}
mD_0\ddot\boldeta=-D_0\nabla_{\boldeta}( V_Q(\boldeta,t)+V(\boldeta))\,.\label{BohmianEL}
\end{align} 
We can now demonstrate how the dynamics of $\boldeta$  are not equivalent to point particle dynamics. In principle, the latter could be obtained by setting a point-like initial density of the type $D_0(\bx_0)=\delta(\bx_0-\bq_0)$ and then integrating the Euler-Lagrange equation over $D_0$.
However, as we have previously commented, this type of initial condition  is not allowed by the structure of the quantum potential. For this reason, asymptotic semiclassical methods are required to properly derive the effects of the quantum potential in a weak limit as $\hbar^2\to0$. For more details, see e.g. \cite{JinLi2003}.

Nonetheless, one can proceed as in Section \ref{Sec:1-QHD}, whereby the Newtonian limit neglects the order $O(\hbar^2)$ quantum dispersion term in the Lagrangian \eqref{hydro-Lagr} (or, equivalently, \eqref{hydro-Lagr1})
and varies the remainder. The resulting equation for $\boldeta$ becomes
$
D_0\big(m\ddot\boldeta+\nabla_{\boldeta}V(\boldeta)\big)=0
$.
It is clear that the point particle initial condition $D_0(\bx_0)=\delta(\bx_0-\bq_0)$ is now allowed and thus denoting $\bq(t)=\boldeta(\bq_0,t)$ and integrating over space again yields Newton's Law
$m\ddot\bq+\nabla V_{\bq}(\bq)=0$.

We conclude this section by presenting a visual summary of the geometric 
formulations of QHD that have been considered so far. 

\vspace{1cm}

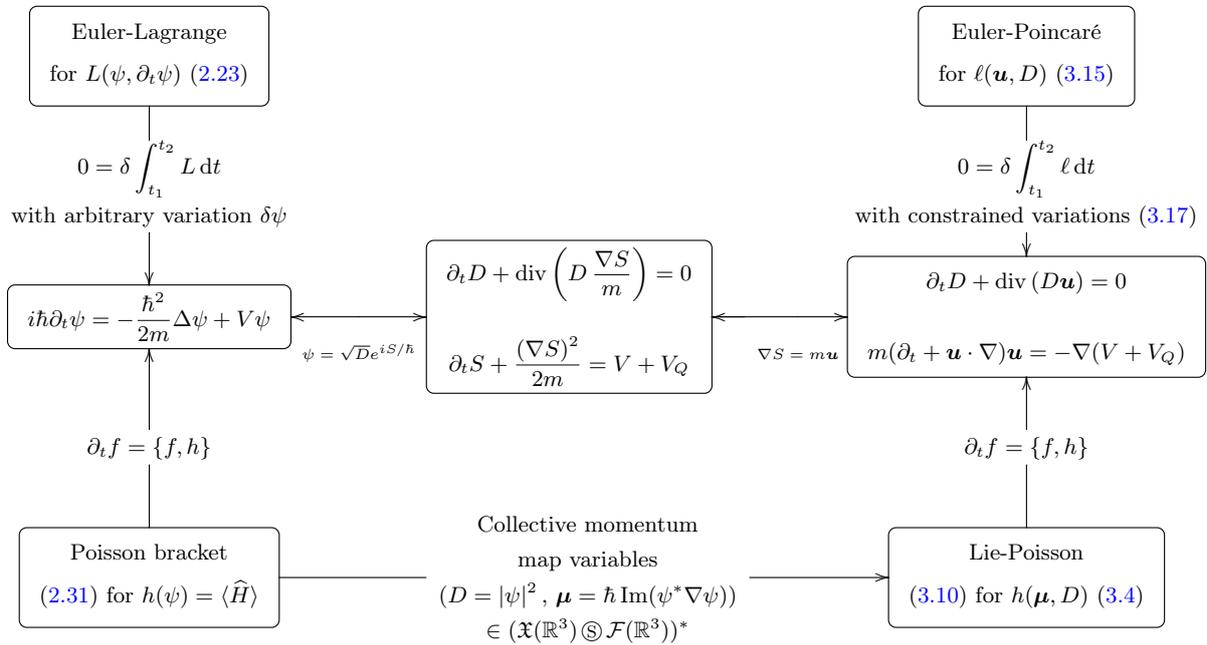
\begin{figure}[h]
\footnotesize\center
\noindent
\resizebox{1.24\textwidth}{!}{\begin{xy}
\hspace{0cm}
\xymatrix{
*+[F-:<3pt>]{
\begin{array}{c}
\vspace{0.1cm}\text{Euler-Lagrange}\\
\vspace{0.1cm} \text{for $L(\psi, \partial_t\psi)$ \eqref{DFStateLagrangian}}
\end{array}
} 
\ar[dd]|{\begin{array}{c}
\displaystyle 0=\delta\int_{t_1}^{t_2} L\,\text{d} t\\
\text{with arbitrary variation $\delta \psi$}
\end{array}}
&  & &&*+[F-:<3pt>]{
\begin{array}{c}
\vspace{0.1cm}\text{Euler-Poincar\'e}\\
\vspace{0.1cm} \text{for $\ell(\bu, D)$ \eqref{hydro-Lagr}}\\
\end{array}
} 
\ar[dd]|{\begin{array}{c}
\displaystyle 0=\delta\int_{t_1}^{t_2} \ell\,\text{d} t\\
\text{with constrained variations \eqref{EPvar}}
\end{array}}\\
&   &&&  \\
*+[F-:<3pt>]{
\begin{array}{c}
\displaystyle i\hbar\partial_t\psi = -\frac{\hbar^2}{2m}\Delta\psi + V\psi
\end{array}
}
\ar[rr]|{\begin{array}{c} \\
\\
\text{\tiny $\psi=\sqrt{D}e^{iS/\hbar}$}\\
\end{array}}
 &   &
*+[F-:<3pt>]{\begin{array}{c}
\displaystyle\partial_t D + \text{div}\left(D\,\frac{\nabla S}{m}\right) = 0\\
\\
\displaystyle\partial_t S +\frac{(\nabla S)^2}{2m}= V+V_Q
\end{array}
} 
\ar[ll]|{\begin{array}{c}
\end{array}}
\ar[rr]|{\begin{array}{c}
\\
\\
\text{\tiny $\nabla S=m\bu$}\\
\end{array}}
&&
*+[F-:<3pt>]{\begin{array}{c}
\partial_t D + \text{div}\left(D\bu\right)=0\\
\\
 m (\partial_t+\bu\cdot\nabla)\bu= -\nabla(V+V_Q)
\end{array}}
\ar[ll]|{\begin{array}{c}
\end{array}}
&  & &&\\
&  & &&\\
*+[F-:<3pt>]{\begin{array}{c}
\vspace{0.1cm}\text{Poisson bracket}\\
\vspace{0.1cm} \text{\eqref{SchrPB} for $h(\psi)=\braket{\widehat{H}}$}\\
\end{array}}
\ar[rrrr]|{\begin{array}{c}\text{Collective momentum}\\
\text{map variables}\\
(D=|\psi|^2\,, \,\bmu = \hbar\,\text{Im}(\psi^*\nabla\psi))\\
\in (\mathfrak{X}(\mathbb{R}^3)\,\circledS\, \mathcal{F}(\mathbb{R}^3))^*\\
\end{array}}
\ar[uu]|{\begin{array}{c}
\partial_t f =\{f, h\}\\
\end{array}}
&  & &&*+[F-:<3pt>]{\begin{array}{c}
\vspace{0.1cm}\text{Lie-Poisson}\\
\vspace{0.1cm}\text{\eqref{LPB} for $h(\bmu, D)$ \eqref{collective1}}
\end{array}}
\ar[uu]|{\begin{array}{c}
\partial_t f =\{f, h\}\\
\end{array}}
}
\end{xy}}
\vspace{.5cm}
\caption{Schematic description of geometric relationships between the Schr\"odinger and hydrodynamic pictures of quantum mechanics.}
\label{figure1}
\end{figure}

\section{Regularised QHD and Bohmion solutions}\label{Sec:2-Bohmions}

In remark \ref{ultra} we discussed some of the problems with the naive $\hbar^2 \to 0$ limit in the QHD Hamiltonian \eqref{collective1}, which is frequently used as the standard approach for allowing classical single particle trajectories. 
In this section we present an alternative solution by introducing a {regularisation method} in the variational principle, allowing for single-particle solutions without taking this limit. It is however the $O(\hbar^2)-$term in QHD which usually prevents
 the existence of particle-like solutions and hence this term must be treated with particular care. 
Specifically, we will introduce a {regularised} QHD Lagrangian which smoothens the fluid variables appearing in the $O(\hbar^2)$ quantum potential 
term so that the $\hbar^2\to0$ limit is no longer 
singular. 
Then, the choice of a singular solution for the fluid density results in a Newtonian equation in which particle dynamics are governed by the regularised 
form of the quantum potential. 

 We begin by defining the spatially smoothened counterpart of $D$,
 \begin{align}
\bar{D} := K*D = \int \!K(\bx,\bs)D (\bs)\,{\de}^3s\,,\label{Smooth}
 \end{align}
 where $K(\bx,\bs)$ is a positive definite, symmetric smoothing kernel which falls off at least exponentially in $|\bx-\bs|$.
 For example one could take the kernel $K(\bx,\bs)$ to be the Green's function of the Helmholtz operator ${(1-\alpha^2\Delta)}$, in which $\alpha$  is a length scale and the limit $\alpha\to 0$  returns the original  hydrodynamic variables,
or another choice could be to take $K(\bx,\bs)$ to be a Gaussian. Replacing this expression in the quantum potential term of \eqref{hydro-Lagr} results in
\begin{align}
\ell(\bu,D)=\int\!\left[\frac{1}{2}mD|\bu|^2-\frac{\hbar^2}{8m}\frac{| \nabla\bar{D}|^2}{\bar{D}}-{D}V\right]\de^3 x
\,.
\label{hydro-RLagr}
\end{align}
This allows us to consider the singular solution for the density
\begin{align}
D(\bx,t)=\sum_{a=1}^{\cal N} w_a\delta(\bx- \boldsymbol{q}_a{(t)})\,,\label{DSING}
\end{align}
(with $\sum_a w_a =1$), which, upon recalling the Lagrange-to-Euler map \eqref{defns}, amounts to 
selecting the initial condition of the same type $D_0(\bx_0)=\sum_{a=1}^{\cal N} w_a\delta(\bx_0- 
\boldsymbol{q}_a{(0)})$, where $\boldsymbol{q}_a(t)=\boldsymbol\eta(\boldsymbol{q}_a{(0)},t)$ and $\boldsymbol\eta({\bx},t)$ is the Lagrangian path such that $\dot{\boldsymbol\eta}(t)=\bu(\boldsymbol\eta(t),t)$. 
We will call these particle-like solutions `Bohmions'.

In turn, this implies that the smoothened variable \eqref{Smooth} is now expressed as
\begin{align}
     \bar{D}(\br,t)=\sum_{a=1}^{\mathcal{N}}w_a K(\br-\bq_a(t))\,,
\end{align} 
so that the {regularised} QHD Lagrangian \eqref{hydro-RLagr} is written as
\begin{align}
L(\{\boldsymbol{q}_a\},\{\dot{\boldsymbol{q}}_a\}) 
=\sum_{a} w_a\bigg(
\frac{m}2|\dot{\boldsymbol{q}}_a|^2 
{
- \frac{\hbar^2}{8m}\int \frac{\sum_{b} w_b \nabla K(\boldsymbol{y}-\boldsymbol{q}_a)\cdot\nabla K(\boldsymbol{y}-\boldsymbol{q}_b)}{\sum_{c} w_c K(\boldsymbol{y}-\boldsymbol{q}_c)} \,\de^3y}
- V(\boldsymbol{q}_a)
\bigg)\,.\label{hydro-RLagrAGAIN}
\end{align}
Then, applying Hamilton's principle for arbitrary variations $\delta\bq_a$ produces the Euler-Lagrange equation
\begin{align}
m\ddot{\boldsymbol{q}}_a=-\nabla_{\bq_a} \left(V(\boldsymbol{q}_a)
-\frac{\hbar^2}{8m} {
\int \frac{\sum_{b} w_b \nabla K(\boldsymbol{y}-\boldsymbol{q}_a)\cdot\nabla K(\boldsymbol{y}-\boldsymbol{q}_b)}{\sum_{c} w_c K(\boldsymbol{y}-\boldsymbol{q}_c)} \,\de^3y }
\right) \,.
\end{align}
Hence, this regularisation procedure results in a form of Newtonian dynamics, in 
which the particle trajectory still feels a force due to the effect of the 
quantum potential, now suitably modified to allow for singular solutions of the type \eqref{DSING}. 
In addition, due to the presence of $\bar{D}$ in the denominator, this approach provides extensive, potentially long-range coupling among the singular particle-like solutions.


\section{Density operators and classical closures}\label{Sec:2-ColdFluid}

So far, the discussion has focused uniquely on the geometric picture of QHD for wavefunctions identifying pure quantum states. 
In this section, we demonstrate how we can generalise the momentum map \eqref{QHDmomap} 
approach for mixed states by working with density operators as described in Section \ref{Sec:1-Postulates}.

To do so, we consider the following Hamiltonian $\text{Diff}(\mathbb{R}^3)$-action 
on ${\rm Her}(\mathscr{H})$ (the space of Hermitian operators on $\mathscr{H}$)
\begin{equation}
\rho(\bx,\bx')\mapsto\frac{\rho(\boldeta^{-1}(\bx),\boldeta^{-1}(\bx'))}{\,\sqrt{\text{det}(\nabla_{\bx}{\boldeta(\bx)}^T\nabla_{\bx'}{\boldeta(\bx')}^T)}\,}
\,,
\label{DiffeoDensityMatrixAction}
\end{equation}
which naturally extends the corresponding action on wavefunctions 
\eqref{half-dens-action}. Recall from Section \ref{Sec:1-Postulates} that $\rho(\bx,\bx')$ 
is the kernel of the density operator as given in Definition \ref{Def:DensityOp}.
Then, the verification detailed in the Appendix \ref{App:HalfDen2}  shows that its infinitesimal generator may be written as
\begin{equation}\label{DiffeoDensityMatrixInfGen}
u(\rho\,)=-\frac{i}{2\hbar}\left[\{\widehat{u}^k,\widehat{P}_k\}_{+\,},\rho\right]\,.
\end{equation}
 From here, using the standard momentum map formula for a group action on a Poisson manifold \eqref{MOMAPGENPOISSON}, one can prove that the corresponding momentum map is $J:{\rm Her}(\mathscr{H})\to \mathfrak{X}^*(\mathbb{R}^3)$, given in matrix element notation as
  \begin{align}
\bJ(\rho\,)=\frac12\{\widehat{P},\rho\,\}_+(\bx,\bx)
=:\boldsymbol{\mu}(\bx)
\,.\label{DiffeoDensityMatrixMomap}
\end{align}
Again see Appendix \ref{App:HalfDen2} for a discussion and full proof. 
In the special case of pure states, one simply verifies that $\rho=\psi\psi^{\dagger}$ recovers the momentum map \eqref{QHDmomap}. However, unlike pure states, in the general case of mixed quantum states, the dynamics of ${\boldsymbol{J}}(\rho)$ cannot be expressed only in terms of $\bmu(\bx)$ and $D(\bx):=\rho(\bx,\bx)$ \cite{Wyatt2006, BuMoHu2007}. 
{Rather, mixed states lead to a multi-fluid system that is obtained by combining the arguments in Section \ref{Sec:2-HD} with the expression in terms of a mixture of pure states \eqref{rhomixture}}.

\begin{remark}[Non-zero vorticity for mixed states]
\hfill \\
Unlike pure states, the fluid velocity corresponding to \eqref{DiffeoDensityMatrixMomap}, $\bu:= m^{-1}\bmu/D$, is no longer an exact 
 differential, instead written as 
 \begin{align}
   \bu(\bx) = \frac{\sum_a w_a \bmu_a(\bx)}{m \sum_a w_a D_a(\bx)}\,,\label{FluidVelocityMixture}
 \end{align}
which follows upon using the expression of $\rho$ as a mixture of pure states \eqref{rhomixture}. Hence, $\bmu_a:= \hbar{\rm Im}(\psi_a^*\nabla\psi_a)$ and $D_a:=|\psi_a|^2$ are the momentum map variables for each pure state 
 $\psi_a$. Then, via Kelvin's theorem, the corresponding hydrodynamic flow preserves a nontrivial circulation $\oint_{c(t)} \bu(\bx)\cdot\de \bx$, for an arbitrary loop $c(t)$ moving with velocity $\bu$ which corresponds to the vorticity dynamics $\partial_t\bomega={\rm curl}(\bu\times\bomega)$, with   $\bomega:=\nabla\times\bu$.
 We will not consider {this explicit form of such fluid velocities in this thesis, instead concluding this chapter by invoking a closure for the density 
 matrix.}
 \end{remark}

In the final part of this chapter, we show that the classical limit of mixed state dynamics (as given by the Liouville-von Neumann equation \eqref{LvN}) can be obtained from an exact closure by allowing for the operator $\rho$ to be sign-indefinite. This should come as no surprise, since classical states violate the uncertainty principle in such a way that the density operator can no longer be positive-definite. The proposed classical closure for the Liouville-von Neumann equation \eqref{LvN} is expressed as
\begin{align}
\rho(\bx,\bx')&=D\Big(\frac{\bx+\bx'}{2}\Big)\exp\!\left[i\frac{m}\hbar 
(\bx-\bx')\cdot\bu\Big(\frac{\bx+\bx'}{2}\Big)\right],
\label{coldfluid}
\end{align}
which upon substitution into the momentum map \eqref{DiffeoDensityMatrixMomap} returns the desired results $\rho(\bx,\bx)=D(\bx)$ and $\boldsymbol{\mu}(\bx)=mD(\bx)\bu(\bx)$, as one can show by a direct verification (see Appendix \ref{App:ColdFluid}). 
\begin{remark}[Cold fluid closure]
  \hfill \\ 
  The expression of the density matrix \eqref{coldfluid} is known as the {\em cold fluid closure}  
  and is more simply expressed in terms of the Wigner-Moyal formulation of quantum mechanics on 
phase space, as described in Appendix \ref{App:Wigner}. Without entering further discussion, we simply comment that in the Wigner-Moyal formalism a quantum state is 
given by the Wigner function
\begin{align}
  W(\bx,\bp)=\frac{1}{(2\pi\hbar)^{3}}\int\rho\left(\bx+\frac{{\by}}{2},\bx-\frac{{\by}}{2}\right) 
  \,e^{-i\hbar^{-1}\bp\cdot\by}\,\de^3y\,,
\end{align}
associated to the density matrix $\rho$. Under this (Wigner) transform, the classical closure \eqref{coldfluid} of the operator $\rho$ returns the phase space Wigner function
\begin{align}
  W(\bx,\bp)= D(\bx)\delta({\bp - m\bu(\bx)})
  \,,
  \label{CF-Wigner}
\end{align}
in which the phase space momentum variable is restricted to be the hydrodynamic momentum field $m\bu(\bx)$.
It is from this expression in terms of the Wigner function that \eqref{coldfluid} derives its name, as the analogous
 expression for the phase space probability density in kinetic theory is referred to as the cold plasma/fluid closure.  
Similar considerations of \eqref{CF-Wigner} have already appeared \cite{JinLi2003} in the context of the semiclassical limit for pure state dynamics as well as in \cite{BuMoHu2007} and references therein in the context of hybrid quantum-classical dynamics.
We emphasise again that the Wigner function in \eqref{CF-Wigner} does not identify a quantum state. This is analogous to what happens for the quantum harmonic oscillator: in this case, the Wigner-Moyal equation coincides with the classical Liouville equation thereby allowing for delta-function solutions returning classical motion. However, delta-function Wigner distributions do not correspond to quantum states, as described in Appendix \ref{App:Wigner}.
\end{remark}

Despite agreeing with the standard expressions for the fluid variables, the cold fluid closure \eqref{coldfluid} returns an alternate form of the total energy $\langle\rho,\widehat{H}\rangle$ 
given by
\begin{align}
\langle\rho,\widehat{H}\rangle=
\int \left[\frac{|\boldsymbol{\mu}|^2}{2mD} + 
  DV\right]\,\de^3 x\,,\label{COLDFLUIDHAM}
\end{align}
coinciding with the QHD Hamiltonian \eqref{collective1} after dropping the $\hbar^2-$term. 
Again see Appendix \ref{App:ColdFluid} for the full derivation.
Thus, the corresponding equations of motion naturally coincide with the classical hydrodynamic limit \eqref{QHDClassicalLimit}  in terms of Newton's Law.
 
In Section \ref{Bohmions} we will make use of the cold-fluid closure in the construction of a  
new density matrix approach to nonadiabatic dynamics, allowing for classical 
nuclear dynamics coupled to quantum electronic dynamics.

\chapter{Geometry of wavefunction factorisations in quantum chemistry}\label{Chap:EFOLD}
Having already introduced the topic of nonadiabatic quantum chemistry in Section \ref{NonadiabaticSection}, 
this chapter follows further results from the author's publication \cite{FoHoTr2019}, focusing on the application of the geometric approach to QHD from Chapter \ref{Chap:QHD}, with the goal of
furthering the understanding of nonadiabatic techniques currently available in 
the quantum chemistry literature.

Section \ref{meanfield-sec} again considers the mean-field factorisation of the molecular 
wavefunction, now rederiving the equations of motion from the Euler-Poincar{\'e} 
formulation of QHD as a simple first application of the momentum map method. 

Following this, Section \ref{Sec:EF} is devoted to the exact factorization of the 
molecular wavefunction. Specifically, exact factorization dynamics are rederived from a variational standpoint, here considering an arbitrary total energy functional. The resulting Euler-Lagrange equations demonstrate more clearly the general 
features of this factorisation and highlight important interpretative aspects of 
the theory. We then proceed to introduce the geometric approach for nuclear QHD analogous to the mean-field case, 
in which the geometric approach helps shed new light on the interesting features of the QHD approach to EF.

\section{Geometric approach to mean-field 
QHD}\label{meanfield-sec}

This section presents the Euler-Poincar{\'e} hydrodynamic approach to the mean-field model \eqref{GeneralMeanField}.
 Although this model fails to retain correlation effects between nuclei and electrons, it is of paramount importance as one of the most common models in nonadiabatic dynamics \cite{MarxHutter2009,deCarvalhoEtAl2014,Tully1998}.
  As we shall see, the geometry of quantum hydrodynamics can be directly applied to this model, thereby leading to the most basic example of hybrid {quantum-classical} dynamics, replicating the result in Section \ref{NonadiabaticSection}, now obtained via a geometric formulation.  
  In addition, the techniques developed in this section will serve as a 
  stepping stone to similar approaches for the exact factorization of the molecular wavefunction, considered in 
  depth in Section \ref{Sec:EF}.

To begin, recall the mean-field ansatz \eqref{GeneralMeanField}
\begin{align}
\Psi(\boldsymbol{r},\boldsymbol{x},t)=\Omega(\boldsymbol{r},t)\psi(\boldsymbol{x},t)
\,,
\label{MFfact}
\end{align}
where as before both $\psi$ and $\Omega$ are {normalised} with respect to the coordinate upon which they depend. 
At this point we expand the molecular Hamiltonian operator \eqref{molecularHamiltonian} as 
\begin{align}
\widehat{H}=-\frac{\hbar^2}{2M}\Delta+\widehat{T}_e+V_I(\br,\bx)
\,,
\label{mario}
\end{align}
so that upon replacing \eqref{MFfact} in the total energy 
$h(\Psi)=\text{Re}\iint\Psi^*\widehat{H}\Psi\,\de^3r\,\de^3x$, as well as employing Dirac's notation for the electronic state so that $\ket{\psi(t)}:=\psi(\bx,t)$, one obtains
\begin{align}
h(\Omega, \psi)= \braket{\psi|\widehat{T}_e\psi} + \int \frac{\hbar^2}{2M}|\nabla\Omega|^2 + |\Omega|^2\braket{\psi|V_I(\br,\bx)\psi}\,\de^3r\,.\label{MFHam}
\end{align}
Then, upon recalling the QHD momentum map \eqref{QHDmomap} and applying it to 
the nuclear wavefunction, therefore denoting $D=|\Omega|^2$ and $\bmu=\bJ(\Omega)=\hbar\text{Im}(\Omega^*\nabla\Omega)$, the mean-field Hamiltonian functional \eqref{MFHam} becomes
\begin{align}
\begin{split}
h(\bmu,D,\psi)=
\langle\psi|\,\widehat{T}_e\psi\rangle+
\int\!\left[\frac{|\bmu|^2}{2MD}+\frac{\hbar^2}{8M}\frac{|\nabla D|^2}{D}+D\langle\psi|V_I(\br,\bx)\psi\rangle\right]\de^3r \,.
\label{MFHamiltonian}
\end{split}
\end{align}
This Hamiltonian functional is a mapping $h:\mathscr{H}_e\times(\mathfrak{X}^*(\Bbb{R}^3)\times \text{Den}(\Bbb{R}^3))\rightarrow \mathbb{R}$, where $\mathfrak{X}^*(\Bbb{R}^3)$ is understood to be the space of {1-form }densities on $\Bbb{R}^3$.
 At this point, as the nuclear and electronic factors are only coupled through the potentials, we may perform the reduced Legendre transform 
 in the nuclear sector according to the procedure outlined in Section \ref{Sec:2-HD}
$
\bu={\delta h}/{\delta \bmu}=M^{-1}{\bmu}/{D}
$,
 whilst using the Dirac-Frenkel type construction for the electronic factor. Hence, we write a mixed {hydrodynamic/phase space} Lagrangian 
 of the form $\ell(\psi,\dot{\psi},\bu,D)$ defined as the mapping $\ell: T\mathscr{H}_e \times(\mathfrak{X}(\Bbb{R}^3)\times\text{Den}(\Bbb{R}^3))\rightarrow 
 \mathbb{R}$.
 The resulting variational principle and equations are presented in the 
 following theorem:
 
 \begin{theorem}[Variational approach to mean-field nuclear QHD]\label{Theorem:MFQHD}\hfill\\
Consider Hamilton's principle $0=\delta\int_{t_1}^{t_2} \ell\,\de t$ applied to the 
Lagrangian
\begin{align}
\ell(\psi,\dot{\psi},\bu,D)=\int\!\left[\frac{1}{2}MD|\bu|^2-\frac{\hbar^2}{8M}\frac{|\nabla D|^2}{D}-D\langle\psi|V_I(\br,\bx)\psi\rangle\right]\de^3r
+\langle\psi,i\hbar\dot\psi-\,\widehat{T}_e\psi\rangle
\,,
\label{MFLagrangian}
\end{align}
with arbitrary variation $\delta \psi$ and constrained Euler-Poincar{\'e} variations 
\eqref{EPvar}
\begin{align*}
\delta D = -{\rm div}(D\bw)\,,\qquad \delta\bu &= 
  \partial_t\bw + (\bu\cdot\nabla)\bw-(\bw\cdot\nabla)\bu\,,
  \end{align*}
   where $\bw := \delta \eta\circ \eta^{-1}$ is arbitrary and 
   vanishing at the endpoints. 
   Together with the Lagrange-to-Euler map \eqref{defns} $D=\eta_*D_0$, 
   this variational principle is equivalent to the equations of motion
   \begin{align}
&\partial_t D + {\rm div}(D\bu)=0\,,\\
&M(\partial_t+\bu\cdot\nabla)\bu=-\nabla V_Q-\langle\psi|\nabla V_I(\br,\bx)\psi\rangle
\,,\label{MF1}
\\
&i\hbar\dot\psi=\left(\,\widehat{T}_e+\int \!DV_I(\br,\bx)\,\de^3r\right)\psi
\label{MF3}
\,.
\end{align}
 \end{theorem}
 \paragraph{Proof:}
 Consider a general Lagrangian of the type $\ell(\psi,\dot{\psi},\bu,D)$. Upon expanding Hamilton's principle and substituting in the variations \eqref{EPvar}
 we obtain
 \begin{align*}
  0 &= \int_{t_1}^{t_2} \Bigg(\left\langle \frac{\delta \ell}{\delta \psi},\delta\psi\right\rangle +\left\langle \frac{\delta \ell}{\delta \dot\psi},\delta\dot\psi\right\rangle + \int \left(\frac{\delta \ell}{\delta \bu}\cdot \Big( \partial_t\bw + \pounds_{\bu}\bw\Big) + \frac{\delta \ell}{\delta D}\Big(-{\rm div}(D\bw)\Big) \right)\,\de^3 r \Bigg) \,\de t\\
  &= \int_{t_1}^{t_2} \Bigg(\left\langle \frac{\delta \ell}{\delta \psi} - \partial_t\left(\frac{\delta \ell}{\delta \dot\psi}\right),\delta\psi\right\rangle + \int \left(-\partial_t\frac{\delta \ell}{\delta \bu} - \pounds_{\bu}\frac{\delta \ell}{\delta \bu}  + D\nabla\frac{\delta \ell}{\delta D}\right)\cdot\bw \,\de^3 r \Bigg) \,\de t  \,.
 \end{align*}
 Recalling $\delta \psi$ and $\bw$ are arbitrary we read off the general equations of motion
\begin{align}
(\partial_t +\pounds_{\bu})\frac{\delta \ell}{\delta \bu}&=D\nabla\frac{\delta \ell}{\delta D}\label{generalMF1}
\,,\\
\frac{\delta \ell}{\delta 
{\psi}}-\partial_t\!\left(\frac{\delta \ell}{\delta 
\dot{\psi}}\right) &= 0\label{generalMF3}
\,,
\end{align}
also recalling the auxilliary advection equation $(\partial_t +\pounds_{\bu})D=0$, which follows immediately from the Lagrange-to-Euler map \eqref{defns}.
Clearly we recognise these as the standard Euler-Poincar{\'e} equations for continua  \cite{HolmEtAl1998} in 
the nuclear variables, along with an Euler-Lagrange equation for the electronic 
state. Then, we compute the variational derivatives associated to the Lagrangian 
\eqref{MFLagrangian}
\begin{align*}
  \frac{\delta \ell}{\delta \psi} &= i\hbar\dot{\psi} - 2\,\widehat{T}_e\psi - 2\int 
  DV_I(\br,\bx)\,\de^3r\,\psi\,,\qquad   \frac{\delta \ell}{\delta \dot\psi} = 
  -i\hbar\psi\\
  \frac{\delta \ell}{\delta \bu} &= MD\bu\,,\qquad \frac{\delta 
  \ell}{\delta D} = \frac{1}{2}M|\bu|^2 - V_Q - 
  \braket{\psi|V_I(\br,\bx)\psi}\,,
\end{align*}
which, upon direct substitution and a little rearrangement, return the desired 
equations \eqref{MF1} and \eqref{MF3}.
 \hfill$\square$\\
 
If the term corresponding to the quantum potential were then simply discarded, following the usual practice in taking the classical restriction of the nuclear dynamics by neglecting $\hbar^2$,
 the reduced mean-field Lagrangian \eqref{MFLagrangian} would become,
\begin{align}\label{joe}
\ell(\psi,\dot{\psi},\bu,D)=\int\!\left[\frac{1}{2}MD|\bu|^2-D\langle\psi|V_I(\br,\bx)\psi\rangle\right]\de^3r
+\langle\psi,i\hbar\dot\psi-\,\widehat{T}_e\psi\rangle
\,.
\end{align}
An analogous result could be obtained by following an alternative procedure which would exploit the cold fluid closure in the density operator formalism for the nuclear dynamics, as indicated in Section \ref{Sec:2-ColdFluid}.
 In this case, one would obtain the same Lagrangian \eqref{joe}, although with non-zero vorticity $\bomega:=\nabla\times\bu\neq0$. 
The resulting fluid equation \eqref{MF1} would lose the quantum force term $\nabla V_Q$ and,
 as explained in  Section \ref{Sec:2-HD},  one can set $D(\br,t)=\delta(\br-\bq(t))$. Then, integrating \eqref{MF1} over space yields classical trajectories 
 and eventually, the corresponding mixed quantum-classical system reads
\begin{align}
M\ddot{\bq}=- \nabla_{\bq}\langle\psi|V_I(\bq,\boldsymbol{x})\psi\rangle
\,,\qquad
i\hbar\dot\psi=\left(\,\widehat{T}_e+V_I(\bq,\bx)\right)\psi
\,,\label{EhrenfestSystem}
\end{align}
in agreement with the (Ehrenfest) mean-field equations presented in Section 
\ref{NonadiabaticSection}.

\begin{remark}[Variational structure of Ehrenfest mean-field model]\hfill\\
  Interestingly, the classical restriction \eqref{EhrenfestSystem} preserves the variational structure, derived from the Lagrangian $L:T\mathbb{R}^3\times T\mathscr{H}_e\to\mathbb{R}$ given by
\begin{align}
  L(\bq,\dot{\bq},\psi,\dot{\psi})= \frac{1}{2} {M}|\dot{\bq}|^2 + 
  \braket{\psi, i\hbar\dot{\psi} - (\,\widehat{T}_e 
  +V_I(\bq,\boldsymbol{x}))\psi}\,,
\end{align}
which has been studied from the geometric perspective of the Heisenberg group in \cite{BonetLuzTronci2015}. 
\end{remark}

 As previously explained, the mean-field model is not satisfactory in many cases, because the {factorisation} \eqref{MFfact} disregards the correlation between nuclei and electrons. 
On the other hand, the exact factorization model \cite{AbediEtAl2012} fully captures the electron-nuclear correlation though the explicit dependence of the electronic factor on the nuclear coordinate.
The remainder of this chapter is devoted to using geometric techniques to better 
understand the EF model as currently employed in the quantum chemistry field.  

\section{Exact factorization of the molecular wavefunction}\label{Sec:EF}

This section begins by rederiving the EF dynamical equations \eqref{EFNuclearEqnSEC1} and \eqref{EFElectronicEqnSEC1}, this time from the DF variational principle. This procedure is first carried out in the general setting of an arbitrary total energy, 
thus producing a new compatibility condition, before specialising to the physical case and throughout commenting on some of the 
finer details of this model. We then proceed to introduce the hydrodynamic interpretation of the nuclear dynamics using the Euler-Poincar{\'e} approach from Section 
\ref{Sec:2-HD}, ultimately resulting in a nuclear fluid equation with a Lorentz force given by the time-dependent Berry curvature. We then consider the classical Newtonian limit as well as corresponding circulation theorem, which itself provides new 
insight into the geometry of the associated Berry phase and its evolution. 

\subsection{General equations of motion}\label{Sec:EF-Sub:1}
In this section, we consider the equations of motion for the exact factorization 
of the molecular wavefunction as outlined in Section \ref{NonadiabaticSection}, first 
 in the general case of an arbitrary Hamiltonian functional 
$h(\Omega, \psi)$, i.e. using the treatment given in Section 
\ref{Sec:1-GeomQM}, before specialising to the physical case of the molecular Hamiltonian.

We commence by recalling the expression for the exact factorization of the molecular 
wavefunction
\begin{align}
  \Psi(t)=\Omega(\br,t)\ket{\psi(\br,t)}\,,\label{GenExFact}
\end{align}
in which we again make use Dirac's bra-ket notation for the electronic state so that the natural $L^2$ inner product on $\mathscr{H}_e$ is denoted by
 $
\langle\psi_1|\psi_2\rangle(\br)
=
\psi_1^\dagger\psi_2(\br)
=
\int \!\psi_1^*(\boldsymbol{x};\br)\psi_2(\boldsymbol{x};\br)\,\de^3x
$. In fact, this notation allows all of the results up to Section \ref{Sec:2Level} to be valid for an arbitrary electronic Hilbert space $\mathscr{H}_e$.
Then, in line with the quantum chemistry literature we refer to $\psi$ as the electronic factor and $\Omega$ as the nuclear factor. 
In the factorisation \eqref{GenExFact}, the electronic degree of freedom $\psi$ depends parametrically on the nuclear coordinate $\br$ so that geometrically $\psi$ is understood
as a smooth map $\psi\in \mathcal{F}(\mathbb{R}^3,\mathscr{H}_e)$ from physical space into the Hilbert space $\mathscr{H}_e$. 
As we have seen, the {factorisation} \eqref{GenExFact} also invokes the {\it partial normalisation condition} (PNC)  
$
\|\psi(\br,t)\|^2=\braket{\psi|\psi}=1\,,
$
which as a result of \eqref{GenExFact} ensures that 
$
1={\int \|\Psi(\boldsymbol{r},t)\|^2\de^3r
= \int|\Omega(\br,t)|^2}\de^3r\,,
$
so that $D(\br,t):=|\Omega(\br,t)|^2$ may be interpreted as the nuclear probability density. 

\begin{remark}[Interpretation as nuclear and electronic wavefunctions]\hfill\\
Consider the density matrix associated to \eqref{GenExFact} given by
\begin{align}
  \widehat{\rho}(\br,\br')=\Omega(\br)\Omega^*(\br')\ket{\psi(\br)}\!\!\bra{\psi(\br')}\,,\label{StraightEFMolDensity}
\end{align} 
where we have dropped the time dependence for brevity. Then the corresponding densities on each subspace are obtained by projections as follows
\begin{align}
  &\widehat{\rho}_e= \iint   
  \widehat{\rho}(\br,\br')\delta(\br-\br')\,\de^3r\,\de^3r' = \int 
  |\Omega(\br)|^2 \ket{\psi(\br)}\bra{\psi(\br)}\,\de^3r\,,\label{TrueElectronDensity}\\
 & \rho_n(\br,\br')={\rm Tr}\Big( \widehat{\rho}(\br,\br')\Big)= 
  \braket{\psi(\br')|\psi(\br)}\Omega(\br)\Omega(\br')\label{TrueNuclearDensity}\,,
\end{align}
in which we notice that the PNC does not apply. Hence, in the context of quantum chemistry \cite{AbediEtAl2010,AbediEtAl2012}, this means
the quantities $\Omega$ and $\psi$ are \emph{not true wavefunctions} for the nuclei and electrons (which may not even exist in the presence of {\it decoherence}, i.e. quantum mixing).
 However, we shall continue to refer to them as such, because they retain certain mnemonic relationships. 
\label{A-observ}
At this stage, we shall only emphasise that all the relations above also apply naturally  in the context of the Born-Oppenheimer approximation \cite{IzFr2016}, thereby indicating again that the interpretation of nuclear and electronic motion in terms of genuine wavefunctions needs to be revisited. For example, backreaction effects generated by the presence of $\psi$ in \eqref{TrueNuclearDensity} can lead to \textit{nuclear decoherence effects} since indeed one has $\rho_n^2\neq\rho_n$. This is a general feature of {quantum-classical} coupling \cite{BoGaTr2019}, which in fact erodes purity in both the classical and the quantum subsystems.
\end{remark}

We now utilise the variational approach to derive the dynamical equations and therefore insert \eqref{GenExFact} into the DF Lagrangian 
\eqref{DFStateLagrangian}. In addition, we allow for an arbitrary total energy functional $h(\Omega, \psi)$ and enforce the PNC via a Lagrange multiplier.
 The result is expressed in the following theorem.

\begin{theorem}[Variational approach to exact factorization dynamics]\label{Theorem:EFEL}
  \hfill\\
  The DF Lagrangian associated to the EF ansatz \eqref{GenExFact} has the form $L(\Omega,\partial_t\Omega,\psi,\partial_t\psi, \lambda)$ understood as the mapping $L:T\mathscr{H}_n\times T\mathcal{F}(\mathbb{R}^3,\mathscr{H}_e)\times\mathcal{F}(\mathbb{R}^3)\to 
\mathbb{R}$ and is given by
\begin{align}
L
= {\rm Re}\int\! \Big[i\hbar\|\psi\|^2\Omega^*\partial_t{\Omega}+|\Omega|^2\langle\psi|i\hbar\partial_t{\psi}\rangle+ \lambda\big(\|\psi\|^2 - 1\big) \Big]
\,\de^3r
-h(\Omega,\psi) 
\,.\label{EFDF1}
\end{align}
This gives rise to the corresponding Euler-Lagrange equations 
\begin{align}
    i\hbar\partial_t{\Omega} + \braket{\psi|i\hbar\partial_t{\psi}}\Omega - \frac{1}{2}\frac{\delta h}{\delta \Omega} 
  &=0 \label{EFOmegaequation}\,,\\
  (\mathbbm{1}-\psi\psi^{\dagger})\left(i\hbar\partial_t{\psi} - \frac{1}{2|\Omega|^2}\frac{\delta h}{\delta\psi}\right) &= 
    0\,,
    \label{EFELpsiprojective}\\
  \|\psi\|^2 &= 1\,,\label{PNC}
\end{align}
along with the compatibility condition on the Hamiltonian functional
\begin{align}
{\rm Im}\left\langle \psi \bigg| \frac{\delta h}{\delta \psi} \right\rangle
= {\rm Im}\left(\Omega^*\frac{\delta h}{\delta \Omega} 
\right)\label{EFELcondition}\,.
\end{align}
\end{theorem}
\paragraph{Proof:}
Before commencing we note that this proof fixes minor oversights found in \cite{FoHoTr2019}.
Firstly, the Lagrange multiplier equation enforces the PNC \eqref{PNC} via the Euler-Lagrange equation for $\lambda$. 
Then, directly computing the $\Omega $ Euler-Lagrange equation and using the PNC immediately yields the 
desired nuclear equation \eqref{EFOmegaequation}. Consequently, upon using the nuclear equation to 
derive
\begin{align}
\hbar\partial_t|\Omega|^2={\rm Im}\left(\Omega^*\frac{\delta h}{\delta\Omega}\right)\,,
\end{align}
we can compute the $\psi$ Euler-Lagrange equation to obtain
  \begin{align}
 ( \mathbbm{1}-\psi\psi^{\dagger}) \Big(i\hbar|\Omega|^2\partial_t{\psi}\Big) + \frac{1}{2}\bigg(\Omega^*\frac{\delta h}{\delta \Omega}\bigg) \psi= &\ 
    \frac{1}{2}\frac{\delta h}{\delta \psi} 
  -\lambda\psi\,,\label{lambdapsiequation}
 \end{align}
 expressed in terms of the notation $\psi^{\dagger} \,\cdot := \braket{\psi| 
  \,\cdot\,}$. Then taking the real part of the inner product of this equation with $\psi$, one finds 
 that the Lagrange multiplier is written as
\begin{align}
   \lambda =\frac{1}{2}\text{Re}\left(\left\langle\psi\Bigg| \frac{\delta h}{\delta \psi} \right\rangle- \Omega^*\frac{\delta h}{\delta \Omega}\right)\,,\label{LambdaLagrange}
\end{align}
 whilst taking the imaginary part of the same object yields the compatibility condition
\eqref{EFELcondition}. Upon substituting \eqref{LambdaLagrange} back into the 
the $\psi$ equation \eqref{lambdapsiequation}, one arrives at the electronic equation 
\eqref{EFELpsiprojective}. 
  \hfill$\square$\\
  
Before continuing, we comment on the difference between equation \eqref{EFELcondition} and the usual compatibility condition for the standard Schr{\"o}dinger equation \eqref{CompCond1} from Section \ref{Sec:1-GeomQM}.
In particular, the non-zero right-hand side of \eqref{EFELcondition} arises because the inner product (on the left) is taken only over the electronic degrees of freedom. Hence, equation \eqref{EFELcondition} depends on the nuclear coordinate $\br$
whilst the original expression \eqref{CompCond1} describes a real number.

\begin{remark}[Local phases and gauge freedom]\label{EFlocalphases}
  \hfill \\ 
  One may observe that the exact {factorization} \eqref{GenExFact} is defined only up to compensating local phase shifts of the nuclear and electronic wavefunctions. Namely, the replacements
\begin{align}
\begin{split}
\ket{\psi(\br,t)}\mapsto&\  \ket{\psi'(\br,t)}=e^{-i\hbar^{-1}\theta(\br,t)}\ket{\psi(\br,t)} 
\\
\Omega(\br,t)\mapsto&\  \Omega'(\br,t)=e^{i\hbar^{-1}\theta(\br,t)}\Omega(\br,t) 
\label{EF-gaugefreedom}
\end{split}
\end{align}
leave the EF product wavefunction $\Psi(t)=\Omega(\br,t)\ket{\psi(\br,t)}$ invariant for an arbitrary local phase $\theta(\br,t)$. This is a typical example of {gauge freedom} in a field theory.


The gauge freedom under the compensating local phase shifts in \eqref{EF-gaugefreedom} implies that  
\begin{align}
\langle \psi|i\hbar\partial_t\psi\rangle \mapsto \langle \psi'|i\hbar\partial_t\psi'\rangle=\partial_t\theta +\langle \psi|i\hbar\partial_t\psi\rangle.
\end{align}
Hence, one may choose $\theta$ at will (gauge fixing) so as to accommodate any value of $\langle \psi'|i\hbar\partial_t\psi'\rangle$. For example, one may fix 
$
2|\Omega'|^{2}\langle\psi'|i\hbar\partial_t\psi'\rangle=\left\langle\psi'|({\delta h}/{\delta\psi'})\right\rangle
$, 
so that the $\psi$ equation in \eqref{EFELpsiprojective}  reads 
\begin{align*}
i\hbar\partial_t\psi'=\frac{1}{2|\Omega'|^{2}}\frac{\delta h}{\delta\psi'}\,.
\end{align*}
The same type of gauge was chosen in passing from equation \eqref{ProjectiveSchrodinger} to equation \eqref {GenSchr}, earlier. 

Another convenient choice  consists in fixing $\langle\psi'|i\hbar\partial_t\psi'\rangle=0$, so that the $\psi$ equation in \eqref{EFELpsiprojective} becomes 
\begin{align*}
i\hbar|\Omega'|^{2}\partial_t\psi'=\frac{1}{2}\left(\frac{\delta h}{\delta\psi'}-\left\langle\psi'\bigg|\frac{\delta 
h}{\delta\psi'}\right\rangle\psi'\right)\,.
\end{align*}
As we shall see later in this chapter, $\langle\psi|i\hbar\partial_t\psi\rangle$ can be seen as analogous to the scalar potential in electromagnetism and hence analogously this gauge is called the \emph{temporal gauge} (or {\it Weyl gauge}) and it has been adopted recently in \cite{AbediEtAl2010,AgostiniEtAl2015,SuzukiEtAl2015}. 
Gauge theory will play in an important role in Chapters \ref{Chap:Holonomy} and \ref{Chap:Gamma} of this thesis, but for now we simply remark that gauge theory is also important in other aspects of chemical physics; for example, see \cite{LittlejohnReinsch1997} for applications of gauge theory in molecular mechanics.
\end{remark}

Having investigated the stucture of the general equations of motion for the EF system, we now consider the case of the physical Hamiltonian functional. 
To do so, recall the molecular Hamiltonian \eqref{molecularHamiltonian}
\begin{align}
  \widehat{H} = -\frac{\hbar^2}{2M}\Delta + 
  \widehat{H}_e(\br)\label{molecularHamiltonian2}\,,
\end{align}
in which (as in Section \ref{SectionBOApprox}) $\nabla$ will denote the gradient with respect to the `nuclear' coordinate $\nabla_{\br}$. In addition, $\widehat{H}_e$ is understood as a Hamiltonian operator that acts on the electronic Hilbert space $\mathscr{H}_e$ whilst containing dependence on $\br$ through the interaction potential.
Then, following the analogous procedure as for the BO total energy (Remark \ref{BODFVP}), involving completing the square and integration by parts, substituting this expression into the total energy $h=\text{Re}\int 
\braket{\Psi|\widehat{H}\Psi}\,\de^3r$ yields $h:\mathscr{H}_n\times\mathcal{F}(\mathbb{R}^3,\mathscr{H}_e)\to\mathbb{R}$ 
given by
\begin{align}\label{EFGenHam}
h(\Omega, \psi) =\text{Re}\int \!\bigg[ 
\frac1{2M}\Omega^*(-i\hbar\nabla+{\boldsymbol{\cal A}})^2\Omega+|\Omega|^2\epsilon(\psi,\nabla\psi)
    \bigg]\text{d}^3r\,,
\end{align}
in agreement with \cite{SuzukiEtAl2015,AgostiniEtAl2015,AgostiniEtAl2016} and in which we recall the definition of the generalised Berry connection \eqref{GenBOBerryConnection}
\begin{align*} 
 {\boldsymbol{\cal A}}(\br,t)&:= \braket{\psi|-i\hbar\nabla\psi}\in \Lambda^{1}(\mathbb{R}^3)\,,
 \end{align*}
 as well as the effective electronic potential $\epsilon(\psi,\nabla\psi)$
 \begin{align}
\begin{split}
\epsilon(\psi,\nabla\psi)&:= \langle\psi|\widehat{H}_e\psi\rangle+\frac{\hbar^2}{2M}\|\nabla\psi\|^2 - 
  \frac{|\boldsymbol{\mathcal{A}}|^2}{2M}\\
    &=\braket{\psi|\widehat{H}_e\psi} 
    +\frac{\hbar^2}{2M}
    \left\langle\partial_i\psi,(\mathbbm{1}-\psi\psi^\dagger)\partial_i\psi\right\rangle
    \,. \label{EFepsilonDEF}
\end{split}
\end{align}
As in Section \ref{NonadiabaticSection} we see the interesting feature that the 
Hamiltonian \eqref{EFGenHam} corresponds to that of the nucleus in an effective 
electromagnetic field generated by the electron, in which the time-dependent Berry connection $\boldsymbol{\cal A}$ 
acts as the magnetic vector potential and $\epsilon$ acts as the scalar 
potential. As we shall see, the fact that this effective field now depends on 
time creates an interesting interplay between electronic and nuclear dynamics 
that goes beyond the Born-Oppenheimer approximation \eqref{BOAnsatz}.

At this point, we comment further on the structure of this effective potential as given by the final equality in \eqref{EFepsilonDEF}. 
Indeed, analogous to the BO case, the last term is the trace of the real part of the {quantum geometric tensor} (QGT) \cite{PrVa1980, WilczekShapere1989, ZanardiEtAl2007}
\begin{align}
\label{QGTensor}
Q_{ij}:=\left\langle\partial_i\psi\Big|(\mathbbm{1}-\psi\psi^\dagger)\partial_j\psi\right\rangle=\langle\partial_i\psi|\partial_j\psi\rangle - \hbar^{-2}{\cal A}_i {\cal A}_j
\,,
\end{align}
which is a Hermitian $(0,2)$-tensor field on $\mathbb{R}^3$, where we denote ${\cal A}_j := \braket{\psi|-i\hbar\partial_j\psi}$. 
In fact, as described in further detail in \cite{ZanardiEtAl2007}, $Q_{ij}$ is pulled-back from 
the projective electronic Hilbert space $\mathbb{P}\mathscr{H}_e$ onto $\mathbb{R}^3$ by the mapping $\br \mapsto \psi(\br)$.  

Correspondingly, (up to a factor) the imaginary part of $Q_{ij}$ returns the 2-form given by the Berry curvature {${\cal B}_{ij}:= \partial_i {\cal A}_j - \partial_j {\cal A}_{i\,}\in \Lambda^2(\mathbb{R}^3)$; namely, $2\hbar{\rm Im}(Q_{ij}) = {\cal B}_{ij}$}, whilst the 
real part,
\begin{align}
\label{T=ReQ}
 T_{ij}=\operatorname{Re}(Q_{ij}),
\end{align}
is a Riemannian metric tensor field on $\mathbb{R}^3$, which is the pull-back of the Fubini-Study metric on $\mathbb{P}\mathscr{H}_e$. 
The emergence of the trace of \eqref{QGTensor} in the electron energy in \eqref{EFepsilonDEF} indicates the geometry underlying the present formulation. 
Notice that the interpretation of $ \epsilon(\psi,\nabla\psi)$ in \eqref{EFepsilonDEF} as an effective electronic potential departs slightly from that found in the literature, where this quantity is called the \emph{gauge invariant part of the time-dependent potential energy surface} \cite{AgostiniEtAl2015,AgostiniEtAl2016,SuzukiEtAl2015}. 
As mentioned earlier, we will further investigate the geometric structure of the QGT and present interesting new relations in terms of non-Abelian gauge 
connections in Chapter \ref{Chap:Gamma}. 

\begin{corollary}[EF equations for the physical Hamiltonian functional]\label{Cor:EFPhysEqns}\hfill\\
  Specialising Theorem \ref{Theorem:EFEL} to the physical EF Hamiltonian functional \eqref{EFGenHam}, the general Euler-Lagrange equations \eqref{EFOmegaequation}-\eqref{EFELpsiprojective} 
  become
\begin{align}
  &i\hbar\partial_t{\Omega} = \left(\widehat{T}_n  - \langle \psi | i\hbar\partial_t{\psi} \rangle+\epsilon(\psi,\nabla\psi)\right)\Omega +\frac{1}{2M}\boldsymbol{\cal A}\cdot(\boldsymbol{\cal A}-2i\hbar\nabla)\Omega -\frac{i\hbar}{2M}({\rm div}\boldsymbol{\cal A})\,\Omega\label{EFEL1}\,,\\
  &(\mathbbm{1}-\psi\psi^{\dagger})\left[i\hbar\partial_t\psi +\hbar^2\frac{\Omega^*\nabla\Omega}{M|\Omega|^2}\cdot\nabla\psi-\widehat{H}_e\psi+ \frac{\hbar^2}{2M}\Delta\psi  
  \right]=0\,.
  \label{EFEL2}
\end{align}
These equations agree with those obtained in Section \ref{NonadiabaticSection} 
by direct substitution into the molecular Schr\"odinger equation as well as those found in the literature 
\cite{AbediEtAl2010,AbediEtAl2012}.
\end{corollary}
\paragraph{Proof:}
The variational derivatives of the Hamiltonian \eqref{EFGenHam} can be computed to 
obtain
\begin{align}
\frac{\delta h}{\delta\Omega}=&\ 2(\widehat{T}_n+\epsilon(\psi,\nabla\psi))\Omega+\frac{1}{M}\boldsymbol{\cal A}\cdot(\boldsymbol{\cal A}-2i\hbar\nabla)\Omega 
-\frac{i\hbar}{M}(\text{div}\boldsymbol{\cal 
A})\,\Omega\label{NUCLEAREFDERIV}\,,
\\
\frac{\delta h}{\delta\psi}=&\ 
2|\Omega|^2\widehat{H}_e\psi-2i\hbar^2\frac{\operatorname{Im}(\Omega^*\nabla\Omega)}{M}\cdot\nabla\psi-i\hbar^2\frac{\operatorname{Im}(\Omega^*\Delta\Omega)}{M}\psi- \frac{\hbar^2}{M}\operatorname{div}(|\Omega|^2\nabla\psi)
\label{FunDer1}\,,
\end{align}
at which point direct substitution into the general equations \eqref{EFOmegaequation}-\eqref{EFELpsiprojective} 
regains the above result and proves the corollary.
\hfill$\square$\\

We conclude this section by making the direct relation bewteen the equations \eqref{EFEL1} and \eqref{EFEL2} of Corollary \ref{Cor:EFPhysEqns} with \eqref{EFNuclearEqnSEC1} and \eqref{EFElectronicEqnSEC1} seen earlier in Section \ref{NonadiabaticSection} as well as those found in the recent chemical physics literature \cite{AbediEtAl2010,AbediEtAl2012}.
 Specifically, after expanding the necessary terms, one can use the useful relation
\begin{align}
\begin{split}
\braket{\psi|\Delta\psi} &= \text{Re}   \braket{\psi|\Delta\psi} + i\text{Im}  
  \braket{\psi|\Delta\psi}\\
  &= -\|\nabla\psi\|^2+ \frac{i}{\hbar}\text{div}\boldsymbol{\cal A}\,,
\end{split}
\end{align}
to verify that these sets of equations agree.
Furthermore, one can show that the variational derivatives \eqref{NUCLEAREFDERIV} and \eqref{FunDer1} imply
that the Lagrange multiplier $\lambda$ given by \eqref{LambdaLagrange} vanishes identically (in agreement with \cite{abedi2013response}) and that the compatibility condition \eqref{EFELcondition} is indeed satisfied.

\subsection{Hydrodynamic approach to nuclear dynamics}\label{Sec:EF-SubHydro}
Thus far the EF model has only been considered in terms of the nuclear and 
electronic wavefunctions. At this point, we follow the approach used throughout this thesis,
 employing a QHD description of the nuclear dynamics, now 
utilising the geometric mechanics machinery of Chapter \ref{Chap:QHD}. In 
particular, we remain on the Lagrangian side for the remainder of this chapter, 
whilst the Hamiltonian structure will be considered in Chapter \ref{Chap:EFNEW}.

As usual we begin by defining the momentum map nuclear fluid variables $D:=|\Omega|^2$ and $\bmu:= 
\bJ(\Omega)=\hbar\text{Im}(\Omega^*\nabla\Omega)$. Then, rewriting \eqref{EFGenHam} 
in these collective variables returns
\begin{align}
h(\boldsymbol{\mu},D,\psi)=
\int\left(\frac{1}{2M}\frac{|\boldsymbol{\mu}+ D\boldsymbol{\cal A}|^2}{D}+ \frac{\hbar^2}{8M}\frac{(\nabla
  D)^2}{D}+ D\,\epsilon(\psi,\nabla\psi)  \right)\text{d}^3r\, 
  \label{EFHydroHamiltonian}.
  \end{align}
At this stage, we will treat the quantity $\psi$ in \eqref{EFHydroHamiltonian} as a parametric variable, whose variations will be taken independently of those for $\bmu$ and $D$. 
Hence, analogous to the approach used in the treatment of the mean-field factorisation of Section \ref{meanfield-sec}, we apply the partial Legendre transform to obtain 
\begin{align}
\bu:=\frac{\delta h}{\delta \bmu}=\frac{\bmu+D\boldsymbol{\cal A}}{MD}\,.
\label{EFpartialLegendre}
\end{align}
Then, upon recalling the DF Lagrangian and again enforcing the PNC via a Lagrange multiplier, we can construct an EF Lagrangian as follows
\begin{align}
  \ell(\bu,D,\psi, \partial_t{\psi}, \lambda) &= \int\bigg[ D\braket{\psi, i\hbar\partial_t\psi} + \bu\cdot\bmu + \lambda\Big(\|\psi\|^2-1\Big)\bigg]\,\de^3r - h(\bu, D, 
  \psi)\,,
\end{align}
understood as the mapping $\ell:\mathfrak{X}(\mathbb{R}^3)\times{\rm Den}(\mathbb{R}^3)\times T\mathcal{F}(\mathbb{R}^3,\mathscr{H}_e)\times\mathcal{F}(\mathbb{R}^3)\to 
\mathbb{R}$.
This leads us to the following theorem.
\begin{theorem}[Euler-Poincar{\'e} approach to EF nuclear QHD]\label{Theorem:EFEPEL}
  \hfill \\ The exact factorization 
  system \eqref{EFEL1}-\eqref{EFEL2} can be equivalently derived from the Lagrangian 
$\ell:\mathfrak{X}(\mathbb{R}^3)\times{\rm Den}(\mathbb{R}^3)\times T\mathcal{F}(\mathbb{R}^3,\mathscr{H}_e)\times\mathcal{F}(\mathbb{R}^3)\to 
\mathbb{R}$ with $\ell=\ell(\bu,D,\psi, \partial_t{\psi}, \lambda)$ given by
  \begin{align}
  \ell = \int \bigg[D\left(\frac{1}{2}M |\bu|^2 - \boldsymbol{\cal A}\cdot \bu+\braket{\psi,i\hbar\partial_t{\psi}}-  V_Q -\epsilon(\psi,\nabla\psi)\right)+ \lambda\Big(\|\psi\|^2-1\Big)\bigg]\de^3r 
  \,.\label{EFHydroL1}
\end{align}
After defining the {\it electronic Hamiltonian functional}
\begin{align}
F(D,\psi) := \int \!D \epsilon(\psi,\nabla\psi)\,\de^3r\,,
\label{Fdef}
\end{align} 
Hamilton's principle $0=\delta\int_{t_1}^{t_2}\ell(\bu,D,\psi, \partial_t{\psi}, \lambda)\,\de t$ for arbitrary 
variations $\delta\psi$, $\delta \lambda$, and constrained Euler-Poincar{\'e} variations \eqref{EPvar} for $\bu$ and 
$D$, results in the following equations of motion
\begin{align}
 &\partial_t D + {\rm div}(D\bu)=0\,,\label{Ineedanotherlabel}\\
 & M(  \partial_t + \bu\cdot \nabla)\bu =- \nabla\left(V_Q + \epsilon\right) - \boldsymbol{\cal E}   - \bu \times \boldsymbol{\cal B}\label{EFEPELu1} 
  \,,\\
&(\mathbbm{1}-\psi\psi^{\dagger})\left( i\hbar \partial_t{\psi} + i\hbar \boldsymbol{u}\cdot\nabla\psi - \frac{1}{2D}\frac{\delta F}{\delta \psi}\right) 
  = 0\,,
\label{EFpsiprojective}\\
  &\|\psi\|^2 = 1\,,
\end{align}
where the first equation follows from the Lagrange-to-Euler map \eqref{defns} $D =\eta_* D_0$.  
Here, $V_Q$ denotes the nuclear quantum potential \eqref{BohmPot} for the nuclei, $\boldsymbol{\cal B}:=\nabla\times\boldsymbol{\cal A}$ is the {Berry curvature} and 
\begin{align}
\boldsymbol{\cal  E}:=-\,\partial_t \boldsymbol{\cal A} - 
\nabla\langle{\psi|i\hbar\partial_t\psi}\rangle\,.
\label{ElField-def-psi}
\end{align}
 In addition, 
the variational principle returns the new form of the compatibility condition
\begin{align}
  {\rm Im}\left\langle\psi \Bigg|\frac{\delta F}{\delta \psi} 
  \right\rangle=0\,.  \label{EFEPcondition}
\end{align}
\end{theorem}
\paragraph{Proof:} Up to the Lagrange multiplier (which naturally enforces the PNC), this Lagrangian can be 
considered as of the same type as that appearing in the treatment of the 
mean-field factorisation from Theorem \ref{Theorem:MFQHD}. Hence, we have 
already encountered the general form of the equations for Lagrangians of this type and it remains to compute and substitute in the 
appropriate variational derivatives. These are given by
\begin{align*}
  \frac{\delta \ell}{\delta \bu} = D(M\bu -\boldsymbol{\cal A})\,,\qquad   \frac{\delta \ell}{\delta D} =\frac{1}{2}M |\bu|^2 - \boldsymbol{\cal A}\cdot \bu+\braket{\psi,i\hbar\partial_t{\psi}}-  V_Q 
  - \epsilon\,,\\
  \frac{\delta \ell}{\delta \psi} = i\hbar D\partial_t\psi +2i\hbar D\bu\cdot\nabla\psi + i\hbar{\rm div}(D\bu)\psi - \frac{\delta F}{\delta \psi}+ 
  2\lambda\psi\,,\qquad   \frac{\delta \ell}{\delta \dot\psi} = -i\hbar D 
  \psi\,,
\end{align*}
recalling that $\boldsymbol{\cal A}=\braket{\psi|-i\hbar\nabla\psi}$.
Computing the $\psi$ Euler-Lagrange equation and using the auxilliary equation $\partial_tD +\text{div}(D\bu)=0$ 
(derived from the L-to-E map \eqref{defns}) yields
\begin{align}
  i\hbar D(\partial_t + \bu\cdot\nabla)\psi - \frac{1}{2}\frac{\delta F}{\delta \psi} 
  = -\lambda \psi\,.\label{YETANOTHEREQUATION}
\end{align}
Hence, upon 
taking the inner product with $\psi$, the real part returns the following expression for 
the Lagrange multiplier $\lambda$
\begin{align}
\lambda =   D\Big(\bu\cdot\boldsymbol{\cal A} - \braket{\psi|i\hbar\partial_t\psi}\Big) + 
  \frac{1}{2}\text{Re}\left\langle\psi \Bigg|\frac{\delta F}{\delta \psi} 
  \right\rangle\,,
\end{align}
whilst the imaginary part indeed returns the compatibility condition 
\eqref{EFEPcondition}. In fact, as shown in Appendix \ref{HydroCompCondApp}, upon postulating $h(\Omega)=h(\bmu, D)$
 one can verify that this expression can be derived from the earlier compatibility condition \eqref{EFELcondition} appearing in  
Section \ref{Sec:EF-Sub:1}. 

Having now characterised the Lagrange multiplier, one can follow straightforward calculations and replace $\lambda$ in the $\psi$ equation
\eqref{YETANOTHEREQUATION} to obtain the desired equation for the electronic {wavefunction} 
\eqref{EFpsiprojective}.

To complete this proof, we must compute the fluid equation \eqref{EFEPELu1}. 
This calculation starts from the general Euler-Poincar{\'e} fluid equation \eqref{generalEPFluid1} 
and procedes as follows:
\begin{align*}
  &(\partial_t + \pounds_{\boldsymbol{u}})\frac{\delta \ell}{\delta \boldsymbol{u}} = D\nabla\frac{\delta \ell}{\delta 
  D}\\
  \implies  &(\partial_t + \pounds_{\boldsymbol{u}})(MD\boldsymbol{u}-D\boldsymbol{\cal A} )= D\nabla\left(\frac{1}{2}M|\boldsymbol{u}|^2 - \boldsymbol{\cal A}\cdot 
  \boldsymbol{u} - V_Q + \braket{\psi| i\hbar\dot{\psi}} - \epsilon\right)\\
  \implies &MD (\partial_t + \pounds_{\boldsymbol{u}})\boldsymbol{u} - D(\partial_t + \pounds_{\boldsymbol{u}})\boldsymbol{\cal A} =  
  \frac{1}{2}MD\nabla|\boldsymbol{u}|^2 - D\nabla(\boldsymbol{\cal A}\cdot 
  \boldsymbol{u}) - D\nabla\Big(V_Q + \epsilon+\braket{\psi| 
  i\hbar\dot{\psi}}\Big)\\
  \implies &M (\partial_t + \boldsymbol{u}\cdot\nabla)\boldsymbol{u}  -\Big(\partial_t \boldsymbol{\cal A} + \nabla(\boldsymbol{\cal A}\cdot 
  \boldsymbol{u}) - \boldsymbol{u}\times\boldsymbol{\cal B}\Big)= - \nabla(\boldsymbol{\cal A}\cdot 
  \boldsymbol{u}) - \nabla\Big(V_Q + \epsilon+\braket{\psi| 
  i\hbar\dot{\psi}}\Big)\,.
\end{align*}
Simply cancelling like terms and recalling the definition of the electronic 
electric field \eqref{ElField-def-psi} returns the desired equation 
\eqref{EFEPELu1}.
\hfill$\square$\\

Having expressed the electronic equation in terms of the functional $F$ \eqref{Fdef}, the exact form in terms of the derivative ${\delta F}/{\delta \psi}$ will now be 
considered. Indeed, upon making use of the effective electronic potential $ \epsilon(\psi,\nabla\psi)$ defined in \eqref{EFepsilonDEF}, one 
computes the functional derivative 
\begin{align}
\begin{split}
\frac{\delta F}{\delta \psi}&=D\frac{\partial \epsilon}{\partial \psi} - {\rm div}\left(D\frac{\partial \epsilon}{\partial 
\nabla\psi}\right)\\
&=  2D \widehat{H}_e\psi+\frac{i\hbar}{M}D\boldsymbol{\cal A} \cdot \nabla \psi +\frac{i\hbar}{M}\operatorname{div}\!\big(D(i\hbar\nabla+\boldsymbol{\cal A})\psi\big)\,, 
\label{EPphysicalFderiv}
\end{split}
\end{align} 
whose insertion into \eqref{EFpsiprojective} yields the explicit form of the electronic Schr\"{o}dinger equation
\begin{align}
  (\mathbbm{1}-\psi\psi^{\dagger})\left( i\hbar \partial_t{\psi} + i\hbar  \Big(\boldsymbol{u}-\frac1{2M}\boldsymbol{\cal A}\Big)\cdot\nabla\psi -\widehat{H}_e\psi -\frac{i\hbar}{2MD}\operatorname{div}\!\big(D(i\hbar\nabla+\boldsymbol{\cal A})\psi\big) \right) 
  &= 0 
  \,.
  \label{EFpsiprojectivePhysical}
\end{align}
Furthermore, one can also verify directly that the derivative \eqref{EPphysicalFderiv} 
satisfies the compatibility condition \eqref{EFEPcondition}.

Before continuing, we comment briefly on the fluid equation \eqref{EFEPELu1}. Following 
the analogy with electromagnetism, the Berry curvature $\boldsymbol{\cal B}$ again acts as an effective magnetic field via a Lorentz force term generated by the electrons (in which $\hbar$ appearing in the Berry connection \eqref{GenBOBerryConnection} plays the role of the coupling constant), whilst 
$\boldsymbol{\cal  E}$ defined by \eqref{ElField-def-psi} plays the role of the electric field generated by the electrons. 
In addition, we remark that equation \eqref{ElField-def-psi} can in fact be thought of as simply the temporal component of a larger object on four-dimensional spacetime, which in the theory of electromagnetism corresponds to the field strength tensor.
 This idea, along with generalisations to non-Abelian geometric structures, will be explored in Chapter \ref{Chap:Gamma}. 

We now extend the discussion of the gauge freedom arising due to local phases from Remark \ref{EFlocalphases} to hydrodynamic nuclear picture. As before, the quantity $\langle{\psi|i\hbar\partial_t\psi}\rangle$ (now appearing in the definition of $\boldsymbol{\cal  E}$) can be fixed by selecting a particular gauge. Once again, one can choose the {\it temporal gauge}, $\braket{\psi|i\hbar\partial_t\psi}=0$, however now another convenient choice includes the {\it hydrodynamic gauge} $\braket{\psi|i\hbar\partial_t\psi}=\boldsymbol{\cal A}\cdot\bu$.
In fact, a more explicit expression of  $\boldsymbol{\cal  E}$ can be found by using \eqref{EFpsiprojective}, thereby leading to
\begin{align}
\begin{split}
\boldsymbol{\cal E}&=-\, \partial_t \boldsymbol{\cal A} - \nabla\langle{\psi|i\hbar\partial_t\psi}\rangle
\\
&= - 2 \braket{\nabla\psi,i\hbar\partial_t\psi}\\
&= -2 \left\langle \nabla\psi, -i\hbar\bu\cdot\nabla\psi + \frac{1}{2D}\frac{\delta F}{\delta \psi}\right\rangle 
- 2 \left\langle\psi\Bigg| i\hbar\partial_t\psi + i\hbar\bu\cdot\nabla\psi - \frac{1}{2D}\frac{\delta F}{\delta 
\psi}\right\rangle\braket{\nabla\psi,\psi}\\
&= -2\hbar\, u_j \,\text{Im}\braket{\nabla\psi|\partial_j\psi} -\frac1D\left\langle\nabla\psi,\frac{\delta F}{\delta 
\psi}\right\rangle\\
&= -\, \bu\times\boldsymbol{\cal B}-\frac1D\left\langle\nabla\psi,\frac{\delta F}{\delta \psi}\right\rangle,
\label{ElField}
\end{split}
\end{align}
where we recall the notation $\langle\,\cdot\,,\,\cdot\rangle
= \operatorname{Re}\braket{\,\cdot\, | \,\cdot\,}$ for the real-valued pairing and have used the definition of the Berry curvature $2\hbar\text{Im}\braket{\partial_i\psi|\partial_j\psi}={\cal B}_{ij} = \epsilon_{ijk}{\cal B}_k$, expressed as both a 2-form and pseudovector. 
Clearly, this calculation allows us to rewrite parts of the nuclear fluid equation 
and hence we now present the new full set of equations as follows.

\begin{proposition}[Alternative form of the EF system of equations]\label{Prop:EFNEWSET}
\hfill \\ The EF system as described by equations \eqref{Ineedanotherlabel}-\eqref{EFpsiprojective} given in Theorem \ref{Theorem:EFEPEL} can be equivalently written as
  \begin{align}
     & \partial_t D + {\rm div}(D\bu) =  0\,,
 \label{density}
  \\
&M(  \partial_t  + \bu\cdot \nabla) u_i = 
- \,\partial_i V_Q 
- \langle\psi,(\partial_i \widehat{H}_e) \psi\rangle 
-  \frac{\hbar^2}{MD}  \partial_j\big(DT_{ij}\big)\,,
\label{EFEPELu2} \\
 &  (\mathbbm{1}-\psi\psi^{\dagger})\left( i\hbar \partial_t{\psi} 
 + i\hbar  \Big(\boldsymbol{u}-\frac1{M}\boldsymbol{\cal A}\Big)\cdot\nabla\psi -\widehat{H}_e\psi 
 + \frac{\hbar^2}{2MD}\operatorname{div}\!\big(D\nabla\psi\big) \right) 
  = 0\,,
  \label{EFpsi2}
\end{align}
recalling that $T_{ij}:={\rm Re}(Q_{ij})$ denotes the real part of the quantum geometric tensor from equation 
\eqref{T=ReQ}.
\end{proposition}
\paragraph{Proof:} Following the calculation \eqref{ElField}, the nuclear fluid 
equation \eqref{EFEPELu1} from Theorem \ref{Theorem:EFEPEL} is rewritten as
\begin{align*}
  M(  \partial_t + \bu\cdot \nabla)\bu =- \nabla\left(V_Q + \epsilon\right)  + 
\frac1D\left\langle\nabla\psi,\frac{\delta F}{\delta \psi}\right\rangle\,.
\end{align*}
Then, using the definition of the electronic Hamiltonian functional 
\eqref{Fdef}, we are able to combine the following terms 
\begin{align}
\begin{split}
&\left\langle\nabla\psi,\frac{\delta F}{\delta 
\psi}\right\rangle-D\nabla\epsilon\\
&= \left\langle \nabla\psi, D\frac{\partial \epsilon}{\partial \psi} - \operatorname{div}\left(D\frac{\partial \epsilon}{\partial \nabla \psi}\right)\right\rangle 
-D\left(\braket{\psi,(\nabla \widehat{H}_e)\psi} + \left\langle \frac{\partial \epsilon}{\partial \psi},\nabla \psi\right\rangle +  \left\langle \frac{\partial \epsilon}{\partial \psi_{,\,j}},\nabla \psi_{,\,j}\right\rangle \right)
\\
&=   -D\langle\psi,(\nabla\widehat{H}_e)\psi\rangle
  -\partial_j \left\langle D
    \nabla\psi, \frac{\partial \epsilon}{\partial \psi_{,\,j}} 
    \right\rangle
    \\
    &= -D\langle\psi,(\nabla\widehat{H}_e)\psi\rangle
-    \frac1M  \partial_j\big(\hbar^2D\langle\nabla\psi,\partial_j\psi\rangle-D\boldsymbol{\cal A} 
{\cal A}_j\big),
\end{split}    
    \label{relation}
\end{align}
in which we have employed the additional notation $\psi_{,\,j}$ for $\partial_j\psi$ and recalled that $\epsilon$ depends on the nuclear coordinate $\br$ not only through $\psi$ but also through $\widehat{H}_e(\br)$.
This expression may then be written in components, in terms of $T_{ij}={\rm Re}(Q_{ij})$ from equation \eqref{T=ReQ}, as
\begin{align}
\left\langle \partial_i\psi,\frac{\delta F}{\delta \psi}\right\rangle-D\partial_i \epsilon
    =    \ -D   \langle\psi,(\partial_i\widehat{H}_e) \psi\rangle
-    \frac{\hbar^2}{M} \partial_j \big(DT_{ij}\big)\,.
\label{relation-comp}
\end{align}
We complete the proof by noting that the difference between the projective 
electronic Schr\"odinger equation \eqref{EFpsiprojectivePhysical} and its new form \eqref{EFpsi2} amounts to 
the simplification obtained by expanding the divergence term in former and 
noticing that all terms along $\psi$ are annihilated by the projection operator.
\hfill$\square$\\

It is important to emphasise once again that the system of equations presented in Proposition \ref{Prop:EFNEWSET} are entirely equivalent to the 
coupled Schr\"odinger equations \eqref{EFNuclearEqnSEC1} and \eqref{EFElectronicEqnSEC1} derived simply upon substituting the EF ansatz into the molecular Schr\"odinger equation.
Indeed, whilst this is easily seen for the electronic equation (upon using the definition of $\bu$ in terms of 
$\Omega$), the nuclear hydrodynamic equations \eqref{EFEPELu2} and \eqref{density} can be 
derived from the original nuclear Schr\"odinger equation \eqref{EFNuclearEqnSEC1} by writing $\Omega=\sqrt{D}e^{i S/\hbar}$ and by finding the evolution for $\bu=({\nabla S+\boldsymbol{\cal A}})/M$.

The remainder of this chapter is devoted to the consideration of some additional 
features of EF system as given in Proposition \ref{Prop:EFNEWSET}. Specifically, 
we pay particular attention to the Newtonian limit and circulation dynamics associated to the fluid equation \eqref{EFEPELu2}. 

\begin{remark}[Conservation of momentum considerations]
 \hfill \\ 
 Notice that equation \eqref{EFEPELu2} does \emph{not} conserve the spatial integral of the nuclear  momentum density,
\begin{align}
MD\bu=\bmu+D\boldsymbol{\cal A}=D(\nabla S+\boldsymbol{\cal A})
\,.
\label{tot-momentum-def}
\end{align}
Here, $S$ is the local phase of the wavefunction $\Omega=\sqrt{D}\,e^{iS/\hbar}$ and $D=|\Omega|^2$, while $D\boldsymbol{\cal A}$ is part of the nuclear momentum density.
This non-conservation of the hydrodynamic momentum should come as no surprise. In fact, this is already apparent in the  original system \eqref{EFNuclearEqnSEC1}-\eqref{EFElectronicEqnSEC1} which, instead, conserves the total  momentum
\begin{align}
\int\braket{\Psi|(-i\hbar\nabla +\widehat{P}_e|\Psi}\,\de^3r=\int\!\left(
\bmu+D\boldsymbol{\cal A}+D\langle\psi|\widehat{P}_e\psi\rangle\right)\de^3r
\,.
\label{tot-momentum-redef}
\end{align}
Thus, the total motion has two momentum contributions: one from the nuclear part and the other from the electronic part.
In particular, the momentum density of the nuclei is given by
$
\braket{\Psi|-i\hbar\nabla\Psi}
= 
\bmu+D\boldsymbol{\cal A}
$
where $\boldsymbol{\cal A}$ is the Berry connection defined in equation \eqref{GenBOBerryConnection}, and $\boldsymbol{\mu}:=\boldsymbol{J}(\Omega)=\hbar{\rm Im}(\Omega^*\nabla\Omega)$.
\end{remark}

\subsubsection{Newtonian limit and Lorentz force}
We now turn our attention to the classical Newtonian limit of the nuclear fluid equation \eqref{EFEPELu1}. 
To begin, we consider the exact factorization analog of the quantum Hamiltonian-Jacobi type equation 
\eqref{DSEqn2}, stated in the following proposition.
\begin{proposition}[Nuclear quantum Hamilton-Jacobi equation]
  \hfill \\
  The nuclear hydrodynamic equation \eqref{EFEPELu1} is equivalent to the quantum Hamilton-Jacobi equation
  \begin{align}
{\partial_t S} +\frac{|\nabla S + \boldsymbol{\cal A}|^2}{2M}=\langle\psi|i\hbar\partial_t\psi\rangle
- \epsilon(\psi,\nabla\psi) -V_Q\,,
\end{align}
upon replacing $M\bu=\nabla S+\boldsymbol{\cal A}$.
\end{proposition}
\paragraph{Proof:} Recall the general hydrodynamic Euler-Poincar\'e equation \eqref{generalEPFluid1}. Specialising 
to the EF Lagrangian \eqref{EFHydroL1}, this becomes
\begin{align}
  (\partial_t+\pounds_{\bu})(MD\bu-D\boldsymbol{\cal A})= D\nabla\left(\frac{1}{2}M|\bu|^2 - \bu\cdot\boldsymbol{\cal A} - V_Q - \epsilon + 
  \braket{\psi|i\hbar\partial_t\psi}\right)\,,
\end{align}
which after further manipulation results in the nuclear fluid equation 
\eqref{EFEPELu1}. Instead, we make use of the auxilliary equation $(\partial_t+\pounds_{\bu})D$ 
and use the relation $M\bu=\nabla S+\boldsymbol{\cal A}$ to obtain
\begin{align*}
  \begin{split}
  (\partial_t+\pounds_{\bu})(\nabla S)&= \nabla\left(\frac{|\nabla S+\boldsymbol{\cal A}|^2}{2M} - M^{-1}(\nabla S+\boldsymbol{\cal A})\cdot\boldsymbol{\cal A} - V_Q - \epsilon + 
  \braket{\psi|i\hbar\partial_t\psi}\right)\\
  \implies  (\partial_t+\pounds_{\bu})S&=\frac{|\nabla S+\boldsymbol{\cal A}|^2}{2M} - M^{-1}(\nabla S+\boldsymbol{\cal A})\cdot\boldsymbol{\cal A} - V_Q - \epsilon + 
  \braket{\psi|i\hbar\partial_t\psi}\\
    \implies  \partial_t S+M^{-1}(\nabla S+\boldsymbol{\cal A})\cdot\nabla S&=\frac{|\nabla S+\boldsymbol{\cal A}|^2}{2M} - M^{-1}(\nabla S+\boldsymbol{\cal A})\cdot\boldsymbol{\cal A} - V_Q - \epsilon + 
  \braket{\psi|i\hbar\partial_t\psi}\\
  \implies {\partial_t S} +\frac{|\nabla S + \boldsymbol{\cal A}|^2}{2M}&=\langle\psi|i\hbar\partial_t\psi\rangle
- \epsilon(\psi,\nabla\psi) - V_Q\,,
\end{split}
\end{align*}
as required.
\hfill$\square$\\
\newline
To obtain a classical equation for the nuclear dynamics, at this point we simply follow the chemistry literature and only drop the nuclear quantum potential term, thus removing the barrier to classical single particle solutions as described earlier in 
Section \ref{Sec:1-QHD}. Then, replacing $M\bu=\nabla S+\boldsymbol{\cal A}$ yields the following classical-type Hamilton-Jacobi equation:
\begin{align}
{\partial_t S} +\frac{|\nabla S + \boldsymbol{\cal A}|^2}{2M}=\langle\psi|i\hbar\partial_t\psi\rangle
- \epsilon(\psi,\nabla\psi) \,,
\end{align}
which corresponds to charged particle motion in a Maxwell field, governed by the equation  
\cite{AgostiniEtAl2016}
\begin{align}
M\ddot{\bq}=-\boldsymbol{\cal E}-\dot{\bq}\times\boldsymbol{\cal 
B}-\nabla_{\bq}\epsilon\,,
\label{classical}
\end{align}
where the electonic wavefunction contained within many of these terms is evaluated along the nuclear trajectory $\psi=\psi(\bq(t),t)$. This same result can be obtained via the analogous method employed in Section \ref{Sec:1-QHD}, in which one sets $D(\boldsymbol{r},t)=\delta(\boldsymbol{r}-\bq(t))$ in \eqref{density} thus producing $\dot\bq(t)=\bu(\bq(t),t)$ so that, after multiplying \eqref{EFEPELu1} by $D(\boldsymbol{r},t)$, integration of the delta function over physical space returns the classical equation \eqref{classical} above. 

An important point here is that the customary operation in chemical physics of neglecting the quantum potential term in the Lagrangian \eqref{EFHydroL1} can be problematic. Often, this step would invoke the limit $\hbar^2\to0$. However, here this process would also lead to discarding the terms ${M^{-1}}({\hbar^2}\|\nabla\psi\|^2 - {\boldsymbol{\cal A}^2})/{2}$ in the effective electronic potential \eqref{EFepsilonDEF}, thereby taking the exact-factorization model into the standard mean-field theory as seen by considering equation \eqref{EFEPELu2}.
 This crucial issue will be resolved in Section \ref{Bohmions} by performing a new variant of the exact factorization at the level of the molecular density operator, allowing us to utilise the cold fluid classical closure of Section \ref{Sec:2-ColdFluid} without taking the $\hbar^2\to0$ limit.


\subsubsection{Circulation dynamics for the Berry connection}\label{Berry-frequency}
In this section, we consider the circulation theorem associated to the nuclear 
hydrodynamic equation \eqref{EFEPELu2}. In particular, thanks to the Euler-Poincar{\'e} formulation of the nuclear QHD equations we can write an associated Kelvin-Noether theorem for the circulation of the nuclear fluid which turns out to describe the dynamics of the Berry connection $\boldsymbol{\cal A}(\br,t) := \braket{\psi|-i\hbar\nabla\psi}$.
This result is expressed as follows:
\begin{theorem}[Kelvin-Noether theorem for exact factorization]\label{Abby}
  \hfill \\ The Kelvin-Noether circulation theorem \cite{HolmEtAl1998} corresponding to the hydrodynamic equation \eqref{EFEPELu2} 
  is given by
    \begin{align}
       \frac{\de}{\de t}\oint_{c(t)} \boldsymbol{\cal A} \cdot \de\br
       =  - \oint_{c(t)} \Big( \langle\psi,(\partial_i\widehat{H}_e) \psi\rangle 
            + \frac{\hbar^2}{MD} \partial_j \big(DT_{ij}\big) \Big) \de r^i 
    \,,
    \label{Berry-Frequency}
    \end{align}
    where $c(t)$ is a closed loop moving with the nuclear hydrodynamic velocity $\bu = 
    (\bmu/D+\boldsymbol{\cal A})/M$.
\end{theorem}
\paragraph{Proof:}
 To see this, we write the motion equation \eqref{EFEPELu2} as the Lie derivative of a circulation 1-form, 
\begin{align}
\begin{split}
M\big( \partial_t + \pounds_{\bu} \big) \big( \bu\cdot \de \br \big) 
&= 
-\, \de \left( \frac12 |\bu|^2 + V_Q \right) - \langle\psi,(\de\widehat{H}_e) \psi\rangle
- \,  \frac{\hbar^2}{MD}  \partial_j\big(DT_{ij}\big) \de r^i
\\&=
\de \big( \partial_t S + \bu \cdot \nabla S\big)  
+ \big( \partial_t + \pounds_{\bu} \big) \big( \boldsymbol{\cal A}\cdot \de \br\big)  
\,,\end{split}
\label{transport-mom}
\end{align}
where we have used the relation \eqref{tot-momentum-def} to further evaluate the left-hand side on the second line and $\de$ denotes spatial differential in $\br$. 
Equating the right hand sides of equation \eqref{transport-mom} and integrating around an arbitrary closed loop $c(t)$ moving with the nuclear flow velocity $\bu(\br,t)$ annihilates the exact differential terms producing
\begin{align}
  \begin{split}
  \oint_{c(t)}\big( \partial_t + \pounds_{\bu} \big) \big( \boldsymbol{\cal A}\cdot \de \br\big) 
  &= - \oint_{c(t)} \Big( \langle\psi,(\partial_i\widehat{H}_e) \psi\rangle 
            + \frac{\hbar^2}{MD} \partial_j \big(DT_{ij}\big) \Big) \de r^i\,.
  \end{split}
\end{align} 
The conclusion of the proof comes from noticing that the Lagrangian time derivative becomes the total derivative when taken outside the integral. Geometrically,
 this occurs as follows
 \begin{align}
  \begin{split}  
 \oint_{c(t)}\big( \partial_t + \pounds_{\bu} \big) \big( \boldsymbol{\cal A}\cdot \de \br\big) 
  &= \oint_{c_0} \eta_t^* \big( \partial_t + \pounds_{\bu} \big) \big( \boldsymbol{\cal A}\cdot \de \br\big) 
  \\
  &= \oint_{c_0} \frac{\de}{\de t} \eta_t^* \big( \boldsymbol{\cal A}\cdot \de \br\big) 
= \frac{\de}{\de t} \oint_{c_0} \eta_t^* \big( \boldsymbol{\cal A}\cdot \de \br\big) 
  \\
  &= \frac{\de}{\de t} \oint_{c(t)} \boldsymbol{\cal A}\cdot \de \br 
  \,,
   \end{split}  
 \end{align}
where $c(t)=\eta_t\circ c_0$ and having used the general relation $\frac{\de}{\de t}(\text{Ad}^*_g\mu) = \text{Ad}^*_g(\partial_t + \text{ad}^*_{\dot{g}g^{-1}})\mu$ for any $g \in G$ and $\mu \in \mathfrak{g}^*$, specialised to the case
of $G=\text{Diff}(\mathbb{R}^3)$. 
\hfill$\square$ \\

This means that the nuclear circulation integral $\oint_{c(t)} \boldsymbol{\cal A} \cdot \de\boldsymbol{r}$, interpreted as the Berry phase obtained by integration around a loop moving with the nuclear fluid, is generated dynamically by an interplay between nuclear and electronic properties.
 Likewise, the evolution of the Berry curvature $\boldsymbol{\cal B}:=\nabla\times\boldsymbol{\cal A}$ follows by applying the Stokes theorem to relation \eqref{Berry-Frequency}.   
 Thus, the flux of the Berry curvature through a surface $S$ whose boundary $\partial S = c(t)$ is a closed loop moving with the nuclear fluid satisfies
    \begin{align}
       \frac{\de}{\de t}\int\!\!\!\!\int_{S} B_{ij}\,{\de}r^j\wedge {\de}r^i
       =  - \oint_{\partial S}  \Big( \langle\psi,(\partial_i\widehat{H}_e) \psi\rangle 
            + \frac{\hbar^2}{MD} \partial_j \big(DT_{ij}\big) \Big) \de r^i 
    \,.
    \label{Berry-CFlux}
    \end{align}

Furthermore, as we have seen, the Berry curvature can itself be viewed as simply 
the imaginary part of the quantum geometric tensor, leading to one final form 
of the circulation theorem:

\begin{remark}[Circulation theorem and the quantum geometric tensor]
\hfill \\ 
In terms of the real and imaginary parts of the quantum geometric tensor, $Q_{ij}$ in \eqref{QGTensor}, the circulation theorem \eqref{Berry-Frequency} 
is equivalently written as
    \begin{align}
       2\hbar\frac{\de}{\de t}\int\!\!\!\!\int_{S} {\rm Im}(Q_{ij}) \,{\de}r^j\wedge {\de}r^i
       =  - \oint_{\partial S}  \Big( \langle\psi,(\partial_i\widehat{H}_e) \psi\rangle 
       +  \frac{\hbar^2}{MD} \partial_j \big(D\,{\rm Re}(Q_{ij})\big) \Big) \de r^i 
    \,.
    \label{QGT-Flux}
    \end{align}
    This result \eqref{QGT-Flux} expresses the quantum geometric mechanics of the correlated nuclear and electronic degrees of freedom in the EF model. 
Namely, the nuclear probability density $D$ and expectation of the gradient $\nabla \widehat{H}_e$ are coupled dynamically with the real and imaginary parts of the quantum geometric tensor, $Q_{ij}$.
    \end{remark}

\chapter{New geometric approaches to exact factorization dynamics}\label{Chap:EFNEW}
This chapter completes the presentation of the material included in this thesis from the author's publication 
\cite{FoHoTr2019}, now taking advantage of the power of the geometric approach to provide new developments of nonadiabatic models in quantum chemistry. 

Section \ref{NuclearFrameSection} takes the EF approach beyond existing 
methods by allowing the electronic dynamics to evolve unitarily in the hydrodynamic frame of 
the nuclear motion. This key step allows us to consider a new pure state density matrix 
description of the electronic motion, unveiling interesting Euler-Poincar{\'e} and Lie-Poisson structures 
previously investigated in the study of complex fluids \cite{GBRaTr2012,GayBalmazRatiu2009,GBRaTr2013,Holm2002}. 

Section \ref{Sec:2Level} incorporates material from the author's work \cite{FoTr2020}, specialising the material from the previous section to a two-level system for the electronic state, applied to 
the Jahn-Teller/spin-boson problem. Remarkably, this two-level restriction of EF also encompasses previous hydrodynamic formulations of the Pauli equation \cite{Takabayasi1955} which 
will be considered in detail in Chapter \ref{Chap:Gamma}.

Section \ref{Bohmions} takes the new developments for EF one step further (now returning to material originating in \cite{FoHoTr2019}) by using the mixed state QHD results of 
Chapter \ref{Chap:QHD} to construct an entirely new 
generalisation of the EF method at the level of the molecular density matrix. Furthermore, the regularisation approach from Section \ref{Sec:2-Bohmions} allows for singular 
 solutions that express the entire electron-nuclear dynamics in terms of a 
 countably infinite set of Bohmion solutions.

\section{Electronic dynamics in the nuclear QHD frame}\label{NuclearFrameSection}

In Chapter \ref{Chap:EFOLD}, we have presented the geometric aspects of the exact factorization model which are currently available in the literature.
This section extends this work to consider the evolution of the electron density matrix.
Motivated by the arguments in \cite{AgostiniEtAl2016}, we shall write the electron dynamics in the Lagrangian frame moving with the nuclear hydrodynamic flow. 

Recalling the notation $\mathscr{H}_e$ for the electronic Hilbert space, we begin by introducing the group, $\mathcal{F}(\mathbb{R}^3, \mathcal{U}(\mathscr{H}_e))$,
  of smooth mappings from the physical space into the unitary group of the electronic Hilbert space.
   Then, we make the following evolution ansatz for the electronic {wavefunction}: $\psi(t)=(U\psi_0)\circ\eta^{-1}$, or more explicitly
\begin{align}
  \psi(\boldsymbol{r},t)=  U(\eta^{-1}(\boldsymbol{r}, t),t)\psi_0(\eta^{-1}(\boldsymbol{r}, 
  t))
  \,,
\label{unitary-evol}
\end{align}
where $\eta$ is the nuclear hydrodynamic path obeying $\dot\boldeta=\bu(\boldeta,t)$ and $U(\boldsymbol{r},t)\in \mathcal{F}(\mathbb{R}^3, \mathcal{U}(\mathscr{H}_e))$ is a local unitary operator on $\mathscr{H}_e$.
The evolution ansatz \eqref{unitary-evol} results in the following equation for the time evolution of $\psi$
\begin{align}
 \partial_t\psi +\bu\cdot\nabla\psi&= \xi\psi\,,\label{psitimeevo}
\end{align}
where we have defined the Lie algebra element $\xi := (\partial_t{U}\,U^{-1})\circ\eta^{-1} \in \mathcal{F}(\mathbb{R}^3, \mathfrak{u}(\mathscr{H}_e))$. 
Upon substituting these relations into the Lagrangian \eqref{EFHydroL1}, one 
obtains $\ell:\mathfrak{X}(\mathbb{R}^3)\times{\rm Den}(\mathbb{R}^3)\times\mathcal{F}(\mathbb{R}^3, \mathfrak{u}(\mathscr{H}_e))\times\mathcal{F}(\mathbb{R}^3, \mathscr{H}_e)\to \mathbb{R}$ 
given by
\begin{align}
  \ell(\bu, D, \xi,\psi) = \int \left[\frac{1}{2}MD |\bu|^2 - \frac{\hbar^2}{8M}\frac{(\nabla 
  D)^2}{D} + D\Big(\braket{\psi,i\hbar\xi{\psi}} - 
  \epsilon(\psi,\nabla\psi)\Big)\right]\text{d}^3r\,,\label{EFHydroL2}
\end{align}
in which we notice the explicit dependence on the Berry connection has been absorbed into $\xi$ via the relation
\begin{align}
 \braket{\psi|i\hbar\partial_t\psi} -\bu\cdot\boldsymbol{\cal A}&= \braket{\psi|i\hbar\xi\psi}  
 \,.\label{AnotherRelation}
\end{align}
At this point, one could continue and formulate the Euler-Poincar{\'e} system 
in terms of the variables $(\bu,\xi,D,\psi)$. Instead however, we notice that 
the relation \eqref{AnotherRelation} admits a density matrix description in 
terms of the electronic pure state $\rho:=\psi\psi^{\dagger}$ as 
$\braket{\psi|i\hbar\xi\psi}=\braket{\rho|i\hbar\xi}$. The only remaining barrier to expressing the entire 
electronic dynamics in terms of $\rho$ lies in the effective electronic 
potential $\epsilon(\psi,\nabla\psi)$. We address this in the following proposition.

\begin{proposition}[Density matrix formulation of the effective electronic potential]\label{Prop:EffPotDen}
  \hfill\\
  The effective electronic potential $\epsilon(\psi,\nabla\psi)$ defined by   \eqref{EFepsilonDEF}
  can equivalently be written in terms of the pure state density matrix $\rho=\psi\psi^{\dagger}$ 
  as
  \begin{align}
\epsilon(\psi,\nabla\psi)=\braket{\rho|\widehat{H}_e} + \frac{\hbar^2}{4M}\|\nabla\rho\|^2
\,,
\label{key-formula}
\end{align}
where one defines $\braket{A|B}:=\operatorname{Tr}(A^\dagger B)$ by using the generalised trace and $\|\,\cdot\,\|^2$ denotes the corresponding norm.
\end{proposition}
\paragraph{Proof:}
Whilst the first term involving the electronic Hamiltonian is a standard result in quantum mechanics, 
for the second term we use the following result. By a direct calculation one has 
that
\begin{align}
\begin{split}
\braket{\partial_i\rho|\partial_j\rho} &=\text{Tr}(\partial_i\rho\,\partial_j\rho)
\\
   &= \text{Tr}\Big(((\partial_i \psi)\psi^{\dagger} + \psi (\partial_i\psi^{\dagger}) )((\partial_j \psi)\psi^{\dagger} + \psi (\partial_j\psi^{\dagger}) 
  )\Big)\\
     &= \text{Tr}\Big((\partial_i \psi)\psi^{\dagger}(\partial_j \psi)\psi^{\dagger} +(\partial_i \psi)\psi^{\dagger}\psi (\partial_j\psi^{\dagger})+ \psi (\partial_i\psi^{\dagger})(\partial_j \psi)\psi^{\dagger} + \psi (\partial_i\psi^{\dagger})  \psi (\partial_j\psi^{\dagger}) 
  \Big)\\
  &= \braket{\psi|\partial_i\psi}\braket{\psi|\partial_j\psi} + \braket{\partial_j\psi|\partial_i\psi}+ \braket{\partial_i\psi|\partial_j\psi} + \braket{\partial_i\psi|\psi}\braket{\partial_j\psi|\psi} 
  \\
  &= 2\,T_{ij}
\,,
\end{split}
\label{minus-calc}
\end{align}
where the tensor $T$ with components $ T_{ij}=\text{Re}(Q_{ij})$ is the real part of the quantum geometric tensor from equation \eqref{QGTensor}. 
Hence upon considering the trace of the above calculation, we write that $\|\nabla\rho\|^2 := \text{Tr}(\nabla\rho\cdot\nabla\rho) = 2\text{Tr}(T)$.
The result follows upon recalling the expression for the effective electronic 
energy \eqref{EFepsilonDEF} in terms of $T$.
  \hfill$\square$\\

It is interesting to note relations resembling \eqref{minus-calc} have previously appeared when considering the Fubini-Study metric on the projective Hilbert space in the context of quantum information, see e.g. \cite{FacchiEtAl2010}.

The key formula \eqref{key-formula} in Proposition \ref{Prop:EffPotDen} now
enables us to express the previous EF Lagrangian \eqref{EFHydroL2} fully in terms of the electronic density 
matrix, resulting in a new Euler-Poincar\'e formulation as follows.

\begin{theorem}[Euler-Poincar\'e formulation of EF dynamics]\label{Theorem:EFEPFull}
\hfill \\
  Consider the evolution of the electronic density matrix given by $\rho=(U\rho_0U^{-1})\circ\eta^{-1}$ with time evolution
   written as $(\partial_t + \bu\cdot\nabla)\rho = [\xi, \rho]$, where $\xi:=(\partial_t{U}\,U^{-1})\circ\eta^{-1} $.  
 Then, the EF system as described by equations \eqref{Ineedanotherlabel}-\eqref{EFpsiprojective} given in Theorem \ref{Theorem:EFEPEL} can be equivalently described by the reduced Lagrangian 
 $\ell:\mathfrak{X}(\mathbb{R}^3)\times{\rm Den}(\mathbb{R}^3)\times\mathcal{F}(\mathbb{R}^3,\mathfrak{u}(\mathscr{H}_e))\times\mathcal{F}(\mathbb{R}^3,{\rm Her}(\mathscr{H}_e))\to\mathbb{R}$ 
 given by
    \begin{align}
  \ell(\bu, D, \xi, \rho) =  \int \bigg[\frac{1}{2}MD |\bu|^2-\frac{\hbar^2}{8M}\frac{(\nabla 
  D)^2}{D} + D\bigg(\braket{\rho,i\hbar\xi}  - \braket{\rho|\widehat{H}_e} - \frac{\hbar^2}{4M}\|\nabla\rho\|^2\bigg)\bigg]\de^3r
 \,. \label{EFHydroL3}
\end{align}
Then, applying Hamilton's principle $0=\delta\int_{t_1}^{t_2} \ell(\bu, D, \xi, \rho)\,\de t$ for constrained Euler-Poincar\'e variations \eqref{EPvar} and
\begin{align}
\begin{split}
  \delta \xi &= \partial_t \Upsilon -\bw\cdot\nabla\xi + \bu\cdot\nabla\Upsilon - [\xi,\Upsilon]\,, \\ 
  \delta \rho &= [\Upsilon,\rho]-\bw \cdot\nabla\rho \,,
\end{split}
\label{EFrhovariation}
\end{align}
where $\Upsilon := (\delta U) U^{-1}\circ\eta^{-1}$ and $\bw := \delta\eta\,\circ\,\eta^{-1}$ are arbitrary, returns the system of equations
\begin{align}
    &\partial_t D+{\rm div}(D\bu)=0\,,
\label{final-D-eqn}\\
&  M(\partial_t + \bu\cdot\nabla)\bu   = 
  - \nabla V_Q-\langle\rho,\nabla\widehat{H}_e\rangle -\frac{\hbar^2}{2MD}\partial_j\langle D \nabla\rho,  \partial_j\rho\rangle
  \label{final-u-eqn}
  \,,
  \\
  &
    i\hbar (\partial_t +\bu\cdot\nabla)\rho = [\widehat{H}_e,\rho]+\frac{\hbar^2}{2MD}{\rm div}(D[\rho,\nabla\rho])
\label{finalrhoeq}
    \,.
\end{align}
\end{theorem}
\paragraph{Proof:} We begin by considering the equations of motion corresponding to the entire class of systems admitting a general reduced Lagrangian of the type $\ell(\bu, D, \xi, \rho)$, where $\bu\in \mathfrak{X}(\Bbb{R}^3)$, {$D\in \text{Den}(\Bbb{R}^3)$}, $\xi\in \mathcal{F}(\Bbb{R}^3,\mathfrak{u}(\mathscr{H}_e))$ and $\rho\in \mathcal{F}(\Bbb{R}^3,{\rm Her}(\mathscr{H}_e))$.
Firstly, the new variations \eqref{EFrhovariation} can be derived 
from their counterparts in the absence of fluid motion, $\delta \xi = \partial_t\Upsilon -[\xi,\Upsilon]$ 
and $\delta \rho=[\Upsilon, \rho]$, upon replacing the material derivatives $\partial_t \mapsto \partial_t + \pounds_{\bu}$ and $\delta \mapsto \delta + \pounds_{\bw}$, as is usual in the Euler-Poincar\'e theory of fluid flows \cite{HolmEtAl1998, GBRaTr2013}. 
For a more detailed proof the reader is referred to \cite{GayBalmazRatiu2009}.
Then, an explicit computation expanding Hamilton's principle yields
\begin{align*}
  0 &= \int_{t_1}^{t_2} \int \left(\frac{\delta \ell}{\delta \bu}\cdot \delta\bu + \frac{\delta \ell}{\delta D}\delta D + \left\langle \frac{\delta \ell}{\delta \xi},\delta \xi \right\rangle+ \left\langle \frac{\delta \ell}{\delta \rho},\delta \rho \right\rangle \right)\,\de^3 r\,\de   t\\
  &= \int_{t_1}^{t_2} \int \Bigg(\frac{\delta \ell}{\delta \bu}\cdot (\partial_t+\pounds_{\bu})\bw - \frac{\delta \ell}{\delta D}\,\text{div}(D\bw) + \left\langle \frac{\delta \ell}{\delta \xi}, \Big((\partial_t +\pounds_{\bu})\Upsilon -[\xi,\Upsilon]-\bw\cdot\nabla\xi \Big)\right\rangle\\
  &\qquad\qquad\qquad\qquad\qquad\qquad\qquad\qquad\qquad\qquad\qquad\qquad+ \left\langle \frac{\delta \ell}{\delta \rho}, [\Upsilon, \rho] - \bw\cdot\nabla\rho \right\rangle \Bigg)\,\de^3 r\,\de  
  t\\
  &= \int_{t_1}^{t_2} \int \Bigg( \Bigg(-(\partial_t+\pounds_{\bu})\frac{\delta \ell}{\delta \bu}+D\nabla\frac{\delta \ell}{\delta D}-\left\langle\frac{\delta \ell}{\delta \xi},\nabla\xi \right\rangle-\left\langle\frac{\delta \ell}{\delta \rho},\nabla\rho \right\rangle \Bigg)\cdot \bw \\
  &\qquad\qquad\qquad\qquad\qquad\qquad\qquad+ \left\langle \Bigg(-(\partial_t +\pounds_{\bu})\frac{\delta \ell}{\delta \xi} +\left[\xi,\frac{\delta \ell}{\delta \xi} \right]+\left[\frac{\delta \ell}{\delta \rho},\rho \right]\Bigg), \Upsilon \right\rangle \Bigg)\,\de^3 r\,\de  
  t\,.
\end{align*}
Hence, recalling that $\bw$ and $\Upsilon$ are arbitrary, we can write the general system of equations as  (cf. 
\cite{GBRaTr2012,GayBalmazRatiu2009,GBRaTr2013,Holm2002})
\begin{align}
\begin{split}
&  (\partial_t + \pounds_{\bu})\frac{\delta \ell}{\delta \bu} = - \left\langle 
  \nabla\xi, \frac{\delta \ell}{\delta \xi}
  \right\rangle - \left\langle 
  \nabla\rho, \frac{\delta \ell}{\delta \rho}
  \right\rangle + D\nabla\frac{\delta \ell}{\delta D}
  \,,
\\&    (\partial_t + \pounds_{\bu})\frac{\delta \ell}{\delta \xi} - \left[\xi,   \frac{\delta \ell}{\delta \xi}\right] 
    = \left[\frac{\delta \ell}{\delta \rho},\rho\right]\,,
 \\&    (\partial_t + \pounds_{\bu})D = 0\,,
 \\
&     (\partial_t + \pounds_{\bu})\rho = [\xi,\rho]\,,
\end{split}
\label{EFrhoevol}
\end{align}
having included the auxilliary equations that follow from the definitions $D=\eta_*D_0$ and 
$\rho=(U\rho_0U^{-1})\circ\eta^{-1}$. We then look to specialise these general 
equations to the EF Lagrangian \eqref{EFHydroL3}. For convenience, we rewrite the electronic Hamiltonian functional in \eqref{Fdef} as 
\\
\begin{align*}
  F(D,\rho) = \int D\left(\braket{\rho|\widehat{H}_e} + \frac{\hbar^2}{4M}\|\nabla\rho\|^2\right)\,\de^3r   = \int D\epsilon(\rho,\nabla\rho)\,\de^3r  
  \,,
\end{align*}
so that the variational derivatives are written as follows
\begin{align*}
  &\frac{\delta \ell}{\delta \bu} = MD\bu\,,\qquad  \frac{\delta \ell}{\delta D} 
  =\frac{1}{2}M|\bu|^2 - V_Q + \braket{\rho,i\hbar\xi} -\epsilon(\rho,\nabla\rho)\,,\\
   &\frac{\delta \ell}{\delta \xi} = -i\hbar D \rho\,,\qquad  \frac{\delta \ell}{\delta \rho} = 
   i\hbar D \xi - \frac{\delta F}{\delta \rho}\,.
\end{align*}
Consequently, the fluid velocity equation in \eqref{EFrhoevol} becomes:
\begin{align}
  M(\partial_t + \bu\cdot\nabla)\bu   &= \frac{1}{D}\left\langle \nabla\rho, \frac{\delta F}{\partial \rho} \right\rangle 
  - \nabla(V_Q + \epsilon)
  \,,
\end{align}
which indeed reduces to equation \eqref{EFEPELu2} upon {specialising} to pure states $\rho=\psi\psi^\dagger$. 
Also, we notice the following analogue of relation \eqref{relation}:
\begin{align}
  \begin{split}
\left\langle \nabla\rho, \frac{\delta F}{\partial \rho} \right\rangle -D\nabla\epsilon=&\ \left\langle \nabla\rho, D\frac{\partial \epsilon}{\partial \rho} -{\rm div}\left(D\frac{\partial\epsilon}{\partial\nabla\rho}\right)\right\rangle -D\nabla\epsilon
\\
=& -D\langle\rho,\nabla\widehat{H}_e\rangle
-\partial_j\left\langle D \nabla\rho, \frac{\partial \epsilon}{\partial \rho_{,j}} \right\rangle
\\
=&-D\langle\rho,\nabla\widehat{H}_e\rangle
-\frac{\hbar^2}{2M}\partial_j\left\langle D \nabla\rho, \partial_j\rho \right\rangle,
  \end{split}
\end{align}
thus recovering the fluid equation \eqref{final-u-eqn}, in which we again recognise the presence of the real part of the QGT in the last 
term. Notice that, using the relation for pure states $\langle\rho,\nabla\rho\rangle=0$, one can write the alternative expression of the final term $\partial_j\langle D \nabla\rho, \partial_j\rho \rangle=-\partial_j\langle D \rho, \partial_j\nabla\rho \rangle$. 
Finally, we consider the $\xi$ equation in \eqref{EFrhoevol} and find, via the $D$ and $\rho$ equations, the following simplifying \textit{algebraic relation}
\begin{align}\label{xi-eliminated?}
\left[i\hbar D \xi - \frac{\delta F}{\delta \rho},\rho\right]=0\,.
\end{align} 
Again using that $(\partial_t+\pounds_{\bu})\rho=[\xi,\rho]$ implies the required electronic von Neumann equation,
\begin{align}\label{rhoeqn}
  i\hbar (\partial_t +\bu\cdot\nabla)\rho = \left[\widehat{H}_e-\frac{\hbar^2}{2MD}{\rm div}(D\nabla\rho),\rho\right]
\,,\end{align}
eliminating the need to solve the algebraic relation for $\xi$ and thus completing 
the proof.
\hfill$\square$\\

\begin{remark}[Analogies with complex fluids]\label{EFComplexFluidsAnalogy}
\hfill \\
We take this opportunity to make the connection between the hydrodynamic exact {factorization} system and 
previous investigations of the geometry of liquid crystal flows, as found in \cite{GayBalmazTronci2010,Tronci2012,GBRaTr2012,GayBalmazRatiu2009,GBRaTr2013,Holm2002}. 
In this comparison, the electronic {wavefunction} $\psi(\boldsymbol{r},t)\in \mathcal{F}(\Bbb{R}^3,\mathscr{H}_e)$ is replaced by the director, 
an orientation parameter field $\bn(\boldsymbol{r},t) \in \mathcal{F}(\Bbb{R}^3,S^2)$; the unitary evolution operator $U(\boldsymbol{r},t)\in \mathcal{F}(\Bbb{R}^3,\mathcal{U}(\mathscr{H}_e))$ becomes a rotation matrix $R(\boldsymbol{r},t)\in \mathcal{F}(\Bbb{R}^3,SO(3))$; and one still considers the coupling to the fluid velocity $\bu(\boldsymbol{r},t)$ given by the action of diffeomorphisms $\eta \in {\rm Diff}(\mathbb{R}^3)$. 
Indeed, with these replacements, one has a reduced Lagrangian of the same type, $\ell(\bu,\boldsymbol\xi,D,\bn)$, 
and the resulting Euler-Poincar{\'e} equations are equivalent to those in \eqref{EFrhoevol}. 
In Chapter \ref{Chap:Gamma} this analogy will continue, in which we will introduce a 
non-Abelian gauge connection as an additional dynamical variable again in the same fashion used in complex 
fluids.
\end{remark}

\begin{remark}[Electron decoherence]
\hfill \\
Equation \eqref{rhoeqn} will determine the evolution of the true electron density matrix defined in \eqref{TrueElectronDensity}, 
$
\widehat{\rho}_e(t) := \int D\rho \,\de^3r =: \int \tilde\rho \,\de^3r
$.
Namely,
\begin{align}
i\hbar \frac{\de}{\de t}{\widehat{\rho}}_e(t) 
= 
\int \left[D\widehat{H}_e-\frac{\hbar^2}{2M}{\rm div}(D\nabla\rho),\rho\right] 
\,\de^3r\,.
\label{rho-e-dyn}
\end{align}
This result implies that spatially uniform pure initial states (such that $\widehat{\rho}_e\,^2=\widehat{\rho}_e$) become mixed states as time proceeds. Thus, in agreement with, e.g., \cite{MinEtAl2015}, the exact {factorization} model captures dynamical electronic decoherence effects (that is, quantum state mixing) from pure initial states; since the density matrix evolution is no longer unitary.
\end{remark}


Having worked solely on the Lagrangian side so far, deriving these equations from an Euler-Poincar{\'e} 
variational principle whose general form has been previously seen in the geometric mechanics literature (see Remark \ref{EFComplexFluidsAnalogy}),
we now proceed by considering the complimentary picture on the Hamiltonian side.

\begin{theorem}[Lie-Poisson structure of the EF system]\label{Theorem:LPEF}
  \hfill \\
  When written in terms of the variables
  \begin{align}
\bm:=MD\bu\,,\qquad\qquad\tilde\rho:=D\rho\,,
\label{MDrho-var}
\end{align}
  the equations \eqref{final-D-eqn}-\eqref{finalrhoeq} possess a Hamiltonian structure in terms of the 
  Lie-Poisson bracket
  \begin{align}
\begin{split}
\{f,g\}(\bm,D,\tilde\rho)=&\,\int\!\bm\cdot\left(\frac{\delta g}{\delta \bm}\cdot\nabla\frac{\delta f}{\delta \bm}-\frac{\delta f}{\delta \bm}\cdot\nabla\frac{\delta g}{\delta \bm}\right)\de ^3 r
\\
&\,
-
\int \!D\left(\frac{\delta f}{\delta \bm}\cdot\nabla\frac{\delta g}{\delta D}-\frac{\delta g}{\delta \bm}\cdot\nabla\frac{\delta f}{\delta D}\right)\de ^3 r
\\
&\,
-\int\left\langle \tilde\rho,\frac{i}{\hbar}\left[\frac{\delta f}{\delta \tilde\rho},\frac{\delta g}{\delta \tilde\rho}\right]+\frac{\delta f}{\delta \bm}\cdot\nabla\frac{\delta g}{\delta \tilde\rho}-\frac{\delta g}{\delta \bm}\cdot\nabla\frac{\delta f}{\delta  \tilde\rho}\right\rangle\de ^3 r
\label{SDP-LPB-EF}\,,
\end{split}
\end{align}
defined on the dual of the Lie algebra $\mathfrak{g}=\mathfrak{X}(\Bbb{R}^3)\,\circledS\, \big(\mathcal{F}(\Bbb{R}^3)\oplus\, \mathcal{F}\big(\Bbb{R}^3,\mathfrak{u}(\mathscr{H}_e)\big)\big)
$ comprising a direct sum of semidirect product actions, corresponding to the Hamiltonian 
\begin{align}
h(\bm,D,\tilde\rho)=
\int\left(\frac{1}{2M}\frac{|\bm|^2}{D} -  \frac{\hbar^2}{8M}\frac{|\nabla D|^2}{D}+ 
\braket{\tilde\rho|\widehat{H}_e} + \frac{\hbar^2}{4M}\frac{\|\nabla\tilde\rho\|^2}{D}\right)\,{\de}^3r  
\,. \label{EFHydroHamiltonian3}
  \end{align}
\end{theorem}
\paragraph{Proof:} To begin we expand $\dot{f}(\bm, D, \tilde{\rho})$ as 
\begin{align*}
  \dot{f}(\bm, D, \tilde{\rho}) = \int \frac{\delta f}{\delta \bm}\cdot \partial_t\bm 
  +  \frac{\delta f}{\delta D}\partial_t D + \left\langle \frac{\delta f}{\delta \tilde{\rho}},\partial_t 
  \tilde{\rho}\right\rangle\,\de^3r\,,
\end{align*}
so that using $\dot{f}
= \{f,h\}$ for the bracket \eqref{SDP-LPB-EF} 
returns the general equations
\begin{align}
  \partial_t \bm + \partial_j\left(\frac{\delta h}{\delta m_j}\bm\right)+ 
  m_j\nabla\frac{\delta h}{\delta m_j} &= -D\nabla\frac{\delta h}{\delta D} - \left\langle  \tilde{\rho}, \nabla\frac{\delta h}{\delta 
  \tilde{\rho}}\right\rangle\,,\\
  \partial_t D + \text{div}\left(D \frac{\delta h}{\delta \bm} \right)&= 0\,,\\
  \partial_t \tilde{\rho} + \text{div}\left(\tilde{\rho} \frac{\delta h}{\delta \bm} 
  \right)&= \frac{i}{\hbar}\left[\tilde{\rho},\frac{\delta h}{\delta 
  \tilde{\rho}}\right]\,.
\end{align}
Rewriting the total energy \eqref{EFHydroHamiltonian} 
in terms of the Hamiltonian variables \eqref{MDrho-var} yields
\begin{align}
h(\bm,D,\tilde\rho)=
\int\left(\frac{1}{2M}\frac{|\bm|^2}{D}+ \frac{\hbar^2}{8M}\frac{(\nabla
  D)^2}{D}+ 
\braket{\tilde\rho|\widehat{H}_e} + \frac{\hbar^2D}{4M}\bigg\|\nabla\bigg(\frac{\tilde\rho}{D}\bigg)\bigg\|^2\right)\,\text{d}^3r  
\,. \label{EFHydroHamiltonian2}
  \end{align}
Upon restricting to {pure quantum states}, so that $\langle\rho,\nabla\rho\rangle=0$ and $\langle\tilde\rho,\nabla\tilde\rho\rangle=D\nabla D$, we find 
the interesting relation
\begin{align}
\frac{\hbar^2D}{4M}\bigg\|\nabla\!\bigg(\frac{\tilde\rho}{D}\bigg)\bigg\|^2=\frac{\hbar^2D}{4M}\bigg\|\frac{\nabla\tilde\rho}{D}-\frac{\nabla D}{D^2}\tilde\rho\bigg\|^2
=
\frac{\hbar^2}{4M}\left(\frac{\|\nabla\tilde\rho\|^2}{D}-\frac{|\nabla D|^2}{D}\right)\,.\label{nablarhocomputation}
\end{align}
Consequently, for pure quantum states, the Hamiltonian \eqref{EFHydroHamiltonian2} 
recovers \eqref{EFHydroHamiltonian3} as required.
  %
Then, upon computing the necessary variational derivatives
\begin{align*}
  \frac{\delta h}{\delta \bm}= \frac{\bm}{MD}\,,\quad \frac{\delta h}{\delta D} 
  = - \frac{|\bm|^2}{2MD^2} - V_Q - 
  \frac{\hbar^2}{4M}\frac{\|\nabla\tilde{\rho}\|^2}{D^2}\,,\quad \frac{\delta h}{\delta 
  \tilde{\rho}}=\widehat{H}_e - 
  \frac{\hbar^2}{2M}\text{div}(D^{-1}\nabla\tilde{\rho})\,,
\end{align*}
the general equations of motion specialise to 
\begin{align}
 \begin{split}
 &\partial_t\bm + ({\rm div}\,\bu)\bm + (\bu\cdot\nabla)\bm
 = 
       D\nabla V_Q-\langle\tilde\rho,\nabla\widehat{H}_e\rangle +\frac{\hbar^2}{2M}\partial_j\left\langle  \tilde\rho,  \nabla\Big(D^{-1}\partial_j\tilde\rho\Big)\right\rangle
  \,,
  \\
  &  i\hbar \big(\partial_t \tilde\rho+{\rm div}(\bu\tilde\rho) \big)= \left[\widehat{H}_e,\tilde\rho\right]+\frac{\hbar^2}{2M}{\rm div}\left(D^{-1}[\tilde\rho,\nabla\tilde\rho]\right)\,,
    \\
&    \partial_t D+{\rm div}(D\bu)  =0\,,
 \end{split}
    \label{final-D-eqn2}
\end{align}
recalling the relation $\bm=MD\bu$. To complete the proof one simply verifies directly that fully changing variables 
back to $(\bu, D, \rho)$ returns the set of equations \eqref{final-D-eqn}, 
making use of the following relation for the gradient of the quantum potential
\begin{align}
  \nabla V_Q = \frac{\hbar^2}{4MD}\partial_j\left(\frac{\partial_j D\,\nabla D}{D}- \nabla(\partial_j 
  D)\right)\,.
\end{align}
\hfill$\square$\\

We conclude this section with the following remarks. Firstly, the above proof 
rectifies the erroneous equation of motion for $\bm$ that was presented in 
\cite{FoHoTr2019} as well as minor oversights in the Poisson bracket \eqref{SDP-LPB-EF}. 
Secondly, the change of variable $\tilde\rho\to i\hbar\tilde\rho$ shows that this bracket is Lie-Poisson on the dual {of the Lie algebra 
$\mathfrak{g}=\mathfrak{X}(\Bbb{R}^3)\,\circledS\, \big(\mathcal{F}(\Bbb{R}^3)\oplus\, \mathcal{F}\big(\Bbb{R}^3,\mathfrak{u}(\mathscr{H}_e)\big)\big)
$, for which} the dual coordinates are $m = \bm\cdot \text{d}\bx \otimes \text{d}^3x \in\mathfrak{X}(\Bbb{R}^3)^* = \Lambda^1(\Bbb{R}^3)\otimes \text{Den}(\Bbb{R}^3)$, $D\in \text{Den}(\Bbb{R}^3)$ and {$i\tilde\rho\in  \mathfrak{u}(\mathscr{H}_e)\otimes\text{Den}(\Bbb{R}^3)$}.
Therefore, the Lie-Poisson bracket may be written more compactly in the usual form as 
\begin{align}
\{f,h\}(m,D,\tilde\rho)= -  \left\langle (m,D,\tilde\rho) \,,\, \left[ \frac{\delta f}{\delta (m,D,\tilde\rho)} \,,\,\frac{\delta h}{\delta (m,D,\tilde\rho)}  \right] \right\rangle_{\br},
\label{LPB-brief}
\end{align}
where in \eqref{LPB-brief} the angle brackets $\langle\,\cdot\,,\,\cdot\,\rangle_{\br}$ denote $L^2$ pairing in the $\br$ coordinates, and the square brackets denote the components of the adjoint action of the semidirect-product Lie algebra $\mathfrak{g}$, whose $\br$-coordinate pairings are given explicitly in equation \eqref{SDP-LPB-EF}.
Finally, we remark that the interesting relation \eqref{nablarhocomputation} demonstrates how the effective potential $\epsilon$ produces terms of the same form as the quantum 
potential when written in terms of the Hamiltonian variable $\tilde\rho$. We will come across this 
relation again in the next section in the context of two-level systems in which it turns out that the analogous relation 
 for the spin density results in an exact cancellation with the explicit form of the quantum potential.

\section{Two-level systems and the spin-boson problem}\label{Sec:2Level}
Having so far considered the general case in which the electronic Hilbert space $\mathscr{H}_e$ is arbitrary, although often
 taken to be $\mathscr{H}_e=L^2(\mathbb{R}^3)$ in the chemistry literature, this section is devoted to 
the specialisation to a two-level system as described in Section \ref{Sec:1-Postulates}, i.e. $\mathscr{H}_e = \mathbb{C}^2$, so that the electronic factor is of the form $\psi \in 
\mathcal{F}(\mathbb{R}^3,\mathbb{C}^2)$. Correspondingly, the Hamiltonian operator on $\mathbb{C}^2$ has the general form
\begin{align}
  \widehat{H}_e=a(\br)\mathbbm{1}+\boldsymbol{b}(\br)\cdot\widehat{\boldsymbol{\sigma}}\,,\label{2levelHam}
\end{align}
where $\mathbbm{1}$ is the $2\times 2$ identity operator and we recall the Pauli matrices defined in Section \ref{Sec:1-Postulates}. 
In addition, the electronic pure state density matrix, defined by $\rho:=\psi\psi^{\dagger}$, can be now expressed in terms of the Pauli matrices using the standard result \eqref{rhoblochvector} to give 
\begin{align}
\rho(\br,t)=\frac{1}{2}\left(\mathbbm{1}+\frac2\hbar{\boldsymbol{s}}(\br,t)\cdot\widehat{\boldsymbol{\sigma}}\right)\,, \label{electronicdensitytospin}
\end{align}
recalling the spin vector $\bs$ defined by $\bs := \hbar\braket{\widehat{\bsigma}}/2$. As a result of the dependence on the nuclear coordinate the spin vector is now understood as the mapping $\br\mapsto\bs(\br,t)$ and hence is an element of $\mathcal{F}(\mathbb{R}^3, \mathbb{R}^3)$. 
Then, due to the above formula, $\bs$ becomes the new dynamical variable describing the electronic motion which, following the reasoning from the previous 
section, we now express in the frame of the nuclear fluid. Then, using relations for the product of Pauli matrices 
from Section \ref{Sec:1-Postulates}, the electronic evolution equation $(\partial_t+\bu\cdot\nabla)\rho = [\xi,\rho]$
becomes
\begin{align}
  (\partial_t+\bu\cdot\nabla)\bs = \bxi\times\bs\,,
\end{align} 
having introduced the new generator $\bxi \in \mathcal{F}(\mathbb{R}^3,\mathbb{R}^3)$ defined by the Lie algebra isomorphism $(\mathbb{R}^3, \,\cdot\,\times\,\cdot\,)$ to $(\mathfrak{su}(2),[\,\cdot\,,\,\cdot\,])$
\eqref{TILDEMAP} applied to the $\mathfrak{su}(2)$ generator $\xi=( \partial_t U \,U^{-1})\circ\eta^{-1}$.
Upon further calculations using the property of Pauli matrices ${\rm Tr}(\sigma_i\sigma_j)=2\delta_{ij}$, we also arrive at the following expressions
\begin{align}
  \braket{\rho|i\hbar\xi}= \bs\cdot\bxi\,,\qquad \|\nabla \rho\|^2 &= \frac{2}{\hbar^2}\|\nabla 
  \bs\|^2\,,\qquad 
  \braket{\rho|\widehat{H}_e}=a+\frac{2}{\hbar}\bs\cdot\boldsymbol{b}\,,
\end{align}
so that for two-level systems the EF system can be formulated as follows.
\begin{corollary}[Euler-Poincar{\'e} EF formulation for electronic two-level systems]\label{Cor:2LevelEP}
  \hfill\\
Specialising the the EF system to a two-level electronic system $\mathscr{H}_e=\mathbb{C}^2$, from Theorem \ref{Theorem:EFEPFull}  
   the Lagrangian \eqref{EFHydroL3} reduces to the mapping $\ell:\mathfrak{X}(\mathbb{R}^3)\times{\rm Den}(\mathbb{R}^3)\times\mathcal{F}(\mathbb{R}^3,\mathbb{R}^3 )\times\mathcal{F}(\mathbb{R}^3,\mathbb{R}^3)\to \mathbb{R}$ 
  given by
  \begin{align}
  \ell(\bu, D, \bxi, \bs) =  \int \bigg[\frac{1}{2}MD |\bu|^2-\frac{\hbar^2}{8M}\frac{|\nabla 
  D|^2}{D} + D\bigg(\bxi\cdot\bs - a-\frac{2}{\hbar}\bs\cdot\boldsymbol{b} - \frac{\|\nabla\bs\|^2}{2M}\bigg)\bigg]\de^3r
 \,. \label{EFHydroLag2Level}
\end{align}
In addition to the constrained fluid variations \eqref{EPvar} we also have that 
\begin{align}
  \delta \bxi = \partial_t\bUpsilon - \bw\cdot\nabla\bxi + \bu\cdot\nabla\bUpsilon 
  -\bxi\times\bUpsilon\,,\qquad \delta\bs = \bUpsilon\times \bs - 
  \bw\cdot\nabla\bs\,,\label{twolevelfluidvariations}
\end{align}
where $\bUpsilon \in \mathcal{F}(\mathbb{R}^3,\mathbb{R}^3)$ is obtained via the Lie algebra isomorphism \eqref{TILDEMAP} applied to $\Upsilon:=(\delta U\,U^{-1})\circ\eta^{-1}\in\mathfrak{su}(2)$.  
Then, applying Hamilton's principle for the arbitrary quantities $\bUpsilon(\br,t)$ and $\bw$ results in the system of equations
\begin{align}
\partial_t D + {\rm div}(D\bu)&= 0\,,\label{Kyle2}\\
  M(\partial_t+\bu\cdot\nabla)\bu&= -\nabla(V_Q+a) -\frac{2}{\hbar}\nabla\boldsymbol{b}\cdot\bs 
  - \frac{\partial_j(D\partial_j\bs\cdot\nabla\bs)}{MD}\,,\label{Nika}\\
  M(\partial_t+\bu\cdot\nabla)\bs &= \bs\times\left(-\frac{2M}{\hbar}\boldsymbol{b}+ 
 \frac{ \partial_j(D\partial_j\bs)}{D}\right)\label{Lola}\,.
\end{align}
\end{corollary}
\paragraph{Proof:} Firstly, both the new Lagrangian \eqref{EFHydroLag2Level} and variations \eqref{twolevelfluidvariations} 
are computed using the isomorphism \eqref{TILDEMAP} applied to their 
counterparts from Theorem \ref{Theorem:EFEPFull}. Then, following analogous steps to the proof of the more general Theorem, Hamilton's principle 
applied to a general Lagrangian of the type $\ell(\bu, D, \bxi, \bs)$ results in 
the following equations 
\begin{align}
   (\partial_t+\pounds_{\bu})\frac{\delta \ell}{\delta \bu} &= D\nabla\frac{\delta \ell}{\delta D} 
      - \frac{\delta \ell}{\delta \bxi}\cdot\nabla\bxi 
   - \frac{\delta \ell}{\delta \bs}\cdot\nabla\bs\,,\\
  (\partial_t+\pounds_{\bu})\frac{\delta \ell}{\delta {\bxi}} &= \bxi\times\frac{\delta \ell}{\delta {\bxi}}
  + \bs\times\frac{\delta \ell}{\delta \bs}\,,
\end{align}
as well as the auxilliary equations $(\partial_t+\pounds_{\bu})D=0$ and $(\partial_t+\pounds_{\bu})\bs = 
{\bxi}\times\bs$. Specialising to the Lagrangian \eqref{EFHydroLag2Level} simply 
requires the calculation of the variational derivatives
\begin{align*}
  &\frac{\delta \ell}{\delta \bu} = MD\bu\,,\qquad \frac{\delta \ell}{\delta D} = 
  \frac{1}{2}M|\bu|^2 - V_Q + \bxi\cdot\bs - a-\frac{2}{\hbar}\bs\cdot\boldsymbol{b} - 
  \frac{\|\nabla\bs\|^2}{2M}\,,\\
  &\frac{\delta \ell}{\delta \bxi} = D\bs\,,\qquad   \quad\,\,\,\frac{\delta \ell}{\delta \bs} = 
  D\bxi - \frac{2}{\hbar}D\boldsymbol{b} - M^{-1}\partial_j(D\partial_j\bs)\,,
\end{align*}
at which point one can again follow the analogous steps in the proof of Theorem 
\ref{Theorem:EFEPFull} to obtain the required equations \eqref{Nika} and \eqref{Lola}.
\hfill$\square$\\

Before continuing, we again acknowledge the presence of the real part of the QGT in the final term 
of \eqref{Nika}, now expressed for a two-level system as 
$T_{ij}=\hbar^{-2}\partial_i\bs\cdot\partial_j\bs$.

At this point, following the introduction of the matrix density $\tilde{\rho}:=D\rho$, we now introduce the analogous object in the two-level restriction, the spin density vector $\tilde{\bs}:=D\bs \in {\rm Den}(\mathbb{R}^3)\otimes\mathcal{F}(\mathbb{R}^3,\mathbb{R}^3)$.
Interestingly, this allows us to write the  counterpart of relation 
\eqref{nablarhocomputation}, now given by
\begin{align}
  \|\nabla\bs\|^2 = \|\nabla\left(D^{-1}\tilde\bs\right)\|^2 
  = \frac{\|\nabla\tilde\bs\|^2}{D^2}-\frac{\hbar^2}{4}\frac{|\nabla 
  D|^2}{D^2}\,,
\end{align}
so that the Lagrangian \eqref{EFHydroLag2Level} is now defined as the mapping $\ell:\mathfrak{X}(\mathbb{R}^3)\times{\rm Den}(\mathbb{R}^3)\times \mathcal{F}(\mathbb{R}^3, \mathbb{R}^3)\times {\rm Den}(\mathbb{R}^3)\otimes\mathcal{F}(\mathbb{R}^3,\mathbb{R}^3)\to \mathbb{R}$ 
given by
\begin{align}\label{EFHDbLagr}
  \ell(\bu, D, \boldsymbol{\xi},\tilde{\boldsymbol{s}} ) = \int\! \left[\frac{1}{2}MD|\bu|^2  
  -D a+ \tilde{\boldsymbol{s}}\cdot\Big(\boldsymbol\xi -\frac{2}{\hbar}\boldsymbol{b}\Big) 
  -\frac{\|\nabla\tilde{\boldsymbol{s}}\|^2}{2MD}\right] \text{d}^3r\,.
\end{align}
Here we notice that the quantum potential term has been absorbed in the Lagrangian so that it no longer appears 
explicitly. Upon computing the new variation
\begin{align}
  \delta \tilde\bs = \bUpsilon\times \tilde\bs - \text{div}(\bw\tilde\bs)\,, 
\end{align}
Hamilton's principle allows us to rewrite the complete system as
\begin{align}
  \partial_tD+\text{div}(D\bu)&= 0\,\label{Kyle}\\
  MD(\partial_t+\bu\cdot\nabla)\bu &=  - 
  D\nabla a-\frac{2}{\hbar}\nabla\boldsymbol{b}\cdot\tilde{\boldsymbol{s}}
+
  M^{-1}\partial_j(\tilde{\boldsymbol{s}}\cdot\nabla(D^{-1}\partial_j\tilde{\boldsymbol{s}}))\,,\label{EFb2}\\
 \hbar M(\partial_t\tilde{\boldsymbol{s}}  + \text{div}(\bu\tilde{\boldsymbol{s}} ))&= \tilde{\boldsymbol{s}} 
 \times \bigg(\hbar\text{div}(D^{-1}\nabla\tilde{\boldsymbol{s}} ) - {2M}\boldsymbol{b}\bigg) 
\label{EFb3}\,,
\end{align}
in which we notice that the quantum potential term correspondingly does not manifest explicitly at the level of the equations. 
Naturally, these equations possess the Lie-Poisson structure analogous to 
Theorem \ref{Theorem:LPEF}.

\begin{corollary}[EF Lie-Poisson structure for two-level electronic dynamics]\label{Cor:2LevelLP}
\hfill \\
In the case of electronic two-level systems the Lie-Poisson structure described in Theorem \ref{Theorem:LPEF} carries over, now defined on the dual of the Lie algebra 
$\mathfrak{g}=\mathfrak{X}(\Bbb{R}^3)\,\circledS\, \big(\mathcal{F}(\Bbb{R}^3)\oplus\, 
\mathcal{F}\big(\Bbb{R}^3,\Bbb{R}^3\big)\big)$.
The Lie-Poisson bracket \eqref{SDP-LPB-EF} becomes
\begin{align}
\begin{split}
\{f,g\}(\bm,D,\tilde\bs)=&\,\int\!\bm\cdot\left(\frac{\delta g}{\delta \bm}\cdot\nabla\frac{\delta f}{\delta \bm}-\frac{\delta f}{\delta \bm}\cdot\nabla\frac{\delta g}{\delta \bm}\right)\de ^3 r
\\
&\,
-
\int \!D\left(\frac{\delta f}{\delta \bm}\cdot\nabla\frac{\delta g}{\delta D}-\frac{\delta g}{\delta \bm}\cdot\nabla\frac{\delta f}{\delta D}\right)\de ^3 r
\\
&\,
-\int \tilde\bs\cdot\left(\frac{\delta f}{\delta \tilde\bs}\times\frac{\delta g}{\delta \tilde\bs}+\frac{\delta f}{\delta \bm}\cdot\nabla\frac{\delta g}{\delta \tilde\bs}-\frac{\delta g}{\delta \bm}\cdot\nabla\frac{\delta f}{\delta  \tilde\bs}\right)\de ^3 
r\,,
\label{LPB-Pauli}
\end{split}
\end{align}
accompanied by the new Hamiltonian
 \begin{align}
   h(D,\bm,\tilde\bs) &= \int \!\left(\frac{|\bm|^2}{2MD} + \frac{\|\nabla \tilde\bs\|^2}{2MD} + \frac{2}{\hbar}\boldsymbol{b}\cdot\tilde\bs +Da \right)
{\rm d}^3r\,.
\end{align}
\end{corollary}
\paragraph{Proof:}
Upon recalling the definition $\bm:=MD\bu$, the equations of motion \eqref{Kyle} - \eqref{EFb3}, 
follow from direct substitution, following the analogous computational steps from the proof of Theorem \ref{Theorem:LPEF}. 
\hfill$\square$\\

We conclude this section by discussing the relevance of the two-level 
restriction of EF dynamics. Indeed, equations \eqref{Kyle} - \eqref{EFb3} describe the closely related spin-boson \cite{LeggettEtAl1987} and Jahn-Teller systems \cite{OBCh93}, 
both of which can be expressed in the general form
\begin{align}
  a(\br)= \frac{1}{2}M\omega^2|\br|^2\,,\qquad \boldsymbol{b}(\br)= 
  \frac{1}{2}\begin{pmatrix}\boldsymbol{C}\cdot\br + E\\0\\ \boldsymbol{D}\cdot\br
  \end{pmatrix}\,,\label{spinbosontype}
\end{align}
where $\boldsymbol{C}$, $\boldsymbol{D}$ and $E$ are time-independent physical constants depending on the nature of the system in question.  All electronic two-level systems coupled to mutliple harmonic oscillators with $\boldsymbol{C}=0$ are known as spin-boson models 
\cite{LeggettEtAl1987} and are used to describe the dissipative 
dynamics of a two-level system coupled to an external bath. In the field of quantum chemistry this Hamiltonian with $E=0$ also captures the Jahn-Teller effect \cite{OBCh93} coupling the electronic two-level system with a
harmonic oscillator model of the nucleus. Specifically, for a single nuclear 
mode (with $\boldsymbol{C}=(0,0,0)^T$, $\boldsymbol{D}=(0,D,0)^T\,$) one has the $E\otimes \beta$ Jahn-Teller effect, 
whilst the extension to two nuclear
coordinates (with $\boldsymbol{C}=(C,0,0)^T$, $\boldsymbol{D}=(0,D,0)^T\,$) captures the $E\otimes \epsilon$ Jahn-Teller effect, which has been of particular interest in quantum chemistry due to the well-known manifestation of geometric phase effects \cite{RequistEtAl2016,RequistEtAl17}.
More generally two-level Hamiltonians similar to \eqref{spinbosontype} describe a whole host of physical phenomena, for example including the Jaynes-Cummings and quantum Rabi models. 

Remarkably, as shown in the author's publication \cite{FoTr2020}, the two-level 
specialisation of exact factorization presented in this section also captures the hydrodynamic interpretation of the Pauli 
equation \cite{BohmSchillerB1955,BohmSchillerTiomnoA1955, Takabayasi1955}, first 
conceived in 1955. This will be considered in detail in Chapter \ref{Chap:Gamma} and demonstrates
an advantage of the geometric approach, here unifying concepts all the way from nonadiabatic quantum chemistry to 
foundational quantum theory.  

Having fully examined the geometry associated to the original exact 
factorization \eqref{GenExFact} of molecular wavefunction, in the next 
section we extend this concept to the level of the molecular density matrix, providing a new generalised
method of describing mixed state nuclear dynamics. 

\section{Density operator factorisation and singular solutions}\label{Bohmions}
Our discussion in Section \ref{Berry-frequency} shows that the only contribution to the circulation of the hydrodynamic flow arises from the Berry connection associated to the electronic function $\psi$. This is due to the fact that the hydrodynamic velocity is written as $M^{-1}\bmu/D=\hbar\operatorname{Im}(\Omega^*\nabla\Omega)/|\Omega|^{2} = M^{-1}\nabla S$ and is therefore an exact differential according to the standard Madelung transform and therefore has zero vorticity.
Whilst we will consider an alternative approach to introducing vorticity in QHD 
via the theory of gauge connections in Chapter \ref{Chap:Holonomy}, here we 
use the methods developed in Section \ref{Sec:2-ColdFluid} in which the zero-vorticity restriction is relaxed by considering density operators.
 Hence, in this section we shall include mixed state dynamics by extending the exact factorization model to a density operator 
 formulation, before implementing the regularisation approach from Section \ref{Sec:2-Bohmions} 
 to allow for finite-dimensional Bohmion singular solutions.

\subsection{New factorisation of the molecular density operator}
In order to generalise the exact factorization \eqref{GenExFact} to density operators, we recall the relation \eqref{StraightEFMolDensity} and extend it to consider a molecular density operator of the form
\begin{align}\label{DMF}
\widehat{\rho}(\br,\br')=\rho_n(\br,\br')\psi(\br)\psi^\dagger(\br')
\,,
\end{align}
for an arbitrary nuclear density matrix $\rho_n$. Here, the `hat' notation on the molecular density matrix refers to its dependence on the electronic degrees of freedom whilst being expressed in terms of the nuclear matrix elements.
The advantage of this new factorisation is that the nuclear factor now accounts 
for more general mixed states via the standard expansion \eqref{rhomixture}
\begin{align}
  \rho_n(\br,\br')=\sum_kw_k\Omega_k(\br)\Omega^*_k(\br')\,,\label{nuclearrhomixture}
\end{align}
reducing to the standard EF of the molecular density matrix 
\eqref{StraightEFMolDensity} at the lowest order truncation.

Naturally, the evolution of the total molecular density matrix is governed by a von Neumann equation \eqref{LvN} featuring the molecular Hamiltonian operator 
\eqref{molecularHamiltonian2}, which, as we saw in Section \ref{Sec:1-GeomQM}, can be derived from a reduced Euler-Poincar{\'e} Lagrangian of the type \eqref{DFLagrDensityMatrix}.
For the molecular density operator this Lagrangian is the mapping $L:{\rm Her}(\mathscr{H})\times\mathfrak{u}(\mathscr{H})\to\mathbb{R}$, where $\mathscr{H}$ denotes the total molecular Hilbert space, and is given by
\begin{align}
L(\,\widehat{\rho}\,, \,\widehat{\xi}\,)= \text{Re}\iint  
  \left\langle 
  {\widehat{\rho}(\br',\br)\Big|i\hbar\widehat{\xi}(\br,\br')-\widehat{H}(\br,\br')}\right\rangle\,\de^3r\,\de^3r'\,.\label{NASA}
\end{align}
Here, we have used the unitary evolution of the molecular density matrix $\widehat\rho(t)=U(t)\widehat\rho(0)U^{-1}(t)$ 
in order to define $\widehat{\xi}:=\partial_t U U^{-1}$ and we understand the angled bracket as the inner product on the space of operators on the electronic Hilbert space, defined by $\braket{\widehat{A}|\widehat{B}}=\text{Tr}(\widehat{A}^{\dagger}\widehat{B})$.

Having characterised the Euler-Poincar{\'e} Lagrangian for the total molecular 
density operator, we now write a variational formulation for the dynamics associated to 
the new factorisation \eqref{DMF}.

\begin{theorem}[Variational formulation for the density matrix factorisation \eqref{DMF}]\label{Theorem:NEWEFDENANSATZFORM}
\hfill \\ Consider the factorisation of the molecular density matrix 
\eqref{DMF} and let $\rho_n= 
U_n\rho_n^{0}U_n^{-1}$,
where $U_n\in\mathcal{U}(\mathscr{H}_n)$ is a time-dependent unitary operator on 
$\mathscr{H}_n=L^2(\Bbb{R}^3)$. Then, the factorisation \eqref{DMF} transforms the Lagrangian for the full molecular density matrix
\eqref{NASA}  into the form
$\ell:{\rm Her}(\mathscr{H}_n)\times\mathfrak{u}(\mathscr{H}_n)\times T\mathcal{F}(\mathbb{R}^3,\mathscr{H}_e)\times\mathcal{F}(\mathbb{R}^3)\to \mathbb{R}$ given by 
\begin{multline}
\ell(\rho_n,{\xi}_n,\psi,\partial_t\psi, \lambda)={\rm Re}\int\!\Big[ 
i\hbar\int\rho_n(\br',\br)\xi_n(\br,\br')\,\de^3r'
+D(\br)\langle\psi|i\hbar\partial_t{\psi}\rangle\\ + \lambda\big(\|\psi\|^2 - 1\big) \Big]\,\de^3 r
-h(\rho_n,\psi) 
\,,
\label{mixednuclearlagrangian}
\end{multline}
where ${\xi}_n:=\partial_t{U}_nU_n^{-1}$ and
the total energy is given by
\begin{align}
  h(\rho_n,\psi) &= \frac{1}{2M}\,\left\langle(\widehat{P}_n+ 
  \widehat{\cal A}\,)^2\Big|\rho_n\right\rangle_{n}+\int\! D(\br)\epsilon(\psi,\nabla\psi)
\,\de^3r\,. \label{EF-HamNuclearRho}
\end{align}
Here, we have denoted the diagonal elements by $D(\br):=\rho_n(\br,\br)$, $\epsilon(\psi,\nabla\psi)$ is the same as in \eqref{EFepsilonDEF}, $\widehat{P}_n$ denotes the nuclear momentum operator, $\widehat{\cal A}(\br,\br')=\boldsymbol{\cal A}(\br)\delta(\br-\br')$, and the angled brackets $\langle\,\cdot\,|\,\cdot\,\rangle_{n}$ denote the inner product of operators on the nuclear Hilbert space $\mathscr{H}_n$ 
expressed in terms of matrix elements as
\begin{align*}
  \left\langle \widehat{A}\,\Big|\widehat{B} \right\rangle_{n}:= \iint 
  \widehat{A}^*(\br,\br')\widehat{B}(\br,\br')\,\de^3\,\de^3r'\,.
\end{align*}
Then, applying Hamilton's principle for arbitrary variations $\delta\psi$ and 
constrained variations
\begin{align}
  \delta \rho_n = [\Upsilon_n, \rho_n]\,,\qquad \delta\xi_n = \partial_t 
  \Upsilon_n + [\Upsilon_n,\xi_n]\,,
\end{align}
where $\Upsilon_n:= \delta {U}_nU_n^{-1}$ is arbitrary, results in the equations 
of motion
\begin{align}
&i\hbar\partial_t\rho_n(\br,\br') = \left[\frac{\delta h}{\delta \rho_n},\rho_n\right](\br,\br') - 
\rho_n(\br,\br')\Big(\braket{\psi|i\hbar\partial_t\psi}\!(\br)-\braket{\psi|i\hbar\partial_t\psi}\!(\br')\Big)\,,\label{nucleardensitymatrixequation}\\
&(\mathbbm{1}-\psi\psi^{\dagger})\left(i\hbar\partial_t\psi - \frac{1}{2D}\frac{\delta h}{\delta 
\psi}\right)=0\,,\label{electronicprojectiveschrodingerAGAIN}\\
&\|\psi\|^2=1\,,\label{PNCAGAIN}
\end{align}
subject to computing the derivatives of the Hamiltonian \eqref{EF-HamNuclearRho}.
\end{theorem}

\paragraph{Proof:} We begin by deriving the reduced Lagrangian \eqref{mixednuclearlagrangian}.  
We do so by considering a sequence of molecular pure states $\{\Psi_k(\br)\}$ and again exploit the general expression of a density matrix as a sum over pure states \eqref{rhomixture} to write
\begin{align}
\widehat{\rho}(\br,\br')=\sum_{k}w_k\Psi_k(\br)\Psi_k^{\dagger}(\br')
\,,
\label{mixture}
\end{align}
where $\sum_k w_k = 1$ and each $\Psi_k$ satisfies a separate (uncoupled) Schr\"odinger equation with Hamiltonian \eqref{molecularHamiltonian2}. 
We then invoke the unitary evolution of each state $\Psi_k(t)=U(t)\Psi_k(0)$ 
in order to define $\widehat{\xi}:=\partial_t U U^{-1}$ so that $\partial_t\Psi_k = 
\widehat{\xi}\Psi_k$. Then, rewriting the Lagrangian \eqref{NASA} we obtain
\begin{align}
L\left(\{\Psi_k\},\{\partial_t\Psi_k\}\right)=\sum_{k} w_k\,\text{Re}\int  
  \left\langle {\Psi_k(\br)\Big|i\hbar\partial_t\Psi_k(\br)-(\widehat{H}\Psi_k)(\br)}\right\rangle\,\de^3r
\,.
\label{MixtureLag1}
\end{align}
At this stage, we consider an electronic state $\psi(\br) \in \mathcal{F}(\mathbb{R}^3, \mathscr{H}_e)$ satisfying the partial normalisation $\|\psi\|^2=1$, and we restrict to  the case
$
\Psi_k(\br)=\Omega_k(\br)\psi(\br)$.
Consequently, the above Lagrangian becomes
\begin{align}
\begin{split}
L\left(\{\Omega_k\},\{\partial_t\Omega_k\},\psi,\partial_t\psi,\lambda\right) =\text{Re}\int\! \bigg[\sum_kw_k\Big(i\hbar\Omega_k^*&\partial_t{\Omega}_k+|\Omega_k|^2\langle\psi|i\hbar\partial_t{\psi}\rangle\Big) 
\\
&+ \lambda\big(\|\psi\|^2 - 1\big) \bigg]
\,\text{d}^3r
-h(\{\Omega_k\},\psi) 
\,.
\label{MixtureLag2}
\end{split}
\end{align}
Here, the Lagrange multiplier enforces $\|\psi\|^2=1$ and the Hamiltonian reads as
$
h(\{\Omega_k\},\psi) =\sum_{k}w_k\int\!\Omega_k^*(\br)\braket{\psi(\br)|\widehat{H}(\Omega_k\psi)(\br)}\,\de^3 r
$.
Then, the new factorisation ansatz \eqref{DMF} is recovered by recalling 
the nuclear mixture \eqref{nuclearrhomixture} and upon restricting $\rho_n$ to undergo its own unitary evolution by writing, for each $k$,
$
\Omega_k(\br,t)=(U_n(t)\Omega_k^{0})(\br)
$, so that $\partial_t\Omega_k = \xi_n\Omega_k$ the Lagrangian can eventually be rewritten to return \eqref{mixednuclearlagrangian}.  
The corresponding equations of motion \eqref{nucleardensitymatrixequation} - \eqref{PNCAGAIN} follow as the Euler-Lagrange 
equations for the electronic state $\psi$ and Lagrange multiplier $\lambda$ along with the Euler-Poincar{\'e} equation 
\eqref{EPdensitymatrixequation} for the nuclear generator $\xi_n$ and the auxilliary equation $\partial_t\rho_n = [\xi_n,\rho_n]$ for nuclear density matrix $\rho_n$ (both introduced in general in Section 
\ref{Sec:1-GeomQM}). The explicit computation follows much the same as the proof 
of Theorem \ref{Theorem:EFEL}.
\hfill$\square$

\begin{remark}[Comparison to the original EF]
  \hfill\\
  The equations of motion in Theorem \ref{Theorem:NEWEFDENANSATZFORM} corresponding to the new EF density matrix ansatz \eqref{DMF} look remarkably similar to those from the standard EF of the molecular wavefunction given in Theorem \ref{Theorem:EFEL}.
  Indeed, the dynamics for $\rho_n$ \eqref{nucleardensitymatrixequation} are identical to those obtained for the pure state $\Omega(\br)\Omega^*(\br')$ emerging from the equation \eqref{EFOmegaequation}. 
  However, in this generalised case, the compatibility condition \eqref{EFELcondition} is now replaced by
\begin{align}
{\rm Im}\left\langle \psi(\br) \bigg| \frac{\delta h}{\delta \psi}(\br) \right\rangle
= 
2\,{\rm Im}\int\!\rho_n(\br',\br)\frac{\delta h}{\delta \rho_n}(\br,\br')\,\de^3 r'.
\end{align}
\end{remark}

Now that the variational principle and corresponding dynamics are completely characterised, we may proceed by restricting nuclear dynamics to undergo classical motion.
 To this purpose, we combine the two approaches described in Sections \ref{Sec:2-Bohmions} and \ref{Sec:2-ColdFluid}, firstly performing the classical closure for the nuclear density operator before applying the regularisation to the resulting Lagrangian.

\subsection{Classical closure}\label{sec:EFCFClosure}

In following the discussion in Section \ref{Sec:2-ColdFluid}, we wish to collectivise the Hamiltonian in terms of the hydrodynamic quantities 
$(\bmu, D)$. This can be achieved by applying the closure \eqref{coldfluid} to $
\rho_n$ in equation \eqref{DMF}, that is 
\begin{align}
\rho_n(\br,\br')=D\left(\frac{\br+\br'}{2}\right)\,\text{exp}\left[i\frac{M}{\hbar}(\br-\br')\cdot \bv\left(\frac{\br+\br'}{2}\right) \right]\label{coldfluidEF}
\,,
\end{align}
with $MD(\br)\bv(\br)=\bmu(\br)$. 
Lengthy but straightforward computations using matrix elements (see Appendix \ref{App:ColdFluid}) eventually take the Hamiltonian \eqref{EF-HamNuclearRho} into the hydrodynamic form
\begin{align}
  h(\bmu,D,\psi) &=
\int \frac{|\bmu+ 
  D\boldsymbol{\cal A}|^2}{2MD} \,\de^3 r+\int\! D\epsilon(\psi,\nabla\psi)
\,\text{d}^3r
\,, \label{EFMixedHydroHamiltonian}
\end{align}
which formally coincides with the Hamiltonian obtained in the wavefunction case \eqref{EFHydroHamiltonian}, except for the quantum potential term. 
Thus, upon performing the reduced Legendre transform, the resulting dynamics take the same form as given in Theorem \ref{Theorem:EFEPEL} from Chapter \ref{Chap:EFOLD}, simply without the contributions from the quantum potential.
Now however, note that the nuclear hydrodynamic variables $(\bu,D)$ are no longer given in terms of a unique wavefunction, so that $M^{-1}\nabla\times(\bmu/D)\neq0$ as can be seen by recalling equation \eqref{FluidVelocityMixture}.
Correspondingly, we have the modified circulation theorem:
\begin{remark}[Modified Kelvin-Noether circulation theorem]
\hfill \\ 
 If the fluid equation \eqref{EFEPELu2} is modified as described above (i.e. without the quantum potential and with an additional non-vanishing vorticity contribution $M^{-1}\nabla\times(\bmu/D)\neq0$)
 the circulation theorem \eqref{Berry-Frequency} is correspondingly modified to read
    \begin{align}
       \frac{\de}{\de t}\oint_{c(t)} \boldsymbol{u} \cdot \de\boldsymbol{r} 
       = - \frac{1}{M} \oint_{c(t)} \Big( \langle\psi,(\partial_i\widehat{H}_e) \psi\rangle 
            + \frac1{MD} \partial_j \big(DT_{ij}\big) \Big) \de r^i 
    \,.
    \label{Berry-Frequency1}
    \end{align}
This means that the dynamics of the nuclear circulation integral $\oint_{c(t)} \boldsymbol{u} \cdot \de\boldsymbol{r}$ is now interpreted as a genuine hydrodynamic Kelvin theorem for the circulation around a loop moving with the nuclear fluid with Eulerian velocity $\boldsymbol{u}:=M^{-1}(\bmu/D+\boldsymbol{\cal A})$. 
\end{remark}

At this point, one follows the treatment in Section \ref{NuclearFrameSection} to express the electron function $\psi$ in the nuclear frame. 
Hence, we arrive at the Lagrangian $\ell:\mathfrak{X}(\mathbb{R}^3)\times{\rm Den}(\mathbb{R}^3)\times\mathcal{F}(\mathbb{R}^3,\mathfrak{u}(\mathscr{H}_e))\times \mathcal{F}(\mathbb{R}^3,{\rm Her}(\mathscr{H}_e))\otimes{\rm Den}(\mathbb{R}^3)\to \mathbb{R}$ 
 given by
\begin{align}
  \ell(\bu, D, \xi, \tilde\rho) &=  \int \bigg(\frac{1}{2}MD |\bu|^2+ \braket{\tilde\rho,i\hbar\xi- \widehat{H}_e} - 
  \frac{\hbar^2}{4MD}\left({\|\nabla{\tilde\rho}\|^2}-{| \nabla D|^2} \right)\bigg)\,\de^3r\,,
  \label{EFMixedHydroLagrangian}
\end{align}
which again coincides with \eqref{EFHydroL3} except for the quantum potential, although a similar potential (the last term) with opposite sign is produced  according to the relation \eqref{nablarhocomputation}.
The corresponding  equations of motion can now be formulated equivalently by the 
Euler-Poincar{\'e} approach as given in Theorem \ref{Theorem:EFEPFull} (with suitable modifications to accomodate the new variable $\tilde\rho=D\rho$) or the 
Lie-Poisson structure of Theorem \ref{Theorem:LPEF} associated to the modified Hamiltonian 
\begin{align}
    h(\bm,D,\tilde{\rho}) &=
\int \left(\frac{|\bm|^2}{2MD}+ \braket{\tilde\rho|\widehat{H}_e} + \frac{\hbar^2}{4MD}\left({\|\nabla\tilde\rho\|^2}-{|\nabla D|^2}\right)
\right)\,\text{d}^3r 
\,. \label{EFMixedHydroHamiltonian2}
\end{align}
In either case, the electronic von Neumann and fluid transport equations do not change, although the force arising from the quantum potential is modified accordingly in the fluid Euler-type equation.

\subsection{New Bohmion closure for nonadiabatic dynamics}\label{sec:BohmionsEF}
We complete the new material presented in this chapter by performing the analogous procedure to that in Section \ref{Sec:2-Bohmions}, now for the above model obtained by factorising the molecular density matrix. 
As before, we {regularise} only the $O(\hbar^2)-$terms on the Lagrangian side.  
However, we notice that, in addition to the last term in the Lagrangian \eqref{EFMixedHydroLagrangian}, which was already regularised in Section \ref{Sec:2-Bohmions}, a further barrier to singular solutions is represented by the  term involving the gradient of $\tilde{\rho}$.
 Hence, we must introduce smoothened version of both of these variables by replacing 
 \begin{align}
   \bar{D}:=K*D = \int K(\br-\bs)D(\bs,t)\,\de^3s\,,\qquad\bar{\rho}:=K*\tilde\rho = \int K(\br-\bs)\tilde{\rho}(\bs,t)\,\de^3s\,. \label{EFSmoothenedvariables}
 \end{align}
 in the $O(\hbar^2)-$terms, so that the Lagrangian now becomes 
 \begin{align}
  \ell(\bu, D, \xi, \tilde\rho) &=  \int \bigg(\frac{1}{2}MD |\bu|^2+ \braket{\tilde\rho,i\hbar\xi-\widehat{H}_e} - 
  \frac{\hbar^2}{4M\bar{D}}\left({\|\nabla{\bar\rho}\|^2}-{| \nabla \bar{D}|^2} \right)\bigg)\,\de^3r\, .
  \label{EFMixedHydroLagrangian2REG}
\end{align}

As before, the advantage of having smoothened these terms in particular now 
comes into play, resulting in the following theorem.
\begin{theorem}[Bohmion singular solutions for regularised EF dynamics]\label{Theorem:CRAZY}
 \hfill \\
  The regularised Lagrangian \eqref{EFMixedHydroLagrangian2REG} admits the 
  `Bohmion' singular solutions
  \begin{align}
  D(\br,t)=\sum_{a=1}^{\mathcal{N}}w_a\delta(\br-\bq_a(t))\,,\qquad
    \tilde{\rho}(\br,t)=\sum_{a=1}^{\mathcal{N}}\varrho_a(t)\delta(\br-\bq_a(t))\,,\label{Andrew}
\end{align}
thus transforming the Lagrangian into the form $L:T\mathbb{R}^{3\mathcal{N}}\times{\rm Her}(\mathscr{H}_e)^{\mathcal{N}}\times\mathfrak{u}(\mathscr{H}_e)^{\mathcal{N}}\to\mathbb{R}$ 
given by
\begin{align}\label{SurelyNearlyThere}
\begin{split}
L(\{\bq_a\},\{\dot{\bq}_a\},\{\varrho_a\},&\{\xi_a\})  =\sum_{a} \Bigg(\frac{Mw_a}2|\dot{\bq}_a|^2+\braket{\varrho_a,i\hbar\xi_a-\widehat{H}_e(\bq_a)}
  \\
  &- \frac{\hbar^2}{4M} \sum_{b} \Big(\langle\varrho_a |\varrho_b\rangle-w_aw_b\Big)  { \int 
  \frac{\nabla K(\br-\bq_a)\cdot \nabla K(\br-\bq_b)}
  {\sum_{c} w_c K(\br-\bq_c)} \,\de^3r }
  \Bigg)
 \,,
\end{split}
\end{align}
where $\xi_a(t):=\big(\partial_tU(\bq_a^{(0)},t)\big)U^\dagger(\bq_a^{(0)},t)$, $\varrho_a(t):=U(\bq_a^{(0)},t)\varrho_a^{(0)}U^\dagger(\bq_a^{(0)},t)$
and we recall that  $\bq_a^{(0)}=\eta^{-1}(\bq(t),t)$. Then, applying Hamilton's principle for arbitrary variations $\delta \bq_a$ and contrained Euler-Poincar{\'e} variations $
\delta \xi_a = \partial_t \Upsilon_a - [\xi_a,\Upsilon_a]
$, 
with $\Upsilon_a :=\big(\delta U(\bq_a^{(0)},t)\big)U^\dagger(\bq_a^{(0)},t)$ arbitrary and vanishing at the 
endpoints, results in the following mixed
quantum-classical system of equations 
\begin{align}
\begin{split}
Mw_a\ddot{\bq}_a &= -\nabla_{\bq_a}\Bigg(\braket{\varrho_a,i\hbar\xi_a-\widehat{H}_e(\bq_a)}
 \\
 &\qquad\qquad\quad - \frac{\hbar^2}{4M} \sum_{b} \Big(\langle\varrho_a |\varrho_b\rangle-w_aw_b\Big)  { \int 
  \frac{\nabla K(\br-\bq_a)\cdot \nabla K(\br-\bq_b)}
  {\sum_{c} w_c K(\br-\bq_c)} \,\de^3r }
  \Bigg)\,,\label{EFBohmionQequations}
  \end{split}\\
  \nonumber \\
i\hbar\dot{\varrho}_a &= 
\big[\widehat{H}_e(q_a),{\varrho}_a\big]
+
\frac{\hbar^2}{4M} \sum_{b} \left[{\varrho_b},{\varrho}_a\right]  {   
\int 
  \frac{\nabla K(\br-\bq_a)\cdot \nabla K(\br-\bq_b)}
  {\sum_{c} w_c K(\br-\bq_c)} \,\de^3r 
}\,.\label{EFBohmionElectronicequations}
\end{align}
\end{theorem}
\paragraph{Proof:} To begin, we comment that the Bohmion solution for $D(\br,t)$ in 
\eqref{Andrew}, as explained in Section \ref{Sec:2-Bohmions}, amounts to selecting the intial density 
$ D_0(\br_0)=\sum_{a=1}^{\mathcal{N}}w_a\delta(\br_0-\bq_a(0))$. Similarly, the form of the electronic 
density matrix $\tilde{\rho}(\br,t)$ in \eqref{Andrew} also comes from selecting an inital condition of the same type
\begin{align}
  \tilde{\rho}_0(\br_0)&=\sum_{a=1}^{\mathcal{N}}\varrho_a(0)\delta(\br_0-\bq_a(0))\,,\label{rho-ini}
\end{align}
so that the evolution of $\tilde\rho$ in terms of the Lagrangian path $\eta$ and the electronic propagator $U(\br_0)$ is given as 
$
\tilde\rho=\eta_*\hat\rho=
\int\! \hat\rho(\br_0,t)\,\delta(\br-{\eta}(\br_0,t))\,\de ^3r_0
$,
{with} 
$
\hat\rho(\br_0,t)=U(\br_0,t)\tilde\rho_0(\br_0)U^\dagger(\br_0,t)
$. In turn, this implies the evolution of the electronic density coefficients, 
$\varrho_a(t):=U(\bq_a^{(0)},t)\varrho_a^{(0)}U^\dagger(\bq_a^{(0)},t)$, and
we set $\varrho_a^{(0)}=\varphi_a^{\scriptscriptstyle (0)}{\varphi_a^{\scriptscriptstyle (0)}}^\dagger$ so that $\varrho_a(t)=\varphi_a(t)\varphi_a^\dagger(t)$ is a projection at all times.

This allows us to evaluate
\[
\int\langle\tilde\rho|i\hbar\xi\rangle\,\de^3 r=
\sum_{a=1}^{\mathcal{N}}\langle\varrho_a|i\hbar\xi_a\rangle
\,,\qquad
\text{where}
\qquad
\xi_a(t):=\big(\partial_tU(\bq_a^{(0)},t)\big)U^\dagger(\bq_a^{(0)},t)
\,,
\]
and we have recalled  $\bq_a^{(0)}=\eta^{-1}(\bq(t),t)$ as well as
$
\xi(\br)=\big(\partial_tU(\eta^{-1}(\br),t)\big)U^\dagger(\eta^{-1}(\br),t)
$.
Then, upon combing the Bohmion solutions \eqref{Andrew} with the regularised quantities \eqref{EFSmoothenedvariables} to obtain
\begin{align*}
   \bar{D}(\br,t)=\sum_{a=1}^{\mathcal{N}}w_a K(\br-\bq_a(t))\,,\qquad
    \bar{\rho}(\br,t)=\sum_{a=1}^{\mathcal{N}}\varrho_a(t) K(\br-\bq_a(t))\,,
\end{align*}
the Lagrangian \eqref{EFMixedHydroLagrangian} is transformed into \eqref{EFMixedHydroLagrangian2REG} 
as required.
While the Euler-Lagrange equations \eqref{EFBohmionQequations} for the trajectories $\bq_a$ are obvious, the electronic dynamics are obtained as Euler-Poincar\'e equations derived from the variational relation
$
\delta \xi_a = \partial_t \Upsilon_a - [\xi_a,\Upsilon_a]
$, 
with $\Upsilon_a :=\big(\delta U(\bq_a^{(0)},t)\big)U^\dagger(\bq_a^{(0)},t)$ arbitrary and vanishing at the endpoints.
 As usual, this is obtained by an explicit calculation using the definition of $\xi_a(t)$, resulting in a set of general equations 
 of the type \eqref{EPdensitymatrixequation} described in Section 
 \ref{Sec:1-GeomQM}. Upon specialising to the Lagrangian \eqref{EFMixedHydroLagrangian2REG} 
 one regains the electronic density coefficient equations \eqref{EFBohmionElectronicequations}.
\hfill$\square$\\

The mixed quantum-classical system of equations \eqref{EFBohmionQequations} and \eqref{EFBohmionElectronicequations} in {Theorem \ref{Theorem:CRAZY}} arising from the new generalisation of EF dynamics is one of the key results relating to applications in quantum chemistry presented in this thesis.
This system is particularly advantageous as it comprises a countably infinite set of equations coupling finite-dimensional nuclear trajectories with an electronic density matrix
evaluated on the nuclear trajectories, thus potentially lifting the so-called `curse 
of dimensions' faced by many-body problems. In addition, the terms proportional to $\hbar^2$ provide extensive, potentially long-range coupling among the singular particle-like solutions due to the presence of $\bar{D}$ in their denominators. 
whilst the limit $\hbar^2\to0$ in these equations is now regular (using the regularisation strategy presented in Chapter \ref{Chap:QHD}).
For future developments, it would now be of interest to examine the solution behaviour of this regularised Bohmion model in the density matrix formulation using computer simulations.

We conclude this chapter with a summary connecting the various geometric formulations of the dynamics corresponding to the original exact factorisation 
ansatz \eqref{GenExFact}. 

Chapter \ref{Chap:EFOLD} focused on the variational point of view, starting with a fully Euler-Lagrange set of equations for both nuclear and electronic dynamics as derived from a Lagrangian $L:T\mathscr{H}_n\times T\mathcal{F}(\mathbb{R}^3,\mathscr{H}_e)\times\mathcal{F}(\mathbb{R}^3)\to 
\mathbb{R}$ in Theorem \ref{Theorem:EFEL}. 
Then, after rewriting the Hamiltonian in terms of the QHD momentum map variables for 
the nuclear dynamics, we used the partial Legendre transform to write a new variational principle in Theorem \ref{Theorem:EFEPEL}. This featured Euler-Poincar{\'e} equations for the nuclear dynamics and Euler-Lagrange equations for the electronic dynamics
as derived from a Lagrangian of the type $\ell:\mathfrak{X}(\mathbb{R}^3)\times{\rm Den}(\mathbb{R}^3)\times T\mathcal{F}(\mathbb{R}^3,\mathscr{H}_e)\times\mathcal{F}(\mathbb{R}^3)\to 
\mathbb{R}$. The resulting equations were then rewritten in an alternative form in Proposition \ref{Prop:EFNEWSET}, highlighting the presence of the QGT in the nuclear fluid equation.

Chapter \ref{Chap:EFNEW} took the EF system beyond the scope of existing 
literature by considering the unitary dynamics of the electronic motion, which, using the advantages of the geometric approach
were expressed in terms of the density operator in the nuclear hydrodynamic frame. This resulted in an entirely new Euler-Poincar{\'e} variational approach for a Lagrangian $\ell:\mathfrak{X}(\mathbb{R}^3)\times{\rm Den}(\mathbb{R}^3)\times\mathcal{F}(\mathbb{R}^3, \mathfrak{u}(\mathscr{H}_e))\times\mathcal{F}(\mathbb{R}^3, \mathscr{H}_e)\to \mathbb{R}$  given in Theorem \ref{Theorem:EFEPFull}. 
Interestingly the same geometric structure had previously been discovered in the geometric approach to liquid crystal 
dynamics \cite{GBRaTr2012,GBRaTr2013}. The corresponding Hamiltonian structure was then expressed in 
Theorem \ref{Theorem:LPEF} which was shown to be Lie-Poisson on the dual of $\mathfrak{X}(\Bbb{R}^3)\,\circledS\, \big(\mathcal{F}(\Bbb{R}^3)\oplus\, \mathcal{F}\big(\Bbb{R}^3,\mathfrak{u}(\mathscr{H}_e)\big)\big)
$.\\
\newline

The sequence of mappings from which these various formulations of EF dynamics have been obtained are compactly expressed in the diagram 
below.


\vspace{1cm}
\begin{figure}[h]
\footnotesize\center
\noindent
\begin{xy}
\hspace{0cm}
\xymatrix{
*+[F-:<3pt>]{
\begin{array}{c}
\vspace{0.1cm}\text{Theorem \ref{Theorem:EFEL} }\\
\vspace{0.1cm}\text{Euler-Lagrange}\\
\vspace{0.1cm} L(\Omega, \partial_t\Omega, \psi, \partial_t\psi, \lambda)\\
\end{array}
} 
\ar[rrrrrdd]|{\begin{array}{c}\text{QHD momentum map}\\
\text{for nuclear factor}\\
D=|\Omega|^2\,, \,\bmu = \hbar\,\text{Im}(\Omega^*\nabla\Omega)\\
\end{array}}
&  & &&&\\
&  & &&&  \\
&  & &&&*+[F-:<3pt>]{
\begin{array}{c}
\vspace{0.1cm}\text{Theorem \ref{Theorem:EFEPEL} }\\
\vspace{0.1cm}\text{Nuclear Euler-Poincar{\'e}}\\
\vspace{0.1cm} \ell(\bu, D, \psi, \partial_t\psi, \lambda)\\
\end{array}}
\ar[llllldd]|{\begin{array}{c}\text{Unitary electronic evolution}\\
\text{in the nuclear QHD frame}\\
 \rho = (U\rho_0U^{-1})\circ\eta^{-1}\\
 \xi = (\partial_t U\,U^{-1})\circ\eta^{-1}\\
 \rho = \psi\psi^{\dagger}
\end{array}}\\
&  & &&&\\
*+[F-:<3pt>]{\begin{array}{c}
\vspace{0.1cm}\text{Theorem \ref{Theorem:EFEPFull}}\\
\vspace{0.1cm}\text{Euler-Poincar{\'e}}\\
\vspace{0.1cm}\ell(\bu, D, \xi, \rho)\\
\end{array}}\ar[rrrrrd]|{\begin{array}{c}\text{Hamiltonian variables}\\
\bm = MD\bu\,,\,\tilde\rho=D\rho
\end{array}}
&  & &&&\\
&  & &&&
*+[F-:<3pt>]{
\begin{array}{c}
\vspace{0.1cm}\text{Theorem \ref{Theorem:LPEF}}\\
\vspace{0.1cm} \text{Lie-Poisson}\\ 
\vspace{0.1cm} h(\bm, D, \tilde\rho)\\
\end{array}
}}
\end{xy}
\vspace{.5cm}
\caption{Schematic description of the various geometric formulations of EF dynamics.}
\label{figure2}
\end{figure}
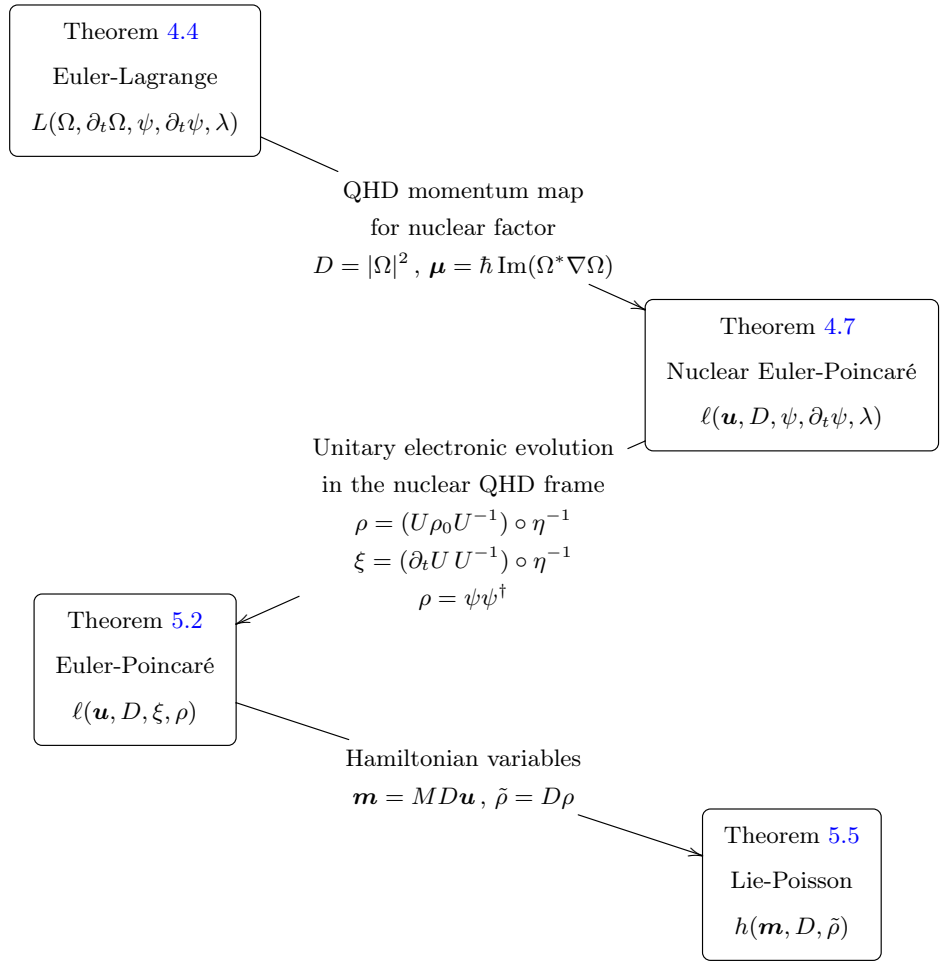

\chapter{Holonomy in quantum hydrodynamics}\label{Chap:Holonomy}

Having previously investigated the geometry of QHD using momentum maps in Chapter \ref{Chap:QHD},
 in this chapter we present an alternative formulation of QHD in terms of dynamical gauge connections following the work of the author's recent publication \cite{FoTr2020}.
 This provides new insights into the role of holonomy in QHD. 

Section \ref{Sec:NEWphase} presents the new formulation of QHD in which a phase connection is the fundamental dynamical variable. This allows for an 
entirely new alternative Euler-Poincar{\'e} reduction procedure for QHD and provides a method for understanding holonomy in this context. In the hydrodynamic picture this approach no longer {constrains} the fluid flow to be irrotational, with 
constant vorticity arising through the non-zero curvature of the connection. In 
addition we see how, upon reconstructing the corresponding Schr\"{o}dinger 
equation, the new connection term (producing non-zero holonomy) appears via minimal coupling as a 
type of vector potential.

Section \ref{Sec:Vortices} considers the possibilty of hydrodynamic vortices in this new setting. Vortex filaments are shown to arise from a specific choice of the phase connection, before being dynamically coupled to the full Schr\"odinger equation via the Rasetti-Regge method \cite{RasettiRegge1975}.

Section \ref{Sec:ConnectionQHDBOApp} uses the techniques of the previous sections to explore a new approach for describing Berry phase effects in Born-Oppenheimer systems. 
After seeing how the Mead-Truhlar method (presented in Section \ref{SectionBOApprox}) can be seen as a special case of the new connection-based description of QHD, we show how the dynamics of single particle nuclear trajectories
can depend on smoothened versions of the Berry connection and PES.  

Section \ref{Sec:Osmotic} concludes this chapter by  
completing the connection picture of QHD and employing an analogous construction for the dynamics of the wavefunction amplitude (in addition to the phase connection).
The corresponding Euler-Poincar{\'e} reduction procedure is then derived and 
the equations of motion are analysed. The interesting new feature of this approach is that it naturally leads to a generalised formulation of the stochastic quantum theory of Nelson 
\cite{Nelson1967}.

\section{Phase connection formulation of QHD}\label{Sec:NEWphase}
Before commencing with the construction of the new connection-based approach to QHD, we develop the introductory discussions of Sections \ref{Graham} and \ref{Sec:1-QHD}.
As we have seen, the QHD equations are only equivalent to the full Schr\"odinger equation provided the quantisation condition \eqref{QHDQuantCirc}  for the circulation holds, arising due to the requirement that the wavefunction be single-valued \cite{Takabayasi1952,Wallstrom1994}. 
This standard picture has recently been endowed a gauge theoretical description in \cite{Spera2016} owing to the fact that this fluid velocity is defined as $\bu := m^{-1}\nabla S$ via the Madelung transform 
\eqref{StandardMadelung}. Then, the quantised circulation condition \eqref{QHDQuantCirc} is rewritten 
as
\begin{align}
  \oint_{c_0} \de S = 2\pi\hbar n\,,\label{QHDHolonomy}
\end{align}
where $n \in \mathbb{Z}$, whose exponential is interpreted as the holonomy associated to the connection $\de S = \nabla S \cdot \de\bx$. Whilst an overview of the geometry of principal fibre bundles is given in Appendix \ref{App:Principalbundles}, here we explain the geometry of equation \eqref{QHDHolonomy} as follows.
 Firstly, in writing the Madelung transform $\psi=\sqrt{D}\,\text{exp}({iS/\hbar})$, we effectively make the decomposition $\mathbb{C}=\mathbb{R}^+\times \mathcal{U}(1)$.
  We then consider the trivial principal bundle $\mathbb{R}^3\times 
\mathcal{U}(1)$ on which the object $i\hbar^{-1}{\rm d}S$ can be considered as a $\mathfrak{u}(1)$-valued connection 1-form corresponding to the phase of the wavefunction. 
We will therefore often refer to such objects as `phase connections' throughout 
this chapter.
Then, whilst we consider a closed loop $c_0:[0,1]\to\mathbb{R}^3$, the 
corresponding path in the bundle may not be closed. The failure to close is 
known as the holonomy and is measured by the exponential of the loop integral \eqref{QHDHolonomy}.
 The quantisation of the holonomy arises due to the fact that $S$ can be considered a multi-valued function, by which we mean that the replacement
\begin{align}
  S(\bx)\to S(\bx) + 2\pi\hbar n \,,
\end{align}
in the Madelung transform \eqref{StandardMadelung} leaves the total wavefunction $\psi$ invariant. Then, the condition \eqref{QHDHolonomy} is non-trivial, that is $n\neq 0$, whenever the curve $c_0$ encloses regions in which $S$ is multi-valued (recall equation \eqref{MultivaluedS} from Section \ref{Sec:1-QHD}).
 In the bundle picture, the curvature of the connection vanishes everywhere except at those points in the base for which $S$ is not single-valued. The holonomy is then discretised, with the exact value depending on the winding number of the loop surrounding the singularity.
 Examples when this occurs are given by the presence of vortices, essentially isolated regions of non-zero vorticity, a topic which has been studied extensively \cite{Takabayasi1952, Takabayasi1983, BialynickiBirula1971, BialynickiBirulaSliwa2000}. 
 In Section \ref{Sec:Vortices} we will consider the possibility of hydrodynamic vortices of a classical nature, coupled with the quantum fluid.

The geometric formulation of QHD presented in Chapter \ref{Chap:QHD} relies on the existence of smooth invertible Lagrangian fluid paths $\eta\in\operatorname{Diff}(\Bbb{R}^3)$, so that $\bu(\bx,t)=\dot{\boldsymbol\eta}(\bx_0,t)$. 
Thus, this description assumes that the phase $S$ is single-valued thereby leading to zero circulation and vorticity. In other words, the previous geometric formulation of QHD does not capture non-zero holonomy as described above.
This chapter demonstrates how holonomy in QHD can still be described by purely geometric arguments using the alternative approach of gauge connections.

\subsection{Euler-Poincar{\'e} approach to phase connection dynamics}\label{Subsec:PhaseConnection}
Having presented the interpretation of the quantised circulation condition \eqref{QHDHolonomy} as the holonomy arising from a singular connection on a trivial bundle,
 we now construct an alternative approach to QHD in which non-trivial holonomy is 
built-in as an initial condition through a new dynamical connection. 

To begin, we use a variant of the Madelung transform \eqref{StandardMadelung}. However, rather than the usual 
 exponential form of the phase factor, we instead use the decomposition
\begin{align}
  \psi(\bx,t)=\sqrt{D(\bx,t)}\,\theta(\bx,t)\label{OurPolar}\,,
  \qquad\text{with}\quad
  \theta\in 
\mathcal{F}(\mathbb{R}^3,\mathcal{U}(1))
\,.
\end{align}
By writing explicitly the $\mathcal{U}(1)$ factor $\theta$  we avoid using the exponential map which is not injective and work only with single-valued functions. 
Furthermore, this expression for the wavefunction has the advantage of writing both terms as Lie group elements, allowing us to use the tools of geometric mechanics \cite{MaRa2013}.
The relation  $\theta^*=\theta^{-1}$ allows us to rewrite the Dirac-Frenkel Lagrangian \eqref{DSLagrangian} 
in terms of $(D, \theta, \partial_t\theta)$ as
  \begin{align}
  L = \int\!  \bigg[i\hbar D \theta^{-1}\partial_t\theta - \frac{\hbar^2}{2m}\Big(\big|\nabla \sqrt{D}\big|^2 + D|\nabla\theta|^2\Big) - DV\bigg]\text{d}^3x 
  \label{Lagrangian2}\,,
\end{align}
understood as the mapping $L:{\rm Den}(\mathbb{R}^3)\times T\mathcal{F}(\mathbb{R}^3,\mathcal{U}(1))\to 
\mathbb{R}$.
Then, following standard procedures in geometric mechanics, we let the phase factor $\theta(\bx,t)$  evolve according to the $\mathcal{U}(1)$ action 
\begin{align}
  \theta(\bx,t)=\Theta(\bx,t)\theta_0(\bx)\label{ThetaEvo}\,,
    \qquad\text{with}\quad
    \Theta\in 
\mathcal{F}(\mathbb{R}^3,\mathcal{U}(1))\,,
\end{align}
which in turn allows us to rewrite the 
time derivative as
\begin{align}
  \partial_t\theta = (\partial_t\Theta\,\Theta^{-1})\theta =: {\zeta}\theta\label{thetadot}\,,
     \qquad\text{where}\quad
     {\zeta}\in \mathcal{F}(\mathbb{R}^3,\mathfrak{u}(1))\,,
\end{align}
so that ${\zeta}(\bx,t)$ is a purely imaginary function as $\mathfrak{u}(1)\cong i \mathbb{R}$. 
It is this standard geometric mechanics approach, introducing the evolution on a Lie group, which leads us to the key step in this new construction, evaluating the gradient of $\theta$ by introducing the 
connection $\bnu$ as follows:
\begin{align}
\begin{split}
   \nabla\theta &= \nabla\Theta\, \theta_0 + \Theta\nabla\theta_0\\
  &= \nabla\Theta\,\Theta^{-1}\theta - \Theta{\bnu}_0\theta_0\\
  &= -(- \nabla\Theta\,\Theta^{-1} + {\bnu}_0)\theta =: -{\bnu}\theta\,,
\end{split}\label{NuConnectionDefinition}
\end{align}
in which we have defined ${\bnu}_0:=-\theta_0^{-1}\nabla\theta_0$. 
Here, we have introduced the connection 
${\nu}=\bnu\cdot\de\bx\in \Lambda^1(\mathbb{R}^3, \mathfrak{u}(1)):=\mathcal{F}(\mathbb{R}^3,\mathfrak{u}(1))\otimes\Lambda^1(\mathbb{R}^3)$, belonging to the space of $\mathfrak{u}(1)$-valued 
1-forms on $\mathbb{R}^3$.
Similar approaches have been used to introduce a connection in 
the geometric approach to liquid 
crystal dynamics \cite{GayBalmazRatiu2009,GayBalmazTronci2010,GBRaTr2012,GBRaTr2013,Holm2002,Tronci2012}. 
Indeed, we present the general formulation for gauge connections in mechanical systems of the type $\nabla n = -\bgamma n$ 
in Chapter \ref{Chap:Gamma}.

\begin{remark}[Trivial vs. non-trivial connections]\label{jack}
\hfill \\ 
As shown in general in Chapter \ref{Chap:Gamma}, any gauge connection introduced in this way must have zero curvature. 
For the phase connection \eqref{NuConnectionDefinition} this is shown by taking the curl of the relation $\nabla\theta=-{\bnu}\theta$ so that
\begin{align}
  \begin{split}
  0 = \nabla\times\nabla\theta &= - \nabla\times(\bnu\theta)\\
  &= - (\nabla\times\bnu)\theta + \bnu \times (\bnu\theta)\\
  &= - (\nabla\times\bnu)\theta \\
  \implies 0 &= \nabla\times\bnu\,.
  \end{split}
\end{align}
In the present approach, we are exploiting this zero curvature case in order to have a final form of the QHD Lagrangian. 
Once variations have been taken in Hamilton's principle, the equations will be allowed to hold also in the case of non-zero curvature. 
This is a common technique used in geometric mechanics to derive new Lagrangians. For example, 
it is used in \cite{BonetLuzTronci2015} to generalise the Dirac-Frenkel Lagrangian \eqref{DFLagrDensityMatrix} to include mixed state dynamics (as mentioned in Section \ref{Sec:1-GeomQM})
 as well as in the study complex fluids \cite{Holm2002,GBRaTr2012,GBRaTr2013}.
\end{remark}

Having introduced the new variables $\zeta$ and $\bnu$, the Lagrangian 
\eqref{Lagrangian2} can now be expressed as $\ell:\mathcal{F}(\mathbb{R}^3,\mathfrak{u}(1))\times\Lambda^1(\mathbb{R}^3, \mathfrak{u}(1))\times {\rm Den}(\mathbb{R}^3)\to \mathbb{R}$ 
given by
\begin{align}
\ell(\zeta, \bnu, D) = \int\!  \bigg[i\hbar D\zeta -\frac{\hbar^2}{2m}\big(\big|\nabla \sqrt{D}\big|^2 + D|\bnu|^2 \big) - DV\bigg]\text{d}^3x \label{Lagrangian2.1}\,,
\end{align}
where we have denoted $|\bnu|^2=\bnu^*\cdot\bnu$. 
Noting that both ${\zeta}$ and ${\bnu}$ 
are purely imaginary, we define their real counterparts 
\begin{align}
 \bar{ \zeta}:={i\hbar}{\zeta}\,,\qquad \bar{\bnu}:={i\hbar}{\bnu}\label{barvariables}\,,
\end{align}
so that $\bar\zeta \in \mathcal{F}(\mathbb{R}^3)$ and $\bar\nu =\bar\bnu(\bx)\cdot \de\bx\in \Lambda^1(\mathbb{R}^3)$.
Then, we can now write an alternative theory of QHD formulated in terms of the phase connection \eqref{NuConnectionDefinition} and derived by Euler-Poincar{\'e} reduction. 

\begin{theorem}[Euler-Poincar{\'e} formulation for phase connections in QHD]\label{Theorem:NEWQHD}
\hfill \\
Written in terms of the new variables \eqref{barvariables}, the transformation \eqref{OurPolar} takes the DF Lagrangian \eqref{DFStateLagrangian} into the reduced Lagrangian 
$\ell:\mathcal{F}(\mathbb{R}^3)\times\Lambda^1(\mathbb{R}^3)\times {\rm Den}(\mathbb{R}^3)\to \mathbb{R}$ 
given by
\begin{align}
  \ell(\bar{\zeta}, \bar{\bnu}, D) = \int\! D \Big( \bar{\zeta}-\frac{|\bar{\bnu}|^2}{2m} - \frac{\hbar^2}{8m}\frac{|\nabla 
  D|^2}{D^2}  - V\Big)\,\de^3x\label{Lagrangian3}\,.
\end{align}
Then, considering the arbitrary variation $\delta D$ and Euler-Poincar{\'e} constrained variations $(\delta\bar\zeta,\delta \bar\bnu) = (\partial_t\bar\eta, -\nabla\bar\eta)$ with arbitrary $\bar\eta:= i\hbar \delta\Theta\,\Theta^{-1}$, 
Hamilton's principle $0=\delta\int_{t_1}^{t_2}  \ell(\bar{\zeta}, \bar{\bnu}, D)\,\de t$
yields the following equations of motion
\begin{align}
  &\partial_tD + {\rm div}(m^{-1}D{\bar\bnu})=0 \label{transport}\,,\\
    &\bar\zeta = \frac{|\bar\bnu|^2}{2m} +V_Q + V 
  \label{xiHamiltonJacobi1}\,,\\
  &\partial_t\bar\bnu +\nabla\bar\zeta= 0\,, \label{NuEvo}
\end{align}
where $V_Q$  is the quantum potential defined by \eqref{BohmPot}. 
\end{theorem}
\paragraph{Proof:} We begin by deriving the Euler-Poincar{\'e} variations explicitly
\begin{align*}
  \delta \bar\zeta = \delta \Big(i\hbar\partial_t\Theta\,\Theta^{-1}\Big) &= 
  i\hbar\Big(\delta\partial_t\Theta\,\Theta^{-1}+ \partial_t\Theta \delta(\Theta^{-1})\Big) 
  \\
  &= i\hbar\Big(\delta\partial_t\Theta\,\Theta^{-1} - \partial_t\Theta\,\Theta^{-2}\delta\Theta\Big) 
  = \partial_t\Big(i\hbar\delta\Theta\,\Theta^{-1}\Big)=:\partial_t\bar\eta\,,\\
  \delta\bar\bnu = \delta\Big(i\hbar(- \nabla\Theta\,\Theta^{-1} + {\bnu}_0)\Big)&= 
  -i\hbar\Big(\delta\nabla\Theta\,\Theta^{-1}+ \nabla\Theta \delta(\Theta^{-1})\Big) 
  \\
  &= -i\hbar\Big(\delta\nabla\Theta\,\Theta^{-1} - \nabla\Theta\,\Theta^{-2}\delta\Theta\Big) 
  = -\nabla\Big(i\hbar\delta\Theta\,\Theta^{-1}\Big)=:-\nabla\bar\eta\,.
\end{align*}
Then, as usual we use these to expand Hamilton's principle
\begin{align*}
  0 &= \delta\int_{t_1}^{t_2}  \ell(\bar{\zeta}, \bar{\bnu}, D)\,\de t = \int_{t_1}^{t_2} 
  \int \left(\frac{\delta \ell}{\delta \bar\zeta}\,\delta\bar\zeta + \frac{\delta \ell}{\delta 
  \bar\bnu}\cdot\delta \bar\bnu+ \frac{\delta \ell}{\delta D}\,\delta 
  D\right)\,\de^3x\,\de t\\
  &=\int_{t_1}^{t_2} 
  \int \left(\frac{\delta \ell}{\delta \bar\zeta}\,\partial_t\bar\eta - \frac{\delta \ell}{\delta 
  \bar\bnu}\cdot\nabla \bar\eta+ \frac{\delta \ell}{\delta D}\,\delta 
  D\right)\,\de^3x\,\de t\\
    &=\int_{t_1}^{t_2} 
  \int \left(\left(-\partial_t\frac{\delta \ell}{\delta \bar\zeta} + {\rm div}\frac{\delta \ell}{\delta 
  \bar\bnu}\right)\cdot\bar\eta+ \frac{\delta \ell}{\delta D}\,\delta 
  D\right)\,\de^3x\,\de t
\end{align*}
 so that upon recalling that $\delta D$ and $\bar\eta$ are arbitrary, we obtain the following general equations of motion
\begin{align}
   \partial_t\left(\frac{\delta \ell}{\delta \bar\zeta}\right) - \text{div}\left(\frac{\delta \ell}{\delta 
  \bar\bnu}\right)&= 0\,,\\
  \frac{\delta \ell}{\delta 
  D}&= 0\,,
\end{align}
along with the auxilliary equation $\partial_t\bar\bnu =-\nabla\bar\zeta $ derived from the definition \eqref{NuConnectionDefinition}. Specialising to the Lagrangian \eqref{Lagrangian3} yields the following variational derivatives
\begin{align}
  \frac{\delta \ell}{\delta \bar\zeta} = D\,,\qquad   \frac{\delta \ell}{\delta \bar\bnu} = 
  -\frac{D}{m}\bar\bnu\,, \qquad \frac{\delta \ell}{\delta D} =\bar{\zeta}-\frac{|\bar{\bnu}|^2}{2m} - V_Q  - 
  V\,,
\end{align}
which upon substitution into the general equations clearly return \eqref{transport} and \eqref{xiHamiltonJacobi1} 
as required.
\hfill$\square$\\

Clearly, equations \eqref{transport} and \eqref{xiHamiltonJacobi1} are the correspondents of the continuity equation \eqref{DSEqn1} and quantum Hamilton-Jacobi equation \eqref{DSEqn2} from the standard QHD theory in Section \ref{Sec:1-QHD},
 now expressed in the connection formalism. Hence, we proceed further by taking the gradient of \eqref{xiHamiltonJacobi1} and use the auxilliary equation \eqref{NuEvo} to eliminate $\bar\zeta$ and simply obtain the coupled  
equations
\begin{align}
  \partial_tD + \text{div}(m^{-1}D{\bar\bnu})&=0\,,\label{transport2}\\
  \partial_t\bar\bnu+\nabla\bigg(\frac{|\bar\bnu|^2}{2m} +V_Q + V \bigg)&=0
\,,
\label{bnueq}
\end{align}
in which we can already see how $\widetilde\bnu:=m^{-1}\bar{\bnu}$ will play the role of a fluid velocity.

\begin{remark}[Lie-Poisson structure I]\label{Rem:LP1}
\hfill \\   
The new QHD equations \eqref{transport2} and \eqref{bnueq} comprise a Lie-Poisson bracket on the dual of the semidirect-product Lie algebra $\mathcal{F}(\Bbb{R}^3)\,\circledS\,\Lambda^1(\Bbb{R}^3)$. Specifically, the Lie-Poisson bracket reads
\begin{equation}
\{f,h\}=\int\left(\frac{\delta h}{\delta\bar\bnu}\cdot\nabla\frac{\delta f}{\delta D}-\frac{\delta f}{\delta\bar\bnu}\cdot\nabla\frac{\delta h}{\delta D}\right)\,{\rm d}^3 x
\,,
\label{AffLPB}
\end{equation}
whilst the Hamiltonian is given by
\[
h(D,\bar\bnu)=\int\!D\bigg(\frac{|\bar{\bnu}|^2}{2m} + \frac{\hbar^2}{8m}\frac{|\nabla 
  D|^2}{D^2}  + V\bigg)\,{\de}^3x
  \,.
\]
 The bracket \eqref{AffLPB} arises from a Lie-Poisson reduction on the semidirect-product group 
 \\
 $\mathcal{F}(\Bbb{R}^3, \mathcal{U}(1))$ $\circledS\,\Lambda^1(\Bbb{R}^3,\mathfrak{u}(1))$. Here, $\Lambda^1(\Bbb{R}^3,\mathfrak{u}(1))$ again denotes the space of differential 1-forms with values in $\mathfrak{u}(1)\simeq i\Bbb{R}$, while the semidirect-product structure is defined by the affine gauge action $\bnu\mapsto\bnu+\Theta^{-1}\nabla\Theta$, where $\Theta\in\mathcal{F}(\Bbb{R}^3, \mathcal{U}(1))$ and $\bnu\in\Lambda^1(\Bbb{R}^3,\mathfrak{u}(1))$. For further details on this type of affine Lie-Poisson reduction, see \cite{GayBalmazRatiu2009,Holm2002}.
\end{remark}
Before continuing, we highlight the three different manifestations of the $\mathcal{U}(1)$ connection 
that have been introduced and will continue to be used throughout this chapter. Firstly, we introduced the $\mathfrak{u}(1)$-valued connection 
$\nu$ via the relation $\nabla\theta=:-\bnu\theta$. Then, its real counterpart $\bar\bnu$
was introduced via $\bar\bnu:= i\hbar\bnu\in\Lambda^1(\Bbb{R}^3)$ which, for  $\theta = e^{iS/\hbar}$, coincides with $\nabla S$.
 Finally, in anticipation of the next section, we define $\widetilde{\bnu}\in\Lambda^1(\Bbb{R}^3)$ 
by performing a further division by the mass, $\widetilde{\bnu}:=i\hbar 
m^{-1}\bnu = m^{-1}\bar\bnu$. This, corresponding to $m^{-1}\nabla S $ in the standard approach, 
will serve as the fluid velocity in the QHD picture.


We also notice how, upon moving back to the standard picture by using the usual exponential form $\theta = e^{iS/\hbar}$ in \eqref{OurPolar},
 $\bar\bnu$ coincides with $\nabla S$ and $\bar\zeta=-\partial_t S$
thus transforming equation \eqref{xiHamiltonJacobi1} 
into the standard phase equation (quantum Hamilton-Jacobi equation) \eqref{DSEqn2} of QHD. However, the crucial difference is that in our approach $\bar\bnu$ is now allowed to have a nontrivial  curvature as can be seen from the curl of \eqref{bnueq}, that is
\begin{align}\label{jeremy}
  \partial_t(\nabla\times\bar\bnu)=0\,.
\end{align}
This relation demonstrates explicitly how, in view of Remark \ref{jack}, the curvature of the connection $\bar\bnu$ need not be trivial (unlike ordinary QHD) but is instead  preserved in time. 
This is one of the main upshots of this more general approach. Indeed, as we shall see, the quantum hydrodynamic velocity $\widetilde{\bnu}$ does not have to be expressed as an exact differential, thereby allowing for non-trivial circulation dynamics. We will develop this observation in the next section.

\subsection{Hydrodynamic equations with vorticity}\label{Sec:PhaseConnectionsSubSec:2}
In order to reconcile these new QHD equations with the standard construction, in this section we consider their hydrodynamic form in terms of the fluid velocity variable $\widetilde{\bnu}$.
\begin{corollary}[Fluid dynamical form of the connection QHD equations]
\hfill\\
The equations of motion \eqref{transport2} and \eqref{bnueq} can be rewritten in 
terms of the fluid velocity $\widetilde\bnu:=m^{-1}\bar\bnu$ as
\begin{align}
  \partial_tD+{\rm div}(D\widetilde{\bnu})&=0\,,\label{BracketEqn1}\\
  m(\partial_t+\widetilde{\bnu}\cdot\nabla)\widetilde{\bnu} &= -\widetilde{\bnu}\times(\nabla\times\bar\bnu) - 
\nabla(V+V_Q)\label{BracketEqn2}\,,
\end{align}
where $\nabla\times\bar\bnu$ is the constant curvature of the connection.
\end{corollary}

The latter equation again clearly demonstrates the importance of not utilizing the exponential form of the phase as we now have the additional Lorentz-force term $-\widetilde{\bnu}\times(\nabla\times\bar\bnu)$, featuring the curvature of $\bar\bnu$, which is  absent in the equation \eqref{QHD2} of standard quantum hydrodynamics.
 Indeed, one sees that this additional term vanishes exactly when $\widetilde{\bnu}$ is a pure gradient. 

As we have seen in Section \ref{Sec:2-HD}, in the Bohmian interpretation the Lagrangian fluid paths are introduced via the relation $\dot{\boldsymbol\eta}(\bx,t)=\widetilde{\bnu}({\boldsymbol\eta}(\bx,t),t)$, so that Bohmian trajectories obey the Lagrangian path equation
\begin{equation}
m\ddot{\boldeta}=-\dot{{\boldeta}}\times\nabla\times\bar\bnu- 
\nabla_{\boldeta}(V+V_Q)\,,
\end{equation}
generalising equation \eqref{BohmianEL}.
We observe that again the non-zero curvature modifies the usual equation of Bohmian trajectories by the emergence of a Lorentz-force term. Notice that this term would also persist in the semiclassical limit after ignoring the quantum potential contributions.

Next, we introduce the sharp isomorphism so that $\widetilde{\bnu}^\sharp= \widetilde{\nu}^j{\text{\bf e}}_j$ denotes the vector field associated to the 1-form $\widetilde{\bnu}\cdot\de\bx$. 
This is induced by the Euclidean metric in the 
fluid kinetic energy term of \eqref{Lagrangian3} and we denote the standard basis of $\mathbb{R}^3$ by ${\text{\bf e}}_j$.
We now consider the fluid circulation dynamics in the 
following theorem.
\begin{theorem}[Kelvin circulation theorem]
  \hfill \\
  The corresponding Kelvin circulation theorem associated to the fluid velocity equation \eqref{BracketEqn2} 
  is given by
  \begin{align}
  \begin{split}
0=  \frac{{\de}}{{\de}t}\oint_{c(t)}\widetilde{\bnu}\cdot{\de}\bx &+  \frac1m
  \oint_{c(t)}\widetilde{\bnu}^\sharp\times(\nabla\times\bar\bnu)\cdot\de\bx
  = \frac{\de}{\de t}\oint_{c_0}\widetilde{\bnu}\cdot\de\bx
  \label{CirculationTheorem}\,,
  \end{split}
\end{align}
for $c(t)$ a loop moving with the fluid velocity $\widetilde{\bnu}^\sharp$ such that 
$c(0)=c_0$.
\end{theorem}
\paragraph{Proof:}  Simply rewriting the fluid equation \eqref{BracketEqn2} in 
terms of the Lie derivative returns
\begin{align}
  m(\partial_t+\pounds_{\widetilde{\bnu}^\sharp})\widetilde{\bnu} = -\widetilde{\bnu}^\sharp\times(\nabla\times\bar\bnu) +\frac{m}{2}\nabla|\widetilde{\bnu}|^2- 
\nabla(V+V_Q)\,.
\end{align}
Then, following the arguments given in the proof of Theorem \ref{Abby}, computing the loop integral and taking the Lagrangian time derivative outside
so that it becomes a total derivative, returns the desired result. The second 
equality of \eqref{CirculationTheorem} follows directly from \eqref{jeremy}.
\hfill$\square$\\

In terms of 
the geometry of principal bundles, the last equality in \eqref{CirculationTheorem} tells us that the holonomy of the connection $\bnu$ must be constant in time.
 Since no singularities are involved and $\bnu$ is assumed to be differentiable, the last equality in \eqref{CirculationTheorem} is an example of nontrivial holonomy which is not discrete, depending smoothly on the choice of loop $c_0$.
\begin{remark}[Helicity conservation]\label{HelicityRemark}
\hfill \\
   Here, we consider the hydrodynamic helicity and its conservation.
To do so, we take the dot product of the second equation in \eqref{NuEvo} with  $\nabla\times\widetilde{\bnu}$ to obtain 
\begin{align*}
 \partial_t (\widetilde{\bnu}\cdot\nabla\times\widetilde{\bnu})&= 
 {-m^{-1}\rm{div}(\zeta\nabla\times\widetilde{\bnu})}\,,
\end{align*}
from which it immediately follows that the hydrodynamic helicity is preserved in time, that is
\begin{align}
\frac{\de}{\de t}\int \widetilde{\bnu}\cdot(\nabla\times\widetilde{\bnu})\,{\rm d}^3{x} &= 0\,.
\end{align}
\end{remark}

Now that the hydrodynamic Bohmian interpretation has been discussed, it is not yet clear how this construction is actually related to the original Schr\"odinger equation of quantum mechanics. This is the topic of the next section.

\subsection{Schr\"odinger equation with holonomy}\label{Subsec:HolonomicSchr}
In this section, we discuss the relation between the new QHD framework given by equations  \eqref{transport2}-\eqref{bnueq} and the original Schr\"odinger equation. As a preliminary step, we consider the Helmholtz decomposition of $\bar\bnu$, that is
\begin{align}
  \bar\bnu(\bx,t) = \nabla s(\bx,t)+\hbar\nabla\times\bbeta(\bx)\label{Helmholtz1}\,,
\end{align}
where $\bbeta$ is a constant function to ensure \eqref{jeremy}. Also, we have added the factor $\hbar$ and fixed the Coulomb gauge $\operatorname{div}\bbeta=0$ for later convenience. Notice that although $s$ appears exactly in the place that $S$ would in the standard Madelung transform from Section \ref{Sec:1-QHD}, here we have used the lowercase letter to emphasise that in this case we consider $s$ as a single-valued function. 
The relation \eqref{Helmholtz1} is reminiscent of similar expressions for the Bohmian velocity $\widetilde{\bnu}$ already appearing in \cite{BohmVigier1954}, although in the latter case these were motivated by stochastic augmentations of standard quantum theory. 
One can verify that upon substituting \eqref{Helmholtz1} into 
the equations of motion \eqref{NuEvo} and \eqref{transport}, the latter become
\begin{align}
 \partial_t s + \frac{|\nabla s+\hbar\nabla\times\bbeta|^2 }{2m}+ 
  V+V_Q&=0
  \,,\label{QHJ+}\\
  \partial_t D + \text{div}\left(D\,\frac{\nabla s+\hbar\nabla\times\bbeta}{m}\right) 
  &=0\,.
  \label{Continuity+}
\end{align}
Here, we have discarded numerical integration factors in the first equation. We recognise that these correspond to the standard Madelung equations in the presence of a magnetic field  $\hbar\Delta\bbeta$.


At this point, we further characterise the 
Helmholtz decomposition \eqref{Helmholtz1}  of the connection $\bnu$ in terms of its defining relation \eqref{NuConnectionDefinition}. In particular, after constructing the Lagrangian \eqref{Lagrangian3}, $\bnu$ is then only defined as the solution of {$\partial_t\bnu=-\nabla\zeta$} in \eqref{NuEvo}. Combining the latter with {$\partial_t\theta=\zeta\theta$} leads to $\partial_t({\bnu}+ \theta^{-1}\nabla\theta)=0$ so that direct integration yields
\begin{align}
  {\bnu}=-\frac{\nabla\theta}{\theta} +i{\bLambda}(\bx)\,,\label{NuHelmholtz}
\end{align}
for some constant real function ${\bLambda}(\bx)$. An immediate calculation shows $
\nabla\times(\theta^{-1}\nabla\theta)=0$ so that $\bLambda$ encapsulates the new non-zero curvature of the connection formulation of QHD.
 Then, upon moving to the real-valued variables \eqref{barvariables}, direct comparison with \eqref{Helmholtz1} yields
\begin{align}
  \nabla s = -{i\hbar}\frac{\nabla\theta}{\theta}\,,\qquad 
\nabla\times\bbeta= -{\bLambda}\label{pbetarelation}\,,
\end{align}
so that, going back to the QHD equations, \eqref{BracketEqn2} can now be 
written in the form
\begin{align}
  m(\partial_t+\widetilde{\bnu}\cdot\nabla)\widetilde{\bnu} = \hbar\widetilde{\bnu}\times\nabla\times{\bLambda} - 
\nabla(V+V_Q)\label{NuFluidWithLambda}
\,.
\end{align}

Now that we have characterised the additional terms due to the presence of non-zero curvature in the Madelung equations, we can now use the expressions above in order to reconstruct the quantum Schr{\"o}dinger equation.

\begin{theorem}[Reconstruction of the Schr\"odinger equation]\label{Theorem:SchrReconstr}
  \hfill \\
Consider the system of equations \eqref{transport}-\eqref{NuEvo} given in Theorem \ref{Theorem:NEWQHD}. Together with the Helmholtz decomposition 
\eqref{NuHelmholtz}, the time evolution of a wavefunction (reconstructed from the polar form \eqref{OurPolar}) results in an associated Schr\"{o}dinger equation 
  \begin{align}
    i\hbar\partial_t\psi &=    
    \left[\frac{(-i\hbar\nabla-\hbar{\bLambda})^2}{2m}+V\right]\psi\label{SchrodingerReconstructEqn}\,,
\end{align}
in which $\hbar{\bLambda}$, corresponding to the constant curvature part of our $\mathcal{U}(1)$ connection, appears via minimal coupling.
\end{theorem}
\paragraph{Proof:}
 Upon denoting $R=\sqrt{D}$ in \eqref{OurPolar}, we compute
\begin{align}
  i\hbar\partial_t\psi  &= i\hbar(\partial_t R\,\theta + R\,\partial_t\theta)\\
   &= \left[-\frac{i\hbar}{m}\left(\frac{\nabla R}{R}\cdot\bar\bnu\right)-\frac{i\hbar}{2m}\text{div}(\bar\bnu)+\frac{|\bar\bnu|^2}{2m} +V_Q\right]\psi 
+ V\psi\,,
\label{prelimSchr}
\end{align}
having used the continuity equation in \eqref{transport} to find $\partial_t R$ and 
\eqref{thetadot} with \eqref{barvariables} and \eqref{xiHamiltonJacobi1} to find 
$\partial_t\theta$.
 At this point, we must manipulate the kinetic energy term to express everything in terms of 
$\psi$. Before continuing, we notice that \eqref{OurPolar} leads to 
$\nabla \psi=(R^{-1}\nabla R+\theta^{-1}\nabla\theta)\psi$,
so that, since $\theta^{-1}\nabla\theta$ is purely imaginary,
\begin{align}
  \frac{\nabla R}{R} = \frac{\text{Re}(\psi^*\nabla\psi)}{|\psi|^2}\,, 
  \qquad
    \frac{\nabla \theta}{\theta} = 
    \frac{i\text{Im}(\psi^*\nabla\psi)}{|\psi|^2}
    \,.
\end{align}
Then, using  \eqref{NuHelmholtz} leads to
$
\bar\bnu =  {\hbar\text{Im}(\psi^*\nabla\psi)}/{|\psi|^2} -\hbar
{\bLambda}$,
so that \eqref{prelimSchr} can now be entirely written in terms of $\psi$. As shown in Appendix \ref{App:Reconstruct}, lengthy calculations yield
\begin{align*}
 \left[-\frac{i\hbar}{m}\left(\frac{\nabla R}{R}\cdot\bar\bnu\right)-\frac{i\hbar}{2m}\text{div}(\bar\bnu)+\frac{|\bar\bnu|^2}{2m} +V_Q\right]\psi
  &= -\frac{\hbar^2}{2m}\Delta\psi + 
  \frac{i\hbar^2}{m}{\bLambda}\cdot\nabla\psi+ 
  \frac{\hbar^2}{2m}|{\bLambda}|^2 \psi\,.
\end{align*}
Hence, putting everything back together we obtain the Schr{\"o}dinger equation 
\eqref{SchrodingerReconstructEqn} as required.
\hfill$\square$\\

As $\hbar{\bLambda}$ appears in the reconstructed Schr{\"o}dinger equation \eqref{SchrodingerReconstructEqn} via minimal coupling, it can be thought of as a type of vector potential (analogous to electromagnetism) in which $\hbar$ plays the role of a coupling constant. 
The role of this vector potential is to incorporate the holonomic effects in quantum dynamics. 
Hence, naturally \eqref{SchrodingerReconstructEqn} coincides with the equation for $\psi=\sqrt{D}e^{is/\hbar}$ in the presence of a magnetic field $\hbar\Delta\bbeta$, as it arises from the new Madelung equations \eqref{QHJ+}-\eqref{Continuity+}.

Having fully characterised the features of the new approach to QHD, we now 
reconnect with the momentum map approach of Chapter \ref{Chap:QHD} in which 
the collective Hamiltonian allowed for both Lie-Poisson and Euler-Poincar{\'e} 
derivations. In fact, having discovered that the novel contribution to the 
non-zero hydrodynamic vorticity arises from the $\bLambda$ term in the curvature of 
$\bnu$, we can now present the complimentary momentum map approach to 
QHD with holonomy.

\begin{proposition}[Lie-Poisson structure II]\label{Prop:LPStructure2}
\hfill \\
Let us define the momentum variable 
\begin{align}\label{HolonomyMomentumDef}
\boldsymbol\mu:=mD\widetilde{\bnu}+ \hbar D\bLambda \in \mathfrak{X}^*(\mathbb{R}^3)\,.
\end{align}
 Then, the hydrodynamic equations \eqref{transport2} and \eqref{NuFluidWithLambda} 
 \begin{align*}
     \partial_tD+{\rm div}(D\widetilde{\bnu})&=0\,,\\
m(\partial_t+\widetilde{\bnu}\cdot\nabla)\widetilde{\bnu} &= \hbar\widetilde{\bnu}\times\nabla\times{\bLambda} - 
\nabla(V+V_Q)\,,
 \end{align*}
 possess an alternative Hamiltonian structure (in addition to that given in Remark \ref{Rem:LP1}) given by the hydrodynamic Lie-Poisson bracket \eqref{LPB}
 \begin{multline*}
\{f,g\}(\bmu,D) 
=  \int \bmu \cdot\left[ \left(\frac{\delta g}{\delta \boldsymbol{\mu}} \cdot \nabla \right)  \frac{\delta f}{\delta \bmu}
-  \left(\frac{\delta f}{\delta \boldsymbol{\mu}} \cdot \nabla \right)  \frac{\delta g}{\delta \bmu} 
\right]\\
+ D \left[  \left(\frac{\delta g}{\delta \boldsymbol{\mu}} \cdot \nabla \right) \frac{\delta f}{\delta D}
- \left(\frac{\delta f}{\delta \boldsymbol{\mu}} \cdot \nabla \right)  \frac{\delta g}{\delta D} 
\right]\,\de^3x\,,
\end{multline*}
defined on the dual of the semidirect product Lie algebra $\mathfrak{X}(\mathbb{R}^3)\,\circledS\, \mathcal{F}(\mathbb{R}^3)$, corresponding to the Hamiltonian 
\begin{align}
h(\bmu, D)= \int\left[ \frac{|\bmu-\hbar D{\bLambda}|^2}{2mD}+\frac{\hbar^2}{8m}\frac{|\nabla 
D|^2}{D}+DV(\boldsymbol{x})\right]\de^3x\,.\label{AltHamiltonianLambda}
\end{align}
\end{proposition}
\paragraph{Proof:} Following previous calculations, using the formula $\dot{f}=\{f,h\}$ we obtain the 
general equations
\begin{align*}
  \partial_t D + \text{div}\left(D \frac{\delta h}{\delta \bmu} \right)= 
  0\,,\qquad    \partial_t \bmu + \partial_j\left(\frac{\delta h}{\delta \mu_j}\bmu\right)+ 
  \mu_j\nabla\frac{\delta h}{\delta \mu_j} = -D\nabla\frac{\delta h}{\delta D}
 \,.
\end{align*}
Then, computing the variational derivatives
\begin{align}
  \frac{\delta h}{\delta \bmu} =\frac{\bmu - \hbar D\bLambda}{mD} = 
  \widetilde\bnu\,,\qquad \frac{\delta h}{\delta D} = -\frac{|\bmu|^2}{2mD^2} + 
  \frac{\hbar^2}{2m}|\bLambda|^2 + V + V_Q\,,
\end{align}
the continuity equation follows immediately whilst the fluid momentum equation 
is derived from the following computation
\begin{align*}
  \partial_t \bmu + \partial_j\left(\widetilde{\nu}_j\bmu\right)+ 
  \mu_j\nabla\widetilde{\nu}_j &= D\nabla\left(\frac{|\bmu|^2}{2mD^2} - 
  \frac{\hbar^2}{2m}|\bLambda|^2 - V - V_Q\right)\\
  \implies   \partial_t \bmu + \partial_j\left(\widetilde{\nu}_j\bmu\right)+ 
  \mu_j\nabla\widetilde{\nu}_j &= D\nabla\left(\frac{1}{2}m|\widetilde{\bnu}|^2 +\hbar\bnu\cdot\bLambda- V - V_Q\right)\\
    \implies   \partial_t \bmu + ({\rm div}\widetilde{\bnu})\bmu+ 
(\widetilde{\bnu}\cdot\nabla)\bmu &= {\hbar D\nu_j\nabla\Lambda_j- D\nabla(V + V_Q)}\,.
\end{align*}
Finally, one regains the necessary fluid velocity equation upon converting fully back 
to $\widetilde{\bnu}$.

\hfill$\square$\\

In fact, we have already encountered this structure for similar Hamiltonians in Chapter \ref{Chap:EFOLD} in which the Berry connection 
appears in the total energy \eqref{EFHydroHamiltonian} via minimal coupling to the nuclear hydrodynamic momentum, analogous to the situation for $\bLambda$ presented above.

At this point we remark that it may be possible to incorporate topological singularities by expressing $\bLambda$ in terms of delta functions 
\cite{Kleinert}. However, this idea will not be considered in the subsequent material, instead retaining the case of a differentiable 1-form $\bLambda$, noting that the presence of topological defects remains an interesting possibility.


To continue, we peform the reduced Legendre transform so that the
equations of motion can be equivalently obtained via an alternative Euler-Poincar{\'e} reduction 
 in addition to that appearing in Theorem \ref{Theorem:NEWQHD}.

\begin{proposition}[Euler-Poincar{\'e} structure II]\label{Prop:EP2Structure}
 \hfill \\
The reduced Legendre transform of the Hamiltonian \eqref{AltHamiltonianLambda} from 
  Proposition \ref{Prop:LPStructure2} introduces the velocity field $\widetilde{\bnu}=\delta h/\delta\bmu \in \mathfrak{X}(\mathbb{R}^3)$, from which the 
 hydrodynamic Lagrangian $\ell:\mathfrak{X}(\mathbb{R}^3)\times{\rm Den}(\mathbb{R}^3)\to \mathbb{R}$ 
 given by
\begin{align}
  \ell(\widetilde{\bnu},D)=\int \bigg[\frac{1}{2}mD |\widetilde{\bnu}|^2+ \hbar D\widetilde{\bnu}\cdot\bLambda-\frac{\hbar^2}{8m}\frac{|\nabla 
D|^2}{D}-DV\bigg]\,\de^3x\,, \label{AlternativeQHDLagrangian}
\end{align}
is constructed via $\ell(\widetilde{\bnu},D)=\int \bmu\cdot\widetilde{\bnu}\,\de^3x - h(\bmu, D)$. Then, the new QHD equations \eqref{transport2} and \eqref{NuFluidWithLambda} follow from applying Hamilton's principle for 
 Euler-Poincar{\'e} variations of the type \eqref{EPvar}. 
\end{proposition}
\paragraph{Proof:} The proof follows analogously to the proof of Theorem \ref{Theorem:EFEPEL} for the exact 
factorization dynamics considered in Chapter \ref{Chap:EFOLD}. Indeed our current fluid equation \eqref{NuFluidWithLambda} takes the form of the nuclear QHD equation
\eqref{EFEPELu1} upon making the replacements $\epsilon\mapsto V$ and $\boldsymbol{\cal A}\mapsto - \hbar\bLambda$, recalling that $\bLambda$ is independent of time.  

\hfill$\square$\\

To close this section, we remark on an important distinction between $\bLambda$ 
and the vector potential of an external magnetic field that manifests for 
multi-particle dynamics.

\begin{remark}[Phase connection vs. magnetic vector potential]
  \hfill \\
  Motivated by the appearance of the phase connection as a minimal coupling term in the Schr\"odinger equation \eqref{SchrodingerReconstructEqn}, it may be useful to  include the effect of an external magnetic field on the quantum system within this new QHD framework. Then, in the case of a spinless unit charge, the Schr\"odinger equation with holonomy reads
\begin{align}\label{modAB}
    i\hbar\partial_t\psi &= \left[\frac{(-i\hbar\nabla- (\hbar{\bLambda}+\bA))^2}{2m}+V\right]\psi\,.
\end{align}
In this instance both ${\bLambda}$ and $\bA$ are  formally equivalent $\mathcal{U}(1)$ gauge connections. Setting  ${\bLambda}=0$ in the Hamiltonian operator of \eqref{modAB} yields the Aharonov-Bohm Hamiltonian, in which case again the magnetic potential has a topological singularity.  However, despite the apparent equivalence between ${\bLambda}$ and $\bA$, they are related to essentially different features: while the holonomy associated to $\bA$ is associated to the properties of the external magnetic field, the holonomy associated to ${\bLambda}$ is intrinsically related to the evolution of the quantum state $\psi$. This specific difference is particularly manifest in the case of two spinless unit charges moving within an external magnetic field. Indeed, in that case the 2-particle wavefunction $\psi(\bx_1,\bx_2,t)$ leads to defining  $\bLambda(\bx_1,\bx_2)$ and $\bA(\bx)$ on different spaces, thereby revealing their essentially different nature. 
At present, it is not clear if this difference plays any role in describing the two-particle {Aharonov-Bohm} effect \cite{Samuelsson}.
\end{remark}

\section{Hydrodynamic vortices in QHD}\label{Sec:Vortices}
 In this section, we consider the how the connection-based approach to QHD presented thus far can be used to capture the presence of vortices in quantum hydrodynamics.
  While the presence of topological vortex singularities in quantum mechanics has been known since the early days, this problem was considered in the context of the Madelung-Bohm formulation by Takabayasi  \cite{Takabayasi1952,Takabayasi1983} and later by Bialynicki-Birula  \cite{BialynickiBirula1971,BialynickiBirulaSliwa2000}.
  In standard hydrodynamics, the vorticity 2-form $\omega$ is the differential of the Eulerian velocity field so that in our case $\bomega:=\nabla\times\widetilde{\bnu}$.
   Then, upon using \eqref{NuHelmholtz}, one has
 \begin{equation}\label{vorticity}
  \bomega(\bx) = -\frac{\hbar}{m}\nabla\times\bLambda(\bx)\in \Lambda^2(\mathbb{R}^3)\,.
\end{equation}
In this section we wish to introduce the presence of non-quantised hydrodynamic vortices in Schr\"odinger quantum mechanics. To this purpose,
we consider a hydrodynamic vortex filament of the following form:
\begin{definition}[Vortex filament \cite{HolmStechmann2004}]
  A {\bf vortex filament} is a distribution of vorticity 
  supported on a curve $\bR(\sigma, t)\in \mathbb{R}^3$ as,
\begin{equation}
   \bomega(\bx) =\int \bR_{\sigma}\,\delta(\bx-\bR(\sigma))\,{\rm d}\sigma\,\label{vortexfilament}
\end{equation}
where $\sigma$ is a 
fixed {parameterisation} of the curve and $\bR_{\sigma}:=\partial\bR/\partial\sigma$ is the vector tangent to the curve and equal to the vorticity at that point.
\end{definition}
   It is well known that three-dimensional vortex filaments of the type \eqref{vortexfilament} cannot be  quantised, as shown in \cite{GoMeSh87, GoMeSh91}.
 Here,  in order to avoid problems with boundary conditions, here we consider the simple case of vortex rings.
\begin{corollary}[Schr{\"o}dinger equation for a hydrodynamic vortex filament]\label{Cor:SchrVort}
\hfill \\
 For a vortex filament of the form 
  \eqref{vortexfilament}, the reconstructed Schr{\"o}dinger equation \eqref{SchrodingerReconstructEqn} from Theorem \ref{Theorem:SchrReconstr} depends explicitly on the vortex position, given by
\begin{align}
    i\hbar\partial_t\psi &= \frac{1}{2m}\left(-i\hbar\nabla-{m}\nabla\times \int \! G(\bx-\bR)\,\bR_\sigma\,\de \sigma\right)^{\!2}\psi+V\psi\label{SchrodingerVortices}\,,
\end{align}
where $G(\bx-\by)=-|\bx-\by|^{-1}/(4\pi)$ is the convolution kernel (Green's function) for the inverse Laplace operator $\Delta^{-1}$. 
\end{corollary}
\paragraph{Proof:}This proof is essentially the same as that for the {\it Biot-Savart law} 
\cite{Saffman1992}. To begin, we take the curl of the equation for the vorticity 
\eqref{vorticity}, resulting in
\begin{align}
 \begin{split}
 -\frac{m}{\hbar}\nabla\times\bomega &= \nabla\times(\nabla\times\bLambda)\\
 &= \nabla\text{div}\bLambda - \Delta \bLambda\\
 \implies \frac{m}{\hbar}\nabla\times\bomega &= \Delta \bLambda\,,
\end{split}
\end{align}
where in the last line we have used that $\text{div}\bLambda = 
-\text{div}(\nabla\times\bbeta)=0$. We now have a vector Poisson equation 
which can be solved using the standard theory of Green's functions as
\begin{align}
  \bLambda(\bx)=  \frac{m}{\hbar}\int 
  G(\bx-\by)\nabla_{\by}\times\bomega(\by)\,\de^3y\,,
\end{align}
where the Green's function given above is the solution to the equation $\Delta G(\bx-\by) = 
\delta(\bx-\by)$. From here, one integrates by parts and then simply rearranges this expression as follows
\begin{align}
  \begin{split}\label{p-vort}
     \bLambda(\bx)&=  -\frac{m}{\hbar}\int 
  \nabla_{\by}G(\bx-\by)\times\bomega(\by)\,\de^3y\\
  &=\frac{m}{\hbar}\int 
  \nabla_{}G(\bx-\by)\times\bomega(\by)\,\de^3y\\
  &= \frac{m}{\hbar}  \nabla\times\int 
G(\bx-\by)\bomega(\by)\,\de^3y \\
 &= \frac{m}{\hbar}  \nabla\times\int 
G(\bx-\by)\int \bR_{\sigma}\,\delta(\by-\bR(\sigma))\,\text{d}\sigma\,\de^3y \\
&= \frac{m}{\hbar}  \nabla\times\int 
G(\bx-\bR(\sigma)) \bR_{\sigma}\,\text{d}\sigma\,,
  \end{split}
\end{align}
as required, having used that $\nabla G(\bx-\by) = -\nabla_{\by}G(\bx-\by)$ in the second 
equality. We complete this proof by presenting the explicit form of $\bLambda$, found using the explicit Green's function 
and computing the curl to obtain
\begin{align}
  \bLambda(\bx) = -\frac{m}{4\pi\hbar}\int 
  \bR_{\sigma}\times\frac{\bx-\bR}{|\bx-\bR|^3}\,\de\sigma\,.\label{LambdaVortexFilament}
\end{align}
\hfill$\square$\\

\begin{remark}[Holonomy for a vortex filament]
\hfill \\
For a vortex filament of the type \eqref{vortexfilament}, the holonomy around a fixed loop $c_0$ 
takes the particular form as the exponential of 
\begin{align}
-\hbar\oint_{c_0}\bLambda\cdot{\rm d}\bx =  m \int_{S_0}\left(\int\bR_{\sigma}\,\delta(\bx-\bR(\sigma)) {\rm d}\sigma\right)\cdot{\rm 
 d}\boldsymbol{S}\,.
\end{align}
Here, we have used Stokes' theorem to write the equality, where $S_0$ is a surface such that its boundary defines the loop $\partial S_0 =: 
c_0$. 
\end{remark}

Up to this point we have considered only the case of constant vorticity, consistent with the new Euler-Poincar{\'e} derivation of QHD. However, the appearance of $\bLambda$ in the Schr\"odinger equation \eqref{SchrodingerReconstructEqn} naturally invokes the thought of including the motion of such vortex filaments, dynamically coupled to the quantum motion.
The dynamics of vortex filaments were first given a Hamiltonian formulation by Rasetti and Regge \cite{RasettiRegge1975} who introduced a non-canonical Poisson bracket in the context of superfluids in 1975, resulting in many further developments, see e.g. \cite{Holm03, HolmStechmann2004, PennaSpera1989, PennaSpera1992, KuznetsovRuban1998, KuznetsovRuban2000,Volovik2006}.
In particular, the Rasetti-Regge Lagrangian for the self-induced motion of vortex filaments can be written as
\begin{align*}
  L(\bR,\partial_t{\bR}) = \frac{1}{3}\int\partial_t\bR\cdot\bR \times\bR_{\sigma} \,{\rm d}\sigma 
  - h(\bR)\,,
\end{align*}
where the Hamiltonian functional $h(\bR)$ must satisfy the consistency relation $ \bR_\sigma\cdot{\delta h}/{\delta \bR}=0$.
This condition has been shown to be valid for any Hamiltonian of the form $h=h(\bomega)$  \cite{Holm03, HolmStechmann2004}, which we prove explicitly in Appendix \ref{App:RR}.

Hence, in our case, we can exploit the Rasetti-Regge approach to let the hydrodynamic vortex filament \eqref{vortexfilament} move (so that $\bR=\bR(\sigma,t)$) in the modified Schr\"odinger equation 
\eqref{SchrodingerVortices}, as described in the following theorem.

\begin{theorem}[Dynamics of a coupled Schr\"odinger-vortex system]\label{Theorem:VORTICES}
  \hfill\\
Consider the the Rasetti-Regge-Dirac-Frenkel (RRDF) Lagrangian 
  $L:T\mathcal{F}(\mathbb{R},\mathbb{R}^3)\times TL^2(\mathbb{R}^3)\to \mathbb{R}$ 
  given by
  \begin{align}
\begin{split}
L(\bR,\partial_t{\bR},\psi,\partial_t\psi) = \frac{1}{3}&\int\partial_t\bR\cdot\bR \times\bR_{\sigma} \,{\rm d}\sigma+\operatorname{Re}\int \Bigg(\!i\hbar\psi^*\partial_t\psi
\\ &- \psi^* \bigg[\frac1{2m}{\bigg(-i\hbar\nabla-{m}\nabla\times \int \!G(\bx-\bR)\,\bR_\sigma\, \de \sigma\bigg)^{\!2}}+V\bigg]\psi\Bigg)\,{\rm d}^3{x}  
\,.\label{RRDFLag}
\end{split}
\end{align}
Here, the second line identifies the Hamiltonian functional $h(\bR,\psi)$ satisfying $ \bR_\sigma\cdot{\delta h}/{\delta \bR}=0$.
Then, the corresponding Euler-Lagrange equations of motion return the coupled 
system
\begin{align}
\partial_t{\bR}=&\,{\hbar}\left({\rm Im}(\psi^*\nabla\psi)-|\psi|^2\bLambda\right)\!\big|_{\bx=\bR\,}
+\kappa\bR_\sigma\,,\label{MOVINGVORTEX}
\\
i\hbar\partial_t\psi=&\,\frac1{2m}{(-i\hbar\nabla-\hbar\bLambda)^{2}}\psi+V\psi\,,
\end{align}
where $\bLambda=\bLambda(\bx,t)$ corresponds to the time-dependent generalisation of \eqref{p-vort} and $\kappa$ is an arbitrary quantity. 
\end{theorem}
\paragraph{Proof:} Clearly, the Euler-Lagrange equation for $\psi$ returns the 
Schr\"odinger equation \eqref{SchrodingerVortices} as required.  Upon considering an arbitrary 
Hamiltonian, one can compute 
the derivatives for the vortex dynamics to obtain
\begin{align}
  \frac{\delta L}{\delta \bR} = \frac{2}{3}\bR_{\sigma}\times\partial_t\bR + 
  \frac{1}{3}\bR\times\partial_t\bR_\sigma - \frac{\delta h}{\delta 
  \bR}\,,\qquad  \frac{\delta L}{\delta \partial_t\bR}= 
  \frac{1}{3}\bR\times\bR_\sigma\,.
\end{align}
In turn, the corresponding Euler-Lagrange equation is given by 
\begin{align}
\bR_\sigma\times\frac{\partial\bR}{\partial t}=\frac{\delta h}{\delta \bR}\,,
\end{align}
so that upon computing explicitly
\begin{align}
  \frac{\delta h}{\delta \bR}=-\frac{m}{\hbar}\bR_\sigma\times\frac{\delta h}{\delta \bLambda}\bigg|_{\bx=\bR}
={\hbar}\bR_\sigma\times\left(\operatorname{Im}(\psi^*\nabla\psi)-|\psi|^2\bLambda\right)\!\Big|_{\bx=\bR\,}\,,\label{DerivativeLambda}
\end{align}
we regain the desired vortex equation \eqref{MOVINGVORTEX}.
\hfill$\square$\\

We notice that the first term on the right hand side of equation \eqref{MOVINGVORTEX} can be rewritten as the hydrodynamic momentum $\bmu - \hbar D\bLambda$ 
(where now $\bmu$ again denotes the momentum map \eqref{QHDmomap}) evaluated at the vortex position $\bx=\bR$). Hence we conclude that this approach invokes a natural description in which the vortex motion is dragged around by the quantum fluid 
flow, whilst in turn the vorticity of the quantum fluid is driven by the filament. 

Before continuing we take a moment to summarise the key results of this chapter thus 
far. 

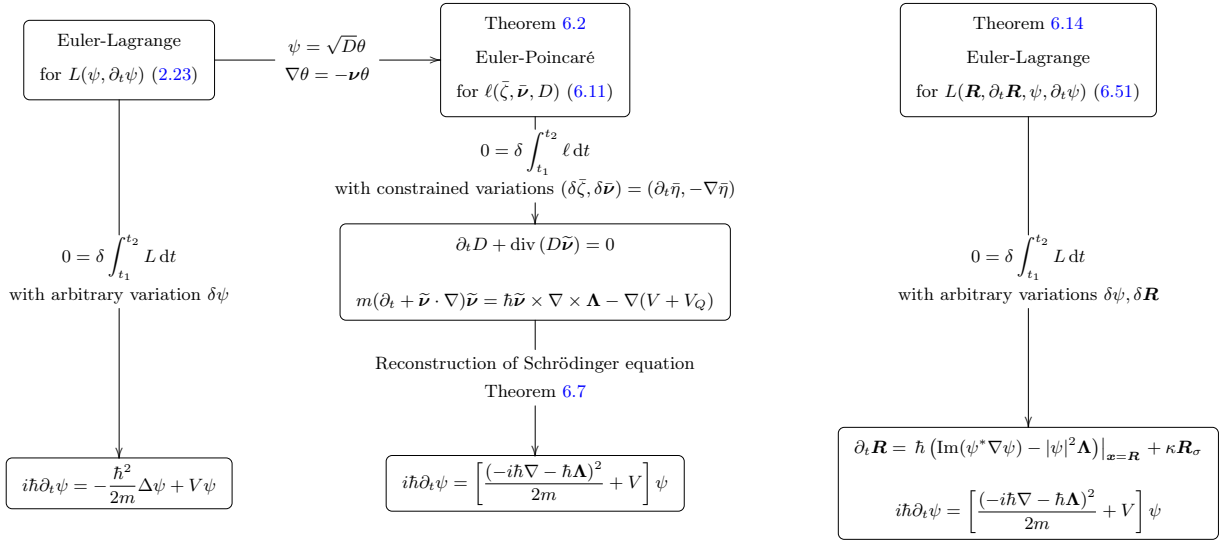
\begin{figure}[h]
\footnotesize\center
\noindent
\resizebox{1.2\textwidth}{!}{\begin{xy}
\hspace{0cm}
\xymatrix{
*+[F-:<3pt>]{
\begin{array}{c}
\vspace{0.1cm}\text{Euler-Lagrange}\\
\vspace{0.1cm} \text{for $L(\psi, \partial_t\psi)$ \eqref{DFStateLagrangian}}
\end{array}
} 
\ar[rr]|{\begin{array}{c}
\psi = \sqrt{D}\theta\\
\nabla\theta = -\boldsymbol{\nu}\theta
\end{array}}
\ar[dddd]|{\begin{array}{c}
\displaystyle 0=\delta\int_{t_1}^{t_2} L\,\text{d} t\\
\text{with arbitrary variation $\delta \psi$}
\end{array}}
&  &*+[F-:<3pt>]{
\begin{array}{c}
\vspace{0.1cm}\text{Theorem \ref{Theorem:NEWQHD}}\\
\vspace{0.1cm}\text{Euler-Poincar{\'e}}\\
\vspace{0.1cm} \text{for $\ell(\bar\zeta, \bar{\boldsymbol{\nu}}, D)$ \eqref{Lagrangian3}}
\end{array}
} 
\ar[dd]|{\begin{array}{c}
\displaystyle 0=\delta\int_{t_1}^{t_2} \ell\,\text{d} t\\
\text{with constrained variations $(\delta\bar\zeta,\delta \bar\bnu) = (\partial_t\bar\eta, -\nabla\bar\eta)$}
\end{array}}
 && *+[F-:<3pt>]{
\begin{array}{c}
\vspace{0.1cm}\text{Theorem \ref{Theorem:VORTICES}}\\
\vspace{0.1cm}\text{Euler-Lagrange}\\
\vspace{0.1cm} \text{for $L(\boldsymbol{R}, \partial_t\boldsymbol{R}, \psi, \partial_t\psi)$ \eqref{RRDFLag}}
\end{array}
} \ar[dddd]|{\begin{array}{c}
\displaystyle 0=\delta\int_{t_1}^{t_2} L\,\text{d} t\\
\text{with arbitrary variations $\delta \psi, \delta\bR$}
\end{array}}\\
&   &&&  \\
 &   &
*+[F-:<3pt>]{\begin{array}{c}
\partial_t D + \text{div}\left(D\widetilde{\boldsymbol{\nu}}\right)=0\\
\\
 m (\partial_t+\widetilde{\boldsymbol{\nu}}\cdot\nabla)\widetilde{\boldsymbol{\nu}}= \hbar\widetilde{\boldsymbol{\nu}}\times\nabla\times\boldsymbol{\Lambda}-\nabla(V+V_Q)
\end{array}}
\ar[dd]|{\begin{array}{c}
\text{Reconstruction of Schr\"odinger equation}\\
\text{Theorem \ref{Theorem:SchrReconstr}}
\end{array}}
&&
&  & &&\\
&  & &&\\
*+[F-:<3pt>]{
\begin{array}{c}
\displaystyle i\hbar\partial_t\psi = -\frac{\hbar^2}{2m}\Delta\psi + V\psi
\end{array}
}
&  &*+[F-:<3pt>]{
\begin{array}{c}
\displaystyle   
    i\hbar\partial_t\psi =    
    \left[\frac{(-i\hbar\nabla-\hbar\boldsymbol{\Lambda})^2}{2m}+V\right]\psi
\end{array}}
&&*+[F-:<3pt>]{\begin{array}{c}
\partial_t{\bR}=\,{\hbar}\left({\rm Im}(\psi^*\nabla\psi)-|\psi|^2\boldsymbol{\Lambda}\right)\!\big|_{\bx=\bR\,}
+\kappa\bR_\sigma\\
\\
\displaystyle  i\hbar\partial_t\psi =    
    \left[\frac{(-i\hbar\nabla-\hbar\boldsymbol{\Lambda})^2}{2m}+V\right]\psi
\end{array}}
}
\end{xy}}
\vspace{.5cm}
\caption{Schematic summary of key results from Sections \ref{Sec:NEWphase} and \ref{Sec:Vortices}.}
\label{figure3}
\end{figure}

The diagram is explained as follows. The left hand side represents the derivation of the Schr\"odinger equation as the Euler-Lagrange equation associated to the Dirac Frenkel Lagrangian. The arrow over to the middle column represents the use of the Lie group form of the Madelung transform \eqref{OurPolar} as well as the $\mathfrak{u}(1)$-connection 
via equation \eqref{NuConnectionDefinition}, allowing us to construct a new alternative Euler-Poincar{\'e} Lagrangian for QHD. The subsequent derivation of the QHD equations given in Theorem \ref{Theorem:NEWQHD} takes us down the middle column.
Recall that the resulting QHD equations of motion allow for the connection (fluid velocity) to 
possess non-zero constant curvature which, when expressed in terms of the Helmholtz decomposition \eqref{Helmholtz1}, manifests as an additional Lorentz force 
term $\hbar\widetilde{\bnu}\times\nabla\times{\bLambda}$ in the fluid equation \eqref{NuFluidWithLambda}. The next arrow down represents Theorem \ref{Theorem:SchrReconstr} 
in which an associated Schr\"odinger equation was reconstructed from the QHD equations and features the component of the connection $\hbar\bLambda$ 
via minimal coupling. The third column represents the passage to Section \ref{Sec:Vortices}, in which $\bLambda$ 
is written such that the vorticity corresponds to a vortex filament. This leads 
to Theorem \ref{Theorem:VORTICES} in which we postulate a new Lagrangian that 
couples the motion of the vortex filament \cite{RasettiRegge1975} to the Schr\"odinger equation. The final arrow indicates the Euler-Lagrange 
derivation of the coupled system.

In the next section we consider a first application of this new geometric approach to QHD for the BO factorisation from Section \ref{SectionBOApprox}.
Specifically, the incorporation of vortices in QHD may be useful in cases where the curvature of the Berry connection is given by a delta function, for example at the point of conical intersections between potential energy surfaces \cite{Kendrick2003}. 

\section{Application to Born-Oppenheimer molecular dynamics}\label{Sec:ConnectionQHDBOApp}
Motivated by the importance of geometric phase effects in quantum chemistry \cite{BohmBoyaKendrick1991,Kendrick2003, RyabinkinIzmaylov2013,RyabinkinEtAl2017,RyabinkinEtAl2014,GheribEtAl2015}, this section applies the formalism outlined so far in Chapter \ref{Chap:Holonomy} to the field of adiabatic molecular dynamics. 
In particular, we will apply the new approach to holonomy in QHD to the Born-Oppenheimer factorisation 
as given in Section \ref{Sec:BOClassicalNuclear}. 

To begin, recall that the BO ansatz 
of the molecular wavefunction $\Psi(\br,t)=\Omega(\br,t)\phi(\br)$ results in 
the DF Lagrangian 
  \begin{align*}
L={\rm Re}\int \Omega^*\left(i\hbar\partial_t\Omega-\frac{(-i\hbar\nabla+ \boldsymbol{\cal A})^2}{2M}\Omega - 
\epsilon(\phi,\nabla\phi)\Omega\right)\,\de^3r\,,
  \end{align*}
as described in Remark \ref{BODFVP}. We then proceed by writing the nuclear wavefunction as \eqref{OurPolar} so that $\Omega(\br,t)=\sqrt{D(\br,t)}\theta{(\br,t)}$.
Then, in the case of a real-valued electronic wavefunction (so that the Berry connection vanishes), following the process given in Section \ref{Sec:NEWphase} results in the nuclear QHD equations
\begin{align}\label{ourMT1}
  {\partial_t D}+\text{div}(D\widetilde\bnu)&= 0
  \,,
\\
 M (\partial_t+\widetilde{\bnu}\cdot\nabla)\widetilde{\bnu}   &= 
\hbar  \widetilde\bnu\times\nabla\times\bLambda - \nabla\bigg(E+\frac{\hbar^2}{2M}\|\nabla\phi\|^2+  V_Q\bigg)\,.
\label{ourMT2}
\end{align}
We notice that in the case when the gauge potential $\bLambda$ is singular, these equations are equivalent to those appearing in the Mead-Truhlar method of adiabatic molecular dynamics \cite{MeadTruhlar1979}.
Specifically, upon replacing $\widetilde{\bnu}=M^{-1}\nabla S$ and $\hbar\bLambda = -\nabla\alpha$ (where $\alpha$ is the multi-valued phase introduced by equation \eqref{MTalphaequation}) 
one recovers the QHD version of the molecular Aharonov-Bohm effect as described in Section 
\ref{Sec:BOClassicalNuclear}.

Despite the large body of work on the topological nature of the Berry phase 
arising due to the presence of conical intersections, essentially topological defects, in BO systems, 
the physical consistency of these singularities has been recently questioned by Gross and collaborators \cite{MinEtAl2014,RequistEtAl2016}.
 In their work, it is argued that the emergence of these topological structures is intrinsically associated to the particular type of adiabatic model arising from the BO factorisation ansatz. Indeed, the results in \cite{MinEtAl2014,RequistEtAl2016} and following papers show that these type of singularities are absolutely absent in the exact case of nonadiabatic dynamics.
This leads to the question of whether alternative approaches to adiabatic dynamics can be obtained in order to avoid dealing with conical intersections.
Notice that the absence of these defects does not imply the absence of a geometric phase.
Indeed, as the Berry connection is not generally vanishing in nonadiabatic dynamics, this leads to nontrivial holonomy which in turn does not arise from topological singularities.
In this context, a gauge connection associated to hydrodynamic vortices as in Section \ref{Sec:Vortices} may be representative of an alternative molecular Aharonov-Bohm effect in which vortex singularities are not quantised and thus produce a geometric phase depending on the integration loop. 
In this case, one could drop the quantum potential in \eqref{ourMT2} and select the single particle solution $D(\br,t)=\delta(\br-\bq(t))$ in \eqref{ourMT1}.
However, this approach may leave a $\delta-$like Lorentz force in the nuclear trajectory equation, thereby leading to major difficulties. 
The latter may be overcome by finding appropriate closures at the level of Hamilton's principle. For example, one could use a Gaussian wavepacket method, such as that presented in the appendix of the author's publication \cite{FoTr2020}.
 However, here we adopt a method  inspired by previous work in plasma physics \cite{Ho83} and geophysical fluid dynamics \cite{Ho86}.

 To do so, we recall Proposition \ref{Prop:EP2Structure} and consider the alternative hydrodynamic Lagrangian of the type \eqref{AlternativeQHDLagrangian}, underlying equations \eqref{ourMT1}-\eqref{ourMT2}:
\begin{align}
\ell(D,\widetilde\bnu)=\int\!D\left(\frac12 {M}|\widetilde\bnu|^2+\hbar\widetilde\bnu\cdot\bLambda-V_Q-\epsilon\right)\de^3 r\,.
\end{align}
Here, the Eulerian variables $D(\br,t)$ and $\widetilde\bnu(\br,t)$ are related to the  Lagrangian fluid path $\boldsymbol\eta(\br,t)$ (Bohmian trajectory) by the relations $\dot{\boldsymbol\eta}(\br,t)=\widetilde\bnu({\boldsymbol\eta}(\br,t),t)$ and $D({\boldsymbol\eta}(\br,t),t)\,\de^3{\eta}(\br,t)= D_0(\br)\,\de^3r$. 
If the Bohmian trajectory is given by ${\boldsymbol\eta}(\br,t)=\br+\bq(t)$, i.e. restricting to pure translations, then $\widetilde\bnu(\br,t):=\dot{\boldsymbol\eta}(\br_0,t)=\dot{\bq}(t)$ and  the Lagrange-to-Euler map \eqref{defns} 
becomes
\begin{align}
D(\br,t):=\int D_0(\br_0)\delta(\br-\boldeta(\br_0,t))\,\de^3r_0 = 
  D_0(\br-\bq(t))\,.
\end{align} 
Then, since in this case {$\int \! DV_Q\,\de^3 r=\text{const.}$}, the Lagrangian $\ell(D,\widetilde\bnu)$ becomes
\begin{align}
L(\bq,\dot\bq)=\frac{M}2|\dot\bq|^2+\int\!D_0(\br-\bq)\big[\hbar\dot\bq\cdot\bLambda(\br)-\epsilon(\br)\big]\,\de^3 r 
\label{Jeff}
\,.
\end{align}
Here, $D_0$ is typically a Gaussian distribution and we recall that we are considering the case of a real electronic wavefunction, so that $\boldsymbol{\cal A}=0$ in the definition of $\epsilon$ \eqref{effectiveelecpot}.
 Then, one obtains the  Euler-Lagrange equation
\begin{equation}\label{particles+vortices}
M\ddot\bq=\hbar\dot\bq\times\nabla\times\int \!D_0(\br-\bq)\bLambda(\br)\,\de^3 r
-  \nabla\!\int \!\epsilon(\br)D_0(\br-\bq)\,\de^3 r
\,,
\end{equation}
where $\bLambda$ is given as in \eqref{p-vort}. We see that the nuclear density acts as a convolution kernel regularising both the connection $\bLambda$ and the potential energy surface appearing in $\epsilon$. For example, while here we are considering non-quantised hydrodynamic vortices, this method could be used to regularise topological singularities arising from conical intersections. Then, the non-quantised geometric phase and the regularised potential energy surface read
\[
-\hbar\oint_{c_0}\int \,D_0(\br'-\br)\bLambda(\br')\,\de^3 r'\cdot\de\br
\,,\qquad\qquad 
\int\,D_0(\br'-\br)E(\br')\,\de^3 r'
\,,
\]
respectively. We complete this section by discussing the application of Theorem \ref{Theorem:VORTICES} 
to adiabatic molecular dynamics.
\begin{remark}[Hydrodynamic vortices in adiabatic dynamics]
\hfill \\
In the present context of BO adiabatic dynamics, the self-consistent vortex evolution may be included upon constructing a RRDF Lagrangian of the type \eqref{RRDFLag}.
Specifically, consider the replacement
 \begin{align}
   L(\bq,\dot{\bq})\to L(\bq,\dot{\bq})+\frac13 \int\!\partial_t\bR\cdot\bR \times\bR_{\sigma} \,{\rm d}\sigma\,, 
 \end{align}
 for the new BO Lagrangian \eqref{Jeff}. Whilst we would again recover the nuclear trajectory equation \eqref{particles+vortices}, in this case the vortex evolution equation would read
  \begin{align}
  \partial_t\bR=MD_0(\bR-\bq)\dot\bq +\kappa\bR_\sigma\,.
\end{align}
It remains an interesting possibilty for future research to understand whether 
these models may help for modelling purposes in quantum chemistry.
\end{remark}

\section{Amplitude connection and osmotic velocity}\label{Sec:Osmotic}

This final section of Chapter \ref{Chap:Holonomy} completes the holonomy formulation of QHD by constructing an analogous connection for the dynamics of amplitude of the wavefunction.
In particular, taking inspiration from the physics literature on QHD, we shall develop the geometry underlying the so-called {\it osmotic velocity} defined as 
\begin{align*}
\bu:=\frac{\hbar}{2m} \frac{\nabla D}{D} = \frac{\hbar}{2m}\nabla\ln D\,,
\end{align*}
 in which $m^{-1}\hbar/2$ is known as the {\it quantum diffusion coefficient} \cite{Wyatt2006, Nelson1967}.
 As mentioned for example in \cite{Spera2016}, the osmotic velocity can be regarded as a trivial gauge connection for scaling transformation $\mathcal{F}(\mathbb{R}^3,\mathbb{R}^+)$. 
 This section develops this point further by allowing for the non-trivial case, following the procedure applied to the phase in previous sections. 
 We note that for the rest of this section we will return to working with the 
 amplitude of the wavefunction $R:=|\psi|=\sqrt{D}$, so that the osmotic velocity 
 above becomes $\bu = \hbar/m \,\nabla\ln R$. We separate the subsequent material 
 under different headings for ease of reading.

\subsubsection{Variational principle and equations of motion}\label{Sec:OsmoticSubsec:1}
To begin recall the group polar form of the wavefunction \eqref{OurPolar}, now written as $\psi = R\theta$. 
In addition to the phase evolution $\theta=\Theta\theta_0$ \eqref{ThetaEvo}, we 
allow the amplitude of the wavefunction to evolve under the group of scaling transformations as follows
\begin{align}
  R(\bx,t)=\lambda(\bx,t)R_0(\bx)\,,
\end{align}
where $\lambda \in \mathcal{F}(\mathbb{R}^3, \mathbb{R}^+)$.
Hence, performing the analogous procedures to those used for the phase connection in Section \ref{Subsec:PhaseConnection}, we have that
\begin{align}
  \partial_t R = \partial_t \lambda \,\lambda^{-1} \,R =: \tau R\,,
  \qquad\text{where}\quad \tau \in 
  \mathcal{F}(\mathbb{R}^3)\,,
\end{align}
and we can introduce a connection via
\begin{align}
 \begin{split} 
 \nabla R &= \nabla\lambda R_0 + \lambda \nabla R_0\\
  &= \nabla\lambda\,\lambda^{-1}\,R - \lambda \bupsilon_0 R_0\\
  &= -(-\nabla\lambda\,\lambda^{-1} + \bupsilon_0)R=: -\bupsilon R\,,
  \qquad\text{where}\quad\bupsilon \in 
  \Lambda^{1}(\mathbb{R}^3)\label{AmplitudeConnectionDef}\,.
 \end{split} 
\end{align}
Again, following the previous  discussion, we can see how at this stage $\nabla\times\bupsilon=0$. 
However, this relation will be relaxed, after constructing the Lagrangian, in the remainder of this section by applying the approach described in Remark \ref{jack}.

Having introduced these new variables we can now formulate QHD in terms of both connections.

\begin{theorem}[Euler-Poincar{\'e} formulation for QHD connections]\label{Theorem:EP2Connections}
  \hfill \\
  Written in terms of the variables $(R,\bupsilon,\bar\zeta,\bar\bnu)$, the transformation \eqref{OurPolar} takes the DF Lagrangian \eqref{DFStateLagrangian} into the reduced Lagrangian
   $\ell:\mathcal{F}(\mathbb{R}^3,\mathbb{R}^+)\times \Lambda^{1}(\mathbb{R}^3)\times \mathcal{F}(\mathbb{R}^3)\times\Lambda^{1}(\mathbb{R}^3)\to \mathbb{R}$ 
  given by
    \begin{align}
  \ell(R,\bupsilon,\bar\zeta,\bar\bnu) =  \int R^2 \Big( \bar\zeta-\frac{|\bar\bnu|^2}{2m} - \frac{\hbar^2}{2m}|\bupsilon|^2  - V\Big)\,{\de}^3x \label{Lagrangian4}\,.
\end{align}
Upon applying Hamilton's principle for Euler Poincar{\'e} variations 
\begin{align*}
  \delta R = \chi R\,,\qquad \delta\bupsilon = 
  -\nabla\chi\,,\qquad\delta\bar\zeta = \partial_t\bar\eta\,,\qquad 
  \delta\bar\bnu = -\nabla\bar\eta\,,
\end{align*}
where $\chi := \delta\lambda \,\lambda^{-1}$ and $\bar\eta := i\hbar\delta\Theta\,\Theta^{-1}$ 
are both arbitrary, we obtain the following system of equations in
$(R,\bar\bnu,\bupsilon)$:
\begin{align}
    \partial_t(R^2) &=-{\rm div}\left(R^2\frac{\bar\bnu}{m}\right)\,,\label{transport3}\\
\partial_t\bar\bnu + \frac{1}{2m}\nabla|\bar\bnu|^2 &= -\nabla\big(V+V_Q(R,\bupsilon)\big)\,,\label{dave}\\
m\partial_t\bupsilon &= \nabla\left(\frac{\nabla R}{R}\cdot 
\bar\bnu\right)+\frac{1}{2}\Delta\bar\bnu + 
\frac{1}{2}\nabla\times(\nabla\times\bar\bnu)\,,\label{David}
\end{align}
where the quantum potential is now expressed in terms of the gauge connection $\bupsilon$ 
as
\begin{align}
  V_Q(R, \bupsilon) = \frac{\hbar^2}{2m}\left(|\bupsilon|^2 + \frac{{\rm div}(R^2\bupsilon)}{R^2} 
  \right)\,.\label{ConnectionQuantPot}
\end{align}
\end{theorem}
\paragraph{Proof:} The Lagrangian \eqref{Lagrangian4} is easily derived from the QHD Lagrangian \eqref{Lagrangian3} 
from Section \ref{Sec:NEWphase} upon using the new expression for the amplitude 
connection. 
Before deriving the equations of motion for this explicit Lagrangian, we proceed in the most general case considering a 
Lagrangian of the form $\ell=\ell(R,\tau,\bupsilon, \theta, \bar\zeta,\bar\bnu)$, that is, 
depending on all dynamical variables introduced in this approach. Hence, we compute the 
following Euler-Poincar{\'e} variations \cite{MaRa2013},
\begin{align*}
  \delta R = \chi R\,, \qquad\delta \tau = \partial_t\chi\,, \qquad\delta\bupsilon = 
  -\nabla\chi\,, \\
  \delta \theta = -\frac{i}{\hbar}\bar\eta\,\theta\,,\qquad \delta\bar\zeta = \partial_t\bar\eta\,,\qquad 
  \delta\bar\bnu = -\nabla\bar\eta\,,
\end{align*}
again with $\chi := \delta\lambda \,\lambda^{-1}$ and $\bar\eta:= i\hbar\delta\Theta\,\Theta^{-1}$ 
both arbitrary. As usual we expand Hamilton's principle using these variations and simple applications of integration by parts yields the following general set of 
equations
\begin{align*}
  \partial_t\left(\frac{\delta \ell}{\delta\tau}\right) &= \text{div}\left(\frac{\delta \ell}{\delta \bupsilon}\right) 
  + R\frac{\delta \ell}{\delta R}\,,\\
  \partial_t\left(\frac{\delta \ell}{\delta\bar\zeta}\right) &= \text{div}\left(\frac{\delta \ell}{\delta \bar\bnu}\right) 
  + \frac{i}{\hbar}\theta^{-1}\frac{\delta \ell}{\delta 
  \theta}\,,
\end{align*}
along with the auxiliary equations
\begin{align*}
  \partial_t R &= \tau R\,,\qquad
  \quad\partial_t\bupsilon = - \nabla\tau\,,\\
    \partial_t \theta &= -\frac{i}{\hbar}\bar\zeta\theta\,,\qquad
  \partial_t\bar\bnu = - \nabla\bar\zeta\,.  
\end{align*}
Now, going back to our QHD Lagrangian \eqref{Lagrangian4}, we specialise the 
first two equations to get
\begin{align}
  \bar\zeta - \frac{|\bar\bnu|^2}{2m} -\frac{\hbar^2}{2m}\left(|\bupsilon|^2 + \frac{\text{div}(R^2\bupsilon)}{R^2} \right)- V &= 0 
  \label{xiHamiltonJacobi2}\,,\\
    \partial_t(R^2) + \text{div}\left(R^2\frac{\bar\bnu}{m}\right)&=0\,,
\end{align}
which clearly correspond with \eqref{xiHamiltonJacobi1} and \eqref{transport} found in Section \ref{Subsec:PhaseConnection}, only this time allowing us to identify a new 
expression for the quantum potential \eqref{ConnectionQuantPot}. At this point, we close the set of equations by writing the equation for the 
evolution of $\bupsilon$ in terms of $R$ and $\bar\bnu$. Noting that $ \partial_t\bupsilon = -\nabla(R^{-1}\partial_t{R})$, we use the transport equation to 
obtain
\begin{align*}
m\partial_t\bupsilon = \nabla\left(\frac{\nabla R}{R}\cdot 
\bar\bnu\right)+\frac{1}{2}\Delta\bar\bnu + 
\frac{1}{2}\nabla\times(\nabla\times\bar\bnu)\,.
\end{align*}
Then, as usual taking the gradient of the phase equation \eqref{xiHamiltonJacobi2} and 
using the auxilliary equation for $\nabla\bar\zeta$,  we regain the required set of equations. 
\hfill$\square$\\

Before continuing, we make a few comments regarding this theorem. Firstly, like for Theorem \ref{Theorem:NEWQHD}, we now postulate this Lagrangian so that $\bupsilon$ 
is only defined as the solution to the general set of equations above which, as we shall again see, allows it to possess constant non-trivial curvature. 
However, as the equations of motion remain the same (up to the form of the quantum potential) after 
introducing the connection $\bupsilon$, we see how $\bar\bnu$ and $\bupsilon$ play dramatically different roles in terms of the dynamics. 
In fact, as we shall see in Section \ref{SchrodingerReconstructionv2} their non-trivial curvatures enter the 
Schr{\"o}dinger equation as a vector and scalar potential respectively. 

At this point it is again natural to introduce the fluid velocity variable $\widetilde\bnu:=m^{-1}\bar\bnu$ 
as defined in previous sections. However, before doing so we consider the 
Helmholtz decomposition for $\bupsilon$, allowing us to characterise the non-zero constant 
curvature of the amplitude connection.

\subsubsection{Helmholtz decomposition}\label{Sec:OsmoticSubsec:2}
At this stage, we wish to apply the Helmholtz decomposition for $\bupsilon$ which, as we shall see, will eventually allow us to decouple the system from the amplitude $R$.
 To do so, we recall that $\partial_t(R^{-1}\nabla R + \bupsilon)=0$ which allows us to write $\bupsilon = - R^{-1}{\nabla R}+ \bkappa(\bx)$ for $\bkappa(\bx)$ an arbitrary constant vector of the integration.
  Upon using the natural logarithm to express the first term as a pure gradient, we see that this is in fact exactly the Helmholtz decomposition.
   Without loss of generality we simply redefine the constant function $\bkappa(\bx)$ as a curl and choose the factor of $1/2$ for later convenience so that 
we write the more usual expression
\begin{align}
  \bupsilon = - \nabla \ln{R} + \frac{1}{2}\nabla\times\bkappa\,.\label{john}
\end{align}

\begin{remark}[Exponential form of the amplitude]\label{Rem:EXPAMP}
  \hfill \\
  The form of the Helmholtz decomposition \eqref{john}, in particular the first 
  term in which there is no ambiguity in using the formula for the derivative of the logarithm, suggests writing the original wavefunction in the form $\psi=e^{\mathcal{R} 
  }
  \theta$, utilising the exponential form for the amplitude. Then, considering 
  this replacement $R= e^{\mathcal{R} 
  }$, equation \eqref{john} would read $\bupsilon=-\nabla {\mathcal{R} 
  }$ (omitting the non-zero curvature term). Such approaches have been considered in \cite{Nelson1967}  
  in which a stochastic theory of Brownian motion is shown to give rise to equations entirely equivalent to QHD. We will elaborate further on the connections with this work later on. 
\end{remark}

Using this new expression for $\bupsilon$ allows us to further evaluate the quantum potential and eliminate any dependence on $R$, writing it solely as
\begin{align}
  V_Q(\bupsilon) = \frac{\hbar^2}{2m}\Big(-|\bupsilon|^2 +\text{div}\bupsilon+ \bupsilon \cdot (\nabla\times\bkappa)
  \Big)\label{quantumpotentialsigma}\,.
\end{align}
Notice that we can select the constant vector $\bkappa(\bx)$ as an 
inital condition on $\bupsilon$ as we wish. For example, choosing $\bkappa(\bx) = (\nabla\times\bupsilon) \times 
\boldsymbol{e}_3$, where $\boldsymbol{e}_3$ is a unit basis vector in $\mathbb{R}^3$ (or equivalently any vector constant in both time and space), allows us to rewrite  
the expression for the quantum potential once more as
\begin{align}
  V_Q(\bupsilon) = \frac{\hbar^2}{2m}\Big(-|\bupsilon|^2 +\text{div}(\bupsilon- \bupsilon \times\bkappa )
  \Big)\,.
\end{align}

%

\subsubsection{Fluid velocity, holonomy and helicity}\label{Sec:OsmoticSubsec:3}
Again motivated by the form of the system of equations, here we will work with the fluid velocity expression of the phase connection, $\widetilde\bnu$.
 However, unlike section \ref{Sec:PhaseConnectionsSubSec:2}, the introduction of $\bupsilon$ allows us to go further and express the system solely in terms of the connections $\widetilde\bnu$ and $\bupsilon$ 
 as follows:
 \begin{proposition}[Closed system for the QHD connections]
   \hfill\\ Utilising the Helmholtz decomposition \eqref{john} for the amplitude 
   connection decouples the dynamical equations for the connections $\bar\bnu$ and $\bupsilon$ presented in Theorem \ref{Theorem:EP2Connections}  from the 
   transport equation and hence the amplitude itself, resulting in a closed system 
   for $(\bar\bnu, \bupsilon)$ given by,
   \begin{align}
(\partial_t+\widetilde\bnu\cdot\nabla)\widetilde\bnu &= -\widetilde\bnu\times(\nabla\times\widetilde\bnu) - 
  m^{-1}\nabla\left(V+\frac{\hbar^2}{2m}\Big(-|\bupsilon|^2 +{\rm div}\bupsilon+ \bupsilon \cdot (\nabla\times\bkappa)
  \Big)\right)\label{separatednuequation}\,,\\
  \partial_t\bupsilon &= -\nabla(\bupsilon\cdot\widetilde\bnu) + \frac{1}{2}\nabla\left(\widetilde\bnu\cdot\nabla\times\bkappa\right)+\frac{1}{2}\Delta\widetilde\bnu + 
\frac{1}{2}\nabla\times(\nabla\times\widetilde\bnu)\label{separatedsigmaequation}\,.
\end{align}
 \end{proposition}
 \paragraph{Proof:}
 To see this, we simply replace the new expression \eqref{quantumpotentialsigma} for the quantum potential $V_Q(\bupsilon)$ in 
the equation for $\bar{\bnu}$ \eqref{dave} and also replace the relation \eqref{john} into 
the $\bupsilon$ equation, resulting in the equations \eqref{separatednuequation} and \eqref{separatedsigmaequation} as given above, both of which no longer depending on the amplitude $R$. 

\hfill$\square$\\

We now turn our attention to the holonomy of the new amplitude connection $\bupsilon$. As was the case for the phase connection,  it follows from the auxilliary equation $\partial_t \bupsilon = -\nabla\tau$ that 
\begin{align}
\frac{\de}{\de t} \oint_{c_0}\bupsilon\cdot\text{d}\bx = 0\,,
\end{align} 
for a fixed loop $c_0$. As before, geometrically this corresponds to a constant holonomy, with the value amounting to an initial condition.
We also note that we now have the additional helicity
\begin{align}
  H_{\bupsilon} &:= \int \bupsilon\cdot(\nabla\times\bupsilon)\,{\rm d}^3x\,,
  \end{align}
  which, using the equivalent computation as in Remark \ref{HelicityRemark}, can be shown to be constant in time. In 
  more generality we notice that the helicity corresponding to the sum of the connections  $\bupsilon + \widetilde\bnu \in 
  \Lambda^{1}(\mathbb{R}^3)$ given by
  \begin{align}
  H_{\bupsilon+\widetilde\bnu} &:= \int (\bupsilon+\widetilde\bnu)\cdot\nabla\times(\bupsilon+\widetilde\bnu)\,{\rm d}^3x\,,
  \end{align}
is also preserved in time. This can be seen by expanding and again using the 
same technique of taking the dot product of one of the auxilliary equations with the necessary curvature to show that each component (including both cross helicities) is constant in time.

\subsubsection{Comparison with the equations of Nelson \cite{Nelson1967}}\label{Sec:OsmoticSubsec:4}
  As mentioned in Remark \ref{Rem:EXPAMP}, in Chapter 15 of \cite{Nelson1967} a QHD-type system is derived using a stochastic approach to Brownian motion on the atomic scale. The resulting equations  
  for the `osmotic velocity' $\bu$ and `current velocity' $\bv$ are given by  
  \begin{align}
  \partial_t\bv &= \frac{\hbar}{2m}\Delta\bu + \frac{1}{2}\nabla|\bu|^2- \frac{1}{2}\nabla|\bv|^2 
  - m^{-1}\nabla V\,,\label{Nelson1}\\
  \partial_t\bu &= -\frac{\hbar}{2m}\Delta\bv - \nabla(\bu\cdot\bv)\,,\label{Nelson2}
\end{align}
for $V$ some external potential (equations (15.3) 
  in \cite{Nelson1967}). The work proceeds by showing the equivalence 
with the Schr\"odinger equation upon defining $\nabla \mathcal{R}=m\hbar^{-1}\bu$ and $\nabla \mathcal{S}=m\hbar^{-1}\bv$ 
and using the change of variable $\psi=e^{\mathcal{R}+i\mathcal{S}}$. The 
relation between these calligraphic amplitude and phase variables and the 
counterparts we have introduced earlier are given by
\begin{align}
  R = e^{\mathcal{R}},\qquad s = \hbar\mathcal{S}\,.
\end{align}
Once again, with the regard to the phase variables, we stress that in the connection 
approach presented in this thesis $s$ is single-valued whilst $\mathcal{S}$, arising from the exponential 
form of the wavefunction, is not.

Naturally, there is much similarity between Nelson's equations for $\bv$ and $\bu$ and our equations \eqref{separatednuequation} and \eqref{separatedsigmaequation}. 
To make the explicit connection, we rewrite our equations in the following form
\begin{align*}
\partial_t\widetilde\bnu &= -\frac{\hbar^2}{2m^2}\nabla\left(-|\bupsilon|^2 +\text{div}\bupsilon+ \bupsilon \cdot (\nabla\times\bkappa)\right)-\frac{1}{2}\nabla|\widetilde{\bnu}|^2 - 
  m^{-1}\nabla V\,,\\
  \partial_t\bupsilon &= \frac{1}{2}\nabla\text{div}(\widetilde{\bnu})-\nabla(\bupsilon\cdot\widetilde\bnu) + \frac{1}{2}\nabla\left(\widetilde\bnu\cdot\nabla\times\bkappa\right)\,.
\end{align*}

Then, one recalls that Nelson's $\bu$ and $\bv$ are both defined as pure gradients. Indeed in the case of $\bkappa=0$, corresponding to zero curvature of the connection $\bupsilon$, we could replace the operator $\nabla\text{div}$ with the Laplacian $\Delta$ so that our equations now read
\begin{align*}
\partial_t\widetilde\bnu &= -\frac{\hbar^2}{2m^2}\Delta\bupsilon+\frac{\hbar^2}{2m^2}\nabla|\bupsilon|^2 -\frac{1}{2}\nabla|\widetilde{\bnu}|^2 - 
  m^{-1}\nabla V\,,\\
  \partial_t\bupsilon &= \frac{1}{2}\Delta\widetilde{\bnu}-\nabla(\bupsilon\cdot\widetilde\bnu) \,.
\end{align*}
Upon changing variables according to $\bv = \widetilde\bnu$ and $\bu = -\hbar m^{-1}\bupsilon$ the equations can be shown equivalent to Nelson's \eqref{Nelson1} and \eqref{Nelson2}.  
Ultimately, this means that whilst in \cite{Nelson1967} the QHD equations have been expressed in terms of variables that are essentially our connections $\widetilde\bnu$ and $\bupsilon$, they, as always, were themselves written as pure gradients and thus do not contain any of the terms corresponding to non-trivial curvature. 
The possibility that the dynamical connections governing QHD possess non-trivial constant curvature is the main new feature of the approach presented in this chapter and is yet to be known whether this may result in a measurable physical phenomena or prove useful for modelling purposes. 

\subsubsection{Reconstruction of the Schr{\"o}dinger equation}\label{SchrodingerReconstructionv2}
We conclude this section with the reconstruction of a Schr{\"o}dinger equation analogous to Section \ref{Subsec:HolonomicSchr}. This can be achieved from the system of equations for the gauge 
constructions from Theorem \ref{Theorem:EP2Connections}. 
\begin{theorem}[Reconstruction of the Schr{\"o}dinger equation II]
  \hfill \\
  Consider the system of equations \eqref{transport3}-\eqref{David} given in Theorem \ref{Theorem:EP2Connections}. Together with the \\
  Helmholtz decompositions for each connection 
\eqref{NuHelmholtz} and \eqref{john}, the time evolution of a wavefunction (reconstructed from the polar form \eqref{OurPolar}) results in an associated Schr\"{o}dinger equation 
\begin{align}
  i\hbar\partial_t\psi &=\left[\frac{(-i\hbar\nabla-\hbar\bLambda)^2}{2m}+\left(V +\frac{\hbar^2}{8m}|\nabla\times\bkappa|^2\right)\right]\psi  \,.\label{RECONSTRUCTAGAIN}
\end{align}
\end{theorem}
\paragraph{Proof:} The full computational details of this proof can be found in 
Appendix \ref{App:Reconstruct}. To begin, we follow the same computation as in Section \ref{Subsec:HolonomicSchr} 
(now using the variable $\bnu = -i\hbar^{-1}\bar\bnu$)
\begin{align}
  i\hbar\partial_t\psi  &= \left[\frac{\hbar^2}{m}\left(\frac{\nabla R}{R}\cdot\bnu\right)+\frac{\hbar^2}{2m}\text{div}(\bnu)-\frac{\hbar^2}{2m}|\bnu|^2 +V_Q(\bupsilon)\right]\psi 
+ V\psi\,.
\end{align}
Interestingly, upon using the Helmholtz decomposition for $\bupsilon$ \eqref{john}, we can write the kinetic term (in parentheses) solely in terms of ${\bnu}$ 
and $\bupsilon$ to obtain the following form
\begin{align*}
  \frac{\hbar^2}{2m}\bigg(-|\bupsilon+{\bnu}|^2 + \text{div}(\bupsilon+{\bnu}) + 
  (\bupsilon+{\bnu})\cdot\nabla\times\bkappa\bigg)\,.
\end{align*}
It is also very important to note that although (as we shall see below) $\bupsilon + {\bnu}$ is complex, the expression $|\bupsilon + {\bnu}|^2$ is not the complex inner product, rather it simply denotes the scalar product $|\bupsilon + {\bnu}|^2=(\bupsilon + {\bnu})\cdot(\bupsilon + {\bnu})$, which may be complex in general.
 At this point, as before, we see that we must manipulate the kinetic energy term to get back to 
$\psi$. Before continuing, we recall the following relations 
\begin{align}
  \frac{\nabla R}{R} = \frac{\text{Re}(\psi^*\nabla\psi)}{|\psi|^2}\,, 
  \qquad
    \frac{\nabla \theta}{\theta} = 
    \frac{i\text{Im}(\psi^*\nabla\psi)}{|\psi|^2}\,,\qquad
     \frac{\nabla \psi}{\psi}=  \frac{\nabla R}{R} +
    \frac{\nabla \theta}{\theta} \,,
\end{align}
which, using our Helmholtz expressions \eqref{john} and \eqref{Helmholtz1}, imply 
that
\begin{align}
    \bupsilon = -\frac{\text{Re}(\psi^*\nabla\psi)}{|\psi|^2} + 
 \frac{1}{2} \nabla\times\bkappa\,,\qquad
\bnu =  -\frac{i\text{Im}(\psi^*\nabla\psi)}{|\psi|^2} +i\bLambda\,,
\end{align}
so that together we have
\begin{align*}
 \bupsilon +\bnu &= -\frac{\nabla\psi}{\psi} + 
 \frac{1}{2} \nabla\times\bkappa+ i\bLambda\,.
\end{align*}
Using this last equation enables us to rewrite the kinetic term in terms of 
$\psi$ as
\begin{align}
  -|\bupsilon+{\bnu}|^2 + \text{div}(\bupsilon+{\bnu}) + 
  (\bupsilon+{\bnu})\cdot\nabla\times\bkappa
  &= -\frac{\Delta \psi}{\psi} + 2i 
  \frac{\nabla\psi}{\psi}\cdot\bLambda + 
  |\bLambda|^2 + 
  \frac{1}{4}|\nabla\times\bkappa|^2\label{kinetictermsimplified}\,,
\end{align}
(see Appendix \ref{App:Reconstruct}) which substituting back into our original expression results in 
the desired Schr\"odinger equation \eqref{RECONSTRUCTAGAIN}.

\hfill$\square$\\

Looking closer at the reconstructed Schr\"odinger equation \eqref{RECONSTRUCTAGAIN}, we see that $\bLambda$ once again plays the same role as in Section \ref{Subsec:HolonomicSchr} as a vector potential, whilst $\nabla\times\bkappa$, corresponding to the curvature of our $\mathbb{R}^+$ connection $\bupsilon$, appears as a scalar potential term.
 Indeed from equation \eqref{quantumpotentialsigma} we see how this term can be understood a remnant of the quantum potential in the case of non-zero curvature for $\bupsilon$.

\chapter{Non-Abelian gauge connections}\label{Chap:Gamma}
In the previous chapter we introduced Abelian connections for the phase and amplitude dynamics of the wavefunction
which provided an alternative formulation of quantum hydrodynamics.
In this chapter, we consider the role of non-Abelian connections 
in quantum mechanics.

In Section \ref{Sec:PauliQHD} we consider the hydrodynamic formulation of the 
Pauli equation. Serving as another manifestation of two-level exact 
factorization dynamics from Section \ref{Sec:2Level}, this system is introduced here in preparation for the subsequent section.
The momentum map QHD formulation of Chapter \ref{Chap:QHD} is used
to formulate a Kelvin-Noether theorem and make relevant remarks relating to the previous literature on this topic. 

In Section \ref{Sec:NonAbelianPauliConnection} the unitary dynamics of the Pauli 
spinor are considered from the perspective of the spin vector, evolving under the action of the rotation group $SO(3)$. 
Following a similar construction as for the QHD connection, this allows 
the first introduction of a non-Abelian connection in this thesis, from which the corresponding Euler-Poincar{\'e} 
equations of motion and subsequent interesting features act as a stepping stone for the remainder of this chapter.  

Section \ref{Sec:GeneralTheory} presents a corresponding general theory for introducing gauge 
connections in mechanical systems. In this approach, the non-zero 
curvature relation (previously seen for both the Abelian QHD and non-Abelian spin vector connections) is used only to construct a reduced Lagrangian, from which 
the resulting Euler-Poincar{\'e} equations allow for more general connections without such constraints. 

Then, in Section \ref{Sec:NonAbelianQuantumSystems} the general theory is 
applied to the action of the full unitary group on a Hilbert space, resulting in 
a non-Abelian connection relevant in a wide variety of quantum systems. This connection
 is shown to generalise that used for the Pauli spin vector from Section \ref{Sec:NonAbelianPauliConnection} 
 as well as recover the Berry connection as its $\mathcal{U}(1)$ projection. This section concludes by utilising this new connection to provide new 
 perspectives on the quantum geometric tensor and associated highly geometric objects.
 
 The chapter concludes with Section \ref{Sec:GammaEF} in which the non-Abelian 
 quantum connection is used to provide further insights into the nature of exact factorization 
 dynamics. In this setting the connection can be viewed as a type of correlation 
 observable accounting for the explicit electron-nuclear coupling that 
 distinguishes EF from mean-field systems. A more general Euler-Poincar{\'e} formulation of 
 the dynamics is presented featuring the non-flat version of the connection, 
 before specialising to the zero-curvature case in order to show the equivalence 
 with the system derived in Section \ref{NuclearFrameSection}.


\section{Hydrodynamic formulation of the Pauli equation}\label{Sec:PauliQHD}
Following the successful hydrodynamic description of the Schr\"odinger equation by Madelung 
in 1926 \cite{Madelung1926,Madelung1927}, later work focused on extending this description 
to the Pauli equation (recall equation \eqref{PauliEquation} from Section \ref{Sec:1-Postulates}), i.e. to include quantum systems with spin. 
Almost simultaneously two papers \cite{BohmSchillerTiomnoA1955, Takabayasi1955} 
by independent authors appeared in 1955 doing exactly this, in both cases 
defining the fluid velocity 
\begin{align*}
  \bu =   
  \frac{\hbar\text{Im}\braket{\Psi|\nabla\Psi}}{m\braket{\Psi|\Psi}}-\frac{\bA}{m}\,.
\end{align*}
Unlike the previous (spinless) QHD, the intrinsic spin nature of the Pauli equation manifests in the hydrodynamic picture as
a non-zero vorticity. This new component of the vorticity was named the {\it Takabayasi vector} in \cite{Bialynicki-Birula1995Weyl} 
and is in fact an earlier manifestation of the celebrated Mermin-Ho relation \cite{MerminHo1976} which has often appeared in geometric approaches to physics 
\cite{Holm2002}. In addition, as we shall see in Remark \ref{Rem:MHo}, we make the new link demonstrating that this expression is also equivalent to the Berry 
curvature.

The primary purpose of this opening section is to provide a geometric formulation of Pauli QHD in preparation for the introduction of a dynamical 
non-Abelian connection in Section \ref{Sec:NonAbelianPauliConnection}. We do this by employing the 
momentum map approach to QHD presented in Section \ref{Sec:2-HD}, which turns out to express Pauli QHD as an additional manifestation of the two-level exact factorization dynamics considered in Chapter 
\ref{Chap:EFNEW}. Through this analogy with exact factorization dynamics, this 
section also provides new intepretations for the circulation of the Pauli fluid flow and Takabayasi vector. 


To begin, recall the dynamics of a spin-½ particle, {governed} by a  spinor wavefunction $\Psi \in L^2(\mathbb{R}^3)\otimes\mathbb{C}^2 $ satisfying the Pauli equation 
\eqref{PauliEquation}, as described in Section \ref{Sec:1-Postulates}. Inspired 
by the previous works \cite{BohmSchillerTiomnoA1955,BohmSchillerB1955, Takabayasi1955, Bialynicki-Birula1995Weyl, Spera2016} on the hydrodynamic interpretation of the Pauli equation, we 
write the spinor wavefunction $\Psi(\bx,t)$ as the product
\begin{align}
\Psi(\bx,t)=\Omega(\bx,t)\ket{\psi(\bx,t)} \in L^2(\mathbb{R}^3)\otimes \mathcal{F}({\mathbb{R}^3,\mathbb{C}^2})\,,\label{Poetry}
\end{align}
where $\Omega \in L^2(\mathbb{R}^3)$ is usually expressed in polar form $\Omega = \sqrt{D}e^{i\hbar^{-1}S}$ and $\psi \in 
\mathcal{F}({\mathbb{R}^3,\mathbb{C}^2})$ such that $\psi^{\dagger}(\bx)\psi(\bx)=1$. Remarkably, we recognise this decomposition, dating back to 1955 \cite{Takabayasi1955}, as
the exact factorization \eqref{GenExFact} from nonadiabatic quantum 
 chemistry and hence can immediately apply the results of Chapter \ref{Chap:EFNEW}, in particular Section \ref{Sec:2Level} specialising to two-level systems. 

To do so, we first recall that the Hamiltonian for spinor wavefunctions \eqref{PauliEquation} 
reads
 \begin{align*}
   \widehat{H}=\frac{(-i\hbar\nabla-\bA)^2}{2m} - 
  \frac{\hbar}{2m}\bB\cdot\widehat{\boldsymbol{\sigma}} + V\,,
 \end{align*}  
recognising the additional feature of the constant external magnetic field $\bB:= 
\nabla\times\bA$ (compared with the molecular Hamiltonian operator \eqref{molecularHamiltonian2}). Hence, the EF total energy \eqref{EFGenHam} is modified to read
\begin{align*}
  h(\Omega, \psi) = \text{Re}\int \Omega^*\frac{(-i\hbar\nabla - \bA - \boldsymbol{\cal A})^2}{2m}\Omega 
  +|\Omega|^2 \epsilon(\psi,\nabla\psi)\,\de^3x\,,
\end{align*}
where $\boldsymbol{\cal A}=\braket{\psi|-i\hbar\nabla\psi}$ is the usual time-dependent Berry 
connection whilst the effective potential defined by \eqref{EFepsilonDEF} 
is specialised as follows
\begin{align*}
  \epsilon(\psi,\nabla\psi):= V -\frac{\hbar}{2m}\bB\cdot\braket{\psi|\widehat{\bsigma}\psi} 
  +\frac{\hbar^2}{2M}\|\nabla\psi\|^2-\frac{|{\boldsymbol{\cal 
  A}}|^2}{2M}\,,
\end{align*}
corresponding to the replacements 
$  a = V$ and $ \boldsymbol{b}=-M^{-1}{\hbar\bB}/{2}$
in the two-level Hamiltonian \eqref{2levelHam}.
The total energy can then be written in terms of the `nuclear' QHD momentum maps $\bmu := \hbar{\rm Im}(\Omega^{*}\nabla\Omega)$ and $D:=|\Omega|^2$ 
so that performing the reduced Legendre transform leads to a modified version of the fluid velocity \eqref{tot-momentum-def} 
\begin{align}
  \bu = \frac{\bmu+D(\boldsymbol{\mathcal{A}}-\bA)}{MD} = 
  \frac{\hbar\text{Im}\braket{\Psi|\nabla\Psi}}{m\braket{\Psi|\Psi}}-\frac{\bA}{m}\label{PauliVelocity}\,.
\end{align}
Here the second equality expresses the velocity in terms of the 
total spinor wavefunction, thus demonstrating the agreement with literature \cite{BohmSchillerTiomnoA1955,Takabayasi1955, Takabayasi1983, 
Bialynicki-Birula1995Weyl}. 

We then follow the procedure given in Section \ref{NuclearFrameSection}. Namely, writing the `electronic' dynamics for $\psi$ as unitary evolution in the 
nuclear frame (see equation \eqref{unitary-evol}) allows a complete description in 
terms of the projection $\rho=\psi\psi^{\dagger}$ (pure state density matrix). In turn 
$\rho$ can be written on the basis of the Pauli matrices via equation \eqref{electronicdensitytospin}, eventually taking the electronic dynamical variable to the spin vector $\bs$.

Ultimately, a modified version of the Lagrangian \eqref{EFHDbLagr} (specialising $a$ and $\boldsymbol{b}$ as above and including the minimal coupling term with the external magnetic potential $\bA$) now governs the hydrodynamic interpretation of the Pauli equation and is given 
by
\begin{align}\label{EFHDbLagr2}
  \ell(\bu, D, \boldsymbol\xi,{\boldsymbol{s}} ) = \int\! D\bigg(\frac{M}{2}|\bu|^2 + 
  \bu\cdot\bA -{\frac{\hbar^2}{8M}\frac{|\nabla 
  D|^2}{D^2} }- V + {\boldsymbol{s}}\cdot\Big(\boldsymbol\xi +M^{-1}\bB\Big) 
  -\frac{\|\nabla{\boldsymbol{s}}\|^2}{2M}\bigg) \text{d}^3x\,,
\end{align}
understood as the mapping $\ell:\mathfrak{X}(\mathbb{R}^3)\times{\rm Den}(\mathbb{R}^3)\times \mathcal{F}(\mathbb{R}^3, \mathbb{R}^3)\times \mathcal{F}(\mathbb{R}^3,\mathbb{R}^3)\to 
\mathbb{R}$.
Naturally, following the Euler-Poincar\'e derivation given in Section \ref{Sec:2Level}, the equations of motion \eqref{Kyle2} - \eqref{Lola} are also modified accordingly, resulting in the 
hydrodynamic representation of the Pauli equation
\begin{align}
  \partial_tD+ \text{div}(D\bu)&= 0\,, \label{PauliEOM1}\\
 \begin{split}\label{PauliEOM2} M(\partial_t+\bu\cdot\nabla)\bu &=  \bu \times \bB  - 
  \nabla (V+V_Q)+ M^{-1}\nabla\bB\cdot{\boldsymbol{s}}
- \frac{\partial_j(D\partial_j\bs\cdot\nabla\bs)}{MD}\,,
 \end{split}\\
 M(\partial_t + \bu\cdot\nabla){\boldsymbol{s}} &= {\boldsymbol{s}} 
 \times \left(\bB+  \frac{ \partial_j(D\partial_j\bs)}{D}\right) 
\label{PauliEOM3}\,,
\end{align}
in agreement with the results of the original work \cite{Takabayasi1955}. 

Having now derived the hydrodynamic system for the Pauli equation via an Euler-Poincar{\'e} variational principle, we can now consider 
the associated Kelvin-Noether circulation theorem. 

\begin{proposition}[Kelvin-Noether theorem for Pauli QHD]\label{Prop:PauliKN}
\hfill \\
For a loop $c(t)$ moving with the fluid velocity $\bu$,  the Kelvin-Noether circulation theorem corresponding to the fluid equation \eqref{PauliEOM2} reads
\begin{align}
\frac{\de}{\de t}\oint_{c(t)} \boldsymbol{\cal A}\cdot\de\bx &=
-\frac{1}{M}\oint_{c(t)} \Big[\nabla{\boldsymbol{s}}\cdot \big(\bB + D^{-1}{\rm div}(D\nabla{\boldsymbol{s}})\big)\Big]\cdot\de\bx\,.\label{Holmwood}
\end{align} 
\end{proposition}
\paragraph{Proof:} For a Lagrangian of the type $\ell(\bu, D, \boldsymbol\xi,{\boldsymbol{s}} ) 
$ the general equation of motion arising from Hamilton's principle corresponding to the arbitrary vector field $\bw:=\delta\eta\circ\eta^{-1}$  is given by 
(as we recall from the proof of Corollary \ref{Cor:2LevelEP})
\begin{align*}
(\partial_t + \pounds_{\bu})\frac{\delta \ell}{\delta \bu} &= D\nabla \frac{\delta \ell}{\delta D} 
- \frac{\delta \ell}{\delta \bxi}\cdot\nabla\bxi - \frac{\delta \ell}{\delta \bs}\cdot\nabla\bs 
\,.
\end{align*} 
Following the general derivation for continuum theories \cite{HolmEtAl1998}, the 
associated Kelvin-Noether theorem reads
\begin{align}
  \frac{\de}{\de t}\oint_{c(t)} \frac{1}{D}\frac{\delta \ell}{\delta \bu}\cdot\de\bx =- 
  \oint_{c(t)} \frac{1}{D}\left[ \frac{\delta \ell}{\delta \bxi}\cdot\nabla\bxi +\frac{\delta \ell}{\delta 
  \bs}\cdot\nabla\bs \right]\cdot\de\bx\,,\label{TomorrowsMyBirthday}
\end{align}
having annihilated the exact differential term on the right hand side. 
Specialising to the Pauli Lagrangian \eqref{EFHDbLagr2} results in 
\begin{align*}
  \frac{\de}{\de t}\oint_{c(t)} (M\bu+ \bA)\cdot\de\bx  &=- 
\oint_{c(t)} \frac{1}{D}\Big[ D\bs\cdot\nabla\bxi + \big(D\bxi + M^{-1}D\bB + M^{-1}{\rm div}(D\nabla{\bs})\big) \cdot\nabla\bs
  \Big]\cdot\de\bx\\
\implies \frac{\de}{\de t}\oint_{c(t)} (\nabla S + \boldsymbol{\cal A})\cdot\de\bx  &=- 
\oint_{c(t)} \Big[\nabla(\bxi\cdot\bs) + M^{-1}\big(\bB +D^{-1}{\rm div}(D\nabla{\bs})\big)\cdot\nabla\bs
  \Big]\cdot\de\bx\,,
\end{align*}
having used the definition of $\bu$ given by \eqref{PauliVelocity} in the second line and recalling the expression $\bmu = D\nabla S$. Again the exact differential terms vanish, thus completing the proof.  
\hfill$\square$\\

The Kelvin-Noether circulation theorem \eqref{Holmwood} is analogous to the EF circulation \eqref{Berry-Frequency} and correspondingly can be
understood as an expression for the time evolution along the flow of a Berry phase, here associated to the spin degrees of freedom of the Pauli equation.
 Correspondingly, the role of the electronic Hamiltonian in \eqref{Berry-Frequency} is played by the external magnetic field $\bB$ whilst the real part of the quantum geometric tensor 
 is now described in terms of the spin vector. We will consider the role of the QGT for two-level systems further in the next section.
 
 We use the analogy with exact factorization dynamics to make 
 relevant remarks regarding the interpretation of the Takabayasi vector in Pauli QHD. 
 
\begin{remark}[Mermin-Ho relation and Takabayasi vector]\label{Rem:MHo}
\hfill \\
The vorticity of the fluid flow \eqref{PauliVelocity} is given by
\begin{align}
  \nabla\times\bu = M^{-1}\nabla\times (\nabla S + \boldsymbol{\cal A}+\bA) =: M^{-1}\left(\frac{\hbar}{2}\boldsymbol{T} - 
  \bB\right)\,,\label{JLo}
\end{align}
where $\boldsymbol{T}:= \epsilon_{ijk}n_i\nabla n_j\times\nabla n_k$ and $\bn:=2\bs/\hbar$ is the Bloch vector field defined in Section \ref{Sec:1-Postulates}. 
The relation $\boldsymbol{\cal B}=\nabla\times\boldsymbol{\cal A}=\hbar\boldsymbol{T}/2$ 
underlying \eqref{JLo} first appeared in Takabayasi's  work \cite{Takabayasi1955} in 1955, hence later motivating the name {\em Takabayasi vector} \cite{Bialynicki-Birula1995Weyl}, 
and can more explicitly be written as
\begin{align}
\begin{split}
  T_c &= \epsilon_{ijk}\epsilon_{abc}n_i(\partial_an_j)(\partial_bn_k)\\
  &= 2i\epsilon_{abc}\Big(\braket{\partial_b\psi|\partial_a\psi}-\braket{\partial_a\psi|\partial_b\psi}\Big)=\frac{2}{\hbar}{\cal B}_c\,,
\end{split}
\end{align}
relying on the definition $n_i = \braket{\psi|\sigma_i\psi}=\psi^*_a(\sigma_i)_{ab}\psi_b$ and the interesting property between elements of the Pauli matrices
\begin{align*}
  \epsilon_{ijk}(\sigma_i)_{ab}(\sigma_j)_{cd}(\sigma_k)_{ef} = 
  2i\Big(\delta_{af}\delta_{cb}\delta_{ed}-\delta_{ad}\delta_{cf}\delta_{eb}\Big)\,.
\end{align*}
Using the analogy with EF dynamics we are able to identify the Takabayasi vector with the Berry curvature, serving as yet another manifestation of the QGT. However, the same general expression
 is more commonly known as the {\rm Mermin-Ho relation} \cite{MerminHo1976} after its appearance in the context of superfluids two decades after Takabayasi's original work. 
For applications of the Mermin-Ho relation in the more general context of complex fluids, see also \cite{Holm2002}. 
\end{remark}

In this section we have provided a geometric approach to the hydrodynamic formulation of the Pauli equation, which will allow us 
to consider the dynamics of the spin vector in further detail in Section \ref{Sec:NonAbelianPauliConnection}. Specifically, exploiting its own evolution under the rotation group, performing the construction 
analogous to that of Chapter \ref{Chap:Holonomy} allows for the introduction of a non-Abelian gauge connection 
which again enters the formulation as an additional dynamical variable.

\section{Non-Abelian gauge theory of the Pauli equation}\label{Sec:NonAbelianPauliConnection}
Following the introduction of the spin vector for both two-level EF systems in Section 
\ref{Sec:2Level} and the Pauli equation in Section \ref{Sec:PauliQHD}, we now further exploit the underlying geometry by considering its own evolution 
on a Lie group. To do so, recall that the unitary evolution \eqref{unitary-evol} of the electronic factor $\psi$ 
enforces its normalisation for all time. Then, through the transition to the pure 
state density matrix description $\rho=\psi\psi^{\dagger}$ and its subsequent 
evaluation on the Bloch sphere, this normalisation amounts to the magnitude of the spin vector 
remaining $|\bs(t)|^2=\hbar^2/4$ for all time. Hence, analogous to the unitary evolution 
of $\psi$ in the nuclear QHD frame according to \eqref{unitary-evol}, the spin vector $\bs$ must evolve 
under the action of the rotation group as $\bs(t)=(R\bs_0)\circ\eta^{-1}$, 
written more explicitly as
\begin{align}
  \bs(\bx,t) = R(\eta^{-1}(\bx,t),t)\bs_0(\eta^{-1}(\bx,t))\,,\label{rotationevolution}
\end{align}
where $\eta$ is the hydrodynamic path obeying $\dot\boldeta=\bu(\boldeta,t)$ and $R(\bx,t) \in \mathcal{F}(\mathbb{R}^3, SO(3))$ denotes the rotation matrix with 
parametric dependence on the physical coordinate $\bx \in \mathbb{R}^3$. 

Following the strategy underlying Chapter \ref{Chap:Holonomy}, the gateway to the introduction of a gauge connection is by 
considering the gradient of the spin vector as an $\mathbb{R}^3$-valued 1-form, that is
\begin{align}
  \de \bs = \nabla \bs \cdot \de\bx = \partial_j \bs \,\de x^j\,,
\end{align}
where we recall that $\bs \in \mathcal{F}(\mathbb{R}^3, \mathbb{R}^3)$. Then, 
using the evolution under the rotation group \eqref{rotationevolution}, we can introduce a (non-Abelian) 
dynamical gauge connection via the relation $\partial_j \bs  = -\widehat{\gamma}_j\bs$
having defined
\begin{align}
  \widehat{\gamma}_j(t) := \Big(-\partial_j R\,R^{-1} + 
  R\,\widehat{\gamma}_j^{(0)}R^{-1}\Big)\circ\eta^{-1} \in \mathcal{F}(\mathbb{R}^3, 
  \mathfrak{so}(3))\,,\label{PauliConnection}
  \end{align}
so that $\widehat{\gamma}:=\widehat{\gamma}_j\de x^j \in \mathcal{F}(\mathbb{R}^3, 
  \mathfrak{so}(3))\otimes \Lambda^1(\mathbb{R}^3)=:\Lambda^1(\mathbb{R}^3, \mathfrak{so}(3))$ is understood as a 
  Lie-algebra valued 1-form.
For convenience we will subsequently work with the $\mathbb{R}^3$ counterpart to $\widehat{\gamma}_j$, constructed via the {\it hat map} isomorphism
from $\mathfrak{so}(3)\cong \mathbb{R}^3$ \cite{DarrylBook2} so that 
\begin{align}
  \partial_j\bs = -\bgamma_j\times\bs = \bs\times\bgamma_j\,,\label{2LevelConnectionVector}
\end{align}
where  $\bgamma_j \in \mathcal{F}(\mathbb{R}^3, \mathbb{R}^3)$.

\begin{remark}[Zero curvature relation]\label{AnotherRemark}
  \hfill \\
  Having introduced the connection $\bgamma_j$ using the analogous construction to that in 
  Chapter \ref{Chap:Holonomy}, it is natural to ask whether the curvature 
  of this connection is again zero. As we shall see in the next section, any 
  connection (Abelian or non-Abelian) defined by the type of relation $\nabla n = -\bgamma n$ 
  must have zero curvature. For the spin vector, this is a consequence of the
  following calculation:
  \begin{align}
    \begin{split}
   0   &= \partial_i\partial_j \bs - \partial_j\partial_i \bs\\
   &= 
   \partial_i\Big(\bs\times\bgamma_j\Big)-\partial_j\Big(\bs\times\bgamma_i\Big)\\
   &= (\bs\times\bgamma_i) \times \bgamma_j + \bs \times \partial_i\bgamma_j - (\bs\times\bgamma_j) \times \bgamma_i- \bs \times 
   \partial_j\bgamma_i\\
   &= \bs\times\Big(\partial_i\bgamma_j - \partial_j\bgamma_i + 
   \bgamma_i\times\bgamma_j\Big)\,.
    \end{split}
  \end{align}
  Here we identify $\bOmega_{ij}:= \partial_i\bgamma_j - \partial_j\bgamma_i + 
   \bgamma_i\times\bgamma_j$ as the curvature of the non-Abelian connection. Via the hat map isomorphism we can equivalently write that $0=\widehat{\Omega}_{ij}\bs$ so that the evolution of the curvature $\widehat{\Omega}_{ij}(t)=R\widehat{\Omega}_{ij}(0)R^{-1}$ (which follows from \eqref{PauliConnection}) implies that 
$\widehat{\Omega}_{ij}(0) \bs(0) =0$. Since we cannot impose this relation in general, we have that $\widehat{\Omega}_{ij}(0)=0$ which in turn implies $\widehat{\Omega}_{ij}(t)=0$ for all time.
    Note that a precise definition of the curvature of a connection will be given in Section \ref{Sec:GeneralTheory} 
   in which we will also prove the general version of this result.  
\end{remark}

We now consider the insertion of 
the gauge connection \eqref{2LevelConnectionVector} into the reduced Lagrangian for any two-level system as an additional dynamical variable. Following the approach used throughout this 
thesis, we do so in the general case of an arbitrary Hamiltonian functional  $h = h(D,\bs, \bgamma_j)$, making the appropriate modifications for the Pauli equation after first deriving all results in the more general setting. 

\begin{proposition}[Reduced dynamics with a spin vector connection]\label{Prop:EP2LevelGamma}
  \hfill \\
  Consider a general reduced Lagrangian of the form $\ell:\mathfrak{X}(\mathbb{R}^3)\times{\rm Den}(\mathbb{R}^3)\times \mathcal{F}(\mathbb{R}^3,\mathbb{R}^3)\times \mathcal{F}(\mathbb{R}^3,\mathbb{R}^3)\times \Lambda^1(\mathbb{R}^3,\mathbb{R}^3)\to \mathbb{R}$ 
  given by
\begin{align}
  \ell(\bu, D, \bs, \bxi, \bgamma_j) = \int \left[\frac{1}{2}MD|\bu|^2 -\frac{\hbar^2}{8M}\frac{|\nabla 
  D|^2}{D}+ 
  D\bxi\cdot\bs\right]\,\de^3 x - h(D,\bs, \bgamma_j)\,.\label{AnotherLagrangian}
\end{align}
Then, computing the variation
\begin{align}
  \delta \bgamma_j = -\partial_j\bUpsilon + \bUpsilon\times\bgamma_j - 
  (\bw\cdot\nabla)\bgamma_j - \bgamma_i \partial_j w_i\,,\label{Adam}
\end{align}
as well as the Euler-Poincar{\'e} variations \eqref{EPvar} and \eqref{twolevelfluidvariations} (where $\bUpsilon$ and $\bw$ are arbitrary), Hamilton's principle 
yields the equations of motion
\begin{align}
  MD(\partial_t+\bu\cdot\nabla)u_i &= -D\partial_i\left(V_Q+\frac{\delta h}{\delta 
  D}\right)+\partial_i\bs\cdot \frac{\delta h}{\delta \bs}+\partial_i\bgamma_j\cdot \frac{\delta h}{\delta \bgamma_j} 
  - \partial_j\left(\bgamma_i\cdot\frac{\delta h}{\delta 
  \bgamma_j}\right)\,,\label{Yeet}\\
  D(\partial_t+ \bu\cdot\nabla)\bs &= \frac{\delta h}{\delta \bs}\times\bs - 
  \partial_j\frac{\delta h}{\delta 
  \bgamma_j} - \bgamma_j\times\frac{\delta h}{\delta 
  \bgamma_j}\,,\label{Tony}\\
    \partial_t D + {\rm div}(D\bu)&=0\,,\\
  (\partial_t + \bu\cdot\nabla)\bs &= \bxi\times\bs\,,\\
\partial_t\bgamma_j + u_i\partial_i\bgamma_j + \bgamma_i\partial_ju_i &= -\partial_j\bxi + 
  \bxi\times\bgamma_j\,.\label{GammaEvoInProp}
\end{align}
\end{proposition}
\paragraph{Proof:} 
As derived explicitly in \cite{GayBalmazRatiu2009,GayBalmazTronci2010} 
the variation $\delta\bgamma_j$ can be computed directly from \eqref{PauliConnection}.
Once again we show this by simply considering the counterpart of $\bgamma_j$ in the absence of fluid flow 
$\delta\bgamma_j = -\partial_j\bUpsilon - \bgamma_j\times\bUpsilon$ (derived explicitly in the proof of Theorem \ref{Theorem:GammaGenEP}) 
and then replace the material derivative $\delta \mapsto \delta + 
\pounds_{\bw}$. A direct verification using the definition of the Lie derivative 
verifies that $\pounds_{\bw}(\bgamma_j\,\de x^j) = \Big((\bw\cdot\nabla)\bgamma_j + \bgamma_i \partial_j w_i\Big)\,\de 
x^j$.
Then, expanding Hamilton's principle yields
\begin{align*}
  0 &= \int_{t_1}^{t_2}\int\Bigg[ \frac{\delta \ell}{\delta\bu}\cdot\delta \bu + \frac{\delta \ell}{\delta D}\,\delta D + \frac{\delta \ell}{\delta\bs}\cdot\delta \bs  + \frac{\delta \ell}{\delta\bxi}\cdot\delta \bxi  + \frac{\delta \ell}{\delta\bgamma_j}\cdot\delta \bgamma_j\Bigg]\,\de^3 x\,\de  t\\
  &= \int_{t_1}^{t_2}\int\Bigg[ \frac{\delta \ell}{\delta\bu}\cdot(\partial_t + \pounds_{\bu})\bw - \frac{\delta \ell}{\delta D}\,{\rm div}(D\bw)
 + \frac{\delta \ell}{\delta\bs}\cdot(\bUpsilon\times\bs - \bw\cdot\nabla\bs)  \\
  &\qquad\qquad\qquad+ \frac{\delta \ell}{\delta\bxi}\cdot((\partial_t+\pounds_{\bu})\bUpsilon - \bxi\times\bUpsilon  - \bw\cdot\nabla\bxi) \\
  &\qquad\qquad\qquad\qquad\qquad+ \frac{\delta \ell}{\delta\bgamma_j}\cdot(-\partial_j\bUpsilon + \bUpsilon\times\bgamma_j - 
  (\bw\cdot\nabla)\bgamma_j - \bgamma_i \partial_j w_i)\Bigg]\,\de^3 x\,\de t\\
    &= \int_{t_1}^{t_2}\int\Bigg[ -(\partial_t + \pounds_{\bu})\frac{\delta \ell}{\delta\bu}\cdot\bw + D\nabla \frac{\delta \ell}{\delta D}\cdot\bw
 + \bs\times\frac{\delta \ell}{\delta\bs}\cdot\bUpsilon - \left(\frac{\delta \ell}{\delta\bs}\cdot\nabla\bs\right)\cdot\bw  \\
  &\qquad\qquad\qquad- \left((\partial_t+\pounds_{\bu})\frac{\delta \ell}{\delta\bxi}-\bxi\times\frac{\delta \ell}{\delta\bxi}\right)\cdot\bUpsilon - \left(\frac{\delta \ell}{\delta\bxi}\cdot\nabla\bxi\right)\cdot\bw \\
  &\qquad\qquad\qquad\qquad+ \left(\partial_j\frac{\delta \ell}{\delta\bgamma_j}+ \bgamma_j\times\frac{\delta \ell}{\delta\bgamma_j}\right)\cdot\bUpsilon
  +\left(- \partial_i\bgamma_j\cdot\frac{\delta \ell}{\delta \bgamma_j} 
 + \partial_j\left(\bgamma_i\cdot\frac{\delta \ell}{\delta 
  \bgamma_j}\right)\right)\,w_i\Bigg]\,\de^3 x\,\de t\\
\end{align*}
Then, recalling that both $\bw$ and $\bUpsilon$ are arbitrary leads the general equations
\begin{align*}
  \left[(\partial_t + \pounds_{\bu})\frac{\delta \ell}{\delta \bu}\right]_i &= 
  D\partial_i \frac{\delta \ell}{\delta D} - \partial_i\bxi\cdot\frac{\delta \ell}{\delta \bxi}- \partial_i\bs\cdot\frac{\delta \ell}{\delta \bs}- \partial_i\bgamma_j\cdot\frac{\delta \ell}{\delta \bgamma_j} 
 + \partial_j\left(\bgamma_i\cdot\frac{\delta \ell}{\delta 
  \bgamma_j}\right)\,,\\
  (\partial_t + \pounds_{\bu})\frac{\delta \ell}{\delta \bxi} &= \bxi \times\frac{\delta \ell}{\delta 
  \bxi}+ \bs \times\frac{\delta \ell}{\delta 
  \bs} + \partial_j\frac{\delta \ell}{\delta \bgamma_j} +\bgamma_j \times\frac{\delta \ell}{\delta 
  \bgamma_j}\,,
\end{align*}
along with the auxilliary equations $  (\partial_t + \pounds_{\bu})D=0,\,\,
  (\partial_t + \pounds_{\bu})\bs = \bxi\times\bs,\,\,
  (\partial_t + \pounds_{\bu})(\bgamma_j\,\de x^j) = -\de\bxi + 
  \bxi\times\bgamma_j \,\de x^j$. Then, we compute the variational derivatives
\begin{align*}
  \begin{split}
  \frac{\delta \ell}{\delta \bu} = MD\bu\,,&\qquad   \frac{\delta \ell}{\delta D} 
  =\frac{M}{2}|\bu|^2+\bxi\cdot\bs -  V_Q - \frac{\delta h}{\delta D}\,,\qquad   \frac{\delta \ell}{\delta \bxi} = 
  D\bs\,,\\
    &\frac{\delta \ell}{\delta \bs} = D\bxi -   \frac{\delta h}{\delta 
    \bs}\,,\qquad  \frac{\delta \ell}{\delta \bgamma_j} = -\frac{\delta h}{\delta 
    \bgamma_j}\,,
  \end{split}
\end{align*}
which specialises the equations for the fluid momentum and generator 
accordingly.
\hfill$\square$\\

At this point, one may notice that the system of equations presented in Proposition \ref{Prop:EP2LevelGamma} are not closed as we require an equation for $\bxi$ to evaluate 
the evolution of $\bgamma_j$. This can be remedied by expanding the auxilliary equation for $\bs$ as $(\partial_t+ 
\bu\cdot\nabla)\bs=\bxi\times\bs$, which upon replacement in \eqref{Tony} 
results in an algebraic equation for the generator
\begin{align}
D\bxi\times\bs  &= \frac{\delta h}{\delta \bs}\times\bs - 
  \partial_j\frac{\delta h}{\delta 
  \bgamma_j} - \bgamma_j\times\frac{\delta h}{\delta 
  \bgamma_j}\,.\label{XIEQUATION}
\end{align}
Upon rearranging we write
          \begin{align*}
               \left(D \bxi -\frac{\delta h}{\delta \bs}\right)\times \bs &=- \partial_j^{\gamma}\frac{\delta h}{\delta \bgamma_j} \,,         
 \end{align*}
in which we have now introduced the notation for the covariant derivative 
$  \partial_j^{\gamma}:= \partial_j + 
{\rm ad}_{\bgamma_j}$. We can then find a solution for this equation by considering the result of repeated cross products of the RHS with 
$\bs$.  More explicitly, we write that    
\begin{align*}
 \left(\partial_j^{\gamma}\frac{\delta h}{\delta 
  \bgamma_j}\times\bs\right)\times\bs=   - \frac{\hbar^2}{4}\partial_j^{\gamma}\frac{\delta h}{\delta \bgamma_j} 
  + \left(\bs\cdot\partial_j^{\gamma}\frac{\delta h}{\delta 
  \bgamma_j}\right)\bs\,,
\end{align*}
having used that $|\bs|^2=\hbar^2/4$. Hence, provided that the second term on the RHS vanishes, we have that
\begin{align*}
   \left(D \bxi -\frac{\delta h}{\delta \bs} - \frac{4}{\hbar^2}\partial_j^{\gamma}\frac{\delta h}{\delta 
  \bgamma_j}\times\bs \right)\times \bs=0\,,
\end{align*}
from which we can find a solution up to a term parallel to $\bs$.  
This general solution is formalised as follows.       
          
\begin{lemma}\label{Lemma:2levelXiSolution}
  Equation \ref{XIEQUATION} possesses the solution for the generator
  \begin{align}
   D \bxi = \frac{\delta h}{\delta \bs} + \frac{4}{\hbar^2}\partial_j^{\gamma}\frac{\delta h}{\delta \bgamma_j}\times\bs + \beta\bs\,,\label{Boris}
  \end{align}
  where $\beta$ is an arbitrary real parameter, provided the Hamiltonian functional satisfies the compatibility condition
  \begin{align}
    \bs\cdot\partial_j^{\gamma}\frac{\delta h}{\delta \bgamma_j}=0\,.\label{Johnson}
  \end{align} 
\end{lemma}

As we shall see, we note that this condition \eqref{Johnson} is only satisfied by the Hamiltonian for the Pauli equation in the case of zero curvature, shown explicitly in the  
proof of Proposition \ref{Prop:PauliQHDGamma}. Having found a solution for the generator, we can now write the closed system of equations in the variables $(D, \bu, \bs, 
\bgamma_j)$.
\begin{corollary} The system of equations \eqref{Yeet}-\eqref{GammaEvoInProp} 
given in Proposition \ref{Prop:EP2LevelGamma} are closed by the generator solution 
\eqref{Boris}, resulting in the closed set of equations (choosing $\beta=0$)
\begin{align}
  \partial_t D + {\rm div}(D\bu)&= 0\,,\\
     MD(\partial_t+\bu\cdot\nabla)u_i &= -D\partial_i\left(V_Q+\frac{\delta h}{\delta 
  D}\right)+\partial_i\bs\cdot \frac{\delta h}{\delta \bs}+\bOmega_{ij}\cdot\frac{\delta h}{\delta 
  \bgamma_j} - \bgamma_i\cdot \partial_j^{\gamma}\frac{\delta h}{\delta \bgamma_j} \,,\label{FLUIDEQNPREV}\\
  D(\partial_t+ \bu\cdot\nabla)\bs &= \frac{\delta h}{\delta \bs}\times\bs - \partial_j^{\gamma}\frac{\delta h}{\delta \bgamma_j}\,,\label{Pikachu}\\
  \partial_t\bgamma_j + u_i\partial_i\bgamma_j + \bgamma_i\partial_ju_i&= -\partial_j^{\gamma}\left(\frac{1}{D}\frac{\delta h}{\delta \bs} + \frac{4}{\hbar^2 D}\partial_j^{\gamma}\frac{\delta h}{\delta \bgamma_j}\times\bs \right)\label{Onyx}
  \,,
\end{align}
where $\bOmega_{ij}:=\partial_i\bgamma_j - \partial_j\bgamma_i + 
\bgamma_i\times\bgamma_j$ is the curvature of the $\mathfrak{so}(3)$ connection \eqref{PauliConnection}.
\end{corollary}
\paragraph{Proof:} The final two terms of the fluid equation \eqref{Yeet} are rewritten 
as follows
\begin{align*}
  \partial_i\bgamma_j\cdot\frac{\delta h}{\delta \bgamma_j} - \partial_j\left(\bgamma_i\cdot\frac{\delta h}{\delta 
  \bgamma_j}\right)&=   (\partial_i\bgamma_j - \partial_j\bgamma_i)\cdot\frac{\delta h}{\delta \bgamma_j} - \bgamma_i\cdot\partial_j\frac{\delta h}{\delta 
  \bgamma_j}\\
  &= (\bOmega_{ij} - \bgamma_i\times\bgamma_j)\cdot\frac{\delta h}{\delta \bgamma_j} - \bgamma_i\cdot\partial_j\frac{\delta h}{\delta 
  \bgamma_j}\\
  &= \bOmega_{ij} \cdot\frac{\delta h}{\delta \bgamma_j} - \bgamma_i\cdot\left(\partial_j\frac{\delta h}{\delta 
  \bgamma_j}+ \bgamma_j\times\frac{\delta h}{\delta 
  \bgamma_j}\right)\,.
\end{align*}
Then, the generator solution (with $\beta = 0$) \eqref{Boris} is simply 
substituted into the connection equation \eqref{GammaEvoInProp} to complete the 
set of equations.
\hfill$\square$\\

At this point we make the important remark that, analogous to the connection-based approach to QHD of Chapter \ref{Chap:Holonomy},
 these equations have been formulated without invoking any assumptions on the curvature of the connection,
and hence possess the new feature of allowing for non-zero holonomy associated to the connection $\bgamma$. 
From the general theory in Appendix \ref{App:Principalbundles} the holonomy of the connection is 
written as
\begin{align*}
  \mathcal{P} \,\text{exp} \oint_{c(t)} \bgamma_j\,\de x^j\,,
\end{align*}
where $\mathcal{P}$ denotes the path ordering operator. In general, this holonomy is a dynamical quantity whose evolution, in principle, can be derived from equation 
\eqref{Onyx}. However, due to the nature of the path ordering operation this evolution equation cannot easily be written. 
 Instead, the connection equation \eqref{Onyx} also allows us to consider the evolution of the related quantity, the $\bgamma$-circulation \cite{Holm2002, GayBalmazRatiu2009}:
\begin{proposition}[Circulation theorem for the connection]\label{Charizard}
  \hfill \\ Recall the reduced system of equations for two-level EF dynamics 
  presented in Proposition \ref{Prop:EP2LevelGamma}. Then, the time evolution of the 
  circulation of the connection is given by
  \begin{align}
    \begin{split}
    \frac{\de}{\de t}\oint_{c(t)}\bgamma_i\,\de x^i &= \oint_{c(t)} 
    \bxi\times\bgamma_i \,\de x^i\\
    &=\oint_{c(t)} \frac{1}{D}\left(\frac{\delta h}{\delta \bs} + \frac{4}{\hbar^2}\partial_j^{\gamma}\frac{\delta h}{\delta \bgamma_j}\times\bs\right)\times\bgamma_i \,\de x^i\,,
      \end{split}\label{Ditto}
  \end{align}
  where we have again selected the $\beta=0$ solution of the generator equation \eqref{Boris} and $c(t)$ is a loop moving with the `nuclear' fluid flow $\bu$. 
\end{proposition}
\paragraph{Proof:} This proposition can be easily proven upon considering the auxilliary equation for the evolution of the connection. 
Indeed, simply taking the integral of \eqref{GammaEvoInProp} $(\partial_t+\pounds_{\bu})(\bgamma_j \,\de x^j)= -\de\bxi + (\bxi\times\bgamma_j)\,\de x^j$ around the loop $c(t)$ 
yields
\begin{align*}
  \oint_{c(t)} (\partial_t+\pounds_{\bu})(\bgamma_j\de x^j) =   \oint_{c(t)} 
  \bxi\times\bgamma_j\,\de x^j\,.
\end{align*}
Then, once again using the standard method (provided more explicitly in the proof of Theorem \ref{Abby}), we 
take the Lagrangian time derivative outside of the integral and substitute in the solution for the generator \eqref{Boris} (with $\beta=0$) to complete the 
proof.
\hfill$\square$\\


Next, due to the Euler-Poincar\'e derivation of the fluid equation \eqref{FLUIDEQNPREV}, 
we can write the associated Kelvin-Noether circulation theorem, depending explicitly on the non-zero curvature 
of the connection.

\begin{proposition}[Kelvin-Noether circulation theorem]\label{Theorem:TIRED2}
  \hfill \\
  The fluid equation \eqref{FLUIDEQNPREV} possesses the Kelvin-Noether circulation 
  theorem
  \begin{align}\label{KNTheoremwithCurv}
    \frac{\de}{\de t}\oint_{c(t)} \boldsymbol{\cal A}\cdot\de\bx= -\oint_{c(t)}\frac{1}{D}\left(\bgamma_i\cdot\partial_j^{\gamma}\frac{\delta h}{\delta \bgamma_j} -\bOmega_{ij}\cdot\frac{\delta h}{\delta \bgamma_j}- \partial_i\bs\cdot \frac{\delta h}{\delta \bs} \right)\de 
    x^i\,,
  \end{align}
  for a loop $c(t)$ moving with the fluid flow $\bu$. 
\end{proposition}
\paragraph{Proof:} As usual, we rewrite \eqref{FLUIDEQNPREV} in terms of the Lie 
derivative as
\begin{align*}
      M(\partial_t+\pounds_{\bu})(\bu\cdot\de \bx) -\frac{M}{2}\de |\bu|^2 &= -\de\left(V_Q+\frac{\delta h}{\delta 
  D}\right)+\frac{1}{D}\left(\partial_i\bs\cdot \frac{\delta h}{\delta \bs}+\bOmega_{ij}\cdot\frac{\delta h}{\delta 
  \bgamma_j} - \bgamma_i\cdot \partial_j^{\gamma}\frac{\delta h}{\delta \bgamma_j} \right)\de x^i\,.
\end{align*}
Then taking the integral around a loop $c(t)$ moving with the fluid kills all 
terms given as exact differentials leading to
\begin{align*}
  \oint_{c(t)}M(\partial_t+\pounds_{\bu})(\bu\cdot\de \bx) &=   -\oint_{c(t)}\frac{1}{D}\left(\bgamma_i\cdot\partial_j^{\gamma}\frac{\delta h}{\delta \bgamma_j} -\bOmega_{ij}\cdot\frac{\delta h}{\delta \bgamma_j}- \partial_i\bs\cdot \frac{\delta h}{\delta \bs} \right)\de x^i\,.
\end{align*}
We then take the Lagrangian time derivative outside of the integral as explained 
in the proof of Theorem \ref{Abby} so that upon replacing $M\bu\cdot\de\bx = \de S + \boldsymbol{\cal A}\cdot\de\bx$ 
we have
  \begin{align*}
    \frac{\de}{\de t}\oint_{c(t)} \boldsymbol{\cal A}\cdot\de\bx= -\oint_{c(t)}\frac{1}{D}\left(\bgamma_i\cdot\partial_j^{\gamma}\frac{\delta h}{\delta \bgamma_j} -\bOmega_{ij}\cdot\frac{\delta h}{\delta \bgamma_j}- \partial_i\bs\cdot \frac{\delta h}{\delta \bs} \right)\de 
    x^i\,.
  \end{align*}
\hfill$\square$\\
Analogous to the result of 
Theorem \ref{Abby}, this above equation can be understood as describing the time evolution along the flow of a Berry connection 
associated to the spin degrees of freedom of the two-level system.

In the fluid equation \eqref{FLUIDEQNPREV} and the corresponding Kelvin-Noether 
theorem \eqref{KNTheoremwithCurv}, we see how the new feature of non-zero 
curvature appears as an additional force influencing the fluid flow. 
However, due to the presence of $\partial_i\bs$ in the equations, one is naturally led to also consider the special case of zero curvature (i.e. $\bOmega_{ij}=0$), in which we reinstate the relation $\partial_j\bs=\bs\times\bgamma_j$ and introduce new connection terms. 
In this case, the fluid equation \eqref{FLUIDEQNPREV} can in fact be rewritten 
as
\begin{align}
    MD(\partial_t+\bu\cdot\nabla)u_i &= -D\partial_i\left(V_Q+\frac{\delta h}{\delta 
  D}\right)-\bgamma_i\cdot\left(\bs\times \frac{\delta h}{\delta \bs} +\partial_j^{\gamma}\frac{\delta h}{\delta \bgamma_j} \right)\,.\label{FLUIDEQN}
\end{align}

Correspondingly, the Kelvin-Noether theorem of Proposition \ref{Theorem:TIRED2} 
takes the particular form given in the following corollary.
\begin{corollary}[Kelvin-Noether circulation theorem for a flat connection]\label{Theorem:TIRED}
  \hfill \\
  The zero curvature fluid equation \eqref{FLUIDEQN} possesses the Kelvin-Noether circulation 
  theorem
  \begin{align}
    \frac{\de}{\de t}\oint_{c(t)} \bs\cdot\bgamma_i\,\de x^i = -\oint_{c(t)}\frac{1}{D}\bgamma_i\cdot\left(\bs\times \frac{\delta h}{\delta \bs} +\partial_j^{\gamma}\frac{\delta h}{\delta \bgamma_j} \right)\de 
    x^i\,,
  \end{align}
  for a loop $c(t)$ moving with the fluid flow $\bu$.
\end{corollary}
\paragraph{Proof:} 
The first part of the proof is analogous to that of proposition \ref{Theorem:TIRED2}. Whilst the RHS clearly follows from the new form of the fluid momentum equation 
\eqref{FLUIDEQN}, the LHS requires a proof of the equation ${\cal 
 A}_i=\bs\cdot\bgamma_i$. As will be discussed in Remark \ref{Rem:connectionConnections} from Section 
 \ref{Sec:NonAbelianQuantumSystems}, this follows from the introduction of 
 an additional connection $\partial_i\psi = -\bar{\gamma}_i\psi$ 
  where $\bar{\gamma}_i \in \mathcal{F}(\mathbb{R}^3, \mathfrak{su}(2))$, corresponding to the $\mathfrak{so}(3)$ connection $\bgamma_i$. This is seen as follows.
  For pure state density matrices $\rho=\psi\psi^{\dagger}$, 
we have that $\partial_i\rho = [\rho, \bar{\gamma}_i]$, at which point we write both operators $\rho \in  \mathcal{F}(\mathbb{R}^3, {\rm Her}(\mathbb{C}^2))$ and $\bar\bgamma_i$ on the basis of Pauli matrices via equations \eqref{rhoblochvector} and \eqref{TILDEMAP} respectively to recover $\partial_i\bs = \bs\times\bgamma_i$. 
  Then, one simply computes 
  \begin{align*}
    {\cal A}_i&=\braket{\psi|-i\hbar\partial_i\psi}=\braket{\psi|i\hbar\bar\gamma_i\psi}\\
    &=\braket{\rho|i\hbar\bar{\gamma}_i}=    i\hbar{\rm Tr}(\rho\bar{\gamma_i})\\
&= i\hbar{\rm Tr}\left(\frac{1}{2}\left(\mathbbm{1}+\frac{2}{\hbar}\bs\cdot\widehat{\bsigma}\right)\cdot-\frac{i}{2}({\bgamma_i}\cdot\widehat{\bsigma})\right)\\
&= i\hbar\left(-\frac{i}{4}(\gamma_i)_b {\rm Tr}(\sigma_b) - \frac{i}{2\hbar}s_a(\gamma_i)_b{\rm 
Tr}(\sigma_a\sigma_b)\right)\\
&= \frac{1}{2}s_a(\gamma_i)_b{\rm Tr}(\sigma_a\sigma_b)\\
 &= \bs\cdot\bgamma_i\,,
  \end{align*}
  making use of the standard relations ${\rm Tr}(\sigma_a)=0$ and ${\rm Tr}(\sigma_a\sigma_b)=2\delta_{ab}$.
\hfill$\square$\\

Notice how the Berry connection is here expressed for a two-level system as ${\cal A}_i:=\braket{\psi|-i\hbar\partial_i\psi}=\bs\cdot\bgamma_i$ and hence can be understood as the projection of the $\mathfrak{su}(2)$ connection $\bgamma_i$ in the direction of the spin vector $\bs$. 
 We will develop this point in more generality in Section \ref{Sec:NonAbelianQuantumSystems}.

We complete this section by specialising these results for including the connection \eqref{2LevelConnectionVector} in two-level EF systems 
for an arbitrary Hamiltonian functional to the specific case of the Pauli equation from Section \ref{Sec:PauliQHD}. 
Recall the reduced form of the Pauli QHD Lagrangian $\ell(\bu, D, \boldsymbol\xi,{\boldsymbol{s}})$ given by 
\eqref{EFHDbLagr2}:
\begin{align*}
  \ell(\bu, D, \boldsymbol\xi,{\boldsymbol{s}} ) = \int\! D\bigg(\frac{M}{2}|\bu|^2 + 
  \bu\cdot\bA -\frac{\hbar^2}{8M}\frac{|\nabla 
  D|^2}{D}- V + {\boldsymbol{s}}\cdot\Big(\boldsymbol\xi +M^{-1}\bB\Big) 
  -\frac{\|\nabla{\boldsymbol{s}}\|^2}{2M}\bigg) \text{d}^3x\,.
\end{align*}
Upon introducing the $\mathbb{R}^3$-valued connection 1-form \eqref{2LevelConnectionVector}, we have that 
\begin{align}
  \begin{split}
  \|\nabla \bs\|^2&=\partial_j\bs\cdot\partial_j\bs \\&= (\bs\times\bgamma_j)\cdot(\bs\times\bgamma_j) 
  = |\bs|^2|\bgamma_j|^2 - (\bs\cdot\bgamma_j)^2 \\&= \frac{\hbar^2}{4}|\bgamma_j|^2 - 
  (\bs\cdot\bgamma_j)^2\,,
  \end{split}
\end{align}
which is now more reminiscent of the original expression for the effective electronic potential in terms of the real part of the quantum 
geometric tensor (QGT) \eqref{QGTensor}. Motivated by this, we first consider how the form of the entire QGT 
is affected by the introduction of the connection, before returning to the 
specific application to Pauli QHD.

\begin{proposition}[Two-level quantum geometric tensor in terms of the connection]
  \hfill\\
  Recall the quantum geometric tensor (QGT) defined by \eqref{QGTensor}
    \begin{align*}
Q_{ij}:=\left\langle\partial_i\psi\Big|(\mathbbm{1}-\psi\psi^\dagger)\partial_j\psi\right\rangle=\langle\partial_i\psi|\partial_j\psi\rangle - \hbar^{-2}{\cal A}_i {\cal A}_j
\,.
\end{align*} 
Upon the introduction of the gauge connection \eqref{2LevelConnectionVector} the 
QGT takes on the form
\begin{align}
  Q_{ij} = \frac{1}{4}\bgamma_i \cdot\bgamma_j - \hbar^{-2}(\bs\cdot\bgamma_i)(\bs\cdot\bgamma_j) 
  + \frac{i}{2\hbar}\bs\cdot\bgamma_i\times\bgamma_j\,,\label{QGTAGAIN}
\end{align}
so that the metric tensor and Berry curvature are correspondingly expressed as
\begin{align}
  T_{ij}&:= {\rm Re}(Q_{ij})=\frac{1}{4}\bgamma_i \cdot\bgamma_j - 
  \hbar^{-2}(\bs\cdot\bgamma_i)(\bs\cdot\bgamma_j)\,,\label{GammaMetricTensor2level}\\
    {\cal B}_{ij}&:=2\hbar{\rm Im}(Q_{ij})=\bs\cdot\bgamma_i\times\bgamma_j\,.\label{GammaBerryCurv2Level}
\end{align}
\end{proposition}
\paragraph{Proof:} Analogous to the method employed in the proof of Corollary 
\ref{Theorem:TIRED}, we introduce an $\mathfrak{su}(2)$ connection for electronic states via $\partial_i\psi = -\bar{\gamma}_i\psi$ 
  where $\bar{\gamma}_j \in \mathcal{F}(\mathbb{R}^3, \mathfrak{su}(2))$. Upon writing the pure state density matrix $\rho=\psi\psi^{\dagger}$, this 
  takes the QGT into the form
  \begin{align*}
    Q_{ij}=\braket{\rho|\bar\gamma_i}\braket{\rho|\bar\gamma_j}-\braket{\rho|\bar\gamma_i\bar\gamma_j}\,,
  \end{align*}
  which is itself is interesting result stated explicitly and in more generality in Proposition 
  \ref{Prop:QGTCov}. From here, one uses the standard decompositions on the 
  basis of Pauli matrices as described in Section \ref{Sec:1-Postulates}
  \begin{align*}
    \rho=\frac{1}{2}\left(\mathbbm{1} + 
    \frac{2}{\hbar}\bs\cdot\widehat{\bsigma}\right)\,,\qquad \bar\gamma_i = 
    -\frac{i}{2}\bgamma_i\cdot\widehat{\bsigma}\,,
  \end{align*}
and substitute directly into the expression for the QGT. As shown in the proof of Corollary \ref{Theorem:TIRED} we have that $\braket{\rho|\bar\gamma_i}=-i\hbar^{-1}\bs\cdot\bgamma_i$ which can be used in the first term. 
For the second term we compute
\begin{align*}
  \braket{\rho|\bar\gamma_i\bar\gamma_j} &= {\rm 
  Tr}(\rho\bar\gamma_i\bar\gamma_j)\\
  &= {\rm 
  Tr}\left(\frac{1}{2}\left(\mathbbm{1} + 
    \frac{2}{\hbar}\bs\cdot\widehat{\bsigma}\right)\cdot
    -\frac{i}{2}(\bgamma_i\cdot\widehat{\bsigma})\cdot
    -\frac{i}{2}(\bgamma_j\cdot\widehat{\bsigma})\right)\\
    &=  {\rm 
  Tr}\left(-\frac{1}{8}(\gamma_i)_b(\gamma_j)_c\sigma_b\sigma_c - \frac{1}{4\hbar}s_a(\gamma_i)_b(\gamma_j)_c\sigma_a\sigma_b\sigma_c\right)\\
   &=  -\frac{1}{8}(\gamma_i)_b(\gamma_j)_c {\rm Tr}(\sigma_b\sigma_c) - \frac{1}{4\hbar}s_a(\gamma_i)_b(\gamma_j)_c{\rm Tr}(\sigma_a\sigma_b\sigma_c)\\
&= -\frac{1}{4}\bgamma_i\cdot\bgamma_j - 
\frac{i}{2\hbar}\bs\cdot\bgamma_i\times\bgamma_j\,,
\end{align*}
using the trace relations for the Pauli matrices $ {\rm Tr}(\sigma_b\sigma_c) = 2\delta_{bc}$ and $ {\rm Tr}(\sigma_a\sigma_b\sigma_c) = 2i\epsilon_{abc}$. Simply putting these expressions together  
returns the desired form of the QGT in terms of the $\mathfrak{so}(3)$ 
connection \eqref{QGTAGAIN}.
\hfill$\square$\\

This proposition provides a new expression for the quantum geometric tensor for a two-level system written in terms of the $\mathfrak{so}(3)$ connection \eqref{PauliConnection}.
In this form the real and imaginary parts separate explicitly allowing us to simply read off the corresponding new relations for the Berry curvature and metric 
tensor. Furthermore, following the discussion in Remark \ref{Rem:MHo}, equation 
\eqref{GammaBerryCurv2Level} provides the expression for the Mermin-Ho relation in terms of the gauge connection in agreement with \cite{Holm2002}. 
As discussed in the proof, we will discover a more general version of 
this result in Section \ref{Sec:NonAbelianQuantumSystems} in which the QGT is 
written in terms of a connection derived from the full unitary group on Hilbert 
space $\mathcal{U}(\mathscr{H})$. As we shall see this will lead to the new interpretation of the QGT featuring the connection as the more fundamental geometric object. 

Having characterised the form of the QGT in terms of the connection, we now turn our attention back to the physical application of Pauli QHD.
We can now see how the Pauli QHD Lagrangian \eqref{EFHDbLagr2} specialises the general two-level Lagrangian \eqref{AnotherLagrangian} by the addition of the minimal coupling term $D\bu\cdot\bA$ and the choice of physical Hamiltonian
\begin{align}
  h(D,\bs,\bgamma_j)=\int D\left(V - M^{-1}\bB\cdot \bs + \frac{\hbar^2}{8M}|\bgamma_j|^2 - 
  \frac{(\bs\cdot\bgamma_j)^2}{2M}\right)\,\de^3x\,.\label{PauliQHDHamiltonianWithGamma}
\end{align}
Before formulating the dynamical system of equations, we first check whether this Hamiltonian satisfies the compatibility condition \eqref{Johnson}. Recall that this is required for \eqref{Boris} to be a valid solution to the generator equation which allows us to close the system.
\begin{lemma}[Zero curvature connection for Pauli QHD]\label{Lemma:PauliZeroCurv}
  \hfill \\
  The Hamiltonian functional \eqref{PauliQHDHamiltonianWithGamma} for Pauli QHD satisfies the 
  compatibility condition \eqref{Johnson} provided 
  \begin{align}
    \partial_j\bs\cdot\bgamma_j=0\,.\label{Dakota}
  \end{align}
  This equation implies that $\bOmega_{ij}=\partial_i\bgamma_j-\partial_j\bgamma_i +\bgamma_i\times\bgamma_j 
  =0$, i.e. that the curvature vanishes and the connection is flat.
\end{lemma}
\paragraph{Proof:}
An explicit calculation yields the variational derivative
\begin{align*}
\frac{\delta h}{\delta \bgamma_j} = \frac{\hbar^2}{4M}D\bgamma_j - 
  \frac{D}{M}(\bs\cdot\bgamma_j)\bs = \frac{D}{M}(\bs^T\bs - 
  \bs\bs^T)\bgamma_j\,,
\end{align*}
where the last equality highlights the equivalent expression in terms of the projection $\bs^T\bs - \bs\bs^T$. Here, $\,\cdot\,^T$ denotes the transpose operation so that $\bs^T\bs = |\bs|^2=\hbar^2/4$. Then, the corresponding covariant divergence
is computed to be
\begin{align*}
  \partial_j^{\gamma} \frac{\delta h}{\delta \bgamma_j} = 
    M^{-1}(\bs^T\bs - \bs\bs^T)\partial_j(D\bgamma_j) - \frac{D}{M}(\bs\cdot\bgamma_j)(\partial_j\bs-\bs\times\bgamma_j) - 
    \frac{D}{M}(\partial_j\bs\cdot\bgamma_j)\bs\,,
\end{align*} 
from which we calculate directly the required expression.
\begin{align*}
  \bs\cdot \partial_j^{\gamma} \frac{\delta h}{\delta \bgamma_j} &= 
    M^{-1}(\bs^T\bs - \bs\bs^T)\partial_j(D\bgamma_j)\cdot\bs - \frac{D}{M}(\bs\cdot\bgamma_j)(\partial_j\bs-\bs\times\bgamma_j)\cdot\bs - 
    \frac{D}{M}(\partial_j\bs\cdot\bgamma_j)|\bs|^2\\
    &= -\frac{\hbar^2 D}{4M}(\partial_j\bs\cdot\bgamma_j)\,.
\end{align*}
Hence, we clearly see that equation \eqref{Dakota} is required for this expression to 
vanish. Then, combined with $\bs\cdot\partial_j\bs=0$, \eqref{Dakota} implies 
that $\partial_j\bs=c (\bs\times\bgamma_j)$, up to a scale factor $c \in \mathbb{R}$. 
From here, the calculation given in Remark \ref{AnotherRemark} again implies that the connection is flat, i.e. has  
zero curvature $\bOmega_{ij}=0$.
\hfill $\square$ \\

We are now in a position to write the full system of equations for Pauli QHD as they arise from the Euler-Poincar{\'e} reduction of Proposition \ref{Prop:EP2LevelGamma}.

\begin{proposition}[Gauge connection formulation for Pauli QHD]\label{Prop:PauliQHDGamma}
 \hfill \\ Recall the hydrodynamic formulation of the Pauli equation as derived from the Lagrangian \eqref{EFHDbLagr2}. Upon introducing the gauge connection \eqref{PauliConnection}, the Lagrangian 
is transformed into the mapping $\ell:\mathfrak{X}(\mathbb{R}^3)\times{\rm Den}(\mathbb{R}^3)\times \mathcal{F}(\mathbb{R}^3,\mathbb{R}^3)\times \mathcal{F}(\mathbb{R}^3,\mathbb{R}^3)\times \Lambda^1(\mathbb{R}^3,\mathbb{R}^3)\to \mathbb{R}$  given by
\begin{align}\label{EFHDbLagr2AGAINAGAINAGAIN}
  \ell= \int\! D\bigg(\frac{1}{2}M|\bu|^2 + 
  \bu\cdot \bA - V + {\boldsymbol{s}}\cdot\Big(\boldsymbol\xi +M^{-1}\bB\Big) 
  -\frac{\hbar^2}{8M}|\bgamma_j|^2 - \frac{(\bs\cdot\bgamma_j)^2}{2M}\bigg) 
  {\rm d}^3x\,.
\end{align}
Then, Hamilton's principle for constrained variations \eqref{EPvar}, \eqref{twolevelfluidvariations} and \eqref{Adam}, returns the equations of motion
 \begin{align}
  \partial_t D + {\rm div}(D\bu)&= 0\,,\\
    M(\partial_t+\bu\cdot\nabla)u_i &= -\partial_i\left(V+V_Q\right)+ \big(\bu\times\bB\big)_i+\frac{1}{M}\partial_i\bB\cdot\bs -\frac{\hbar^2}{MD}\partial_j(DT_{ij})\,,\label{FLUIDEQNAGAIN}\\
   M(\partial_t+ \bu\cdot\nabla)\bs &= \bs\times\bB + (\bs\cdot\bgamma_j)\bs\times\bgamma_j-\frac{1}{D}\Big(\bs^T\bs-\bs\bs^T\Big)\partial_j(D\bgamma_j)\,,\label{2levelConnectionPauliSpinDynamicsZEROCURV}\\
 M(\partial_t\bgamma_i + u_j\partial_j\bgamma_i + \bgamma_j\partial_iu_j)&=\partial_i^{\gamma}\left(\bB + (\bs\cdot\bgamma_j)\bgamma_j + \frac{\partial_j D}{D} \bs\times\bgamma_j + \bs\times\partial_j\bgamma_j \right)\label{2LEVELZEROCURVGAMMAEQUATION}
\end{align}
where we recall the new expression for the metric tensor $T_{ij}$ as given by 
\eqref{GammaMetricTensor2level} and $\,\cdot\,^T$ denotes the transpose operation so that $\bs^T\bs = |\bs|^2=\hbar^2/4$.
\end{proposition}
\paragraph{Proof:} The direct proof of this result follows analogously to the general case 
given in Proposition \ref{Prop:EP2LevelGamma} as well as the subsequent step 
in solving the generator equation as described in \ref{Lemma:2levelXiSolution}. 
Hence, rather than repeating calculations, we explain the suitable modifications that apply for the treatment of the Pauli equation. 
Firstly, the Lagrangian \eqref{EFHDbLagr2AGAINAGAINAGAIN} modifies its general two-level 
counterpart \eqref{AnotherLagrangian} with the addition of the minimal coupling term $D\bu\cdot\bA$, arising from the additional external magnetic field.
In addition, recall from Lemma \ref{Lemma:PauliZeroCurv} that the connection is flat so that $\bOmega_{ij}=0$. Hence we have a modified version of the general (zero curvature) fluid equation \eqref{FLUIDEQN}, now 
taking the form
\begin{align*}
      MD(\partial_t+\bu\cdot\nabla)u_i &= -D\partial_i\left(V_Q+\frac{\delta h}{\delta 
  D}\right)+D\big(\bu\times\bB\big)_i -\bgamma_i\cdot\left(\bs\times \frac{\delta h}{\delta \bs} +\partial_j^{\gamma}\frac{\delta h}{\delta \bgamma_j} \right)
  \,,
\end{align*}
featuring the additional Lorentz force term, whilst the fluid continuity equation as well as \eqref{Pikachu}-\eqref{Onyx} remain unchanged. Specialising to the physical Hamiltonian functional 
\begin{align*}
  h(D,\bs,\bgamma_j)=\int D\left(V - M^{-1}\bB\cdot \bs + \frac{\hbar^2}{8M}|\bgamma_j|^2 - 
  \frac{(\bs\cdot\bgamma_j)^2}{2M}\right)\,\de^3x\,,
\end{align*}
corresponding to the Lagrangian \eqref{EFHDbLagr2AGAINAGAINAGAIN}, results in 
the following variational derivatives
\begin{align*}
\frac{\delta h}{\delta D} = V - &M^{-1}\bB\cdot\bs + \frac{\hbar^2}{8M}|\bgamma_j|^2 - 
  \frac{(\bs\cdot\bgamma_j)^2}{2M}\,,\qquad \frac{\delta h}{\delta \bs}=- 
  \frac{D}{M}\Big(\bB + (\bs\cdot\bgamma_j)\bgamma_j\Big)\,,\\
  &\frac{\delta h}{\delta \bgamma_j} = \frac{\hbar^2}{4M}D\bgamma_j - 
  \frac{D}{M}(\bs\cdot\bgamma_j)\bs = \frac{D}{M}(\bs^T\bs - \bs\bs^T)\bgamma_j\,.
\end{align*}
The corresponding covariant divergence of the final derivative now takes on the simpler form using the zero curvature relation
$\partial_j\bs = \bs\times\bgamma_j$ given by
\begin{align*}
  \partial_j^{\gamma} \frac{\delta h}{\delta \bgamma_j} = 
    M^{-1}(\bs^T\bs - \bs\bs^T)\partial_j(D\bgamma_j)\,.
\end{align*} 
Then substituting these derivatives into the fluid equation given above, further algebraic manipulations 
eventually yield
\begin{align*}
  M(\partial_t+\bu\cdot\nabla)u_i = -\partial_i\left(V+V_Q\right)&+ \Big(\bu\times\bB\Big)_i+\frac{1}{M}\partial_i\bB\cdot\bs \\
  &  -\frac{D}{M}\bgamma_j\cdot(\bs^T\bs - \bs\bs^T)\partial_i\bgamma_j - 
  M^{-1}\bgamma_i\cdot(\bs^T\bs - \bs\bs^T)\partial_j(D\bgamma_j)\,.
\end{align*}
From here one can recover the desired equation 
\eqref{FLUIDEQNAGAIN} upon computing explicity the following quantity
\begin{align*}
  -\frac{\hbar^2}{M}\partial_j(DT_{ij})&=  -\frac{\hbar^2}{M}\partial_j\left(D\left(\frac{1}{4}\bgamma_i \cdot\bgamma_j - 
  \hbar^{-2}(\bs\cdot\bgamma_i)(\bs\cdot\bgamma_j)\right)\right)\\
&=  -\frac{\hbar^2}{4M}\Big(\bgamma_i\cdot\partial_j(D\bgamma_j) +D\bgamma_j\cdot\partial_j\bgamma_i\Big) \\
&\qquad+ \frac{1}{M}\Big(D(\bs\cdot\bgamma_j)\partial_j(\bs\cdot\bgamma_i)+ (\bs\cdot\bgamma_i)\partial_j(\bs\cdot D\bgamma_j)\Big)\\
&=  -\frac{\hbar^2}{4M}\Big(\bgamma_i\cdot\partial_j(D\bgamma_j) +D\bgamma_j\cdot(\partial_i\bgamma_j + \bgamma_j\times\bgamma_i)\Big) \\
&\qquad+ \frac{1}{M}\Big(D(\bs\cdot\bgamma_j)(\bs\times\bgamma_j\cdot\bgamma_i+ \bs\cdot\partial_j\bgamma_i)+ (\bs\cdot\bgamma_i)(D\bs\times\bgamma_j\cdot \bgamma_j+ \bs\cdot\partial_j(D\bgamma_j))\Big)\\
&=  -\frac{\hbar^2}{4M}\Big(\bgamma_i\cdot\partial_j(D\bgamma_j) +D\bgamma_j\cdot\partial_i\bgamma_j\Big) \\
&\qquad+ \frac{1}{M}\Big(D(\bs\cdot\bgamma_j)(\bs\cdot\partial_i\bgamma_j)+ 
(\bs\cdot\bgamma_i)(\bs\cdot\partial_j(D\bgamma_j))\Big)\\
&=-\frac{D}{M}\bgamma_j\cdot(\bs^T\bs - \bs\bs^T)\partial_i\bgamma_j - 
  M^{-1}\bgamma_i\cdot(\bs^T\bs - \bs\bs^T)\partial_j(D\bgamma_j)\,,
\end{align*}
making use of the zero curvature relations $\partial_j\bs=\bs\times\bgamma_j$ and $\partial_i\bgamma_j-\partial_j\bgamma_i=\bgamma_j\times\bgamma_i$. 
Next, substituting the necessary variational 
derivatives into the equation for the spin vector \eqref{Pikachu} easily returns 
the required equation \eqref{2levelConnectionPauliSpinDynamicsZEROCURV}. Finally, we consider the equation for the evolution of the connection 
itself. Once again one obtains the desired result simply replacing the variational derivatives 
in the general equation \eqref{Onyx} and performing simple algebraic computations.
\hfill$\square$\\

We remark that that upon inserting the relation $\partial_j\bs = 
\bs\times\bgamma_j$ into 
the fluid equation \eqref{FLUIDEQNAGAIN} and spin vector equation \eqref{2levelConnectionPauliSpinDynamicsZEROCURV} recovers their counterparts \eqref{PauliEOM2} and \eqref{PauliEOM3} from the previous formulation (without $\bgamma$) in Section \ref{Sec:PauliQHD}. 

Finally, we present the particular form of the Kelvin-Noether circulation theorem \ref{Theorem:TIRED} 
as follows:
\begin{corollary}[Kelvin-Noether circulation theorem for Pauli QHD]
 \hfill \\ 
 The Kelvin-Noether circulation theorem associated to the fluid flow \eqref{FLUIDEQNAGAIN} 
  can be derived simply by specialising the general result for two-level systems 
  with the $\mathbb{R}^3$-valued connection from Corollary  \ref{Theorem:TIRED} in 
  the case of the Lagrangian \eqref{EFHDbLagr2AGAINAGAINAGAIN}. This results in
  \begin{align}
\frac{\de}{\de t}\oint_{c(t)} \bs\cdot\bgamma_i \,\de x^i &= \oint_{c(t)} 
\Big(M^{-1}\partial_i\bB\cdot\bs
            - \frac{\hbar^2}{MD} \partial_j \big(DT_{ij}\big) \Big) \de x^i \,,
\end{align}
where $c(t)$ is a loop moving with the fluid velocity $\bu$ and the real part of 
the QGT is now expressed in terms of the non-Abelian connection, as given explicitly in \eqref{GammaMetricTensor2level}. 
\end{corollary}


In this section, we have introduced a gauge connection in two-level EF systems derived from the evolution of the spin 
vector under the rotation group. The subsequent Euler-Poincar{\'e} formulation 
of the dynamics led to a new set of hydrodynamic equations which feature explicit dependence on the curvature of the connection \eqref{PauliConnection}.
The resulting circulation theorem again takes the form of the evolution of a Berry phase. Interestingly, in this new formalism,
the Berry connection can be understood as the $\mathcal{U}(1)$ projection of the non-Abelian connection, whilst new expressions for the two-level quantum geometric
 tensor were unveiled. Finally, these results were specialised to the Pauli 
 equation, providing an alternative hydrodynamic system of equations to those 
 presented in Section \ref{Sec:PauliQHD}.

\section{General theory for connections in mechanical systems with order parameters}\label{Sec:GeneralTheory}

We have now encountered multiple examples of gauge connections in this thesis: Abelian 
 connections for the QHD phase and amplitude from Chapter \ref{Chap:Holonomy} and the non-Abelian 
connection for the spin vector from Section 
\ref{Sec:NonAbelianPauliConnection}. All of these connections have been derived 
using a similar construction which, as described in Remarks \ref{jack} and \ref{AnotherRemark} 
respectively, imply a zero-curvature relation. The goal of this section is to extend this construction   
 to a whole class of models employing the general relation $\nabla n = -\bgamma n$. Here, $n\in\mathcal{F}(\Bbb{R}^3,M)$ is an order parameter field and $\gamma=\bgamma\cdot\de\bx$ is a connection corresponding to a gauge group $\mathcal{F}(\Bbb{R}^3,G)$, where $G$ acts on $M$. 
Nevertheless, even though the relation $\nabla n = -\bgamma n$, which implies zero curvature, is used to construct a Lagrangian,  the resulting equations of motion still allow for a more general non-zero constant curvature that again arises as an initial condition.

For this general approach we consider an order parameter $n \in \mathcal{F}(\mathbb{R}^3,M)$ (where at this point $M$ is an arbitrary manifold) whose evolution is given by an element $g$ of the Lie group 
$\mathcal{F}(\mathbb{R}^3,G)$,
\begin{align}
  {n}(\bx,t)={g}(\bx,t){n}_0(\bx)\,.\label{generalgroupactionOrderParam}
\end{align}
Then, one can construct a gauge connection $\gamma$ from the gradient of ${n}$ using the analogous approach to that found in Sections \ref{Sec:NEWphase} and \ref{Sec:NonAbelianPauliConnection} as 
follows,
\begin{align}
    \begin{split}
  \nabla{n} &= \nabla {g} \,{n}_0 + {g}\nabla{n}_0\\
  &= \nabla {g} \,{g}^{-1} {n} - {g} \bgamma_0{n}_0\\
  &= - (-\nabla {g}\,{g}^{-1}+ {g}\bgamma_0{g}^{-1}){n} := -\bgamma{n}\,.
  \end{split}\label{generalgammaconstruction}
\end{align}
Hence, we have the general relation $\nabla{n} = -\bgamma{n}$.
Here the 1-form $\gamma=\gamma_i\de x^i$ can be considered as a $\mathfrak{g}$-valued connection 1-form $\gamma\in\Lambda^{1}(\mathbb{R}^3, \mathfrak{g})$ (where $\mathfrak{g}$ denotes the Lie algebra of $G$) defined on the trivial principal bundle $\mathbb{R}^3\times 
G$. The geometry of such gauge connections has previously been considered in the context of complex fluids and the reader is referred to \cite{GayBalmazRatiu2009, GayBalmazTronci2010} 
for further details.

In order to prove that a connection constructed in this way must have zero curvature, we first give a general definition of the curvature of a gauge 
connection.
\begin{definition}[Curvature 2-form]\label{Def:Curvature}
  Associated to any gauge connection $\gamma \in \Lambda^{1}(\mathbb{R}^3, \mathfrak{g})$, the corresponding {\bf curvature 2-form} 
  is defined by
  \begin{align}
    \Omega = {\rm d}^{\gamma}\gamma := \de \gamma + \frac{1}{2}[\gamma\wedge\gamma]\in \Lambda^{2}(\mathbb{R}^3, 
    \mathfrak{g})\,.\label{CurvatureDef1}
  \end{align} 
  Here $\de^{\gamma}$ denotes the exterior covariant derivative and the bracket $[\,\cdot\, \wedge \,\cdot\,]$ is regarded
  as a bilinear operation on $\Lambda^{1}(\mathbb{R}^3, \mathfrak{g})$ given by 
  \begin{align*}
   [\alpha \wedge \beta]= [\alpha_i \de x^i \wedge \beta_j \de x^j] := [\alpha_i,\beta_j]\,\de x^i\wedge \de 
   x^j\,,
     \end{align*}
for all $\alpha =\alpha_i \de x^i, \beta=\beta_j \de x^j \in \Lambda^1(\mathbb{R}^3, \mathfrak{g})$.
  Using this we can expand \eqref{CurvatureDef1}, so that upon writing $\Omega = 1/2 \,\Omega_{ij}\de x^i\wedge\de x^j$, we obtain an expression for the curvature in components as 
  \begin{align}       
        \Omega_{ij}:=\partial_i\gamma_j-\partial_j\gamma_i+[\gamma_i,\gamma_j]\,.\label{CurvatureDef2}
  \end{align}
\end{definition}

We can now state the important new result for connections constructed via the 
method described in equation \eqref{generalgammaconstruction}.
\begin{theorem}[General zero curvature relation]\label{Theorem:GenZeroCurv}
\hfill \\ Any gauge connection $\gamma \in \Lambda^1(\mathbb{R}^3,\mathfrak{g})$ satisfying the relation $\de n = -\gamma n$  
  \eqref{generalgammaconstruction} 
  must have zero curvature and is called `flat'. That is, the corresponding curvature 2-form $\Omega \in \Lambda^2(\mathbb{R}^3, \mathfrak{g})$ as defined in Definition \ref{Def:Curvature} must satisfy $\Omega = 0$.
Equivalently, written explicitly in terms of components, this zero curvature relation reads
  \begin{align}
    \Omega_{ij}=\partial_i\gamma_j-\partial_j\gamma_i+[\gamma_i,\gamma_j]=0\,.
  \end{align}
\end{theorem}
\paragraph{Proof:} Writing our defining relation in terms of differential forms as $\text{d}{n}=-\gamma{n}$, we compute the 
following:
\begin{align}
    \begin{split}
  0= \text{d}^2{n} &=  -\text{d}(\gamma_j{n}\,\text{d}x^j)\\
  &= -\partial_{[i}(\gamma_{j]}{n})\,\text{d}x^i\wedge \text{d}x^j\\
  &= -\Big((\partial_{[i}\gamma_{j]}){n} + \gamma_{[j}(\partial_{i]}{n})\Big)\,\text{d}x^i\wedge \text{d}x^j\\
    &= -\Big(\partial_{[i}\gamma_{j]} - \gamma_{[j}\gamma_{i]}\Big){n}\,\text{d}x^i\wedge \text{d}x^j\\
        &= -\frac{1}{2}\Big(\partial_i\gamma_j - \partial_j\gamma_i + \gamma_i\gamma_j - \gamma_j\gamma_i\Big){n}\,\text{d}x^i\wedge \text{d}x^j\\
                &= -\frac{1}{2}\Omega_{ij}{n}\,\text{d}x^i\wedge \text{d}x^j\\
                &= -\Omega{n}
                \,,
  \end{split}\label{ZeroCuvatureGenCalc}
\end{align}
where the square brackets denote  index anti-symmetrisation.
At this point, we observe that since $\gamma(t) := -\de {g}\,{g}^{-1}+ {g}\gamma_0{g}^{-1}$, it follows that 
\begin{align}
\Omega(t)=g\Omega_0 g^{-1}
\end{align}
so that \eqref{ZeroCuvatureGenCalc} implies $0= \Omega_0 n_0$. Thus, we end up in a situation in which either $\Omega_0=0$ thus rendering $\Omega={\rm d}^\gamma\gamma=0$ for all time, or $n_0$ belongs to the kernel of $\Omega_0$, a statement that we cannot impose in general. 
\hfill$\square$\\

Despite this result, we now turn our attention to the particular step in which the gauge connection is introduced in the Lagrangian of a field theory.
As we have seen in Chapter \ref{Chap:Holonomy}, after initially using the type 
of relation $\nabla n = -\bgamma n$ to introduce a dynamical connection in the Lagrangian, this 
restriction can be lifted upon considering the general equations of motion that 
follow. We now consider the analogous procedure for general Lagrangians of the 
type $L=L(n,\dot{n},\nabla n)$, where $n \in \mathcal{F}(\mathbb{R}^3, M)$ (for example,  in the case of the Ericksen-Leslie theory of liquid crystal nematodynamics, we have  $M=S^2$ \cite{GBRaTr2012,GBRaTr2013}).
  According to the previous discussion one can let $n$ evolve under the group $g\in \mathcal{F}(\mathbb{R}^3, G)$, so that $n(t)=g(t)n_0$ as given by \eqref{generalgroupactionOrderParam}. 
  This leads to introducing the generator $\xi:=\partial_t{g}\,g^{-1}\in \mathcal{F}(\mathbb{R}^3, \mathfrak{g})$ such that $\partial_t{n}=\xi n$ and a gauge connection $\gamma\in \Lambda^1(\mathbb{R}^3, \mathfrak{g})$ via equation \eqref{generalgammaconstruction}. These relations can be used to obtain a reduced Lagrangian of the form $\ell=\ell(n, \xi, \bgamma)$.
   Then, applying Hamilton's principle to this new Lagrangian produces a more general set of equations in which $\gamma$ is allowed to have a non-zero curvature (constant if the gauge group is 
   Abelian). This procedure is formalised in the following theorem, specialising the more general treatment found in 
   \cite{GayBalmazRatiu2009}.
 
 \begin{theorem}[Euler-Poincar{\'e} theory for connections in mechanical systems]\label{Theorem:GammaGenEP}
\hfill \\ Consider a mechanical system described by the Lagrangian $L:T\mathcal{F}(\mathbb{R}^3,M)\to \mathbb{R}$ 
given by $L=L(n, \partial_t n, \nabla n)$, where $n=n(\bx,t)\in \mathcal{F}(\mathbb{R}^3, 
M)$. If the order parameter evolves under the transitive group action \eqref{generalgroupactionOrderParam} 
\begin{align*}
    {n}(\bx,t)={g}(\bx,t){n}_0(\bx)\,,\qquad g \in \mathcal{F}(\mathbb{R}^3, 
    G)\,,
\end{align*}
the Lagrangian can be written in the reduced form $\ell: \mathcal{F}(\mathbb{R}^3,M)\times \mathcal{F}(\mathbb{R}^3,\mathfrak{g})\times\Lambda^1(\mathbb{R}^3,\mathfrak{g})\to \mathbb{R}$ 
given by $\ell=\ell(n, \xi, \bgamma)$, where $\xi:= \partial_t g \,g^{-1}\in  \mathcal{F}(\mathbb{R}^3,\mathfrak{g})$ and $\gamma \in \Lambda^1(\mathbb{R}^3,\mathfrak{g})$ 
is defined by equation \eqref{generalgammaconstruction}. Then, applying 
Hamilton's principle for constrained variations
\begin{align}
  \delta n = \Upsilon n\,,\qquad \delta \xi = \partial_t\Upsilon + [\Upsilon, 
  \xi]\,,\qquad \delta \gamma_j = -\partial_j^{\gamma} \Upsilon :=-\partial_j\Upsilon - [\gamma_j,\Upsilon]\,,\label{EPVarGammaSystem}
\end{align}
where $\Upsilon:=\delta g\,g^{-1}\in\mathcal{F}(\mathbb{R}^3,\mathfrak{g})$ is 
arbitrary, results in the following Euler-Poincar{\'e} equations of motion
 \begin{align}
   \partial_t\frac{\delta \ell}{\delta \xi}-\left[\xi, \frac{\delta \ell}{\delta 
   \xi}\right]&=\frac{\delta \ell}{\delta n}\diamond n + \partial_j^{\gamma}\frac{\delta \ell}{\delta 
   \gamma_j}\,,\label{GenEPwithGammaGenEqn}\\
   \partial_t n &= \xi n\,,\\
   \partial_t \gamma_j &= -\partial_j^{\gamma}\xi\,.\label{Hannah}
 \end{align}
 The final two equations are auxilliary 
 equations following from the evolution of $n$ and $\bgamma$ under the group $G$, whilst the first equation must be solved for the generator 
 $\xi$.
  \end{theorem}
 \paragraph{Proof:} We begin this proof by considering the derivation of the 
 variations \eqref{EPVarGammaSystem}. Whilst the first two are standard in Euler-Poincar{\'e} reduction theory (see Appendix \ref{App:GMTheory}), 
 we compute the variation of $\gamma$ directly as follows.
\begin{align*}
   \delta\bgamma &= \delta\Big(-\nabla U\,U^{-1}+U\bgamma_{0}U^{-1}\Big)\\
   &= - (\nabla\delta U)U^{-1} - \nabla U \delta(U^{-1}) + 
   \delta U \bgamma_0U^{-1} + U\bgamma_0\delta(U^{-1})\\
   &= -\nabla(\delta U\,U^{-1}) + \delta U\nabla(U^{-1})- \nabla U \delta(U^{-1})+ 
   \delta U \,U^{-1}\,U\bgamma_0U^{-1} + U\bgamma_0\delta(U^{-1})\\
   &= -\nabla(\delta U\,U^{-1}) - (-\nabla U 
   \,U^{-1}+U\bgamma_0U^{-1})\delta U\,U^{-1} + \delta U\,U^{-1}(-\nabla U 
   \,U^{-1}+U\bgamma_0U^{-1})\\
   &= -\nabla\Upsilon - [\bgamma,\Upsilon]\,,
 \end{align*}
in which we recognise the covariant gradient and have denoted $\bgamma_0 = \bgamma(\bx,0)$. In addition, via the analogous computation one can derive the auxilliary 
 equation $\partial_t\bgamma = -\nabla^{\gamma}\xi$. We complete the proof by 
 deriving the algebraic equation \eqref{GenEPwithGammaGenEqn} for the generator $\xi$. To do so, we 
 expand Hamilton's principle and use the variations \eqref{EPVarGammaSystem}:
 \begin{align*}
   0 &= \delta \int_{t_1}^{t_2}\ell(n, \xi, \bgamma)\,\de t\\
   &= \int_{t_1}^{t_2} \Bigg(\left\langle \frac{\delta \ell}{\delta n}, \delta n \right\rangle +\left\langle \frac{\delta \ell}{\delta \xi}, \delta \xi \right\rangle +\left\langle \frac{\delta \ell}{\delta \gamma_j}, \delta \gamma_j \right\rangle\Bigg)\,\de 
   t\\
    &= \int_{t_1}^{t_2} \Bigg(\left\langle \frac{\delta \ell}{\delta n}, \Upsilon n \right\rangle +\left\langle \frac{\delta \ell}{\delta \xi}, \partial_t \Upsilon -{\rm ad}_{\xi}\Upsilon \right\rangle +\left\langle \frac{\delta \ell}{\delta \gamma_j}, -\partial_j^{\gamma}\Upsilon \right\rangle\Bigg)\,\de 
    t\\
    &= \int_{t_1}^{t_2} \Bigg(\left\langle \frac{\delta \ell}{\delta n}\diamond n, \Upsilon  \right\rangle +\left\langle -\partial_t\frac{\delta \ell}{\delta \xi} - {\rm ad}^*_{\xi}\frac{\delta \ell}{\delta \xi},\Upsilon \right\rangle +\left\langle \partial_j^{\gamma}\frac{\delta \ell}{\delta \gamma_j}, \Upsilon \right\rangle\Bigg)\,\de t
 \end{align*}
 from which we read off the desired equation \eqref{GenEPwithGammaGenEqn} 
 recalling that $\Upsilon$ is arbitrary.
 \hfill$\square$\\

The general Euler-Poincar{\'e} equations given in Theorem \ref{Theorem:GammaGenEP} allow the connection $\gamma$ to possess non-zero curvature (constant if the gauge group is Abelian)
as described by the following corollary. 

 \begin{corollary}[Time evolution of the curvature]\label{Cor:CurvEvo}
   \hfill \\
   Taking the covariant differential $\de^{\gamma}=\de + {\rm ad}_\gamma$ of the evolution equation $\partial_t\gamma=-{\rm d}^{\gamma}\xi$ results in 
   the evolution equation for the curvature $\Omega=\de^{\gamma}\gamma$ given by
\begin{align}
  \partial_t\Omega=[\xi,\Omega]\,.
\end{align}
 \end{corollary}
 \paragraph{Proof:} Applying the covariant differential to both sides of the 
 auxilliary equation $\partial_t\gamma = -\de^{\gamma}\xi$ yields $\partial_t  \de^{\gamma}\gamma = -(\de^{\gamma})^2\xi$ 
 so that, upon recalling Definition \ref{Def:Curvature}, the above result 
 amounts to proving $(\de^{\gamma})^2\xi=[\de^{\gamma}\gamma, \xi]=[\Omega, 
 \xi]$. We do this as follows
 \begin{align*}
   (\de^{\gamma})^2\xi &= \de^{\gamma}(\de^\gamma\xi)\\
   &= \Big(\partial_i+ [\gamma_i, \,\cdot\,]\Big)\de x^i \wedge \Big(\partial_j\xi+ [\gamma_j, \xi]\Big)\de 
   x^j\\
   &= \frac{1}{2}\Big(\partial_i[\gamma_j, \xi] - \partial_j[\gamma_i, \xi] + [\gamma_i, \partial_j\xi] - [\gamma_j,\partial_i\xi] +[\gamma_i,[\gamma_j,\xi]]-[\gamma_j,[\gamma_i,\xi]]\Big)\de x^i \wedge\de 
   x^j\\
      &= \frac{1}{2}\Big([\partial_i\gamma_j- \partial_j\gamma_i, \xi] + [[\gamma_i,\gamma_j],\xi]\Big)\de x^i \wedge\de 
   x^j\\
   &= \left[\frac{1}{2}\Omega_{ij}\de x^i \wedge\de 
   x^j, \xi\right] = [\Omega, \xi]\,.
 \end{align*}
 Here we have used the Jacobi identity for the Lie bracket on 
 $\mathfrak{g}$ to obtain the penultimate line and recalled that $\Omega = 1/2\,\Omega_{ij}\de x^i \wedge\de 
   x^j.$
 \hfill$\square$\\

We conclude the material in this section with a couple of remarks. The first concerning the relation between $\de n$ 
and $\gamma n$ in the general case of non-zero curvature and the second discussing the extension of this theory to flowing order parameter systems, i.e. including coupling to an underlying fluid motion.
\begin{remark}\label{Rem:Heatwave}
  Now that the connection $\gamma$ is defined purely as the solution to the 
  equation $\partial_t\gamma = -\de^{\gamma}\xi$ (hence allowing for non-zero curvature), it is natural to 
  now ask how the differential $\de n$ can be expressed in terms of the 
  connection. To do so we notice the equation
  \begin{align}
(    \partial_t-\xi)\Big({\rm d}n+\gamma n\Big)= 0\,,\label{Jamie}
  \end{align}
  which follows by a direct computation using the two auxilliary equations of 
  Theorem \ref{Theorem:GammaGenEP}. 
This equation implies the evolution of the initial condition by
\begin{align}
  {\rm d}n(t)+\gamma(t) n(t) = g(t)\Big({\rm d}n_0+\gamma_0 n_0\Big)\,,\qquad \text{{\rm with} $g\in \mathcal{F}(\mathbb{R}^3, G)$}\,,
\end{align}
 so that allowing for a non-trivial connection simply amounts to the failure of the initial condition $\nabla n_0\neq-\bgamma_0 n_0$ and offers a general method for constructing defect theories whose defect topology does not depend on time.
\end{remark}

\begin{remark}[Extension to flowing order parameter systems]
  \hfill \\
  Whilst the material in this section has been presented in its simplest form, a natural extension is to include coupling to an underlying fluid flow. 
  This is the domain of {\rm complex fluids} \cite{Holm2002,GayBalmazRatiu2009} and has many applications, for example 
  in studying flowing liquid crystal dynamics \cite{GBRaTr2012,GBRaTr2013}. 
  This extension naturally requires the introduction of the diffeomorphism group on physical space 
  ${\rm Diff}(\mathbb{R}^3)$ which we recall consists of smooth invertible maps $\eta:\bx\mapsto\boldeta(\bx)$. 
  In this case, the Euler-Poincar{\'e} theory of Theorem \ref{Theorem:GammaGenEP} 
  specialises as follows. The evolution of the order parameter \eqref{generalgroupactionOrderParam} is 
  replaced by $n(t)=(gn_0)\circ \eta^{-1}$, resulting in the corresponding modifications to the definitions of the generator and 
  gauge connection
  \begin{align*}
    \xi= (\partial_t g\,g^{-1})\circ \eta^{-1}\,,\qquad \bgamma=(-\nabla g\,g^{-1}+g\bgamma_0g^{-1})\circ 
    \eta^{-1}\,.
  \end{align*}
  Upon deriving the new variations (amounting to replacing the material derivatives $\delta\mapsto\delta+\pounds_{\bw}$ and $\partial_t\mapsto\partial_t+\pounds_{\bu}$) 
 Hamilton's principle applied to $\ell(n, \xi, \bgamma)$ results in the system of 
 equations
  \begin{align}
(   \partial_t+\pounds_{\bu})\frac{\delta \ell}{\delta \xi}-\left[\xi, \frac{\delta \ell}{\delta 
   \xi}\right]&=\frac{\delta \ell}{\delta n}\diamond n + \partial_j^{\gamma}\frac{\delta \ell}{\delta 
   \gamma_j}\,,\\
(   \partial_t+\pounds_{\bu}) n &= \xi n\,,\\
(   \partial_t+\pounds_{\bu}) \gamma &= -\de^{\gamma}\xi\,,
 \end{align}
 corresponding to \eqref{GenEPwithGammaGenEqn}-\eqref{Hannah} in Theorem 
 \ref{Theorem:GammaGenEP}. This approach will be used later in this chapter
 to introduce a non-Abelian connection into the hydrodynamic exact 
  factorization system of Section \ref{NuclearFrameSection}.
\end{remark}

We now apply the results of this section to construct a general non-Abelian connection for a entire class of quantum mechanical systems, resulting in interesting new 
interpretations for many of the geometric objects encountered so far in this thesis.

\section{A $\mathfrak{u}(\mathscr{H})$ connection for quantum systems}\label{Sec:NonAbelianQuantumSystems}
Having presented the general theory for introducing dynamical gauge connections in systems with order parameters in the previous 
section, we now consider the application of this approach to quantum mechanics. We consider this direction motivated by the exact factorization \eqref{GenExFact}, in which we can consider the electronic wavefunction (which depends on the nuclear coordinate)
as an order parameter $\psi(\br,t)\in \mathcal{F}(\mathbb{R}^3, \mathscr{H}_e)$. Then, its unitary 
 evolution via $\psi(\br,t) = 
U(\br,t)\psi_0(\br)$ provides the gauge group from which we can introduce a connection of the form \eqref{generalgammaconstruction}.
Motivated by this possibility, in this section we consider the general case of non-Abelian connections in quantum systems associated to the 
full unitary group action on a Hilbert space, before returning to the specific application in the EF system in Section \ref{Sec:GammaEF}.

Hence, consider a normalised quantum state $\psi(\bx)$
that depends smoothly on an external parameter $\bx\in\mathbb{R}^3$. 
Here, $\psi$ can be understood as beloging to $\mathcal{F}(\mathbb{R}^3, \mathscr{H})$, the space of smooth mappings from $\mathbb{R}^3$ into $\mathscr{H}$, with the latter denoting the Hilbert space in consideration.
The normalisation of such states is written as $\braket{\psi(\bx)|\psi(\bx)}=1$, where the angled bracket denotes the inner product on $\mathscr{H}$, so that the normalisation is true for all $\bx \in \mathbb{R}^3$.
Notice that we have encountered many such systems throughout this thesis\footnote{Quantum systems that depend on some external parameter are often considered in the physics literature, see for example \cite{Berry1984, Simon1983, PrVa1980, WilczekZee1984}.}. 
We then consider the evolution of $\psi$ on the unitary group of the Hilbert space $\mathcal{U}(\mathscr{H})$ as 
follows
\begin{align}
  \psi(\bx,t)=U(\bx,t)\psi_0(\bx)\,,\qquad U(\bx,t)\in  \mathcal{F}(\mathbb{R}^3, 
  \mathcal{U}(\mathscr{H}))\,.\label{ParametricUnitaryEvo}
\end{align} 
Following the general theory from Section \ref{Sec:GeneralTheory}, this allows us to evaluate derivatives
 with respect to the parameter more explicitly as, considering $\nabla\psi(\bx)$ as the components of a 1-form, we have $\nabla\psi(\bx)\cdot\de\bx = \partial_j\psi(\bx)\,\de 
 x^j$. Hence, via the computation \eqref{generalgammaconstruction} we can write
\begin{align}
\begin{split}\partial_j\psi(\bx,t)
  &= -\gamma_j(\bx,t)\psi(\bx,t)\,,\\
\gamma_j(\bx,t)&=-\partial_j U(\bx,t)U^{-1}(\bx,t) + U(\bx,t)\gamma^0_j(\bx) 
  U^{-1}(\bx,t)\,,
\end{split}\label{FullUnitaryConnection}
\end{align}
where $\gamma_j(\bx) \in \mathcal{F}(\mathbb{R}^3, \mathfrak{u}(\mathscr{H}))$ is a mapping from physical space into the $\mathfrak{u}(\mathscr{H})$, the Lie-algebra 
of the unitary group on $\mathscr{H}$. Hence, each $\gamma_j$ is a skew-Hermitian 
operator and we can write the $\mathfrak{u}(\mathscr{H})$-valued 1-form (gauge connection)
$  \gamma= \gamma_j(\bx)\,\de x^j$. In physics $\gamma$ is described as the magnetic vector potential of a Yang-Mills (non-Abelian) field.

\begin{remark}[Relation to the $\mathfrak{su}(2)$ connection for the spin vector]\label{Rem:connectionConnections}
  \hfill \\
  Recall the $\mathfrak{su}(2)$ connection defined by \eqref{PauliConnection} for the spin vector from Section 
  \ref{Sec:NonAbelianPauliConnection}. We can now see how this is simply the two-level 
  restriction of the connection \eqref{FullUnitaryConnection}. Specifically, upon writing the density matrix as $\rho=\psi\psi^{\dagger}$, a simple calculation 
  using \eqref{FullUnitaryConnection}
  yields that $\partial_j\rho = [\rho, \gamma_j]$. Then, upon restricting to two-level systems so that $\rho$ and $\gamma_j$ are operators on
  $\mathbb{C}^2$, we transform this equation by writing these operators in terms of the Pauli matrices as follows
  \begin{align*}
    \partial_j\left(\frac{1}{2}\left(\mathbbm{1}+ \frac{2}{\hbar}\bs\cdot\widehat\bsigma\right)\right) &= \left[\frac{1}{2}\left(\mathbbm{1}+ \frac{2}{\hbar}\bs\cdot\widehat\bsigma\right), -\frac{i}{2}\bgamma_j\cdot\widehat\bsigma\right]\\
    \implies \hbar^{-1}\partial_j\bs\cdot\widehat\bsigma &= -\frac{i}{2\hbar} 
    s_{\ell}(\gamma_j)_k [\sigma_{\ell},\sigma_k]\\
    &= \hbar^{-1}\epsilon_{\ell k m} s_{\ell}(\gamma_j)_k \sigma_m\\
    &= \hbar^{-1}(\bs\times\bgamma_j)\cdot\widehat\bsigma\,,
  \end{align*} 
  thus regaining the previously used formula $\partial_j\bs=\bs\times\bgamma_j$.
 Hence, via the $\mathfrak{su}(2)\cong \mathbb{R}^3$ Lie-algebra isomorphism, we 
see how the new connection \eqref{FullUnitaryConnection} generalises that used 
in Section \ref{Sec:NonAbelianPauliConnection}.
\end{remark}

Having introduced the new connection \eqref{FullUnitaryConnection} associated to 
the action of the full unitary group on the Hilbert space, we now consider some 
of the interesting new perspectives on quantum mechanics that it unveils. We note that throughout this section we freely make use of the original definition of the connection $\nabla \psi = -\bgamma \psi$ which (recalling Theorem \ref{Theorem:GenZeroCurv}) automatically invokes the zero curvature relation. 

Firstly, motivated by the desire for physical intuition,
 we define a Hermitian counterpart to $\gamma$ via
\begin{align}
 \widetilde{\gamma}:= -i\hbar\gamma \label{GammaTildeDefn}\,,
\end{align}
so that $ \widetilde{\gamma}$ may be understood an observable. Then, the defining relation \eqref{FullUnitaryConnection}
can be rewritten as the Schr\"odinger type equation $i\hbar\nabla\psi=\widetilde{\gamma}\psi$ in which $\widetilde{\gamma}$ plays the role of a spatial counterpart to a Hamiltonian operator. 
Furthermore, we once again consider the celebrated Berry connection \cite{Berry1984}, which itself is a connection on the trivial principal bundle 
$\mathcal{U}(1)\times\mathbb{R}^3$.
 Upon introducing $\gamma$, we now have that
\begin{align}
  {\cal A}_j := \braket{\psi|-i\hbar\partial_j\psi} = -\braket{\widetilde{\gamma}_j}\,,\label{BerryConnectionIsTheExpectation}
\end{align}  
where we have introduced the notation $\braket{\widehat{A}}=\braket{\psi|\widehat{A}|\psi}$ for the pointwise expectation value of an operator, $\widehat{A}$.
 Hence, we see that the Berry connection can be understood as simply the $\mathcal{U}(1)$ projection of $\widetilde{\gamma}$.
 
 \begin{remark}[Finite-dimensional projections and the Wilczek-Zee connection]
   \hfill \\
Given that the Berry connection is understood as the $\mathcal{U}(1)$ projection of $\gamma$, it is natural to consider the projections onto other finite-dimensional subgroups of $\mathcal{U}(\mathscr{H})$.
In fact, the whole class of such projections onto $\mathcal{U}(\mathcal{N})$ are known as Wilczek-Zee connections \cite{WilczekZee1984} 
and arise in the case of degenerate Hamiltonian operators.  Considering a single 
energy level $E$, we say a quantum Hamiltonian operator (parametrically dependent on external parameter $\bx \in \mathbb{R}^3$) is $\mathcal{N}$-fold 
degenerate if
\begin{align*}
  \widehat{H}(\bx)\psi_a(\bx)=E(\bx)\psi_a(\bx)\,,\qquad a\in 
  \{1,\,\cdots,\mathcal{N}\}\,,
\end{align*}
with the eigenvectors normalised such that $\braket{\psi_a|\psi_b}=\delta_{ab}$. 
Following the analogous procedure to that described in Section \ref{SECBOBerryPhase} (invoking the adiabatic theorem), it can be shown that the geometric phase is given by 
the path-ordered exponential of the loop integral of
\begin{align*}
  {\cal A}_{ab} := \braket{\psi_a|-i\hbar\de \psi_b}\in \Lambda^1(\mathbb{R}^3,\mathfrak{u}(\mathcal{N}))\,, 
\end{align*}
which is the {\emph Wilczek-Zee connection} on the trivial bundle $\mathbb{R}^3\times\,\mathcal{U}(\mathcal{N})$\cite{WilczekZee1984,ChruscinskiJamiolkowski2012,BohmEtAl2003}.
Then, intoducing the new $\mathfrak{u}(\mathscr{H})$-connection \eqref{FullUnitaryConnection} 
we have that
\begin{align*}
  {\cal A}_{ab} = \braket{\psi_a|-i\hbar\de \psi_b} = 
  i\hbar\braket{\psi_a|\gamma|\psi_b}\,,
\end{align*}
and we see how the Wilczek-Zee connection is contained within $\gamma$. Further generalisations of quantum geometric phases not corresponding to adiabatic 
evolution \cite{AharonovAnandan1987} or even closed loops in the base space \cite{SamuelBhandari1988} have also been 
discovered, and it would be of interest to clarify what role $\gamma$ plays in 
relation to the corresponding connections.
 \end{remark}

 The previous relations suggest how $\gamma$ opens up interesting new possibilties for the interpretation of exisiting 
 geometric objects in quantum mechanics. As we have previously seen throughout this thesis, another object  
 of significance in both quantum physics and applications in quantum chemistry is the 
 quantum geometric tensor \eqref{QGTensor}. In order to pursue this direction further, we first give the following defintion.

\begin{definition}[Covariance]\label{Def:Cov} The {\bf covariance} of two Hermitian operators (observables) $\widehat{A}$ 
  and $\widehat{B}$ is defined as
  \begin{align}
    {\rm cov}(\widehat{A},\widehat{B}):= 
\braket{\widehat{A}\widehat{B}}-\braket{\widehat{A}}\braket{\widehat{B}}\,.
  \end{align}
The variance of $\widehat{A}$ from Definition \ref{Def:Variance} is then regained as simply the covariance of $\widehat{A}$ with itself, i.e. $  \sigma_{{A}}^2\equiv{\rm 
cov}(\widehat{A},\widehat{A})$.
\end{definition}

This allows us to write the following proposition expressing the QGT in terms of the connection \eqref{FullUnitaryConnection}.

\begin{proposition}[Quantum geometric tensor as the covariance of the connection]\label{Prop:QGTCov}
  \hfill \\
  Recall the quantum geometric tensor (QGT) defined by \eqref{QGTensor}
  \begin{align*}
Q_{ij}:=\left\langle\partial_i\psi\Big|(\mathbbm{1}-\psi\psi^\dagger)\partial_j\psi\right\rangle=\langle\partial_i\psi|\partial_j\psi\rangle - \hbar^{-2}{\cal A}_i {\cal A}_j
\,.
\end{align*} 
Expressed in terms of the $\mathfrak{u}(\mathscr{H})$ connection 
\eqref{FullUnitaryConnection}, the QGT is equivalently written as
 \begin{align}
\begin{split}
Q_{ij} &= \braket{\gamma_i}\braket{\gamma_j} - \braket{\gamma_i\gamma_j}\\
&= \hbar^{-2}\Big(\braket{\widetilde{\gamma_i}\widetilde{\gamma_j}}-\braket{\widetilde{\gamma_i}}\braket{\widetilde{\gamma_j}} 
  \Big)=\hbar^{-2}{\rm cov}(\widetilde{\gamma_i}, \widetilde{\gamma_j})\,,
  \end{split}
  \label{QGTCovariance}
\end{align}
where we have given expressions with both $\gamma$ and $\widetilde{\gamma}$ for completeness. 
\end{proposition}
\paragraph{Proof:} The result follows simply by applying the definition of the 
connection \eqref{FullUnitaryConnection} into the QGT as follows
\begin{align*}
  Q_{ij}&= \langle\partial_i\psi|\partial_j\psi\rangle - 
  \braket{\partial_i\psi|\psi}\braket{\psi|\partial_j\psi}\\
  &=  \langle\gamma_i\psi|\gamma_j\psi\rangle - 
  \braket{\gamma_i\psi|\psi}\braket{\psi|\gamma_j\psi}\\
  &=  \langle\psi|\gamma_i^{\dagger}\gamma_j\psi\rangle - 
  \braket{\psi|\gamma_i^{\dagger}\psi}\braket{\psi|\gamma_j\psi}\\ 
    &=   \braket{\psi|\gamma_i\psi}\braket{\psi|\gamma_j\psi}-\langle\psi|\gamma_i\gamma_j\psi\rangle \,.
\end{align*}
Then, upon introducing the Hermitian counterpart \eqref{GammaTildeDefn} one obtains 
the equation \eqref{QGTCovariance}. 

\hfill$\square$\\

Hence, recognising the 
Definition \ref{Def:Cov} for the covariance in the second line, we see that the introduction of the gauge connection allows for the interpretation of the QGT as the covariance of the operators $\widetilde{\gamma_i}$ 
and $\widetilde{\gamma_j}$ (up to a factor of $\hbar^{-2}$). It follows that the 
diagonal elements of the QGT (which must be real as $Q$ is Hermitian) are given 
by
\begin{align*}
  {\rm no \,\,sum}\qquad Q_{jj} = \braket{\gamma_j}^2 - \braket{\gamma_j^2}= \hbar^{-2}\Big(\braket{\widetilde{\gamma_j}^2}-\braket{\widetilde{\gamma_j}}^2 
\Big)=\hbar^{-2}\sigma_{\widetilde{\gamma_j}}^2\,,
\end{align*}
which naturally returns the variance of 
$\widetilde{\gamma}_j$ (again up to the $\hbar^{-2}$ factor). This tells us some new information: not only are the diagonal elements of the QGT real, they must also be positive.

Now that the QGT can be understood as a covariance when expressed in terms of the connection,  we recall that the metric tensor (Fubini-Study) and Berry curvature  
are given by the real and imaginary parts of the QGT respectively, and hence ask what 
  form these will take in terms of the new $\mathfrak{u}(\mathscr{H})$ connection.
  
  \begin{corollary}[Connection expression for the metric tensor and Berry curvature]\label{Cor:QGTOffshoots}
    \hfill \\
    The Fubini-Study metric tensor and Berry curvature can be derived from the real and imaginary parts 
    of the QGT as follows
      \begin{align*}
    T_{ij}&= {\rm Re}(Q_{ij}) = \langle\partial_i\psi,\partial_j\psi\rangle - \hbar^{-2}{\cal A}_i 
    {\cal A}_j\,,\\
     {\cal B}_{ij}&= 2\hbar{\rm Im}(Q_{ij}) = \partial_i {\cal A}_j - \partial_j {\cal A}_i\,.
  \end{align*}
  When expressed in terms of the connection $\gamma$ defined by 
  \eqref{FullUnitaryConnection}, these objects can be re-expressed in the 
  following forms
  \begin{align}
  T_{ij}&= \braket{\gamma_i}\braket{\gamma_j}-\frac{1}{2}\braket{\{\gamma_i,\gamma_j\}}
    = \hbar^{-2}\Big(\frac{1}{2}\braket{\{\widetilde{\gamma_i},\widetilde{\gamma_j}\}}-\braket{\widetilde{\gamma_i}}\braket{\widetilde{\gamma_j}}\Big)   
    \label{MetricTensorGammaTilde}\,,\\
{\cal B}_{ij} &= i\hbar\braket{[\gamma_i,\gamma_j]}
= - 
\frac{i}{\hbar}\braket{[\widetilde{\gamma_i},\widetilde{\gamma_j}]}\label{BerryCurvGammaTilde}\,.
\end{align}
  \end{corollary}

The knowledge that the quantum geometric tensor represents a covariance of 
Hermitian operators dates back to its discovery in \cite{PrVa1980}. 
This feature has subsequently reappeared \cite{VenutiZanardi2007}, with several works even constructing a skew-Hermitian operator analogous to \eqref{FullUnitaryConnection}
and obtaining results analogous to  Proposition \ref{Prop:QGTCov} and Corollary \ref{Cor:QGTOffshoots} \cite{CarolloValentiSpagnolo2020,ZanardiEtAl2007,KoSeMePo2017}.
Despite such developments, these previous works use such constructions purely as a tool to probe the structure of the QGT and do not consider the geometric significance of these operators.  
The work presented in this section not only independently arrives at analogous results, but crucially provides a new interpretation, identifying the fundamental geometric object underlying such expressions as a $\mathfrak{u}(\mathscr{H})$-valued connection 1-form given by \eqref{FullUnitaryConnection}. 
We now take the existing ideas further by using the Hermitian counterpart of the connection to consider {uncertainty relations that can} be derived from Proposition \ref{Prop:QGTCov}.

These last two equations \eqref{MetricTensorGammaTilde} and \eqref{BerryCurvGammaTilde} take a remarkably similar form to the uncertainty relations given in Section \ref{Sec:1-Postulates}.
When delving into the details, this actually comes as no suprise as both the defintions of $T$ and ${\cal B}$ and the derivation of the uncertainty relation involves separating a complex expression 
for the covariance of Hermitian operators into real and imaginary parts. 
Hence we now simply specialise the general relations in Theorem \ref{Theorem:SchrUR} and Corollary \ref{Cor:RobertsonUR} to provide novel uncertainty relations for the QGT.

\begin{proposition}[Schr\"odinger uncertainty relation for the QGT]\label{Theorem:QGTUR}
  \hfill \\
  Following the expression of QGT in terms of the $\mathfrak{u}(\mathscr{H})$ connection 
  as in Proposition \ref{Prop:QGTCov}, the Schr\"odinger uncertainty relation 
  \eqref{SchrodingerUR} results in the equivalent expressions (all which have no sum over the repeated indices):
\begin{align}   
\begin{split}   
 \sigma_{\widetilde{\gamma_i}}^2\sigma_{\widetilde{\gamma_j}}^2 &\geq \left|{\frac {1}{2}}\langle \{{ {\widetilde{\gamma_i}}},{ {\widetilde{\gamma_j}}}\}\rangle -\langle { {\widetilde{\gamma_i}}}\rangle \langle { {\widetilde{\gamma_j}}}\rangle \right|^{2}+\left|{\frac {1}{2i}}\langle [{ {\widetilde{\gamma_i}}},{ {\widetilde{\gamma_j}}}]\rangle 
    \right|^{2} \,,\\
{T_{ii}T_{jj}} &\geq 
   |T_{ij}|^2 + \left|\frac{1}{2\hbar}{\cal B}_{ij} \right|^2  \,,\\
   Q_{ii}Q_{jj}&\geq|Q_{ij}|^2\,.
\end{split} \label{SchrodingerQRelations}
\end{align}
\end{proposition}

\begin{corollary}[Robertson uncertainty relation for the QGT]\label{Cor:QGTUR}
  \hfill \\
  Analogous to the general case in which Corollary \ref{Cor:RobertsonUR} follows 
  as a weaker version of Theorem \ref{Theorem:SchrUR}, we can write the corresponding 
  weaker uncertainty relation for the QGT from Theorem \ref{Theorem:QGTUR} as (again with no summation over the repeated indices)
  \begin{align}
 \begin{split}
 \sigma_{\widetilde{\gamma_i}}\sigma_{\widetilde{\gamma_j}} &\geq 
  \frac{1}{2}\left|\braket{[\widetilde{\gamma_i},\widetilde{\gamma_j}]}\right|\,,\\
  \hbar\sqrt{T_{ii}T_{jj}}&\geq\frac{1}{2}|{\cal B}_{ij}|\,,\\
\sqrt{Q_{ii}Q_{jj}} &\geq    |{\rm Im}(Q_{ij})|\,.
\end{split}\label{RobertsonQRelations}
\end{align}
\end{corollary}

To the best of the author's knowledge such constraints on the QGT have not been seen before. However, the exact geometric interpretation of these uncertainty relations and their implications for the QGT 
remains an open question and is of future research interest. Before closing this 
section, we briefly remark on one further interesting possibility that arises 
from the introduction of the connection \eqref{FullUnitaryConnection}.

\begin{remark}[Spacetime formalism]\label{Rem:Spacetime}
  \hfill \\
The gauge connection \eqref{FullUnitaryConnection} 
 derives from the unitary evolution \eqref{ParametricUnitaryEvo} of a parametrically dependent quantum state. 
Throughout this thesis this unitary evolution has been used to derive a 
reduced description for the dynamics of quantum systems via the equation
\begin{align}
  \partial_t\psi(\bx,t)=\xi(\bx,t)\psi(\bx,t)\,,\qquad \text{with}\quad \xi:=\partial_t U\,U^{-1} \in \mathcal{F}(\mathbb{R}^3, 
  \mathfrak{u}(\mathscr{H}))\,.\label{generalunitarygenerator}
\end{align}
A further interesting possibility consists of unifying the gauge 
connection $\gamma$ with the generator $\xi$ into four-dimensional spacetime 
connection $\gamma_{\mu}=(\xi, \bgamma)$. Pursuing this idea naturally leads to 
four-dimensional counterparts of the Berry connection and curvature, analogous 
to the covariant formulation of electromagnetism, whilst for the non-Abelian 
connection natural links with Yang-Mills theories in physics emerge. These ideas are not relevant to the discussion of the remainder of this thesis and hence preliminary findings are reported in Appendix \ref{App:Spacetime}.   
There potentially remain further features within this approach which also could 
form interesting avenues for future work.
\end{remark}

This section has introduced a $\mathfrak{u}(\mathscr{H})$ connection for general quantum systems that depend on an external parameter. 
It has been shown to encapsulate both the celebrated Berry connection and Wilczek-Zee connections as well as 
give a new geometric foundation for the QGT. To conclude this chapter, we
shall consider the application of such a connection to the exact factorization 
model of quantum chemistry.


\section{Application to nonadiabatic dynamics}\label{Sec:GammaEF}

In this final section, we consider the application of the non-Abelian connection 
introduced in Section \ref{Sec:NonAbelianQuantumSystems} to nonadiabatic quantum chemistry. 
We continue from the exact factorization approach with electronic dynamics expressed in the nuclear hydrodynamic frame, hence expanding upon the material presented in Section \ref{NuclearFrameSection}. 

Recall the EF of the molecular wavefunction \eqref{GenExFact} 
$\Psi(t)=\Omega(\br,t)\ket{\psi(\br,t)} \in L^2(\mathbb{R}^3)\otimes\mathcal{F}(\mathbb{R}^3, \mathscr{H}_e)$ with partial normalisation condition (PNC) $\|\psi(\br,t)\|^2=1$. 
The PNC must be valid for all time and hence the 
electronic factor was allowed to evolve under unitary transformations via 
$\psi(t)=(U\psi_0)\circ\eta^{-1}$ as given in equation \eqref{unitary-evol}, with $\eta$ the nuclear Lagrangian path such that $\dot\boldeta=\bu(\boldeta,t)$. 
This then allowed for the introduction of the generator \eqref{psitimeevo}
\begin{align*}
 \partial_t\psi +\bu\cdot\nabla\psi&= \xi\psi\,,
\end{align*}
where $\xi := (\partial_t{U}\,U^{-1})\circ\eta^{-1} \in \mathcal{F}(\mathbb{R}^3, \mathfrak{u}(\mathscr{H}_e))$. 
Clearly, the electronic dynamics are exactly of the type described in Section \ref{Sec:NonAbelianQuantumSystems} and hence we can immediately apply the relevant results. 
Specifically, we introduce a non-Abelian electronic connection via
\begin{align}
  \partial_j{\psi(\br,t)}=-\gamma_j(\br,t){\psi(\br,t)}\,,\label{EFGamma}
\end{align} 
where $\bgamma\cdot\de \br = \gamma_j\,\de r^j\in \Lambda^1(\mathbb{R}^3, 
\mathfrak{u}(\mathscr{H}_e))$ and $\gamma_j = (-\partial_j U\,U^{-1}+U\gamma_j^0 \,U^{-1})\circ\eta^{-1}\in \mathcal{F}(\mathbb{R}^3, \mathfrak{u}(\mathscr{H}_e))$.
In addition recall that in Chapter \ref{Chap:EFNEW} the electronic dynamics of the EF system were expressed in terms of  
the pure state density matrix via 
$\rho=\psi\psi^{\dagger}$. In this section, we again utilise this approach and 
note that \eqref{EFGamma} becomes the relation $\partial_j\rho=[\rho,\gamma_j]$. {Interestingly, we note that a similar operator has previously been introduced in the study of geometric phases in molecular dynamics \cite{Mead1992}. However, the geometric nature of this operator as a connection on a principal bundle had previously not been fully 
realised.}

Before continuing, we also recall Theorem \ref{Theorem:EFEPFull}. In particular, we rearrange the Lagrangian \eqref{EFHydroL3} 
to read
    \begin{align}
  \ell(\bu, D, \xi, \rho) =  \int \bigg[\frac{1}{2}MD |\bu|^2-\frac{\hbar^2}{8M}\frac{(\nabla 
  D)^2}{D} + D\braket{\rho,i\hbar\xi-\widehat{H}_e} \bigg]\,\de^3 r- G(D,\rho) \,,\label{Sertse}
\end{align}
where we have introduced the functional $G:{\rm Den}(\mathbb{R}^3)\times\mathcal{F}(\mathbb{R}^3,{\rm Her}(\mathscr{H}_e))\to \mathbb{R}$ 
given by 
\begin{align}
  G(D, \rho):=\frac{\hbar^2}{4M} \int D\|\nabla\rho\|^2\, 
  \de^3r\,,\label{Gfunct}
\end{align}
and 
$\|\nabla\rho\|^2=\braket{\partial_j\rho|\partial_j\rho}$. In this form, we notice how it is only the presence of this final term $G(D,\rho)$
that the EF system differs formally from mean field type Lagrangians as seen in Section \ref{meanfield-sec}.
In addition, it is clear that the connection \eqref{EFGamma} will only manifest through this functional and hence we simply apply the result of Corollary \ref{Cor:QGTOffshoots} 
to present the new result:

\begin{proposition}[Correlation functional]\label{Prop:GammaG}
  \hfill \\
Introducing the gauge connection \eqref{EFGamma} takes the functional $G(D, \rho)$ defined by equation \eqref{Gfunct} into the functional
$\bar{G}:{\rm Den}(\mathbb{R}^3)\times\mathcal{F}(\mathbb{R}^3,{\rm Her}(\mathscr{H}_e))\times\Lambda^1(\mathbb{R}^3, \mathfrak{u}(\mathscr{H}_e))\to \mathbb{R}$ given by
\begin{align}
\bar{G}(D, \rho, \gamma_j)
  &=  \frac{\hbar^2}{2M}\int D
\Big(\braket{\gamma_j}^2 - \braket{\gamma_j^2}\Big)\,\de^3r\,.\label{GfunctGamma}
\end{align}
Here, we have used that $\rho=\psi\psi^{\dagger}$ and adopted the notation $\braket{\widehat{A}}\equiv{\rm Tr}(\rho\widehat{A})$ for the expectation.
\end{proposition}
  \paragraph{Proof:} The proof simply relies of the introduction of the 
  connection \eqref{EFGamma} and using the pure state property of the electronic 
  density matrix. A direct computation for the last term yields
  \begin{align*}
    \|\nabla\rho\|^2&=\braket{\partial_j\rho|\partial_j\rho}\\
    &= \braket{[\rho,\gamma_j]|[\rho,\gamma_j]}\\
    &=     \text{Tr}\Big((\rho\gamma_j-\gamma_j\rho)^{\dagger}(\rho\gamma_j-\gamma_j\rho)\Big)\\
    &=   \text{Tr}\Big(\rho\gamma_j\rho\gamma_j-\gamma_j\rho^2\gamma_j -\rho\gamma_j^2\rho+\gamma_j\rho\gamma_j\rho\Big)\\
    &=   2\text{Tr}\Big(\rho\gamma_j\rho\gamma_j-\gamma_j^2\rho^2\Big)\\
    &=   2\text{Tr}\Big(\psi\psi^{\dagger}\gamma_j\psi\psi^{\dagger}\gamma_j-\gamma_j^2\psi\psi^{\dagger}\psi\psi^{\dagger}\Big)\\
        &=     
        2\Big(\braket{\psi|\gamma_j\psi}\braket{\psi|\gamma_j\psi}-\braket{\psi|\gamma_j^2\psi}\Big)\,.
\end{align*}
  \hfill$\square$\\

\begin{remark}[Connection as a correlation observable]\label{Rem:CorObs}
  \hfill \\
  Following from Proposition \ref{Prop:GammaG}, we introduce the Hermitian 
  counterpart to the connection, $\widetilde{\gamma}_j:=-i\hbar\gamma_j$ so that 
  the effective electronic potential \eqref{EFepsilonDEF} reads
  \begin{align*}
  \epsilon(\psi,\widetilde\gamma)
  &=\braket{\widehat{H}_e}  + 
  \frac{\braket{\widetilde{\gamma}_j^2} - \braket{\widetilde{\gamma}_j}^2}{2M} = \braket{\widehat{H}_e}  +\frac{1}{2M}\sum_{j=1}^3\sigma_{\widetilde{\gamma}_j}^2\,,
\end{align*}
recalling the notation for the variance from Definition \ref{Def:Variance}.

Hence, if the variances of the connection $\widetilde\gamma$ are zero, the 
Lagrangian \eqref{Sertse} formally coincides with its mean-field counterpart. 
We therefore propose the interpretation of the Hermitian operator $\widetilde{\gamma}$ as a {\rm correlation 
observable}, which we hope may prove to be a usual measure of the strength of electron-nuclear 
interactions in nonadiabatic quantum chemistry. Analogously we will call $\bar{G}(D, \rho, \gamma_j)$ a {\rm correlation functional}.  
\end{remark}

Having seen how the $\mathfrak{u}(\mathscr{H}_e)$ connection provides an 
 new method for analysing electron-nuclear coupling, we formulate the dynamics of the exact 
 factorization system including the connection as an additional dynamical 
 variable. Following the analogous approach to Section \ref{Sec:NonAbelianPauliConnection} we commence considering an arbitrary functional $\bar{G}$, before specialising to the physical case to complete this section.
 
 \begin{proposition}[Euler-Poincar{\'e} formulation of EF dynamics with connection]\label{Prop:EFwithGamma}
   \hfill \\
Recall the reduced Lagrangian \eqref{EFHydroL3} from Theorem 
\ref{Theorem:EFEPFull}. Introducing the $\mathfrak{u}(\mathscr{H}_e)$ connection \eqref{EFGamma} takes this Lagrangian into the mapping 
  $\ell:\mathfrak{X}(\mathbb{R}^3)\times{\rm Den}(\mathbb{R}^3)\times\mathcal{F}(\mathbb{R}^3,\mathfrak{u}(\mathscr{H}_e))\times\mathcal{F}(\mathbb{R}^3,{\rm Her}(\mathscr{H}_e))\times \Lambda^1(\mathbb{R}^3, \mathfrak{u}(\mathscr{H}_e))\to\mathbb{R}$ 
 given by
    \begin{align}
  \ell(\bu, D, \xi, \rho, \gamma_j) =  \int \bigg[\frac{1}{2}MD |\bu|^2-\frac{\hbar^2}{8M}\frac{(\nabla 
  D)^2}{D} + D\braket{\rho,i\hbar\xi-\widehat{H}_e}
\bigg]\de^3r - \bar{G}(D, \rho, \gamma_j)
 \,,\label{EFHydroL4}
\end{align}
where $\bar{G}$ is considered an arbitrary correlation functional $\bar{G}:{\rm Den}(\mathbb{R}^3)\times\mathcal{F}(\mathbb{R}^3,{\rm Her}(\mathscr{H}_e))\times\Lambda^1(\mathbb{R}^3, \mathfrak{u}(\mathscr{H}_e))\to 
\mathbb{R}$.
Then, applying Hamilton's principle $0=\delta\int_{t_1}^{t_2} \ell(\bu, D, \xi, \rho,\gamma_j)\,\de t$ for constrained Euler-Poincar\'e variations \eqref{EPvar} and
\begin{align}
\begin{split}
  \delta \xi &= \partial_t \Upsilon -\bw\cdot\nabla\xi + \bu\cdot\nabla\Upsilon - [\xi,\Upsilon]\,, \\ 
  \delta \rho &= [\Upsilon,\rho]-\bw \cdot\nabla\rho \,,\\
  \delta \gamma_j &= -\partial_j\Upsilon +[\Upsilon, \gamma_j]-(\bw\cdot\nabla)\gamma_j - \gamma_i\partial_jw_i 
\end{split}\label{EFGammavariations}
\end{align}
where $\Upsilon := (\delta U) U^{-1}\circ\eta^{-1}$ and $\bw := \delta\eta\,\circ\,\eta^{-1}$ are arbitrary, returns the system of equations
\begin{align}
    \begin{split}
    M(\partial_t+\bu\cdot\nabla)u_i&=-\partial_i \left(V_Q+\frac{\delta \bar{G}}{\delta D}\right)-\braket{\rho,(\partial_i\widehat{H}_e)}  \\
    &\qquad+ \frac{1}{D}\left(\left\langle \partial_i\rho, \frac{\delta \bar{G}}{\delta \rho} \right\rangle+ \left\langle \partial_i\gamma_j, \frac{\delta \bar{G}}{\delta \gamma_j} \right\rangle-\partial_j\left\langle \gamma_i,\frac{\delta \bar{G}}{\delta 
    \gamma_{j}}
    \right\rangle\right) \,, 
    \end{split}\label{Yeet2}\\
    \left[i\hbar\xi - \widehat{H}_e - \frac{1}{D}\frac{\delta \bar{G}}{\delta \rho},\rho\right] &= \frac{1}{D}\partial_j^{\gamma}\frac{\delta \bar{G}}{\delta \gamma_{j}} \label{xiequationwithgamma}\,,\\
    \partial_t D+{\rm div}(D\bu)&=0\,,\\
  (\partial_t+\bu\cdot\nabla)\rho &= [\xi,\rho] \,, \\
\partial_t\gamma_j + u_i\partial_i\gamma_j + \gamma_i\partial_ju_i  &= -\partial_j^{\gamma}\xi \label{Bros}\,,
\end{align}
where we recall the covariant derivative notation $\partial_j^{\gamma}\,\cdot\, = \partial_j\,\cdot\, + \,[\gamma_j, \,\cdot\,]\,.$
 \end{proposition}

\paragraph{Proof:} Once again, see \cite{GayBalmazRatiu2009,GayBalmazTronci2010} 
for complete derivations of the variations \eqref{EFGammavariations}.
Then, expanding Hamilton's principle and recalling that $\Upsilon$ and $\bw$ are arbitrary yields the general equations
\begin{align*}
  &\left[(\partial_t + \pounds_{\bu})\frac{\delta \ell}{\delta \bu}\right]_i = 
  D\partial_i \frac{\delta \ell}{\delta D} - \left\langle\partial_i\xi,\frac{\delta \ell}{\delta \xi}\right\rangle- \left\langle\partial_i\rho,\frac{\delta \ell}{\delta \rho}\right\rangle- \left\langle\partial_i\gamma_j,\frac{\delta \ell}{\delta \gamma_j} 
  \right\rangle
 + \partial_j\left\langle\gamma_i,\frac{\delta \ell}{\delta 
  \gamma_j}\right\rangle\,,\\
 & (\partial_t + \pounds_{\bu})\frac{\delta \ell}{\delta \xi} - \left[\xi,\frac{\delta \ell}{\delta 
  \xi}\right]= \left[\frac{\delta \ell}{\delta 
  \rho},\rho\right] + \partial_j^{\gamma}\frac{\delta \ell}{\delta \gamma_j} \,,
\end{align*}
along with the auxilliary equations $  (\partial_t + \pounds_{\bu})D=0,\,\,
  (\partial_t + \pounds_{\bu})\rho = [\xi,\rho],\,\,
  (\partial_t + \pounds_{\bu})(\gamma_j\de x^j) = -\de^{\gamma}\xi$. Then, we compute the variational 
  derivatives of the Lagrangian \eqref{EFHydroL4}
\begin{align*}
  \begin{split}
  \frac{\delta \ell}{\delta \bu} = MD\bu\,,&\qquad   \frac{\delta \ell}{\delta D} 
  =\frac{M}{2}|\bu|^2-  V_Q +\braket{\rho,i\hbar\xi-\widehat{H}_e} - \frac{\delta \bar{G}}{\delta D}\,,\qquad   \frac{\delta \ell}{\delta \xi} = 
-i\hbar D\rho\,,\\
    &\frac{\delta \ell}{\delta \rho} = i\hbar D\xi -   D\widehat{H}_e-\frac{\delta \bar{G}}{\delta 
    \rho}\,,\qquad  \frac{\delta \ell}{\delta \gamma_j} = -\frac{\delta \bar{G} }{\delta 
    \gamma_j}\,,
  \end{split}
\end{align*}
which specialises the equations for the fluid momentum and generator 
accordingly.
\hfill$\square$\\

Naturally, this system is the generalisation of the two-level dynamics given in Proposition \ref{Prop:EP2LevelGamma}.
Hence, the remainder of this section closely follows the material and generalises the results 
of Section \ref{Sec:NonAbelianPauliConnection}.
Firstly, we must again find a solution to the generator equation \eqref{xiequationwithgamma} in order to close the system. 

\begin{lemma}\label{Lemma:GammaXiSolution}
  Equation \ref{xiequationwithgamma} possesses the solution for the generator
  \begin{align}
   i\hbar \xi = \widehat{H}_e + \frac{1}{D}\frac{\delta \bar{G}}{\delta \rho} + \frac{1}{D}\left[\partial_j^{\gamma}\frac{\delta \bar{G} }{\delta \gamma_j}, \rho\right] + \beta\rho\,,\label{Boris2}
  \end{align}
  where $\beta$ is an arbitrary real parameter, provided the correlation functional satisfies the compatibility condition
  \begin{align}
\partial_j^{\gamma}\frac{\delta \bar{G} }{\delta \bgamma_j}=\left\{\partial_j^{\gamma}\frac{\delta \bar{G} }{\delta \bgamma_j},\rho\right\}\,.\label{Johnson2}
  \end{align} 
  Here, we have employed the notation $\{\widehat{A},\widehat{B}\}=\widehat{A}\widehat{B}+\widehat{B}\widehat{A}$ 
  for the anticommutator of operators.
\end{lemma}
\paragraph{Proof:}
One can easily see that the solution \eqref{Boris2} generalises the 
corresponding two-level solution \eqref{Boris}. We verify this solution 
explicitly as follows.
\begin{align*}
      \left[i\hbar\xi - \widehat{H}_e - \frac{1}{D}\frac{\delta \bar{G}}{\delta \rho},\rho\right] &= \left[\widehat{H}_e + \frac{1}{D}\frac{\delta \bar{G}}{\delta \rho} + \frac{1}{D}\left[\partial_j^{\gamma}\frac{\delta \bar{G} }{\delta \gamma_j}, \rho\right] + \beta\rho- \widehat{H}_e - \frac{1}{D}\frac{\delta \bar{G}}{\delta \rho},\rho\right]\\
&=\frac{1}{D}\left[ \left[\partial_j^{\gamma}\frac{\delta \bar{G} }{\delta \gamma_j}, \rho\right] ,\rho\right]\\
&= \frac{1}{D}\left[\partial_j^{\gamma}\frac{\delta \bar{G} }{\delta \gamma_j}\psi\psi^{\dagger}-\psi\psi^{\dagger}\partial_j^{\gamma}\frac{\delta \bar{G} }{\delta \gamma_j},\psi\psi^{\dagger}\right]\\
&= \frac{1}{D}\left(\partial_j^{\gamma}\frac{\delta \bar{G} }{\delta \gamma_j}\psi\psi^{\dagger} - 2\psi\psi^{\dagger}\partial_j^{\gamma}\frac{\delta \bar{G} }{\delta \gamma_j}\psi\psi^{\dagger}+ \psi\psi^{\dagger}\partial_j^{\gamma}\frac{\delta \bar{G} }{\delta 
\gamma_j}\right)\\
&= \frac{1}{D}\left(\left\{\partial_j^{\gamma}\frac{\delta \bar{G} }{\delta \gamma_j},\psi\psi^{\dagger}   \right\} - 2\left\langle\psi\bigg|\partial_j^{\gamma}\frac{\delta \bar{G} }{\delta \gamma_j}\bigg|\psi 
\right\rangle\psi\psi^{\dagger}\right)\\
&=\frac{1}{D}\left(\left\{\partial_j^{\gamma}\frac{\delta \bar{G} }{\delta \gamma_j},\rho   \right\} - 2\left\langle\rho\bigg|\partial_j^{\gamma}\frac{\delta \bar{G} }{\delta 
\gamma_j}\right\rangle\rho\right)\,,
\end{align*}
where we have used the pure state expression $\rho=\psi\psi^{\dagger}$. At this 
point, we notice that we have the important relations
\begin{align*}
  \text{Tr}\left(\partial_j\frac{\delta \bar{G} }{\delta \gamma_j}\right) = \left\langle\rho\bigg|\partial_j^{\gamma}\frac{\delta \bar{G} }{\delta 
\gamma_j}\right\rangle=0\,,
\end{align*}
which follow from taking the traces of the generator equation \eqref{xiequationwithgamma} 
as well as \eqref{xiequationwithgamma} after multiplying by $\rho$. Hence, \eqref{Boris2} 
is indeed a solution to \eqref{xiequationwithgamma} provided the compatibility 
condition \eqref{Johnson2} holds.
\hfill$\square$\\

Having found a solution for the generator, we can now write the closed system of equations in the variables $(D, \bu, \rho, 
\gamma_j)$.
\begin{corollary} The system of equations \eqref{Yeet2}-\eqref{Bros} 
given in Proposition \ref{Prop:EFwithGamma} are closed by the generator solution 
\eqref{Boris2}, resulting in the closed set of equations (choosing $\beta=0$)
\begin{align}
  \partial_t D + {\rm div}(D\bu)&= 0\,,\\
   \begin{split}
    MD(\partial_t+\bu\cdot\nabla)u_i&=-D\partial_i \left(V_Q+\frac{\delta \bar{G}}{\delta D}\right)-D\braket{\rho,(\partial_i\widehat{H}_e)}
    \\    &\qquad\qquad
    + \left\langle \partial_i\rho, \frac{\delta \bar{G}}{\delta \rho} \right\rangle+ \left\langle \Omega_{ij}, \frac{\delta \bar{G}}{\delta \gamma_j} \right\rangle-\left\langle \gamma_i,\partial^{\gamma}_j\frac{\delta \bar{G}}{\delta 
    \gamma_{j}}
    \right\rangle \label{GAMMAEFFLUIDEQUATION}\,, 
    \end{split}\\ 
  i\hbar(\partial_t+ \bu\cdot\nabla)\rho &= \left[\widehat{H}_e + \frac{1}{D}\frac{\delta \bar{G}}{\delta \rho} , \rho\right]+ \frac{1}{D}\partial_j^{\gamma}\frac{\delta \bar{G} }{\delta \gamma_j}\,,\label{Erica}\\
i\hbar(\partial_t\gamma_j + u_i\partial_i\gamma_j + \gamma_i\partial_ju_i )&= -\partial_j^{\gamma}\left(\widehat{H}_e + \frac{1}{D}\frac{\delta \bar{G}}{\delta \rho} + \frac{1}{D}\left[\partial_j^{\gamma}\frac{\delta \bar{G} }{\delta \gamma_j}, \rho\right] \right)
  \,.\label{Laura}
\end{align}
where $\Omega_{ij}:=\partial_i\gamma_j - \partial_j\gamma_i + 
[\gamma_i,\gamma_j]$ is the curvature of the $\mathfrak{u}(\mathscr{H}_e)$ connection \eqref{EFGamma}.
\end{corollary}
\paragraph{Proof:} Analogous to the two-level system of Section \ref{Sec:NonAbelianPauliConnection}, the final two terms of the fluid equation \eqref{Yeet} are rewritten 
as follows
\begin{align*}
\left\langle \partial_i\gamma_j, \frac{\delta \bar{G}}{\delta \gamma_j} \right\rangle-\partial_j\left\langle \gamma_i,\frac{\delta \bar{G}}{\delta 
    \gamma_{j}}
    \right\rangle&= \left\langle \partial_i\gamma_j - \partial_j\gamma_i,  \frac{\delta \bar{G}}{\delta \gamma_j}\right\rangle 
    - \left\langle \gamma_i, \partial_j\frac{\delta \bar{G}}{\delta \gamma_j} 
    \right\rangle\\
    &= \left\langle \Omega_{ij}-[\gamma_i,\gamma_j],  \frac{\delta \bar{G}}{\delta \gamma_j}\right\rangle 
    - \left\langle \gamma_i, \partial_j\frac{\delta \bar{G}}{\delta \gamma_j} 
    \right\rangle\\
        &= \left\langle \Omega_{ij}, \frac{\delta \bar{G}}{\delta \gamma_j}\right\rangle 
    - \left\langle \gamma_i, \partial_j\frac{\delta \bar{G}}{\delta \gamma_j} + \left[\gamma_j,\frac{\delta \bar{G}}{\delta \gamma_j}\right]
    \right\rangle
\end{align*}
where in the final line we have used the definition of the pairing $\braket{A,B}={\rm Re}({\rm Tr}(A^{\dagger}B))$ to rearrange the commutator term.
Then, the generator solution (with $\beta = 0$) \eqref{Boris2} is simply 
substituted into the connection equation \eqref{Bros} to complete the 
set of equations.
\hfill$\square$\\

We once again emphasise that in the above equations, no assumption has been made on the 
curvature of the connection. Following the results of Section \ref{Sec:GeneralTheory} the above system may allow for non-trivial curvature of $\gamma$ and consequently non-zero holonomy. 
Analogous to Section \ref{Sec:NonAbelianPauliConnection} we consider the related quantity, the circulation of the connection along the flow \cite{Holm2002,GayBalmazRatiu2009}.

\begin{proposition}[Circulation theorem for the connection]\label{Charizard2}
  \hfill \\ Recall the reduced system of equations 
  presented in Proposition \ref{Prop:EFwithGamma}. Then, the time evolution of the circulation of the connection \eqref{EFGamma} is given by
  \begin{align}
    \begin{split}
    \frac{\de}{\de t}\oint_{c(t)}\gamma_i\,\de r^i &= \oint_{c(t)} 
    [\xi,\gamma_i ]\,\de r^i\\
    &= \oint_{c(t)} 
    \left[-\frac{i}{\hbar}\left(\widehat{H}_e + \frac{1}{D}\frac{\delta \bar{G}}{\delta \rho} + \frac{1}{D}\left[\partial_j^{\gamma}\frac{\delta \bar{G} }{\delta \gamma_j}, \rho\right]\right),\gamma_i \right]\,\de r^i\,,
    \end{split}\label{Ditto2}
  \end{align}
where we have again selected the $\beta=0$ solution of the generator equation \eqref{Boris2} and $c(t)$ is a loop moving with the nuclear fluid flow $\bu$.
\end{proposition}
\paragraph{Proof:}
As for the two-level analog, this proposition can be easily proven upon considering the auxilliary equation for the evolution of the connection. 
Simply taking the integral of $(\partial_t+\pounds_{\bu})(\gamma_j\,\de x^j) = -\de\xi + [\xi,\gamma_j]\,\de x^j$ around the loop $c(t)$ moving with the fluid flow  
and passing the Lagrangian time derivative outside of the integral (as explained in Theorem \ref{Abby}) returns the first equality. The second equality follows from the insertion of the generator solution \eqref{Boris2} (with $\beta =0$) .
\hfill $\square$\\

Once again, due to the Euler-Poincar\'e derivation of the fluid equation \eqref{GAMMAEFFLUIDEQUATION}, we can write the associated Kelvin-Noether circulation theorem, depending explicitly on the non-zero curvature of the connection.
\begin{theorem}[Kelvin-Noether circulation theorem]
\hfill \\
The fluid equation possesses the Kelvin-Noether circulation theorem
\begin{align}
\frac{\de}{\de t}\oint_{c(t)}\boldsymbol{\cal A}\cdot\de \br = 
-\oint_{c(t)}\left(\braket{\rho, \partial_i\widehat{H}_e} + \frac{1}{D}\left(\left\langle \gamma_i,\partial^{\gamma}_j\frac{\delta \bar{G}}{\delta 
    \gamma_{j}}
    \right\rangle-\left\langle \partial_i\rho, \frac{\delta \bar{G}}{\delta \rho} \right\rangle- \left\langle \Omega_{ij}, \frac{\delta \bar{G}}{\delta \gamma_j} \right\rangle\right)\right)\,\de 
r^i\,,
\end{align}
for a loop $c(t)$ moving with the fluid flow $\bu$.
\end{theorem}
\paragraph{Proof:} To begin we rewrite \eqref{GAMMAEFFLUIDEQUATION} in terms of the Lie 
derivative as
\begin{align*}
      M(\partial_t+\pounds_{\bu})(\bu\cdot\de \br)  &= -\de\left(V_Q+\frac{M}{2}|\bu|^2+\frac{\delta \bar{G} }{\delta 
  D}\right) - \braket{\rho,\de \widehat{H}_e}\\
  &\qquad\qquad-\frac{1}{D}\left(\left\langle \gamma_i,\partial^{\gamma}_j\frac{\delta \bar{G}}{\delta 
    \gamma_{j}}
    \right\rangle-\left\langle \partial_i\rho, \frac{\delta \bar{G}}{\delta \rho} \right\rangle- \left\langle \Omega_{ij}, \frac{\delta \bar{G}}{\delta \gamma_j} \right\rangle\right)\,\de 
r^i\,.
\end{align*}
Then taking the integral around a loop $c(t)$ moving with the fluid velocity $\bu$ annihilates the exact differentials terms leading to
\begin{align*}
  \oint_{c(t)}M(\partial_t+\pounds_{\bu})(\bu\cdot\de \br) &=  -\oint_{c(t)}\Bigg(\braket{\rho, \partial_i\widehat{H}_e} \\
  &\qquad\qquad+ \frac{1}{D}\left(\left\langle \gamma_i,\partial^{\gamma}_j\frac{\delta \bar{G}}{\delta 
    \gamma_{j}}
    \right\rangle-\left\langle \partial_i\rho, \frac{\delta \bar{G}}{\delta \rho} \right\rangle- \left\langle \Omega_{ij}, \frac{\delta \bar{G}}{\delta \gamma_j} \right\rangle\right)\Bigg)\,\de 
r^i\,.
\end{align*}
We then take the Lagrangian time derivative outside of the integral as explained 
in the proof of Theorem \ref{Abby} so that upon replacing $M\bu\cdot\de\br = \de S + \boldsymbol{\cal A}\cdot\de\br$ 
we have
  \begin{align*}
    \frac{\de}{\de t}\oint_{c(t)} \boldsymbol{\cal A}\cdot\de\br= -\oint_{c(t)}\left(\braket{\rho, \partial_i\widehat{H}_e} + \frac{1}{D}\left(\left\langle \gamma_i,\partial^{\gamma}_j\frac{\delta \bar{G}}{\delta 
    \gamma_{j}}
    \right\rangle-\left\langle \partial_i\rho, \frac{\delta \bar{G}}{\delta \rho} \right\rangle- \left\langle \Omega_{ij}, \frac{\delta \bar{G}}{\delta \gamma_j} \right\rangle\right)\right)\,\de 
r^i\,,
  \end{align*}
as required.
\hfill$\square$\\

This result provides a generalisation of Theorem \ref{Abby} expressing 
the time evolution of the Berry phase along the fluid flow, now with the additional term arising from the curvature of the connection.

Despite the general possibility of incorporating holonomy effects associated to a non-flat connection, we will now specialise to the case of zero curvature, allowing the usage of the relations $\partial_j\rho=[\rho,\gamma_j]$ and $\partial_i\gamma_j - \partial_j\gamma_i + [\gamma_i,\gamma_j]=0$.
In this case, the fluid equation \eqref{GAMMAEFFLUIDEQUATION} can again be rewritten as
\begin{align}
    M(\partial_t+\bu\cdot\nabla)u_i&=-\partial_i \left(V_Q+\frac{\delta \bar{G}}{\delta D}\right)-\braket{\rho,(\partial_i\widehat{H}_e)} + \frac{1}{D}\left\langle \gamma_i, \left[\rho,\frac{\delta \bar{G}}{\delta \rho}\right] -\partial_j^{\gamma}\frac{\delta \bar{G}}{\delta \gamma_j} \right\rangle\,. \label{GAMMAEFFLUIDEQUATIONAGAIN}
\end{align}
The associated Kelvin-Noether theorem is modified as follows.

\begin{theorem}[Kelvin-Noether circulation theorem for a flat connection]
\hfill \\
The fluid equation \eqref{GAMMAEFFLUIDEQUATIONAGAIN} possesses the Kelvin-Noether circulation theorem
\begin{align}
\frac{\de}{\de t}\oint_{c(t)}i\hbar\braket{\rho|\gamma_i} \de r^i = 
-\oint_{c(t)}\left(\braket{\rho, \partial_i\widehat{H}_e} + \frac{1}{D}\left\langle \gamma_i, \left[\frac{\delta \bar{G}}{\delta \rho},\rho\right] +\partial_j^{\gamma}\frac{\delta \bar{G}}{\delta \gamma_j} \right\rangle\right)\,\de 
r^i\,,\label{KNTheoremEFNOCURV}
\end{align}
for a loop $c(t)$ moving with the fluid flow $\bu$.
\end{theorem}
\paragraph{Proof:} To begin we rewrite \eqref{GAMMAEFFLUIDEQUATIONAGAIN} in terms of the Lie 
derivative as
\begin{align*}
      M(\partial_t+\pounds_{\bu})(\bu\cdot\de \br)  &= -\de\left(V_Q+\frac{M}{2}|\bu|^2+\frac{\delta \bar{G} }{\delta 
  D}\right) - \braket{\rho,\de \widehat{H}_e}-\frac{1}{D}\left\langle \gamma_i, \left[\frac{\delta \bar{G}}{\delta \rho},\rho\right]  +\partial_j^{\gamma}\frac{\delta \bar{G}}{\delta \gamma_j} \right\rangle\,\de 
r^i\,.
\end{align*}
Then taking the integral around a loop $c(t)$ moving with the fluid velocity $\bu$ annihilates the exact differentials terms leading to
\begin{align*}
  \oint_{c(t)}M(\partial_t+\pounds_{\bu})(\bu\cdot\de \br) &=   -\oint_{c(t)}\left(\braket{\rho, \partial_i\widehat{H}_e} + \frac{1}{D}\left\langle \gamma_i, \left[\frac{\delta \bar{G}}{\delta \rho},\rho\right]  +\partial_j^{\gamma}\frac{\delta \bar{G}}{\delta \gamma_j} \right\rangle\right)\,\de 
r^i\,.
\end{align*}
We then take the Lagrangian time derivative outside of the integral as explained 
in the proof of Theorem \ref{Abby} so that upon replacing $M\bu\cdot\de\br = \de S + \boldsymbol{\cal A}\cdot\de\br$ 
we have
  \begin{align*}
    \frac{\de}{\de t}\oint_{c(t)} \boldsymbol{\cal A}\cdot\de\br= -\oint_{c(t)}\left(\braket{\rho, \partial_i\widehat{H}_e} + \frac{1}{D}\left\langle \gamma_i, \left[\frac{\delta \bar{G}}{\delta \rho},\rho\right]  +\partial_j^{\gamma}\frac{\delta \bar{G}}{\delta \gamma_j} \right\rangle\right)\,\de 
r^i\,.
\end{align*}
Finally, now that we have restricted to the consider zero curvature, we can make 
use of the relation $\partial_i\psi = -\gamma_i \psi$ and hence rewrite the 
electronic Berry connection as 
${\cal A}_i=\braket{\psi|-i\hbar\partial_i\psi}=i\hbar\braket{\rho|\gamma_i}$.

\hfill$\square$\\

The remainder of this section will make 
use of the explicit form of the correlation functional \eqref{GfunctGamma}. We 
first check whether it satisfies the compatibility condition \eqref{Johnson2}.

\begin{lemma}[Zero curvature connection for exact factorization]\label{Lemma:EFZeroCurv}
  \hfill \\
  The functional \eqref{GfunctGamma}  satisfies the 
  compatibility condition \eqref{Johnson2} provided 
  \begin{align}
    \partial_j\rho = [\rho, \gamma_j]\,.\label{Dakota2}
  \end{align}
As we have seen this equation implies $\Omega_{ij}=\partial_i\gamma_j-\partial_j\gamma_i +[\gamma_i,\gamma_j] 
  =0$, i.e. that the curvature vanishes and the connection is flat.
\end{lemma}
\paragraph{Proof:}
An explicit calculation yields the variational derivative
\begin{align*}
\frac{\delta \bar{G}}{\delta \gamma_j} = \frac{\hbar^2}{2M}D\Big(\{\rho, 
\gamma_j\}-2\braket{\gamma_j}\rho\Big)\,.
\end{align*}
Then, the corresponding covariant divergence
is computed to be
\begin{align*}
  \partial_j^{\gamma} \frac{\delta \bar{G}}{\delta \gamma_j} &= \frac{\hbar^2}{2M}\Big(\{\rho, 
\partial_j(D\gamma_j)\}-2\braket{\partial_j(D\gamma_j)}\rho\Big) 
-\frac{\hbar^2}{M}D \braket{\gamma_j}\Big(\partial_j\rho + [\gamma_j,\rho]\Big) \\
&\qquad\qquad+\frac{\hbar^2}{2M}D \{\partial_j\rho+[\gamma_j, \rho], \gamma_j\}
- \frac{\hbar^2}{M}D\braket{\partial_j\rho|\gamma_j}\rho\,.
\end{align*} 
From here we compute the compatibility condition \eqref{Johnson2} directly. 
Lengthy algebraic manipulations show that the first two terms above satisfy the 
condition directly whilst the third and fourth require the zero curvature relation $\partial_j\rho= [\rho, 
\gamma_j]$.

\hfill $\square$ \\

We are now in a position to write the full system of equations for exact factorization as they arise from the Euler-Poincar{\'e} reduction 
including the dynamics of the flat gauge connection.

\begin{proposition}[Explicit EF system with a non-Abelian gauge connection]\label{Prop:FinalEFwithGamma}
  \hfill \\
  The system of equations that follow from Proposition \ref{Prop:EFwithGamma} upon 
  using the explicit form of the correlation functional \eqref{GfunctGamma} and 
  under the assumption of zero curvature are given by
  \begin{align}
    &\partial_t D +{\rm div}(D\bu)=0\,,\\
    &M(\partial_t + \bu\cdot\nabla)u_i = - \partial_i V_Q 
    - \braket{\rho, \partial_i\widehat{H}_e} 
    -\frac{\hbar^2}{MD}\partial_j\left(D\Big(\braket{\gamma_i}\braket{\gamma_j}-\frac{1}{2}\braket{\{\gamma_i,\gamma_j\}}\Big)\right)\,,
\\
 &i\hbar(\partial_t+\bu\cdot\nabla)\rho = 
  \left[\widehat{H}_e, 
  \rho\right]+\frac{\hbar^2}{2MD}\Bigg(D\Big[2\braket{\gamma_j}\gamma_j-\gamma_j^2,\rho\Big]+ 
  \{\rho, \partial_j(D\gamma_j)\} - 2\braket{\partial_j(D\gamma_j)}\rho\Bigg)\,,\\
&i\hbar(\partial_t\gamma_j + u_i\partial_i\gamma_j + \gamma_i\partial_ju_i )=-\partial_j^{\gamma}\left(\widehat{H}_e+\frac{\hbar^2}{2M}\left(2\braket{\gamma_j}\gamma_j-\gamma_j^2+\frac{1}{D}\left[\partial_j(D\gamma_j),\rho\right]\right)\right)\,.
  \end{align}
\end{proposition}

\paragraph{Proof:} We begin by computing the variational derivatives of the 
correlation function \eqref{GfunctGamma}
\begin{align*}
\bar{G}(D, \rho, \gamma_j)
  &=  \frac{\hbar^2}{2M}\int D
\Big(\braket{\gamma_j}^2 - \braket{\gamma_j^2}\Big)\,\de^3r\,.
\end{align*}
These are given by
\begin{align*}
  \frac{\delta \bar{G}}{\delta D}= 
  \frac{\hbar^2}{2M}\Big(\braket{\gamma_j}^2-\braket{\gamma_j^2}\Big)\,,\quad   \frac{\delta \bar{G}}{\delta \rho}= 
  \frac{\hbar^2 D}{2M}  \Big(2\braket{\gamma_j}\gamma_j - \gamma_j^2\Big)\,,\quad
    \frac{\delta \bar{G}}{\delta \gamma_j}= \frac{\hbar^2 D}{2M}\Big(\{\rho, \gamma_j\} 
    -2\braket{\gamma_j}\rho\Big)\,.
\end{align*}
Recall that the corresponding covariant divergence is given by
\begin{align*}
 \partial_j^{\gamma} \frac{\delta \bar{G}}{\delta \gamma_j} &= \frac{\hbar^2}{2M}\Big(\{\rho, 
\partial_j(D\gamma_j)\}-2\braket{\partial_j(D\gamma_j)}\rho\Big)\,.
\end{align*}
Lengthy computations substituting these derivatives into the general fluid 
equation \eqref{GAMMAEFFLUIDEQUATIONAGAIN}, electronic equation \eqref{Erica} and 
connection equation \eqref{Laura} yield the desired results.
\hfill$\square$\\

We remark that the set of equations presented in Proposition \ref{Prop:FinalEFwithGamma} 
are equivalent to those reported in the chemistry literature as given in 
Corollary \ref{Cor:EFPhysEqns}. This can be shown explicitly by using the defining relation $\partial_j\rho = [\rho, \gamma_j]$ 
as well as the pure state expression of the electronic density matrix 
$\rho=\psi\psi^{\dagger}$. In addition, the 
Kelvin-Noether circulation theorem \eqref{KNTheoremEFNOCURV} is specialised by 
to the physical correlation functional \eqref{GfunctGamma} returning the 
previously derived expression from Theorem \ref{Abby}. Naturally, the real part of the QGT is expressed in terms of the connection via equation \eqref{MetricTensorGammaTilde}. 

However, by incorporating the dynamics of the gauge 
connection, these equations would be able to capture the motion of any topological singularities that may exist in the nonadiabatic regime.
In particular, as described in Remark \ref{Rem:Heatwave}, such defects would arise through the initial conditions and are preserved in time, in this case both dragged around by the nuclear fluid  
and evolve according the unitary evolution of the electronic motion. The 
physical nature of singularities in nonadiabatic dynamics is still under 
investigation \cite{MinEtAl2014, RequistEtAl2016}.

In this final section we have applied the construction of the non-Abelian connection from Section \ref{Sec:NonAbelianQuantumSystems} to the exact factorization model of quantum chemistry previously examined in Chapters \ref{Chap:EFOLD} and \ref{Chap:EFNEW}.
This provided a new perspective on electron-nuclear correlations in which the 
so-called `gauge invariant part of the time-dependent potential energy 
surface' \cite{AgostiniEtAl2015} is given the highly geometric interpretation relating to the variance of the 
connection. A more general EF system was derived from an Euler-Poincar\'e formulation in Proposition \ref{Prop:EFwithGamma} 
in which the connection was {not} constrained to have zero curvature, potentially 
allowing for non-zero holonomy in future studies. Finally, upon restricting back 
to zero curvature, a reformulation of the physical system presented in Theorem \ref{Theorem:EFEPFull} was given, featuring the flat connection as an additional dynamical variable.

\chapter{Conclusions and perspectives}\label{Chap:Conclusions}

This thesis has employed the tools of geometric mechanics to investigate
quantum hydrodynamics and its application to current problems in theoretical chemistry. 
As a result new models for describing nonadiabatic quantum dynamics have been developed, whilst 
the use of a geometric approach has unveiled enlightening new perspectives across many aspects of quantum chemistry. 
This thesis has also developed the use of gauge connections in mechanical systems, resulting in a new geometric framework for QHD
as well as the discovery of insightful theoretical developments arising from a novel non-Abelian connection, applicable to a wide variety of quantum systems.
We now summarise in more detail the key contributions made by this thesis. 

\section{Summary of key results}
\subsubsection{Quantum hydrodynamics}
Quantum hydrodynamics has played a central role in this thesis with new results developed across a 
wide range of topics. 

In Chapter \ref{Chap:QHD} a regularised Euler-Poincar{\'e} 
Lagrangian for QHD is introduced  \eqref{hydro-RLagr} in which the quantum potential term is modified by smoothened fluid density. In turn, this allowed for singular solutions 
of the form
\begin{align*}
D(\bx,t)=\sum_{a=1}^{\cal N} w_a\delta(\bx- \boldsymbol{q}_a{(t)})\,,
\end{align*}
which we called `Bohmions'. This takes the regularised Lagrangian into a new 
Lagrangian \eqref{hydro-RLagrAGAIN} defined on {$T\mathbb{R}^{3\mathcal{N}}$} and the resulting trajectory equations evolve under the force generated from a regularised version of the  
quantum potential. This regularisation strategy provides a new method for 
obtaining single particle trajectories without taking the $\hbar^2\to 0$ limit and neglecting the effects due to the quantum 
potential. 

Chapter \ref{Chap:QHD} also extended the previous geometric 
treatment of QHD to mixed quantum states. In particular, the momentum map $J:{\rm Her}(\mathscr{H})\to \mathfrak{X}^*(\mathbb{R}^3)$ 
given by
  \begin{align*}
\bJ(\rho\,)=\frac12\{\widehat{P},\rho\,\}_+(\bx,\bx)
=:\boldsymbol{\mu}(\bx)
\,,
\end{align*}
was introduced, however the total energy Hamiltonian is no longer collective. 
This issue is circumvented by the introduction of the cold fluid closure \eqref{coldfluid} which, although not a true quantum state, 
returns a collective Hamiltonian $h(\bmu, D)$ without the quantum potential term. This geometric approach hence provides an alternative method for neglecting the 
quantum potential, again without invoking any limiting process on $\hbar$.     

In Chapter \ref{Chap:Holonomy} an entirely new formulation of QHD was presented, 
based on the theory of gauge connections. Specifically, upon adopting a 
technique used previously in the geometric study of complex fluids \cite{Holm2002,GayBalmazRatiu2009,GBRaTr2012,GBRaTr2013}, the $\mathcal{U}(1)$ 
gauge connection $\bnu$ was introduced, replacing the role usually played by the phase in the 
standard approach to QHD. This resulted in the alternative Euler-Poincar{\'e} 
reduction process for QHD presented in Theorem \ref{Theorem:NEWQHD}. 
Crucially, within the new theory, the connection $\bar\bnu$ is not expressed 
as an exact differential, instead allowed constant non-zero curvature through 
the Helmholtz decomposition
\begin{align*}
  \bar\bnu(\bx,t) = \nabla s(\bx,t) - \hbar\bLambda(\bx)\,.
\end{align*}
From a hydrodynamic perspective this connection also acts as the fluid velocity $\widetilde{\bnu}$, with the non-zero curvature endowing the flow with constant vorticity 
$\bomega=-\hbar m^{-1}\nabla\times\bLambda$, appearing in the fluid velocity equation as an additional Lorentz force term
\begin{align*}
    m(\partial_t+\widetilde{\bnu}\cdot\nabla)\widetilde{\bnu} = \hbar\widetilde{\bnu}\times\nabla\times{\bLambda} - 
\nabla(V+V_Q)\,.
\end{align*}
 
 In Section \ref{Subsec:HolonomicSchr} 
the new QHD equations were used to reconstruct the Schr\"odinger equation in which 
this additional non-flat connection term appears as a vector potential via minimal coupling 
as
\begin{align*}
      i\hbar\partial_t\psi &=    
    \left[\frac{(-i\hbar\nabla-\hbar{\bLambda})^2}{2m}+V\right]\psi\,.
\end{align*}
 This new connection-based formulation of QHD generalises the 
standard picture through this additional term and allows for non-trivial constant holonomy. 

One particular choice for $\bLambda$ results in a vortex filament form for the 
vorticity \eqref{vortexfilament}. In Section \ref{Sec:Vortices} this picture was 
extended to consider the dynamics of such vortex filaments using the method of 
Rasetti and Regge \cite{RasettiRegge1975}. This led to the new system of equations 
appearing in Theorem \ref{Theorem:VORTICES}
coupling quantum evolution to classical hydrodynamic vortex dynamics, for which 
it remains to be seen if there are practical applications. 

Finally in Section \ref{Sec:Osmotic} 
the connection approach was further extended to include the amplitude dynamics. 
The resulting equations for the phase and amplitude connections decoupled from the amplitude 
itself resulting in a generalised version of the stochastic theory of Nelson 
\cite{Nelson1967}.

\subsubsection{Quantum chemistry}
The original motivation that led to the work in this thesis was the prospect of 
applying geometric methods in quantum chemistry. The main developments have been 
based on the so-called exact factorization of the molecular wavefunction 
\cite{AbediEtAl2012}.

Chapter \ref{Chap:EFOLD} examined the geometry of the EF approach as it is currently implemented 
in the chemistry literature, focusing only on the use of variational techniques. Theorem \ref{Theorem:EFEL} derived the nuclear and electronic dynamics from a Lagrangian 
$L:T\mathscr{H}_n\times T\mathcal{F}(\mathbb{R}^3,\mathscr{H}_e)\times\mathcal{F}(\mathbb{R}^3)\to 
\mathbb{R}$ in which the electronic normalisation was enforced via a Lagrange 
multiplier and the total energy was left arbitrary, leading to a new 
compatibility condition \eqref{CompCond1}. The physical energy functional was 
then analysed and the presence of the quantum geometric tensor 
identified. 

Upon invoking the geometric approach to nuclear QHD (as described in Section \ref{Sec:2-HD}), a new variational principle was presented in Theorem \ref{Theorem:EFEPEL}. This featured Euler-Poincar{\'e} equations for the nuclear dynamics coupled to Euler-Lagrange equations for the electronic 
dynamics,
with the presence of the QGT again highlighted upon writing a new alternative expression for the EF system
\begin{align*}
     & \partial_t D + {\rm div}(D\bu) =  0\,,
  \\
&M(  \partial_t  + \bu\cdot \nabla) u_i = 
- \,\partial_i V_Q 
+ \langle\psi,(\partial_i \widehat{H}_e) \psi\rangle 
-  \frac{\hbar^2}{MD}  \partial_j\big(DT_{ij}\big)\,,\\
 &  (\mathbbm{1}-\psi\psi^{\dagger})\left( i\hbar \partial_t{\psi} 
 + i\hbar  \Big(\boldsymbol{u}-\frac1{M}\boldsymbol{\cal A}\Big)\cdot\nabla\psi -\widehat{H}_e\psi 
 + \frac{\hbar^2}{2MD}\operatorname{div}\!\big(D\nabla\psi\big) \right) 
  = 0\,,
\end{align*}
as given in  in Proposition \ref{Prop:EFNEWSET}.
Naturally, the Euler-Poincar{\'e}  formulation of the hydrodynamic nuclear equations resulted in a corresponding Kelvin-Noether theorem which 
was shown to account for the evolution of the electronic Berry phase via
    \begin{align*}
       \frac{\de}{\de t}\oint_{c(t)} \boldsymbol{\cal A} \cdot \de\br
       =  - \oint_{c(t)} \Big( \langle\psi,(\partial_i\widehat{H}_e) \psi\rangle 
            + \frac{\hbar^2}{MD} \partial_j \big(DT_{ij}\big) \Big) \de r^i 
    \,,
    \end{align*}
from Theorem \ref{Abby}. This equation was then rewritten solely in terms of the 
geometric tensor and electronic Hamiltonian in equation \eqref{QGT-Flux}, again 
demonstrating the highly geometric structure underlying EF dynamics.

Chapter \ref{Chap:EFNEW} used additional techniques in geometric mechanics to 
take the EF beyond the scope of the existing literature.  In particular, Section 
\ref{NuclearFrameSection} considered the unitary evolution of the electronic 
factor which naturally ensures the normalisation for all times. Crucially the geometric approach provided a natural method for expressing these
dynamics in the nuclear hydrodynamic frame as
\begin{align*}
  \psi(\boldsymbol{r},t)=  U(\eta^{-1}(\boldsymbol{r}, t),t)\psi_0(\eta^{-1}(\boldsymbol{r}, 
  t))
  \,,
\end{align*}
where $\eta(\br,t)$ is the Lagrangian path of the nuclear flow and $U$ is the unitary operator acting on the electronic Hilbert space. 
This approach, when combined with the result of Proposition 
\ref{Prop:EffPotDen}, allows the EF Lagrangian to be written in terms of the 
electronic density matrix $\rho=\psi\psi^{\dagger}$ in a new reduced form as $\ell:\mathfrak{X}(\mathbb{R}^3)\times{\rm Den}(\mathbb{R}^3)\times\mathcal{F}(\mathbb{R}^3,\mathfrak{u}(\mathscr{H}_e))\times\mathcal{F}(\mathbb{R}^3,{\rm Her}(\mathscr{H}_e))\to\mathbb{R}$ 
. Theorem \ref{Theorem:EFEPFull} provides the resulting Euler-Poincar{\'e} variational principle, which carries the same underlying geometric structure 
as previously seen in the study of liquid crystal flows \cite{GBRaTr2012,GBRaTr2013}, 
and results in a new formulation of the EF system given by
\begin{align*}
    &\partial_t D+{\rm div}(D\bu)=0\,,\\
&  M(\partial_t + \bu\cdot\nabla)\bu   = 
  - \nabla V_Q-\langle\rho,\nabla\widehat{H}_e\rangle -\frac{\hbar^2}{2MD}\partial_j\langle D \nabla\rho,  \partial_j\rho\rangle
  \,,
  \\
  &
    i\hbar (\partial_t +\bu\cdot\nabla)\rho = [\widehat{H}_e,\rho]+\frac{\hbar^2}{2MD}{\rm div}(D[\rho,\nabla\rho])
    \,.
\end{align*}
This geometric picture was completed by Theorem \ref{Theorem:LPEF} in which these 
equations are shown to possess a Hamiltonian structure given by the Lie-Poisson 
bracket \eqref{SDP-LPB-EF} defined on the dual of the Lie algebra $\mathfrak{g}=\mathfrak{X}(\Bbb{R}^3)\,\circledS\, \big(\mathcal{F}(\Bbb{R}^3)\oplus\, \mathcal{F}\big(\Bbb{R}^3,\mathfrak{u}(\mathscr{H}_e)\big)\big)
$.

Section \ref{Sec:2Level} specialised the above construction to two-level 
electronic systems in which $\mathscr{H}_e=\mathbb{C}^2$. Correspondingly the electronic dynamics 
were described by the spin vector $\bs(\br)\in \mathcal{F}(\mathbb{R}^3,\mathbb{R}^3)$ and, 
upon using the Lie algebra isomorphism $\mathfrak{su}(2)\cong\mathbb{R}^3$, the 
Euler-Poincar{\'e} and Lie-Poisson structures were specialised in Corollaries \ref{Cor:2LevelEP} 
and \ref{Cor:2LevelLP} respectively. This two-level restriction captures both spin-boson \cite{LeggettEtAl1987} and 
Jahn-Teller systems \cite{OBCh93} as well as encompassing the (previously unrelated) hydrodynamic approach to the Pauli 
equation \cite{Takabayasi1955} as described in Section \ref{Sec:PauliQHD}.

Chapter \ref{Chap:EFNEW} concluded by presenting an entirely new factorisation 
ansatz \eqref{DMF} for the molecular density matrix, 
\begin{align*}
\widehat{\rho}(\br,\br')=\rho_n(\br,\br')\psi(\br)\psi^\dagger(\br')
\,,
\end{align*}
generalising the usual EF approach \eqref{GenExFact} by 
allowing for more general mixed nuclear states. After deriving the corresponding 
equations of motion in Theorem \ref{Theorem:NEWEFDENANSATZFORM}, a sequence of 
geometric approximations led to a new mixed quantum-classical system for 
describing nonadiabatic phenomena. Specifically, we employed the cold fluid 
closure for the nuclear density in \eqref{coldfluidEF} and applied the reduced Legendre transform 
to obtain an Euler-Poincar{\'e} Lagrangian expressed in terms of nuclear QHD variables 
that no longer correspond to an irrotational flow. The penultimate step required performing a 
regularisation procedure analogous to that presented in Chapter \ref{Chap:QHD}, 
smoothening the $O(\hbar^2)$ terms in the Lagrangian (thus requiring the introduction of a regularised electronic density matrix). 
The final result is Theorem \ref{Theorem:CRAZY} in which the Bohmion singular solutions
  \begin{align*}
  D(\br,t)=\sum_{a=1}^{\mathcal{N}}w_a\delta(\br-\bq_a(t))\,,\qquad
    \tilde{\rho}(\br,t)=\sum_{a=1}^{\mathcal{N}}\varrho_a(t)\delta(\br-\bq_a(t))\,,
\end{align*}
 result in the new Lagrangian \eqref{SurelyNearlyThere} defined on $L:T\mathbb{R}^{3\mathcal{N}}\times{\rm Her}(\mathscr{H}_e)^{\mathcal{N}}\times\mathfrak{u}(\mathscr{H}_e)^{\mathcal{N}}\to\mathbb{R}$
and ultimately give the equations of motion
\begin{align*}
\begin{split}
Mw_a\ddot{\bq}_a &= -\nabla_{\bq_a}\Bigg(\braket{\varrho_a,i\hbar\xi_a-\widehat{H}_e(\bq_a)}
 \\
 &\qquad\qquad\quad - \frac{\hbar^2}{4M} \sum_{b} \Big(\langle\varrho_a |\varrho_b\rangle-w_aw_b\Big)  { \int 
  \frac{\nabla K(\br-\bq_a)\cdot \nabla K(\br-\bq_b)}
  {\sum_{c} w_c K(\br-\bq_c)} \,\de^3r }
  \Bigg)\,,
  \end{split}\\
  \nonumber \\
i\hbar\dot{\varrho}_a &= 
\big[\widehat{H}_e(q_a),{\varrho}_a\big]
+
\frac{\hbar^2}{4M} \sum_{b} \left[{\varrho_b},{\varrho}_a\right]  {   
\int 
  \frac{\nabla K(\br-\bq_a)\cdot \nabla K(\br-\bq_b)}
  {\sum_{c} w_c K(\br-\bq_c)} \,\de^3r 
}\,.
\end{align*}
The achieves a key goal for modelling in nonadiabatic chemistry. Namely, the 
classical nuclear trajectories (finite-dimensional) are coupled dynamically to the quantum electronic 
dynamics by a countably infinite set of equations in a way that retains both the long-range effects of the quantum 
potential and correlation effects between nuclei and electrons. It is now of significant future interest to see the numerical implementation 
of this model.  

The new connection-based geometric formulation of QHD presented in Chapter \ref{Chap:Holonomy} 
was applied to the Born-Oppenheimer system in Section 
\ref{Sec:ConnectionQHDBOApp}. Upon restricting the nuclear hydrodynamic path to 
evolve under pure translations, we obtained a nuclear trajectory equation \eqref{particles+vortices} whereby the nuclear dynamics are governed by regularised versions of the
potential energy surface and Lorentz force that usually appear in the Mead-Truhlar method. This idea warrants further study 
as the associated Berry phase is no longer a topological quantity, rather a 
true geometric phase in that it is path dependent, which has interesting 
parallels with the recent analogous suggestion for Berry phases in nonadiabatic processes 
    \cite{MinEtAl2014, RequistEtAl2016}.

\subsubsection{Non-Abelian gauge connections}
Inspired by the use of connections in QHD from Chapter \ref{Chap:Holonomy}, Section \ref{Sec:NonAbelianPauliConnection} 
introduced a non-Abelian connection in the EF two-level system as described in Section \ref{Sec:2Level}.
In particular the evolution of the spin vector under the rotation group 
gave rise to an $\mathfrak{so}(3)$ connection defined by equation \eqref{PauliConnection}. 
This led to Proposition \ref{Prop:EP2LevelGamma} which provided an 
Euler-Poincar\'e formulation of EF two-level systems in which the connection possessed
non-trivial curvature
\begin{align*}
  \partial_t D + {\rm div}(D\bu)&= 0\,,\\
     MD(\partial_t+\bu\cdot\nabla)u_i &= -D\partial_i\left(V_Q+\frac{\delta h}{\delta 
  D}\right)+\partial_i\bs\cdot \frac{\delta h}{\delta \bs}+\bOmega_{ij}\cdot\frac{\delta h}{\delta 
  \bgamma_j} - \bgamma_i\cdot \partial_j^{\gamma}\frac{\delta h}{\delta \bgamma_j} \,,\\
  D(\partial_t+ \bu\cdot\nabla)\bs &= \frac{\delta h}{\delta \bs}\times\bs - \partial_j^{\gamma}\frac{\delta h}{\delta \bgamma_j}\,,\\
  \partial_t\bgamma_j + u_i\partial_i\bgamma_j + \bgamma_i\partial_ju_i&= -\partial_j^{\gamma}\left(\frac{1}{D}\frac{\delta h}{\delta \bs} + \frac{4}{\hbar^2 D}\partial_j^{\gamma}\frac{\delta h}{\delta \bgamma_j}\times\bs \right)
  \,.
\end{align*}
The curvature $\bOmega_{ij}$ features explicity in the fluid equation and it is hoped that such equations may eventually be used to describe more general phenomena in which 
holonomy plays a role. The assumption of non-zero curvature was then relaxed, 
resulting in a simplified form of the fluid velocity equation and associated 
Kelvin Noether theorem \ref{Theorem:TIRED}. These results were then specialised 
to describe the hydrodynamic form of the Pauli 
equation from Section \ref{Sec:PauliQHD}. Interesting new expressions for the 
quantum geometric tensor (and hence the Berry curvature and Fubini-Study metric) in terms of the connection were obtained 
\begin{align*}
  Q_{ij} = \frac{1}{4}\bgamma_i \cdot\bgamma_j - \hbar^{-2}(\bs\cdot\bgamma_i)(\bs\cdot\bgamma_j) 
  + \frac{i}{2\hbar}\bs\cdot\bgamma_i\times\bgamma_j\,,
\end{align*}
and the Pauli QHD system of equations were reformulated in Proposition 
\ref{Prop:PauliQHDGamma}.

Section \ref{Sec:GeneralTheory} considered the general theory of introducing 
gauge connections in mechanical systems with order parameters through the use of such relations $\de n = -\gamma 
n$. A new result was presented in Theorem \ref{Theorem:GenZeroCurv}, stating that this relation implies that the connection is flat, i.e. has vanishing curvature, 
and applies to a whole class of similar connections previously introduced in the theory of complex fluids  
\cite{Holm2002,GayBalmazRatiu2009,GBRaTr2012,GBRaTr2013}. The general 
Euler-Poincar{\'e} reduction for such mechanical systems was then presented in 
Theorem \ref{Theorem:GammaGenEP}, specialising the more general theory found in 
\cite{GayBalmazRatiu2009}. Then, despite the previous zero curvature relation, defined solely as the solution to the general Euler-Poincar{\'e} 
equation $ \partial_t \gamma = -\nabla^{\gamma}\xi$, the connection is no longer constrained to be flat, with the 
non-trivial evolution of the curvature given in Corollary \ref{Cor:CurvEvo}.

Section \ref{Sec:NonAbelianQuantumSystems} then applied this construction to the 
unitary evolution of quantum systems, allowing for the introduction of a new $\mathfrak{u}(\mathscr{H})$ 
connection according to $\partial_j\psi(\bx,t)
  = -\gamma_j(\bx,t)\psi(\bx,t)$. Upon defining the Hermitian counterpart $\widetilde{\gamma}=i\hbar\gamma$, this immediately led to the identification of the Berry connection as simply the expectation value
\begin{align*}
  {\cal A}_j := \braket{\psi|-i\hbar\partial_j\psi} = -\braket{\widetilde{\gamma}_j}\,,
\end{align*}  
geometrically corresponding to the $\mathcal{U}(1)$ projection of $\gamma$.
 In proposition \ref{Prop:QGTCov} the quantum geometric tensor was 
given the new interpretation as the covariance of this connection 
 \begin{align*}
Q_{ij} = \hbar^{-2}\Big(\braket{\widetilde{\gamma_i}\widetilde{\gamma_j}}-\braket{\widetilde{\gamma_i}}\braket{\widetilde{\gamma_j}} 
  \Big)=\hbar^{-2}{\rm cov}(\widetilde{\gamma_i}, \widetilde{\gamma_j})\,,
  \end{align*}
in turn, leading to the new uncertainty relations Proposition \ref{Theorem:QGTUR} and 
Corollary \ref{Cor:QGTUR}. As explained in Section \ref{Sec:NonAbelianQuantumSystems} 
similar results have previously been found \cite{CarolloValentiSpagnolo2020,ZanardiEtAl2007,KoSeMePo2017}, however the new interpretation provides a more geometric foundation 
with the fundamental object as the $\mathfrak{u}(\mathscr{H})$ gauge connection. 

Section \ref{Sec:GammaEF} introduced this non-Abelian connection $\gamma_j\de r^j\in \Lambda^1(\mathbb{R}^3, \mathfrak{u}(\mathscr{H}_e))$ as an 
additional dynamical variable in the exact factorization system. This connection 
appears solely through the so-called `gauge invariant part of the time-dependent potential energy 
surface' \cite{AgostiniEtAl2015} which is given a new geometric interpretation as the variance of the 
connection via
  \begin{align*}
  \epsilon
  &=\braket{\widehat{H}_e}  + 
  \frac{\braket{\widetilde{\gamma}_j^2} - 
  \braket{\widetilde{\gamma}_j}^2}{2M}\,,\qquad {\rm with}\,\, \widetilde{\gamma}_j=-i\hbar\gamma_j\,, 
\end{align*}
in Remark \ref{Rem:CorObs}.
 It is exactly when this variance vanishes that the system reduces to the mean-field  model, and hence we proposed that the connection can be understood as a `correlation observable' determining the 
strength of electon-nuclear coupling in nonadiabatic dynamics.
The introduction of the connection also allowed for the Euler-Poincar\'e derivation of a more general EF system in Proposition \ref{Prop:EFwithGamma} 
which may be able to capture non-zero holonomy effects. The original EF system of Theorem \ref{Theorem:EFEPFull} was then re-expressed in terms of the connection.

%
%
%
%

\section{Perspectives}\label{Sec:Perspectives}


Whilst these results have answered and shed new light on many of the questions posed at the beginning 
of this thesis, they have also opened up new questions and interesting directions for future research. 
We conclude this final chapter by summarising some of the key areas that warrant further study. 

\begin{itemize}
  \item Whilst this thesis considers single product factorisations of the 
  molecular wavefunction (Born-Oppenheimer, mean-field, exact factorization), 
  in the more general case of many interacting  molecules it is of great 
  practical and theoretical interest to consider a geometric approach to {\bf multi-particle 
  closures}. For example, one can consider a generalisation of the BO ansatz 
  consisting of a  product of nuclear wavefunctions whilst retaining a single electronic eigenstate. 
  What geometric structures underly such models? Do the results obtained in the single product factorisation case generalise?
  \item Recently, there has been a major development in the theory of mixed quantum-classical 
  dynamics, specifically the discovery of a consistent Hamiltonian theory for the interaction between quantum and classical systems, ultimately leading 
  to the introduction of a hybrid quantum-classical wavefunction \cite{GayBalmazTronci2019}. With the goal 
  of developing applications for this method it would be very interesting to 
  apply the geometric approaches contained in 
  this work to {\bf product factorisations of the hybrid wavefunction}, culminating in interacting quantum and 
  classical subsystems. For example, due to the nature of the hybrid wavefunction, the BO factorisation can be implemented 
  in both directions (parametric dependence of the classical wavefunction on the quantum subsystem and vice versa), potentially resulting in
  different dynamics that may prove advantageous in numerical simulations. 
  \item Another potentially fruitful area of research is the geometric approach to {\bf coherent states in 
quantum chemistry}. Indeed, whilst Gaussian wavepackets are commonly used in 
applications, as demonstrated in an appendix 
  of the author's publication \cite{FoTr2020}, inserting Gaussians of a fixed width within the 
  variational principle leads to the new idea of nuclear dynamics occuring on regularised energy 
  surfaces and in the presence of a smoothened Berry connection. This idea demands further attention
and could redefine the understanding of the nature of the geometric (or topological) phase in adiabatic quantum chemistry. 
\item It would also be of great interest to continue to develop the material on the {\bf non-Abelian connection in quantum systems} established in Chater \ref{Chap:Gamma}.
Much of the material presented in this chapter leads to further avenues that should be pursued. For example, 
following the proposed introduction of a spacetime 
non-Abelian connection in Appendix \ref{App:Spacetime}, one can consider a four-dimensional 
 quantum geometric tensor on spacetime. This object would contain new components 
relating to the covariance between the generator and original (purely spatial) 
connection. Could we formulate additional uncertainty relations between temporal 
and spatial components of the connection? What information would this tell us 
about the wavefunction? Another question may relate to the real part of this spacetime QGT. Does it define
a generalisation of the Fubini-Study metric on Minkowski space?  

\end{itemize}

Finally, it would be of general interest to
 obtain {\bf numerical and experimental validation} of the models presented in this thesis in order to 
 understand their potential application to real-world problems. In particular, 
 how does the Bohmion model of EF given in Section \ref{Bohmions} compare to 
 current simulations? Does the new connection-based approach to QHD of Chapter \ref{Chap:Holonomy}
capture physical phenomena? Is the coupled Schr\"odinger-vortex system of 
practical use? Can the connection formulation of EF dynamics from Section \ref{Sec:GammaEF} result in non-Abelian  
holonomy effects? Is the correlation observable measurable?


\appendix

\chapter{Principles of geometric mechanics}\label{App:GMTheory}
In this appendix, we consider many of the general principles of geometric 
mechanics employed throughout this thesis. This appendix contains material that can be found
in much of the standard geometric mechanics literature and for further details the reader is directed to \cite{MaRa2013, HoScSt2009}.
\section{Lie group actions}
\begin{definition}[Left Lie group action]\label{Def:LeftAction}
  Let $M$ be a manifold and let $G$ be a Lie group. A {\bf left action} of a Lie group $G$ on $M$ is a 
  smooth mapping $\Phi:G\times M\to M$ such that:
  \begin{enumerate}[(i)]
  \item $\Phi(e,x)=x$ $\forall$ $x\in M$.\label{Left1}
  \item $\Phi(g,\Phi(h,x))=\Phi(gh,x)$ $\forall$ $g,h\in G$ and $x\in M$.\label{Left2}
  \item For every $g \in  G$ the map $\Phi_g:M\to M$ defined by $\Phi_g(x):=\Phi(g,x)$ 
  is a diffeomorphism (i.e. smooth and invertible).\label{Left3}
\end{enumerate}
  \end{definition}
\begin{definition}[Right Lie group action]
  The {\bf right action} of $G$ on $M$ satisfies \eqref{Left1} and \eqref{Left3} 
  from Definition \ref{Def:LeftAction}, while \eqref{Left2} is replaced by:
  \begin{align}\label{Right2}
{   \Phi(g,\Phi(h,x))=\Phi(hg,x)}\quad\forall\, g,h\in G\quad \text{and} \quad x\in M\,.
  \end{align}
\end{definition}
  \paragraph{Concatenation notation:} We write $gx$ for the left action $\Phi(g,x)$ and think of 
  a group element $g$ acting on $x$ (from the left). Then \eqref{Left2} becomes 
  $g(hx)=(gh)x$. Similarly, we write $xg$ for the right action $\Phi(g,x)$ and 
  \eqref{Right2} becomes $(xh)g=x(hg)$. 
  \begin{remark}[Left vs right actions]\label{Rem:LeftRight}
    Any left action $(g,x)\mapsto gx$ naturally 
produces a right action by $(g,x)\mapsto g^{-1}x$. With this in mind, throughout 
this thesis we have chosen to work with left actions. However in some cases using the right action 
can result in easier computations, for example in Appendix \ref{App:HalfDen}. 
  \end{remark}
  
  \begin{definition}[Group orbit]
    Let $G$ be a Lie group acting on a manifold $M$. For any point $x\in M$, the 
    subset
    \begin{align}
      {\rm Orb}(x):=\{gx:g\in G\}\,,
    \end{align}
    is called the {\bf group orbit} of $G$ through $x$.
  \end{definition}
  
  \begin{definition}[Properties of group actions]
    A group action $\Phi:G\times M\to M$ is said to be
      \begin{enumerate}[(i)]
  \item {\bf transitive}, if there is only one orbit, or equivalently, if $\forall x,y \in M$ $\exists g \in G$ such that 
  $gx=y$.
    \item {\bf free}, if it has no fixed points, that is, $\Phi(g,x)=x \implies 
    g=e$.
      \item {\bf faithful}, if $\forall g \in G$ such that $g\neq e$, $\exists x\in M$ such that $gx\neq x$.
\end{enumerate}
  \end{definition}
  
  \begin{definition}[Tangent map]
      Let $\Phi_g:M\to M$ be a group action on the manifold $M$. The {\bf tangent 
  map} of $\Phi_g$ at $x \in M$ is the map $T_x\Phi_g:T_xM\to T_{\Phi_g(x)}M$, defined by
  \begin{align}
    v\mapsto T_x\Phi_g(v):= \frac{\de}{\de t}\Bigg|_{t=0}\Phi_g(q(t))\,,
  \end{align}
  where $q(t)\in M$ is an arbitrary path such that $q(0)=x$ and $\dot{q}(0)=v \in T_x 
  M$.
  \end{definition}
  
  \begin{definition}[Tangent lift]
  Let $\Phi_g:M\to M$ be a group action on the manifold $M$. The {\bf tangent 
  lift }of $\Phi_g$ is the group action $T\Phi_g:TM\to TM$ on the tangent bundle of $M$, defined by
  \begin{align}
    (x,v)\mapsto T\Phi_g(x,v):=\Big(\Phi_g(x),T_x\Phi_g(v)\Big)\,,
  \end{align} 
  for all $(x,v)\in TM$, where $T_x\Phi_g$ is the tangent map of $\Phi_g$ defined above.
    \end{definition}
  
    \begin{definition}[Cotangent map]
      Let $\Phi_g:M\to M$ be a group action on the manifold $M$. The {\bf cotangent 
  map} of $\Phi_g$ at $x \in M$ is the map $T_x^*\Phi_g:T^*_{\Phi_g(x)}M\to T_x^*M$, defined by
  \begin{align}
\Big\langle T^*_x\Phi_g(\alpha), v \Big \rangle:= \Big\langle \alpha, T_x\Phi_g(v) \Big 
\rangle\,,
  \end{align}
 for all $v\in T_x M$ and $\alpha \in T_{\Phi_g(x)}^*M$, i.e. is the dual of the tangent 
 map.
  \end{definition}

  \begin{definition}[Cotangent lift]
  Let $\Phi_g:M\to M$ be a group action on the manifold $M$. The {\bf cotangent 
  lift }of $\Phi_g$ is the group action $T^*\Phi_g:T^*M\to T^*M$ on the cotangent bundle of $M$, defined by
  \begin{align}
    (x,\alpha)\mapsto T^*\Phi_{g^{-1}}(x,\alpha):=\Big(\Phi_g(x),T^*_{\Phi_g(x)}\Phi_{g^{-1}}(v)\Big)\,,
  \end{align} 
  for all $(x,\alpha)\in T^*M$, where $T^*_{\Phi_g(x)}\Phi_{g^{-1}}$ is the cotangent map (as defined above), now of $\Phi_{g^{-1}}$ which is the mapping $T^*_x M \to T^*_{\Phi_g(x)}M$.
    \end{definition}
  
\begin{definition}[Left translation]
  The map $L_g:G\to G$ given by $L_g:h\mapsto gh=:L_g(h)$ defines a left, transitive and free action of $G$ on itself.
  By Remark \ref{Rem:LeftRight}, $L_{g^{-1}}:h\mapsto g^{-1}h$ defines a right action of $G$ on itself.
\end{definition}

\begin{definition}[Right translation]
  The map $R_g:G\to G$ given by $R_g:h\mapsto hg=:R_g(h)$ defines a right, transitive and free action of $G$ on itself.
  By Remark \ref{Rem:LeftRight}, $R_{g^{-1}}:h\mapsto hg^{-1}$ defines a left action of $G$ on itself.
\end{definition}

\begin{definition}[Conjugation]
  The map $I_g:G\to G$ given by $I_g:h\mapsto ghg^{-1}=(R_g^{-1}\circ L_g)(h)=:I_g(h)$ 
  is called the {\bf conjugation action} of G on itself. Orbits of this action 
  are called {\bf conjugacy classes}.
\end{definition}

\paragraph{Adjoint and coadjoint actions:} The {\bf adjoint representation} of $G$ 
on $\mathfrak{g}$ is obtained by differentiating the conjugation action at the 
identity of the group $e$. Taking an arbitrary curve $h(t)\in G$ such that 
$h(0)=e$, and denoting $\xi = \dot{h}(0)\in T_eG\cong \mathfrak{g}$, we define
\begin{align}
  {\rm Ad}_g\,\xi := \frac{\de}{\de t}\Big|_{t=0}I_g(h(t)) = g\xi g^{-1} \in 
  \mathfrak{g}\,.\label{GAdjoint}
\end{align}

\begin{definition}[Adjoint action of $G$]
  The {\bf adjoint action} of $G$ on $\mathfrak{g}$ is the map ${\rm Ad}: G\times \mathfrak{g}\to\mathfrak{g}$ 
  given by \eqref{GAdjoint}, that is, ${\rm Ad}_g\,\xi =g\xi g^{-1} $.
  The dual map ${\rm Ad}^*:G\times \mathfrak{g}^*\to\mathfrak{g}^*$, defined by
  \begin{align}
    \braket{{\rm Ad}^*_g\, \mu, \xi} = \braket{\mu, {\rm Ad}_g\,\xi}\,,
  \end{align}
  where the pairing is defined on $\mathfrak{g}^*\times\mathfrak{g}$ is called 
  the {\bf coadjoint action} of $G$ on $\mathfrak{g}^*$.
\end{definition}

\begin{definition}[Coadjoint orbit]
  For any Lie group $G$ and any $\mu \in \mathfrak{g}^*$, the orbit of $\mu$ 
  under the coadjoint action of $G$ on $ \mathfrak{g}^*$, given by
  \begin{align}
    {\rm Orb}(\mu):= \{{\rm Ad}^*_{g^{-1}}\, \mu : g\in G\}\,,
  \end{align}
  is called the {\bf coadjoint orbit} of $\mu$.
\end{definition}

\section{Lie algebra actions}
Having seen how a Lie group $G$ acts on manifolds (and thus also on themselves) 
as well as on its Lie algebra by the adjoint and coadjoint actions, here we consider how Lie algebras act on manifolds to produce vector fields.
\begin{definition}[Infinitesimal generator]\label{Def:InfGen}
  Let $g(t)\in G$ be such that $g(0)=e$ and $\dot{g}(0)=\xi\in\mathfrak{g}$. Let 
  $\Phi:G\times M\to M$ be a group action. Then, the {\bf infinitesimal generator} 
  of $\Phi$ corresponding to $\xi$ at $x\in M$ is defined by
  \begin{align}
    \xi_M(x):=\frac{\de}{\de t}\Big|_{t=0} \Phi_{g(t)}(x)\,.
  \end{align}
  This is also known as the {\bf Lie algebra action} of $\xi$ on $M$. The map $\xi_M:M\to TM$ 
  for all $x\in M$ is a vector field associated to $\xi \in \mathfrak{g}$.
\end{definition}

\begin{definition}[Adjoint action of $\mathfrak{g}$]\label{Def:AlgebraAdj}
  Taking $g(t)\in G$ such that $g(0)=e$ and denoting $\xi=\dot{g}(0)\in 
  \mathfrak{g}$, the infinitesimal generator of the adjoint action of $G$ on $\mathfrak{g}$ 
  defines the map ${\rm ad}: \mathfrak{g}\times\mathfrak{g}\to \mathfrak{g}$, 
  where ${\rm ad}:(\xi,\eta)\mapsto {\rm ad}_{\xi}\,\eta$ defined by
  \begin{align}
    {\rm ad}_{\xi}\,\eta:= \frac{\de}{\de t}\Big|_{t=0}{\rm 
    Ad}_{g(t)}\,\eta\,,\qquad \forall \eta\in \mathfrak{g}\,,
  \end{align}
  and is called the {\bf coadjoint action} of $\mathfrak{g}$ on itself. 
  The dual map ${\rm ad}^*:\mathfrak{g}\times\mathfrak{g}^*\to\mathfrak{g}^*$, 
  where ${\rm ad}^*:(\xi,\mu)\mapsto {\rm ad}^*_{\xi}\,\mu$ given by
  \begin{align}
    \braket{{\rm ad}^*_{\xi} \,\mu, \eta} = \braket{\mu, {\rm ad}_{\xi}\,\nu}\label{Def:CoadjAlgAction}
  \end{align}
  is called the {\bf coadjoint action} of $\mathfrak{g}$ on $\mathfrak{g}^*$.
\end{definition}
\begin{remark}[Adjoint action for matrix Lie groups]
  Consider a matrix Lie group $G\subseteq GL(n,\mathbb{C})$. Let $g(t)\in G$ be a curve such that $g(0)=\mathbbm{1}$ 
  and $\dot{g}(0)=\xi \in \mathfrak{g}$. Then using the the adjoint action of $G$ on $\mathfrak{g}$ 
  ${\rm Ad}_g\,\eta = g\eta g^{-1}$, we can compute the adjoint action of $\mathfrak{g}$ 
  on $\mathfrak{g}$ according to Defintion \ref{Def:AlgebraAdj} as follows
  \begin{align}
    \begin{split}
     {\rm ad}_{\eta}\,\xi &= \frac{\de}{\de t}\left[g(t)\xi g^{-1}(t)\right]_{t=0}\\
     &= \dot{g}(0)\xi g^{-1}(0) - g(0)\xi g^{-1}(0)\dot{g}(0)g^{-1}(0)\\
     &= \eta\xi-\xi\eta = [\eta,\xi]\,.
    \end{split}
  \end{align}
  Hence, for matrix Lie groups we see that the adjoint action of $\mathfrak{g}$ 
  on $\mathfrak{g}$ is given by the matrix commutator, i.e. ${\rm ad}_{\eta}\,\xi =[\eta,\xi]$. 
  Simply by using the duality pairing one can also check that the coadjoint 
  action is given by ${\rm ad}^*_{\eta}\,\mu=[\eta^{\dagger},\mu]$.
\end{remark}

\section{Euler-Poincar{\'e} reduction}\label{App:EPReduction}
Here, we present the general approach of Euler-Poincar{\'e} reduction for the Lagrangian description of a 
mechanical system. To begin, consider a system whose dynamical equations of motion 
correspond to the Euler-Lagrange equations as derived from Hamilton's variational principle 
$0 = \delta \int_{t_1}^{t_2} L(q,\dot{q}) \,\de t$ for a Lagrangian defined as the mapping $L:TQ\to \mathbb{R}$, 
where $(q,\dot{q})$ belong to the tangent bundle of the configuration space 
$TQ$.\\

Upon assuming that the generalised coordinate evolves under the action of a Lie 
group $G$ so that we can write $  q(t)=g(t)q(0)$, where $g(t)\in G$, the 
Lagrangian is then defined over the tangent bundle of the Lie group $L:TG\to \mathbb{R}$ 
and is written as $L=L(g,\dot{g})$. At this point, we in fact consider the more 
common scenario in which the Lagrangian on $TG$ depends parametrically on a variable $a_0 \in 
M$, where $M$ is an arbitrary manifold, so that $L_{a_0}:TG\to\mathbb{R}$. We then define the unique function (if it exists)
$L:TG\times M\to \mathbb{R}$  by $L(g,\dot{g},a_0):=L_{a_0}(g,\dot{g})$ for all 
$(g,\dot{g})\in TG$.\\

If the group $G$ is also a symmetry of the system, for example if the extended Lagrangian $L$ 
is right-invariant (under the tangent-lifted action) i.e. $L(g,\dot{g},a_0)=L(gh,\dot{g}h,a_0h)$ for all $h\in G$, then one can select $h=g^{-1}$
so that the Lagrangian is defined on the reduced space $TG/G\times M$. Hence, using that $TG/G\cong T_eG 
=\mathfrak{g}$ and defining the Lie algebra element 
$\xi:=\dot{g}g^{-1}$ ultimately leads to the reduced Lagrangian $\ell:\mathfrak{g}\times M\to \mathbb{R}$ defined by
\begin{align}
\ell(\xi, a):=  L(gg^{-1},\dot{g}g^{-1},a_0g^{-1})\,,
\end{align}
where $a(t):=a_0 g^{-1}(t)$ and is now dynamical. At this stage we apply 
Hamilton's principle to this reduced Lagrangian, noting that arbitrary 
variations $\delta g$ on $L_{a_0}$ now become the constrained variations
\begin{align}
  \delta\xi = \dot{\eta} - {\rm ad}_{\xi}\,\eta\,,\qquad \delta a = -\eta_M \,
  a\,,
\end{align}
where ${\rm ad}$ is the adjoint action given in Defintion \ref{Def:AlgebraAdj} and $\eta_M$ the infinitsimal generator from Definition \ref{Def:InfGen}, thus resulting in the equations of motion
\begin{align}
  \frac{\de}{\de t}\frac{\delta \ell}{\delta \xi}+ {\rm ad}^*_{\xi}\,\frac{\delta \ell}{\delta \xi} 
  = \frac{\delta \ell}{\delta a}\diamond a\,,\qquad \dot{a}=-\xi_M\,a\,.\label{EPGenRight}
\end{align}
These are the {\bf Euler-Poincar{\'e}} equations for a right-invariant 
Lagrangian (see Theorem 3.3 in \cite{HolmEtAl1998}) and are entiely
equiavalent to the Euler-Lagrange equations for $L(g,\dot{g})$ obtained for arbitrary variations $\delta g$. 
Here, the diamond ($\diamond$) notation is defined by the relation
\begin{align}
  \left\langle a\diamond \frac{\delta \ell}{\delta a}, \zeta \right\rangle =   -\left\langle \frac{\delta \ell}{\delta a}\diamond a, \zeta \right\rangle :=   \left\langle \frac{\delta \ell}{\delta a}, \zeta_M\,a 
  \right\rangle\,,\label{DIAMOND}
\end{align}
where the pairing on the left is defined on $\mathfrak{g}^*\times\mathfrak{g}$ 
whilst the pairing on the right is defined on $T_a^*M\times T_aM$. 
We also note that the diamond operation defines the momentum map $J: TM^*\to \mathfrak{g}^*$ which can be understood as associated to the cotangent lifts of Lie algebra actions \cite{DarrylBook2}.
\begin{remark}[Basic Euler-Poincar{\'e} equations]
  In the simpler case when the Lagrangian does not contain any parametric 
  dependence, i.e. we have $L=L(g,\dot{g})$ only, then the reduced Lagrangian is 
  simply $\ell = \ell(\xi)$ and the Euler-Poincar{\'e} equation reduces to its basic 
  form
  \begin{align}
     \frac{\de}{\de t}\frac{\delta \ell}{\delta \xi}= - {\rm ad}^*_{\xi}\,\frac{\delta \ell}{\delta \xi} 
     \,.
  \end{align}
\end{remark}
\begin{remark}[Left-invariant Lagrangians]
The same reduction process holds for left invariant Lagrangians, that is, 
$L(g,\dot{g},a_0)=L(hg,h\dot{g},ha_0)$ for all $h\in G$, so that $\xi:=g^{-1}\dot{g}$ 
with variations $\delta\xi = \dot{\eta}+ {\rm ad}_{\xi}\,\eta$, where $\eta:=g^{-1}\delta 
g$, resulting in the Euler Poincar{\'e} equations
\begin{align}
  \frac{\de}{\de t}\frac{\delta \ell}{\delta \xi}- {\rm ad}^*_{\xi}\,\frac{\delta \ell}{\delta \xi} 
  = \frac{\delta \ell}{\delta a}\diamond a\,,\qquad \dot{a}=-\xi_M\,a\,.
\end{align}
\end{remark}

\section{Lie-Poisson reduction}
As in the standard approach to classical mechanics Lagrangian and Hamiltonian formalisms are related
 by the Legendre transform, so in the geometric approach Euler-Poincar{\'e} reduction can be 
 related to a Hamiltonian counterpart called Lie-Poisson reduction, this time by a reduced Legendre transform. 
 Here, we present the essential details of this construction.\\
 
 Consider a Hamiltonian defined on the cotangent bundle of a Lie group $G$ that again depends parametrically
 on $a_0\in M$, hence given by $H_{a_0}(g,p)$ where $H_{a_0}:T^*G\to 
 \mathbb{R}$. As before we define (if it exists) the unique function $H:T^*G\times M \to \mathbb{R}$ 
 by $H(g,p,a_0):=H_{a_0}(g,p).$ If this extended Hamiltonian is invariant under 
 the (cotangent-lifted) right action, i.e. $H(gh, ph, a_0h)=H(g,p,a_0)$ for all $h\in G$, then once again choosing $h=g^{-1}$
 allows us to write the reduced Hamiltonian $h:\mathfrak{g}^*\times M \to \mathbb{R}$ 
 given by
\begin{align}
  h(\mu, a):= H(gg^{-1}, pg^{-1}, a_0g^{-1})\,,
\end{align}
where again $a(t):=a_0g^{-1}(t)$ and this time we have 
$\mu:=pg^{-1}\in T_e^*G\cong\mathfrak{g}^*$. From here, it is possible to 
derive the Lie-Poisson equations directly \cite{MaRa2013}, however we proceed by 
assuming that the original Hamiltonian was related to a Lagrangian by a (hyperregular) Legendre 
transform. In this case, there exists a reduced Legendre transform $\mathbb{F}\ell:\mathfrak{g}\times M\to \mathfrak{g}^*\times M$ 
so that the reduced Hamiltonian reads
\begin{align}
  h(\mu, a)=\braket{\mu, \xi} - \ell(\xi, a)\,,\qquad \mu:= \frac{\delta \ell}{\delta 
  \xi}\,.
\end{align}
Then, as the reduced Lagrangian is also hyperregular (smooth and invertible), the last definition implies 
that $\xi = \delta h/\delta \mu$, so that applying the Euler-Poincar{\'e} variational 
principle
\begin{align}
  0 = \delta \int_{t_1}^{t_2} \Big(\braket{\mu, \xi} - h(\mu, a)\Big)\,\de t\,,
\end{align}
results in the equations of motion
\begin{align}
  \dot{\mu}= -{\rm ad}^*_{\frac{\delta h}{\delta \mu}}\,\mu - \frac{\delta h}{\delta a}\diamond a\,,\qquad \dot{a}=-\left(\frac{\delta h}{\delta 
  \mu}\right)_Ma \,.
\end{align}
These are called the {\bf Lie-Poisson} equations for a right-invariant 
Hamiltonian are are entirely equivalent to the canonical Hamilton equations in $(q,p)$ 
for $H_{a_0}(q,p)$. 

\begin{remark}[Left-invariant Hamiltonians]
  The analogous construction applies when the Hamiltonian is left-invariant,
   that is $H(hg, hp, ha_0)=H(g,p,a_0)$ for all $h\in G$, in which case the corresponding 
   Lie-Poisson equations read
   \begin{align}
  \dot{\mu}= {\rm ad}^*_{\frac{\delta h}{\delta \mu}}\,\mu - \frac{\delta h}{\delta a}\diamond a\,,\qquad \dot{a}=-\left(\frac{\delta h}{\delta 
  \mu}\right)_Ma \,.
\end{align}
\end{remark}

\begin{remark}[Lie-Poisson bracket]
  Upon using the Lie-Poisson equations for a (left/right)-invariant Hamiltonian in the relation 
  \begin{align}
    \dot{f}(\mu, a)=\left\langle \frac{\delta f}{\delta \mu}, \dot{\mu} 
    \right\rangle+\left\langle \frac{\delta f}{\delta a}, \dot{a} 
    \right\rangle\,,
  \end{align}
  and recalling the standard relation $\dot{f}=\{f,h\}$, it follows that the space $\mathfrak{g}^*\times M$ is a Poisson manifold 
  endowed with the {\bf Lie-Poisson bracket}
  \begin{align}\label{LPBDEF}
    \{f,g\}(\mu, a) = \mp \left\langle \mu, \left[\frac{\delta f}{\delta \mu}, \frac{\delta g}{\delta \mu}\right] \right\rangle 
    +\left\langle \frac{\delta f}{\delta a}\diamond a, \frac{\delta g}{\delta \mu} \right \rangle  
    -\left\langle \frac{\delta g}{\delta a}\diamond a, \frac{\delta f}{\delta \mu} \right \rangle  
    \,,
  \end{align}
  where the $\mp$ corresponds to a (left/right)-invariant Hamiltonian 
  respectively and $[\,\cdot\,,\,\cdot\,]$ denotes the Lie bracket on 
  $\mathfrak{g}$.
\end{remark}


\section{Momentum maps and collectivisation}\label{App:Momaps}
In this section, we introduce the general notation of momentum maps which are 
highly important in geometric mechanics as they underpin the geometry behind all key objects in 
physics.

\begin{definition}[Canonical group action]
Consider a Poisson manifold $(P, 
  \{\,\cdot\,,\,\cdot\,\})$ and Lie group $G$. Then, the group action $\Phi:G\times P\to P$ 
  is {\bf canonical} if it satisfies
  \begin{align}
    \Phi_g^*\{f,h\}= \{\Phi^*_g(f),\Phi_g^*(h)\}\,, \qquad \forall f,h \in 
    \mathcal{F}(P)\,,\qquad\forall \,g \in G\,,
  \end{align}
  where $\Phi_g^*$ defines the pull-back of the group action.
\end{definition}

\begin{definition}[Momentum map on a Poisson manifold]
  Let $G$ be a Lie group acting canonically (from the left) on a Poisson manifold $(P, 
  \{\,\cdot\,,\,\cdot\,\})$. A map $J:P\to\mathfrak{g}^*$ such that
  \begin{align}
    \{f, \braket{J(z),\xi}\}= \xi_P[f]\,,\qquad \forall f \in P\,,\qquad\forall \,\xi \in 
    \mathfrak{g}\,,\label{MOMAPGENPOISSON}
  \end{align}
  is called a {\bf momentum map}.
  \end{definition}

\begin{remark}[Momentum map on a symplectic vector space]
  In the specific case when the Poisson manifold is given by a symplectic vector 
  space $(V,\omega)$, the above formula for the momentum map $J:V\to\mathfrak{g}^*$ reduces to
  \begin{align}
    \braket{J(z),\xi}=\frac{1}{2}\omega(\xi_V(z),z)\,,\qquad \forall z\in V\,,\qquad\forall\xi \in \mathfrak{g}\,.\label{VSMomapFormula}
  \end{align}
\end{remark}

\begin{definition}[Equivariance]
  A momentum map $J:P\to\mathfrak{g}^*$ associated to the action of a Lie group $G$ 
  on a Poisson manifold $(P, 
  \{\,\cdot\,,\,\cdot\,\})$ is said to be {\bf equivariant} if
  \begin{align}
    J(\Phi_g(z))={\rm Ad}^*_{g^{-1}}(J(z))\,,\qquad\forall z\in P\,,\qquad 
    \forall g \in G\,.
  \end{align}
\end{definition}

\begin{remark}
Equivariant momentum maps are also Poisson maps
and hence define a Poisson isomorphism identifying the Poisson structure on the Poisson manifold
 with the Lie-Poisson bracket on the dual of the Lie algebra of the group:
\begin{align}
(P,  \{\,\cdot\,,\,\cdot\,\}_P) \to (\mathfrak{g}^*, \{\,\cdot\,,\,\cdot\,\}_{\mathfrak{g}^*})\,.
\end{align}
This is exactly what happens in the Lie-Poisson reduction seen earlier.
\end{remark}

\begin{definition}[Hamiltonian group action]
A (left) canonical Lie group action is called {\bf Hamiltonian} if it possesses 
an equivariant momentum map $J$.
\end{definition}

\paragraph{Collectivisation \cite{GuilleminSternberg1980,GuilleminSternberg1990}:}
Consider a (left) Hamiltonian action of a Lie group, with momentum map $J:P\to 
\mathfrak{g}^*$. Then, any function(al) {$h$} on $\mathfrak{g}^*$ gives rise to a 
function(al) $H = h\circ J$ on $P$ which is called a {\bf collective Hamiltonian} 
associated to the group action of $G$. The Hamiltonian $H$ is said to 
{\bf collectivise} through the momentum map $J$.

\chapter{Calculations for diffeomorphism actions}\label{App:HalfDen}
In this appendix we provide the detailed calculations resulting in the formulas 
for the infinitesimal generators and corresponding momentum maps used in 
Section \ref{Sec:2-HD} for wavefunctions and in Section \ref{Sec:2-ColdFluid} for density matrices. 
Throughout this appendix we use the standard definitions from geometric 
mechanics as given in Appendix \ref{App:GMTheory}.
\section{Action on half-densities}
Here we derive explicitly the expressions for the infinitesimal generator \eqref{Diff-action} and momentum map \eqref{QHDmomap} from the half-density action given in Section \ref{Sec:2-HD}.
For simplicity in the calculations that follow we consider the equivalent right 
action to \eqref{half-dens-action} by using the well-known fact that any left action produces a right action by replacing $\boldeta\to\boldeta^{-1}$. 
Hence, we consider the half-density group action from the right
\begin{align}
  \Phi_{\eta}(\psi)= \sqrt{\det \nabla 
  \boldeta(\bx)^T}\psi(\boldeta(\bx))\,.\label{halfDensRightAction}
\end{align}
\subsubsection{Infinitesimal generator}
Here, we compute the infinitesimal generator from the standard defintion, noting that $\boldeta(\bx,0)=\bx$ and 
$\dot\boldeta(\bx,0)=\bu(\bx)$:
\newpage
\begin{align}
  \begin{split}
  u_{\text{Den}^{1/2}}(\psi) 
 &= \frac{\text{d}}{\text{d}t}\Bigg|_{t=0}\sqrt{\det \nabla 
 \boldeta(\bx,t)^T}\psi(\boldeta(\bx,t))\\
 &= \frac{1}{2}(\det \nabla 
 \boldeta^T)^{-1/2}\frac{\text{d}}{\text{d}t}(\det \nabla 
 \boldeta^T)\psi(\boldeta(\bx,t))+ \sqrt{\det \nabla 
 \boldeta^T}\nabla_{\boldeta}\psi \cdot \dot{\boldeta}\Bigg|_{t=0}\\
   &= \frac{1}{2}(\det \nabla 
 \boldeta^T)^{-1/2}\det \nabla\boldeta^T \,\text{Tr}\left((\nabla\boldeta^T)^{-1}(\nabla\dot\boldeta^T)\right)\psi(\boldeta(\bx,t))+ \sqrt{\det \nabla 
 \boldeta^T}\nabla_{\boldeta}\psi \cdot \dot{\boldeta}\Bigg|_{t=0}\\
    &= \frac{1}{2}(\det \nabla 
 \bx^T)^{-1/2}\det \nabla\bx^T \,\text{Tr}\left((\nabla\bx^T)^{-1}(\nabla\bu^T)\right)\psi(\bx)+ \sqrt{\det \nabla 
 \bx^T}\nabla_{\bx}\psi \cdot \bu\\
 &=\frac{1}{2}(\nabla\cdot\bu)\psi+ \nabla\psi\cdot\bu\,,
\end{split}\label{HalfDensRightInfGen}
\end{align}
where we have used Jacobi's formula
\begin{align}
  \frac{\text{d}}{\text{d}t}\det A(t) &= \det A \,\text{Tr}(A^{-1}\dot{A})\,,\label{JacobiFormula}
\end{align}
as well as the facts that $\nabla\bx^T=\mathbbm{1}$ and 
$\text{Tr}(\nabla\bu^T)=\nabla\cdot\bu$. We notice that the last line returns 
minus the expression \eqref{Diff-action}. This is to be expected as we now 
consider the corresponding right action to \eqref{half-dens-action} and is 
compensated for in the general formula for the momentum map in what follows.
\subsubsection{Momentum map}
Finally we compute the corresponding momentum map, according to the vector space formula for a {right action} 
(which is minus that for a left action)
\begin{align}
  \braket{\boldsymbol{J}(\psi),\bu}&= 
  -\frac{1}{2}\omega(u_{\text{Den}^{1/2}}(\psi),\psi)\,,
\end{align}
which, as we have seen, for the standard symplectic form on a Hilbert space \eqref{QMsympform} becomes 
\eqref{QMVSmomap}:
\begin{align}
  \braket{\boldsymbol{J}(\psi),\bu}&= 
  -\hbar\text{Im}\braket{u_{\text{Den}^{1/2}}(\psi)|\psi}\,.
  \end{align}
  Upon using the above equation for the infinitesimal generator, we obtain
  \begin{align}
    \begin{split}
       \braket{\boldsymbol{J}(\psi),\bu}&= 
  -\hbar\text{Im}\braket{u_{\text{Den}^{1/2}}(\psi)|\psi}\\
  &= -\hbar\text{Im}\int \left(\frac{1}{2}(\nabla\cdot\bu)\psi+ \nabla\psi\cdot\bu\right)^*\psi\,\text{d}^3 
  x\\
    &= -\hbar\text{Im}\int \left(-\frac{1}{2}\bu\cdot \nabla|\psi|^2+ \bu\cdot(\psi \nabla\psi^*)\right)\,\text{d}^3 
  x\\
      &= -\hbar\text{Im}\int \left(-\frac{1}{2}\bu\cdot (\nabla\psi^* \psi + \psi^*\nabla\psi)+ \bu\cdot(\psi \nabla\psi^*)\right)\,\text{d}^3 
  x\\
   &= \text{Im}\int \bu\cdot -\frac{\hbar}{2}(\nabla\psi^* \psi - \psi^*\nabla\psi)\,\text{d}^3 
  x\\
     &= \text{Re}\int \bu\cdot \frac{\hbar}{2i}(\psi^*\nabla\psi-\nabla\psi^* \psi)\,\text{d}^3 
  x\,,
    \end{split}
  \end{align}
  thus returning the momentum map $\boldsymbol{J}(\psi) = \hbar\text{Im}(\psi^*\nabla\psi)$ in agreement with \eqref{QHDmomap} as required.
\subsubsection{Operatorial notation}
Many of the calculations above can be expressed more concisely upon adopting the 
operatorial notation used in the final equality of \eqref{Diff-action}. Specifically, 
we denote $\widehat{P}_k:=-i\hbar\partial_k$ for the momentum operator, $\widehat{u}^{\,k\!}$ for the multiplicative operator associated to $u^k(\boldsymbol{x})$ and use the anticommutator notation 
$\{A,B\}_+:=AB+BA$. Then, the infinitsimal generator can be equivalently written as
\begin{align}
    u_{\text{Den}^{1/2}}(\psi) = 
    \frac{i}{2\hbar}\{\widehat{P}_k,\widehat{u}^k\}_+\psi\,.\label{HalfDenInfGenOp}
\end{align}
This can be verified as follows:
\begin{align}
 \begin{split} 
 \frac{i}{2\hbar}\{\widehat{P}_k,\widehat{u}^k\}_+\psi &=   \frac{i}{2\hbar}\left(\widehat{P}_k\widehat{u}^k+ 
  \widehat{u}^k\widehat{P}_k\right)\psi\\
  &= \frac{i}{2\hbar}\left(\widehat{P}_k u^k \psi + 
  \widehat{u}^k(-i\hbar\partial_k\psi)\right)\\
    &= \frac{i}{2\hbar}\left(-i\hbar\partial_k u^k \psi  -i\hbar u^k \partial_k\psi  
  -i\hbar u^k\partial_k\psi\right)\\
  &=\frac{1}{2}\left(\partial_k u^k \psi  + 2 u^k\partial_k\psi\right)\\
  &= \frac{1}{2}(\nabla\cdot\bu)\psi+ 
  \bu\cdot\nabla\psi\,.
\end{split}
   \end{align}
Similarly, the momentum map can be computed in this notation
\begin{align}
  \begin{split}
  \braket{\boldsymbol{J}(\psi),\bu}&=-\hbar\text{Im}\braket{u_{\text{Den}^{1/2}}(\psi)|\psi}\\
  &= 
    -\hbar\text{Im}\left\langle 
    \frac{i}{2\hbar}\{\widehat{P}_k,\widehat{u}^k\}_+\psi\Bigg|\psi\right\rangle\\
    &= \left\langle \widehat{u}^k, 
    \frac{1}{2}\{\widehat{P}_k,\psi\psi^{\dagger}\}_+\right\rangle\,,
  \end{split}
\end{align}
from which we read off the momentum map $\boldsymbol{J}(\psi) = 
\frac{1}{2}\{\widehat{P},\psi\psi^{\dagger}\}_+(\bx,\bx)$. Here, the diagonal matrix elements arise when going from the operator $\widehat{u}^k$ to the vector $\bu$ via the relation $\widehat{u}^k(\bx,\bx')=\bu(\bx)\delta(\bx-\bx')$.
Indeed, this version of the momentum map can be shown to agree with the more explicit expression upon going back to matrix 
elements
\begin{align}
  \begin{split}
    \frac{1}{2}\{\widehat{P},\psi\psi^{\dagger}\}_+(\bx,\bx) &= 
    \frac{1}{2}\Big(\braket{\bx|\widehat{P}|\psi}\!\!\braket{\psi|\bx}+\braket{\bx|\psi}\!\!\braket{\psi|\widehat{P}|\bx}\Big)\\
&=     \frac{1}{2}\Big(\braket{\bx|\widehat{P}|\psi}\!\!\braket{\psi|\bx}+\braket{\bx|\psi}\!\!\braket{\bx|\widehat{P}|\psi}^*\Big)\\
  &=  \frac{1}{2}\Big((-i\hbar\nabla_{}\psi(\bx))\psi^*(\bx)+\psi(\bx)(-i\hbar\nabla_{}\psi(\bx))^*\Big)\\
  &=  \frac{1}{2}\left(-i\hbar\psi^*\nabla \psi+i\hbar\psi\nabla\psi^*\right)\\
  &= -\frac{i\hbar}{2}\left(\psi^*\nabla \psi-\psi\nabla\psi^*\right)\\
  &= \hbar\text{Im}\left(\psi^*\nabla \psi\right)
  \end{split}
\end{align}
as required.

\section{Action on density matrices}\label{App:HalfDen2}
Here, we prove formula \eqref{DiffeoDensityMatrixInfGen} for the infinitesimal generator and \eqref{DiffeoDensityMatrixMomap} for the momentum map. 
Instead of considering the left action  in \eqref{DiffeoDensityMatrixAction}, we simplify the treatment here by again considering the corresponding right action (given by replacing $\boldeta\to\boldeta^{-1}$)
\begin{align}
  \Phi_{\eta}(\rho)= \sqrt{\det \left(\nabla_{\bx} \boldeta(\bx)^T \nabla_{\bx'} 
  \boldeta(\bx')^T\right)}\rho(\boldeta(\bx),\boldeta(\bx'))\label{RightDiffDensityMatrixAction}\,.
\end{align}
Furthermore, to simplify the computations we consider the following unitary representation of this action:
\begin{align}\label{gino}
   \Phi_{\eta}(\rho)=  U_{\eta}\,\rho \,  U_{\eta}^{\dagger}\,,
\end{align}
for the unitary operator (in matrix element notation)
\begin{align}
    U_{\eta}(\bx,\bx')&= \sqrt{\det \nabla_{\bx} \boldeta(\bx)^T}\,\delta(\bx' - 
  \boldeta(\bx))\,.
\end{align}
This can be shown to reproduce \eqref{RightDiffDensityMatrixAction} upon expanding  
$(  U_{\eta}\,\rho \,  U_{\eta}^{\dagger})(\bx,\bx')$ and using the matrix elements explicitly. 
This representation of the action can also be applied for pure states, in which 
case the action \eqref{halfDensRightAction} is written as $\psi \mapsto   U_{\eta}\psi$.

\subsubsection{Infinitesimal generator}
At this point we compute the infinitesimal generator from its definition to obtain
\begin{align}
  \begin{split}
    u(\rho) &= \frac{\text{d}}{\text{d}t}\Bigg|_{t=0}\Phi_{\eta(t)}(\rho)\\
    &=  \frac{\text{d}}{\text{d}t}\Bigg|_{t=0}  U_{\eta}(t)\,\rho 
    \,  U_{\eta}^{\dagger}(t)\\
    &= \Big[\dot{U}_{\boldeta}\,\rho 
    \,  U_{\eta}^{\dagger} +   U_{\eta}\,\rho 
    \,\frac{\text{d}}{\text{d}t}  U_{\eta}^{-1}\Big]_{t=0}\\
    &=\Big[ \dot{U}_{\boldeta}  U_{\eta}^{-1}  U_{\eta}\,\rho 
    \,  U_{\eta}^{\dagger} +   U_{\eta}\,\rho 
    \,\left(-  U_{\eta}^{-1}\dot{U}_{\boldeta}  U_{\eta}^{-1}\right)\Big]_{t=0}\\
    &= \left[\hat\xi, \rho\right]\,,
  \end{split}\label{gino-inf}
\end{align}
as $  U_{\eta}(0)=\mathbbm{1}$ and $\dot{U}_{\boldeta}(0):= \widehat{\xi}$. Again, one can use this formalism in the case of pure states, in which case 
    the infinitesimal generator \eqref{HalfDenInfGenOp} is simply written as $u(\psi)=\widehat{\xi}\psi$, 
    so we can make the identification
    \begin{align}
      \widehat\xi=\frac{i}{2\hbar}\{\widehat{P}_k,\widehat{u}^k\}_+\,.\label{Harry}
    \end{align}
Indeed, we verify this equality explicitly using matrix elements. 

Firstly, for $\widehat\xi$, we consider a curve $\boldsymbol{\eta}(t)\in \text{Diff}(\mathbb{R}^3)$ such that $\boldsymbol{\eta}(0)=\boldsymbol{1}$ and $\dot{\boldsymbol{\eta}}(0)=\boldsymbol{u}$. Then, we 
compute
\begin{align}
  \begin{split}  
 \widehat\xi(\bx,\bx')&= \frac{\text{d}}{\text{d}t}\Bigg|_{t=0}(  U_{\eta}(t))(\bx,\bx')\\
  &=\Bigg[ \frac{\delta(\bx' - 
  \boldeta(\bx,t))}{2\sqrt{\det \nabla_{\bx} \boldeta(\bx,t)^T}} \,\frac{\text{d}}{\text{d}t}
 \big(\!\det \nabla_{\bx} \boldeta(\bx,t)^T\big) + \sqrt{\det \nabla_{\bx} \boldeta(\bx,t)^T}\,\frac{\text{d}}{\text{d}t}\delta(\bx' - 
  \boldeta(\bx,t))\Bigg]_{t=0}\\
    &=\Bigg[\sqrt{\det \nabla_{\bx} \boldeta(\bx,t)^T}\Big(\frac{1}{2} 
  \delta(\bx' - 
  \boldeta(\bx,t))- \dot{\boldeta}(\bx,t)\cdot\nabla_{\bx'}\delta(\bx' - 
  \boldeta(\bx,t)) \Big)\Bigg]_{t=0}\\
    &= \frac{1}{2}\left(\nabla_{\bx}\cdot\bu(\bx)\right)\delta(\bx'-\bx) 
 +\nabla_{\bx}\delta(\bx'-\bx)\cdot \bu(\bx)
 \,,
  \end{split}  
\end{align}
where the third step again uses Jacobi's formula \eqref{JacobiFormula}. 
Equivalently, we compute the matrix elements of $\widehat\xi={i}{\hbar}^{-1}\{\widehat{u}^k,\widehat{P}_k\}_+/2$ as follows
\begin{align}
  \begin{split}  
\frac{i}{2\hbar}\{\widehat{u}^k,\widehat{P}_k\}_+(\bx,\bx') &= \frac{i}{2\hbar}\int 
\widehat{u}^k(\bx,\by)\widehat{P}_k(\by,\bx') + \widehat{P}_k(\bx,\by)\widehat{u}^k(\by,\bx')\,\text{d}^3 
y\\
&= \frac{i}{2\hbar}\int 
i\hbar\delta(\by-\bx')\nabla_{\by_k}\widehat{u}^k(\bx,\by) -i\hbar \delta(\bx - \by) \nabla_{\by_k}\widehat{u}^k(\by,\bx')\,\text{d}^3 
y\\
&=  \frac{i}{2\hbar}\left(i\hbar\nabla_{\bx'_k}\widehat{u}^k(\bx,\bx')-i\hbar  
\nabla_{\bx_k}\widehat{u}^k(\bx,\bx')\right)\\
&=  \frac{i}{2\hbar}\left(i\hbar\nabla_{\bx'_k}\big(u^k(\bx)\delta(\bx-\bx')\big)-i\hbar  
\nabla_{\bx_k}\big(u^k(\bx)\delta(\bx-\bx')\big)\right)\\
&= \frac{1}{2}(\nabla_{\bx}\cdot\bu(\bx))\delta(\bx-\bx') + \bu(\bx)\cdot\nabla_{\bx}\delta(\bx-\bx')
\,,
  \end{split}  
\end{align}
where we have used the matrix elements
\begin{align}
\widehat{u}(\bx,\bx')=\bu(\bx)\delta(\bx-\bx')
\,,\qquad
\widehat{{P}}(\bx,\bx')=-i\hbar\nabla_{\bx}\delta(\bx-\bx')
\,.
\end{align}
Hence, having proven the equality \eqref{Harry}, we can equivalently write the infinitesimal generator for the left action \eqref{DiffeoDensityMatrixAction} as $u(\rho) 
    =- [\widehat\xi, \rho]$ so that 
    \begin{align}
u(\rho)=-\frac{i}{2\hbar}\left[\{\widehat{u}^k,\widehat{P}_k\}_{+\,},\rho\right]\,, 
    \end{align}
    in agreement with \eqref{DiffeoDensityMatrixInfGen}.
    
\subsubsection{Momentum map}
Here, we will prove that \eqref{DiffeoDensityMatrixMomap} given by
\begin{align}
  \boldsymbol{J}(\rho) = \frac{1}{2}\{\widehat{P},\rho\}_+(\bx,\bx)
\end{align} is indeed a momentum map corresponding to the action 
\eqref{DiffeoDensityMatrixAction}.
 However, the computation is very different from that used for wavefunctions earlier. We now have a density matrix $\rho$, for which $i\hbar\rho \in 
\mathfrak{u}(\mathscr{H})^*$, which does not live in a vector space and hence we 
must use the more general formula for a momentum map on a Poisson manifold \eqref{MOMAPGENPOISSON} (having changed sign to account for a right action), 
given by
\begin{align}
   \{F,\braket{\boldsymbol{J}(\rho)|\bu}\} &= -u(F)\,,
\end{align}
where the Poisson bracket on left is the Lie-Poisson bracket \eqref{LPBDens} and the 
on the right we have the infinitesimal generator associated to the right action. 
Expanding both sides yields
\begin{align}
\left\langle \frac{\delta F}{\delta 
   \rho}, \left[\frac{\delta \braket{\boldsymbol{J}(\rho)|\bu}}{\delta \rho},-i\hbar^{-1}\rho\right]
   \right\rangle &= \left\langle \frac{\delta F}{\delta \rho}, -u(\rho) 
   \right\rangle\,,
\end{align}
so that the momentum map $\boldsymbol{J}(\rho)$ must satisfy
\begin{align}
  \left[\frac{\delta \braket{\boldsymbol{J}(\rho)|\bu}}{\delta \rho},i\hbar^{-1}\rho\right] 
  = u(\rho)\, .\label{JAMES}
\end{align}
Hence, we first compute
\begin{align}
  \begin{split}
    \braket{  \boldsymbol{J}(\rho)|\bu} &= 
    \frac{1}{2}\braket{\{\widehat{P}_k,\rho\}_+|\widehat{u}^k}\\
    &=   \left\langle\rho\Bigg|\frac{1}{2}\{\widehat{P}_k,\widehat{u}^k\}_+\right\rangle
  \end{split}
\end{align}
which allows us to easily read off
\begin{align}
  \frac{\delta  \braket{  \boldsymbol{J}(\rho)|\bu}}{\delta \rho} 
  =\frac{1}{2}\{\widehat{P}_k,\widehat{u}^k\}_+\,.
\end{align}
Inserting this into the LHS of our momentum map condition \eqref{JAMES} yields
\begin{align}
  \begin{split}
    \left[ \frac{\delta  \braket{  \boldsymbol{J}(\rho)|\bu}}{\delta \rho} ,i\hbar^{-1}\rho\right] 
    &=  \left[\frac{1}{2}\{\widehat{P}_k,\widehat{u}^k\}_+ ,i\hbar^{-1}\rho\right] \\
    &= \left[\frac{i}{2\hbar}\{\widehat{P}_k,\widehat{u}^k\}_+,\rho\right]=u(\rho) 
    \,,
  \end{split}
\end{align} 
as required, thus proving that $\boldsymbol{J}(\rho)$ is a momentum map.

\section{Semidirect product action}\label{App:HalfDen3}
We conclude this appendix by presenting the explicit proof of momentum map $(\bmu, D)$ associated to the left action of 
the full semidirect product group $\text{Diff}(\mathbb{R}^3)\,\circledS\,\mathcal{F}(\mathbb{R}^3,S^1)$ on the space of half-densities $\text{Den}^{1/2}(\mathbb{R}^3)$ 
given by \eqref{SEMIhalf-dens-action}
\begin{align}
  \Phi_{(\eta, \varphi)}(\psi) &=\frac{1}{\sqrt{\det \nabla 
 \boldeta(\bx)^T}}e^{-i\hbar^{-1}\varphi(\boldeta^{-1}(\bx))}\,\psi(\boldeta^{-1}(\bx))\,.
\end{align}

\subsubsection{Infinitesimal generator}
Here, we compute the infinitesimal generator, noting that we have chosen $\boldeta(\bx,0)=\boldeta^{-1}(\bx,0)=\bx, \dot\boldeta(\bx,0)=\bu(\bx)$ and $\varphi(\bx,0)=0, 
\dot\varphi(\bx,0)=\alpha(\bx)$. Using the standard defintion \eqref{Def:InfGen}, a direct calculation yields 
\begin{align}
  \begin{split}
  u_{\text{Den}^{1/2}}(\psi) &= \frac{\text{d}}{\text{d}t}\Bigg|_{t=0}\Phi_{(\eta_t, \varphi_t)}(\psi)\\
 &= \frac{\text{d}}{\text{d}t}\Bigg|_{t=0}\left[(\det \nabla 
  \boldeta(\bx,t)^T)^{-1/2}e^{-i\hbar^{-1}\varphi(\boldeta^{-1}(\bx,t),t)}\,\psi(\boldeta^{-1}(\bx,t))\right]\\
  &=\Bigg[-\frac{1}{2}(\det \nabla 
  \boldeta^T)^{-3/2}\frac{\text{d}}{\text{d}t}(\det \nabla 
  \boldeta^T)e^{-i\hbar^{-1}\varphi(\boldeta^{-1}(\bx,t),t)}\,\psi(\boldeta^{-1}(\bx,t))\\
  &\qquad+(\det \nabla 
  \boldeta^T)^{-1/2}\cdot -\frac{i}{\hbar}e^{-i\hbar^{-1}\varphi(\boldeta^{-1}(\bx,t),t)}\frac{\text{d}}{\text{d}t}(\varphi(\boldeta^{-1}(\bx,t),t))\,\psi(\boldeta^{-1}(\bx,t))\\
  &\qquad+ (\det \nabla 
  \boldeta^T)^{-1/2}e^{-i\hbar^{-1}\varphi(\boldeta^{-1}(\bx,t),t)} \nabla_{\boldeta^{-1}}\psi\cdot 
  \dot\boldeta^{-1}\Bigg]_{t=0}\\
  &= \Bigg[-\frac{1}{2}(\det \nabla 
  \boldeta^T)^{-3/2} (\det \nabla 
  \boldeta^T) 
  \text{Tr}\left((\nabla\boldeta^T)^{-1}\frac{\text{d}}{\text{d}t}(\nabla\boldeta^T)\right)e^{-i\hbar^{-1}\varphi(\boldeta^{-1}(\bx,t),t)}\psi(\boldeta^{-1}(\bx,t))\\
  &\qquad+ (\det \nabla 
  \boldeta^T)^{-1/2}\cdot 
  -\frac{i}{\hbar}e^{-i\hbar^{-1}\varphi(\boldeta^{-1}(\bx,t),t)}(\nabla_{\boldeta^{-1}}\varphi\cdot\dot\boldeta^{-1}+\dot\varphi)\,\psi(\boldeta^{-1}(\bx,t))\\
  &\qquad+(\det \nabla 
  \boldeta^T)^{-1/2}e^{-i\hbar^{-1}\varphi(\boldeta^{-1}(\bx,t),t)} \nabla_{\boldeta^{-1}}\psi\cdot 
  \dot\boldeta^{-1}\Bigg]_{t=0}\\
    &= \Bigg[-\frac{1}{2}(\det \nabla 
  \boldeta^T)^{-3/2} (\det \nabla 
  \boldeta^T) 
  \text{Tr}\left((\nabla\boldeta^T)^{-1}(\nabla\dot\boldeta^T)\right)e^{-i\hbar^{-1}\varphi(\boldeta^{-1}(\bx,t),t)}\psi(\boldeta^{-1}(\bx,t))\\
  &\qquad+ (\det \nabla 
  \boldeta^T)^{-1/2}\cdot 
  -\frac{i}{\hbar}e^{-i\hbar^{-1}\varphi(\boldeta^{-1}(\bx,t),t)}(\nabla_{\boldeta^{-1}}\varphi\cdot(-\nabla_{\boldeta}\boldeta^{-1}\cdot\dot\boldeta)+\dot\varphi)\,\psi(\boldeta^{-1}(\bx,t))\\
  &\qquad+(\det \nabla 
  \boldeta^T)^{-1/2}e^{-i\hbar^{-1}\varphi(\boldeta^{-1}(\bx,t),t)} \nabla_{\boldeta^{-1}}\psi\cdot 
 (-\nabla_{\boldeta}\boldeta^{-1}\cdot\dot\boldeta)\Bigg]_{t=0}\\
     &= -\frac{1}{2}(\det \nabla 
  \bx^T)^{-3/2} (\det \nabla 
  \bx^T) 
  \text{Tr}\left((\nabla\bx^T)^{-1}(\nabla\bu^T)\right)e^{-i\hbar^{-1}\varphi(\bx,0)}\psi(\bx)\\
  &\qquad+ (\det \nabla 
  \bx^T)^{-1/2}\cdot 
  -\frac{i}{\hbar}e^{-i\hbar^{-1}\varphi(\bx,0)}(\nabla_{\bx}\varphi(\bx,0)\cdot(-\nabla_{\bx}\bx\cdot\bu)+\alpha)\,\psi(\bx)\\
  &\qquad+(\det \nabla 
  \bx^T)^{-1/2}e^{-i\hbar^{-1}\varphi(\bx,0)} \nabla_{\bx}\psi\cdot 
 (-\nabla_{\bx}\bx\cdot\bu)\\
 &= -\frac{1}{2}(\nabla\cdot\bu)\psi -\frac{i}{\hbar}\alpha\psi- \nabla\psi\cdot 
 \bu\,,
\end{split}
\end{align}
in which we have again used Jacobi's formula as well as that $\nabla\bx^T=\mathbbm{1}$ and 
$\text{Tr}(\nabla\bu^T)=\nabla\cdot\bu$. Also, we have 
used that as $\boldeta^{-1}(\boldeta(\bx,t),t)=x$, then
\newpage
\begin{align}
  \begin{split}
\frac{\text{d}}{\text{d}t}[\boldeta^{-1}(\boldeta(\bx,t),t)]&= 0 \\
\implies \nabla_{\boldeta}\boldeta^{-1}\cdot \dot\boldeta + \frac{\text{d}}{\text{d}t}\boldeta^{-1} 
&= 0\\
\implies \frac{\text{d}}{\text{d}t}\boldeta^{-1} &= - \nabla_{\boldeta}\boldeta^{-1}\cdot \dot\boldeta 
\\
\implies\frac{\text{d}}{\text{d}t}\Bigg|_{t=0}\boldeta^{-1} &= - \bu(\bx)  \,.
  \end{split}
\end{align}
\subsubsection{Momentum map}
Here we compute the corresponding momentum map, according to the standard vector space formula 
for a left action on a Hilbert space \eqref{QMVSmomap}. Therefore, using the above equation for the infinitesimal generator, we obtain
  \begin{align}
    \begin{split}
       \braket{\boldsymbol{J}(\psi),\bu}&= 
  \hbar\text{Im}\braket{u_{\text{Den}^{1/2}}(\psi)|\psi}\\
  &= \hbar\text{Im}\int \bigg(-\frac{1}{2}(\nabla\cdot\bu)\psi- \nabla\psi\cdot\bu -\frac{i}{\hbar}\alpha\psi\bigg)^*\psi\,\text{d}^3 
  x\\
    &= \hbar\text{Im}\int \bigg(\frac{1}{2}\bu\cdot \nabla|\psi|^2- \bu\cdot(\psi \nabla\psi^*) + \frac{i}{\hbar}\alpha |\psi|^2\bigg)\,\text{d}^3 
  x\\
      &= \hbar\text{Im}\int \bigg(-\frac{1}{2}\bu\cdot (\nabla\psi^* \psi + \psi^*\nabla\psi)- \bu\cdot(\psi \nabla\psi^*)+ \frac{i}{\hbar}\alpha |\psi|^2\bigg)\,\text{d}^3 
  x\\
   &= \text{Im}\int \bu\cdot -\frac{\hbar}{2}(\nabla\psi^* \psi - \psi^*\nabla\psi) + i\alpha|\psi|^2\,\text{d}^3 
  x\\
     &= \text{Re}\int \bu\cdot \frac{\hbar}{2i}(\psi^*\nabla\psi-\nabla\psi^* \psi)\,\text{d}^3 
  x + \text{Re}\int\alpha |\psi|^2\,\text{d}^3x\,,
    \end{split}
  \end{align}
  thus giving that $\boldsymbol{J}(\psi) =( \hbar\text{Im}(\psi^*\nabla\psi), |\psi|^2)=:(\bmu, D)$ as 
  required.

\chapter{Cold fluid classical closures}\label{App:ColdFluid}
In this appendix we prove some of the results from Sections 
\ref{Sec:2-ColdFluid} and \ref{sec:EFCFClosure} regarding the cold fluid closures for density matrices.
To begin we simply prove that the cold fluid closure \eqref{coldfluid} 
\begin{align}
\rho(\bx,\bx')&=D\Big(\frac{\bx+\bx'}{2}\Big)\exp\!\left[i\frac{m}\hbar 
(\bx-\bx')\cdot\bu\Big(\frac{\bx+\bx'}{2}\Big)\right],
\end{align}
returns 
the expected relations $\rho(\bx,\bx)=D(\bx)$ and 
$\bmu(x):=\{\widehat{P},\rho\}_+(\bx,\bx)/2=mD(\bx)\bu(\bx)$. Whilst the first result 
follows simply observation, the second is proven as follows. Firstly, we simply expand the definition 
to reveal
\begin{align}
  \begin{split}
  \bmu(\bx):=\frac12\{\widehat{P},\rho\}_+(\bx,\bx) &= \frac{1}{2}\Big(\braket{\bx|\widehat{P}\rho|\bx} 
    +\braket{\bx|\rho\widehat{P}|\bx}\Big)\\
    &= \frac{1}{2}\int \Big[\braket{\bx|\widehat{P}|\bx'}\!\!\braket{\bx'|\rho|\bx}+ 
    \braket{\bx|\rho|\bx'}\!\!\braket{\bx'|\widehat{P}|\bx}\Big]\,\de^3x'\\
    &= \frac{\hbar}{2i}\int \Big[\nabla_{\bx}\delta(\bx-\bx')\rho(\bx',\bx)+
\rho(\bx,\bx')\nabla_{\bx'}\delta(\bx'-\bx)\Big]\,\de^3x'\\
    &= \frac{\hbar}{2i}\int \Big[-\nabla_{\bx'}\delta(\bx-\bx')\rho(\bx',\bx)+
\rho(\bx,\bx')\nabla_{\bx'}\delta(\bx'-\bx)\Big]\,\de^3x'\\
    &= \frac{\hbar}{2i} \Big[\nabla_{\bx'}\Big(\rho(\bx',\bx)-
\rho(\bx,\bx')\Big)\Big]_{\bx'=\bx}\,.
  \end{split}
\end{align}
At this point, we introduce the cold fluid closure \eqref{coldfluid} so that
\begin{align}
  \rho(\bx',\bx)-
\rho(\bx,\bx') = -2i 
D\left(\frac{\bx+\bx'}{2}\right)\sin\left(\frac{m}{\hbar}(\bx-\bx')\cdot\bu\left(\frac{\bx+\bx'}{2}\right)\right)\,.
\end{align}
Then, subtituting this expression back into $\bmu$ we obtain
\begin{align}
 \begin{split} 
\bmu &= 
  -\hbar\left[\nabla_{\bx'}\left(D\left(\frac{\bx+\bx'}{2}\right)\sin\left(\frac{m}{\hbar}(\bx-\bx')\cdot\bu\left(\frac{\bx+\bx'}{2}\right)\right)\right)\right]_{\bx'=\bx}\\
  &=  -\hbar\Bigg[\nabla_{\bx'}D\left(\frac{\bx+\bx'}{2}\right)\,\sin\left(\frac{m}{\hbar}(\bx-\bx')\cdot\bu\left(\frac{\bx+\bx'}{2}\right)\right)+D\left(\frac{\bx+\bx'}{2}\right)\cdot\\
  & \cos\left(\frac{m}{\hbar}(\bx-\bx')\cdot\bu\left(\frac{\bx+\bx'}{2}\right)\right)\nabla_{\bx'}\left(\frac{m}{\hbar}(\bx-\bx')\cdot\bu\left(\frac{\bx+\bx'}{2}\right)\right) \Bigg]_{\bx'=\bx}\\
    &= mD(\bx)\bu(\bx)\,,
    \end{split}
\end{align}
as required, having used the result
\begin{align}
  \nabla_{\bx'}\left((\bx-\bx')\cdot\bu\left(\frac{\bx+\bx'}{2}\right)\right) = 
  (x_j-x'_j)\nabla_{\bx'}u_j\left(\frac{\bx+\bx'}{2}\right) - 
  \bu\left(\frac{\bx+\bx'}{2}\right)\,.\label{Jonesy}
\end{align}

Next, we prove that the cold fluid closure for nuclear density matrix \eqref{coldfluidEF} transforms the generalised exact factorization Hamiltonian from \eqref{EF-HamNuclearRho} into the form 
\eqref{EFMixedHydroHamiltonian}.
Specifically, we only need to prove that 
\begin{align}
  \frac{1}{2M}\,\left\langle(\widehat{P}_n+ 
  \widehat{\cal A}\,)^2\Big|\rho_n\right\rangle = \int \frac{|\bmu+ 
  D\boldsymbol{\cal A}|^2}{2MD} \,\de^3 r\,.\label{Jimmy}
\end{align}
Notice that, upon specialising to the standard QHD case, the derivation below can be applied to the work in Section \ref{Sec:2-ColdFluid} thus proving that the cold fluid closure \eqref{coldfluid} results in the Hamiltonian \eqref{COLDFLUIDHAM}.  
To proceed, we work term-by-term. Firstly, we compute  
$\braket{\widehat{P}_n^2|\rho}$ for an arbitrary density matrix, expressed in 
polar form as $\rho(\br,\br')=\widetilde{D}(\br,\br')\,\text{exp}[i\hbar^{-1}\Phi(\br,\br')]$. 
Then,
\begin{align}
  \braket{\widehat{P}_n^2|\rho}&= \iint \rho(\br',\br)\cdot 
  -\hbar^2\Delta_{\br'}\delta(\br-\br')\,\de^3r\,\de^3r'= -\hbar^2\int \Big[\Delta_{\br}\rho(\br,\br')\Big]_{\br'=\br}\,\de^3r\,.\label{Mike}
\end{align}
Hence, we compute
\begin{align}
  \Delta_{\br}\left(\widetilde{D}(\br,\br')e^{i\hbar^{-1}\Phi(\br,\br')}\right) = 
  \left(\Delta_{\br}\widetilde{D} + \frac{2i}{\hbar}\nabla_{\br}\widetilde{D}\cdot \nabla_{\br}\Phi + \frac{i}{\hbar}\widetilde{D}\Delta_{\br}\Phi - 
  \hbar^{-2}\widetilde{D}|\nabla_{\br}\Phi|^2\right)e^{i\hbar^{-1}\Phi(\br,\br')}\,.
\end{align}
Upon substituting this expression back into the kinetic energy, noting that the 
middle two terms vanish as they are purely imaginary, we obtain
\begin{align}
    \braket{\widehat{P}_n^2|\rho} = \int \left[-\hbar^2 \Delta_{\br}\widetilde{D} + \widetilde{D}|\nabla_{\br}\Phi|^2 \right]_{\br'=\br}\,\de^3r\,.
\end{align}
Returning back to the cold fluid closure we compute
\begin{align}
  \left[\Delta_{\br}{D}\left(\frac{\br+\br'}{2}\right)\right]_{\br'=\br}&= \frac{1}{4}\left[\Delta_{\frac{1}{2}\left({\br+\br'}\right)}{D}\left(\frac{\br+\br'}{2}\right)\right]_{\br'=\br} 
  = \frac{1}{4}\Delta_{\br}D(\br)\,,
\end{align}
and use \eqref{Jonesy}, specialising the kinetic energy to
\begin{align}
    \braket{\widehat{P}_n^2|\rho} = \int M^2 D |\bv|^2\,\text{d}^3r= \int 
    \frac{|\bmu|^2}{D}\,\de^3r\,,
\end{align}
where boundary terms are assumed to vanish. Next, we evaluate the term
\begin{align}
  \begin{split}
  \braket{\widehat{P}_n\widehat{\cal A}|\rho_n}&= \iint\rho(\br',\br)(\widehat{P}_n\widehat{\cal 
  A})(\br,\br')\,\de^3r\,\de^3r'\\
&= \iint\rho(\br',\br)\int\widehat{P}_n(\br,\bs)\widehat{\cal A}(\bs,\br')\,\de^3s\,\de^3r\,\de^3r'\\
&= \iint\rho(\br',\br)\int-i\hbar\nabla_{\br}\delta(\br-\bs)\cdot\boldsymbol{\cal A}(\bs)\delta(\bs-\br')\,\de^3s\,\de^3r\,\de^3r'\\
&= -i\hbar\iint\rho(\br',\br)\nabla_{\br}\delta(\br-\br')\cdot\boldsymbol{\cal A}(\br')\,\de^3r\,\de^3r'\\
&= i\hbar\iint\nabla_{\br}\rho(\br',\br)\cdot\boldsymbol{\cal A}(\br')\delta(\br-\br')\,\de^3r\,\de^3r'\\
&= i\hbar\int \left(\frac{1}{2}\nabla_{\br} D(\br) - \frac{iM}{\hbar}D(\br) \bv(\br) \right)\cdot\boldsymbol{\cal 
A}(\br)\,\de^3r\,,
  \end{split}
\end{align}
where in the penultimate line we have used the cold fluid closure \eqref{coldfluid} 
to expand the derivative and then integrate over the remaining delta function. 
The analogous calculation yields
\begin{align}
    \braket{\widehat{\cal A}\widehat{P}_n|\rho_n}&=-i\hbar\int \left(\frac{1}{2}\nabla_{\br} D(\br) + \frac{iM}{\hbar}D(\br) \bv(\br) \right)\cdot\boldsymbol{\cal 
A}(\br)\,\de^3r\,,
\end{align}
so that together we have
\begin{align}
   \braket{\widehat{\cal A}\widehat{P}_n+\widehat{P}_n\widehat{\cal A} |\rho_n} 
   = 2\int MD\bv\cdot\boldsymbol{\cal A}\,\de^3r =2\int \bmu\cdot\boldsymbol{\cal A}\,\de^3r  \,.
\end{align}
For the final term, we compute
\begin{align}
  \begin{split}
  \braket{\widehat{\cal A}^2|\rho_n}&=\iint\rho(\br',\br)(\widehat{\cal A}\widehat{\cal 
  A})(\br,\br')\,\de^3r\,\de^3r'\\
&= \iint\rho(\br',\br)\int\widehat{\cal A}(\br,\bs)\widehat{\cal A}(\bs,\br')\,\de^3s\,\de^3r\,\de^3r'\\
&= \iint\rho(\br',\br)\int\boldsymbol{\cal A}(\br)\delta(\br-\bs)\cdot\boldsymbol{\cal 
A}(\bs)\delta(\bs-\br')\,\de^3s\,\de^3r\,\de^3r'\\
&= \iint\rho(\br',\br)\boldsymbol{\cal A}(\br)\cdot\boldsymbol{\cal 
A}(\br')\delta(\br-\br')\,\de^3r\,\de^3r'\\
&= \int D(\br)|\boldsymbol{\cal A}(\br)|^2\,\de^3r\,.
  \end{split}
\end{align}
Hence, all together we have that 
\begin{align}
\left\langle(\widehat{P}_n+ 
  \widehat{\cal A}\,)^2\Big|\rho_n\right\rangle = \int \frac{|\bmu+ 
  D\boldsymbol{\cal A}|^2}{D} \,\de^3 r\,,
\end{align}
which after dividing by $2M$ returns \eqref{Jimmy} as required.

\chapter{Overview of quantum mechanics in phase space}\label{App:Wigner}
The formalism for quantum mechanics in phase space goes by many names including the Wigner-Weyl, Wigner-Moyal or just simply the Wigner 
formulation. Despite the apparent conflict with the canonical commutation 
relation \eqref{CCR} and the corresponding Heisenberg uncertainty principle, a 
consistent phase space formulation was slowly developed by several independent works over the course of the 20th century. 
Whilst this formulation appears only briefly in this thesis, this appendix 
provides a brief overview of the phase space approach for the interested reader with further details found in \cite{ZaFaCu2005, Wyatt2006} and references therein. 

We organise this material in line with the principles of quantum mechanics from 
Section \ref{Sec:1-Postulates} before considering some additional relevant 
features of the theory.

\subsubsection{I: States} The state of a quantum system is given by a 
quasiprobability density $W(\bx,\bp)$ known as the {\it Wigner function} 
which, given a density matrix description of the same state in terms of $\rho$, is 
defined as 
\begin{align}
  W(\bx,\bp):=\frac{1}{(2\pi\hbar)^{3}}\int\rho\left(\bx+\frac{{\by}}{2},\bx-\frac{{\by}}{2}\right) 
  \,e^{-i\hbar^{-1}\bp\cdot\by}\,\de^3y\,.\label{WignerFunction}
\end{align}
The Wigner function satisfies the normalisation condition 
\begin{align}
  \iint W(\bx,\bp)\,\de^3x\,\de^3p = \text{Tr}(\rho)= 1\,, \label{WignerNorm}
\end{align}
and for pure states $\rho = \psi\psi^{\dagger}$ can be expressed as
\begin{align}
  W(\bx,\bp):=\frac{1}{(2\pi\hbar)^{3}}\int\psi^*\left(\bx+\frac{{\by}}{2}\right)\psi\left(\bx-\frac{{\by}}{2}\right) 
  \,e^{-i\hbar^{-1}\bp\cdot\by}\,\de^3y\,.
\end{align}
so that, upon making use of the general result $\int \exp(i\hbar^{-1}\bp\cdot\bs)\,\de^3s=(2\pi\hbar)^3\delta(\bs)$, the normalisation \eqref{WignerNorm} becomes the familiar $\int |\psi(\bx)|^2\,\de^3x = 
1$.

\subsubsection{II: Observables} The above definition of the Wigner function is in fact a specific example of the more general
{\it Wigner transform}, taking $\widehat{A}$, a Hermitian operator on a Hilbert space, to its 
counterpart real-valued function on phase space, $A(\bx,\bp)$. Specifically, the Wigner transform is the map $\mathcal{W}:\widehat{A}\mapsto A(\bx,\bp)$ 
given by equation \eqref{WignerFunction}, simply replacing $\rho$ and $W(\bx,\bp)$ 
with $\widehat{A}$ and $A(\bx,\bp)$ respectively. The inverse of the Wigner transform $\mathcal{W}^{-1}:A(\bx,\bp)\mapsto \widehat{A}$ is known 
as the {\it Weyl map}.

The expectation of an observable $A(\bx,\bp)$ in the state determined by the Wigner function $W(\bx,\bp)$ 
is given by
\begin{align}
  \braket{A}_W := \iint W(\bx,\bp)A(\bx,\bp)\,\de^3x\,\de^3p\,.
\end{align}

\subsubsection{III: Dynamics} The time evolution of the Wigner 
function $W(\bx,\bp)$ is given by the {\it Wigner-Moyal equation}
\begin{align}
  \partial_t W(\bx,\bp) = \{\!\{H(\bx,\bp),W(\bx,\bp)\}\!\}\,,
\end{align}
also referred to as the {\it quantum Liouville equation}. Here, $H(\bx,\bp)$ is the Hamiltonian on phase space, given by the Wigner transform of the corresponding Hamiltonian operator, 
$H(\bx,\bp)=\mathcal{W}(\widehat{H})$. This equation also introduces the {\it Moyal bracket} $\{\!\{\,\cdot\,,\,\cdot\,\}\!\}$ which is defined 
by \\
$\{\!\{A(\bx,\bp),B(\bx,\bp)\}\!\}:= -i\hbar^{-1}\mathcal{W}([\widehat{A},\widehat{B}])$, where $A(\bx,\bp)=\mathcal{W}(\widehat{A})$ and $B(\bx,\bp)=\mathcal{W}(\widehat{B})$.

\subsubsection{Relation to classical statistical mechanics}
The above formulation is a phase space correspondent to classical statistical 
mechanics in which the Liouville density is replaced by the Wigner function and 
the Liouville equation becomes the Wigner Moyal equation. Indeed this analogy can be further understood upon computing the Moyal bracket more explicitly, 
resulting in
\begin{align}
  \{\!\{A(\bx,\bp),B(\bx,\bp)\}\!\} =\{A(\bx,\bp),B(\bx,\bp)\}_c + \mathcal{O}(\hbar^2)\,, 
\end{align} 
which we recognise as the canonical Poisson bracket $\{\,\cdot\,,\,\cdot\,\}_c$ on phase space plus higher 
order derivative terms organised by even powers of $\hbar$. It follows that if either of the two phase space functions is quadratic in $(\bx,\bp)$,
 then the Moyal bracket reduces to the classical Poisson bracket and the Wigner-Moyal equation coincides with the classical Liouville equation. 
 
 Despite the similarities, there are naturally several important differences 
 between the quantum and classical phase space theories as described below.
\subsubsection{Purely quantum features}
Unlike the Liouville density, the Wigner function \eqref{WignerFunction} is 
a quasiprobability distribution, meaning it is not positive definite and can attain negative values. However, it can be shown that regions  
  for which $W(\bx,\bp)$ is negative are bounded (relating to $\hbar$ as manifestations of the uncertainty principle) and hence 
  become positive if smoothened over larger scales.

 In addition, unlike classical probability densities, the magnitude of the Wigner function is bounded such that $|W(\bx,\bp)|\leq 
  1/\pi\hbar$. This restricts quantum states being given by delta-functions,
   which are allowed in the classical setting and correspond to single 
  particle trajectories.
\\
\newline
We conclude this appendix with a comment on the geometry of the phase space 
approach to quantum mechanics. Indeed as detailed in \cite{BBMo91} the Moyal 
bracket defines a Lie algebra for phase space functions which in turn gives rise to a Lie-Poisson structure on the space of functionals. 
In fact, this can be derived explicitly by performing the Wigner transform of the Lie-Poisson bracket \eqref{LPBDens} 
from Section \ref{Sec:1-GeomQM}.

\chapter{Hydrodynamic form of the EF compatibility condition}\label{HydroCompCondApp}
In this appendix, we prove that the compatibility condition \eqref{EFEPcondition}, 
obtained in Section \ref{Sec:EF-SubHydro} using the Euler-Poincar\'e hydrodynamic approach to 
the nuclear dynamics in exact factorization, corresponds to the
Euler-Lagrange condition \eqref{EFELcondition}, after changing variables using chain 
rule calculations and applying to the Hamiltonian functional 
\eqref{EFHydroHamiltonian}.

To begin, let $h(\Omega)=h(D,\boldsymbol{\mu})$ where $D=|\Omega|^2$ and $\boldsymbol{\mu}= 
-\hbar\text{Im}(\Omega^*\nabla\Omega)$. Then,
\begin{align*}
\delta h(D,\boldsymbol{\mu})&= \left\langle \frac{\delta h}{\delta D}, \delta D \right\rangle 
+ \left\langle \frac{\delta h}{\delta \boldsymbol{\mu}}, \delta \boldsymbol{\mu} 
\right\rangle\\
&=\left\langle \frac{\delta h}{\delta D}, \delta \Omega^* \, \Omega + \Omega^*\,\delta\Omega \right\rangle 
+ \left\langle \frac{\delta h}{\delta \boldsymbol{\mu}}, -\frac{\hbar}{2i}\left(\delta\Omega^*\,\nabla\Omega + \Omega^*\,\nabla\delta\Omega - \delta\Omega\,\nabla\Omega^* - \Omega\,\nabla\delta\Omega^*\right) 
\right\rangle\\
&= \left\langle 2\frac{\delta h}{\delta D}\Omega, \delta \Omega \right\rangle 
+ \left\langle -\frac{\hbar}{i}\frac{\delta h}{\delta \boldsymbol{\mu}}\cdot\nabla\Omega, \delta \Omega 
\right\rangle+ \left\langle -\frac{\hbar}{i}\text{div}\left(\frac{\delta h}{\delta \boldsymbol{\mu}}\Omega\right), \delta \Omega 
\right\rangle\\
&=\left\langle 2\frac{\delta h}{\delta D}\Omega +i\hbar\left(2\frac{\delta h}{\delta \boldsymbol{\mu}}\cdot\nabla\Omega + \text{div}\left(\frac{\delta h}{\delta \boldsymbol{\mu}}\right)\Omega\right), \delta \Omega 
\right\rangle\\
&= \left\langle \frac{\delta h}{\delta \Omega}, \delta \Omega \right\rangle =\delta 
h(\Omega)\,,
\end{align*}
thus resulting in the transformation
\begin{align}
  \frac{\delta {h}}{\delta \Omega} &= 2\frac{\delta {h}}{\delta D}\Omega +i\hbar\left(2\frac{\delta {h}}{\delta \boldsymbol{\mu}}\cdot\nabla\Omega + \text{div}\left(\frac{\delta {h}}{\delta 
  \boldsymbol{\mu}}\right)\Omega\right)\,.
\end{align}
Hence, we can evalute the LHS of \eqref{EFELcondition} to obtain
\begin{align*}
    \text{Im}\left(\Omega^*\frac{\delta h}{\delta \Omega}\right) &=   \text{Im}\left(\Omega^*\left(2\frac{\delta {h}}{\delta D}\Omega +i\hbar\left(2\frac{\delta {h}}{\delta \boldsymbol{\mu}}\cdot\nabla\Omega + \text{div}\left(\frac{\delta {h}}{\delta 
    \boldsymbol{\mu}}\right)\Omega\right)\right)\right)\\
    &= \text{Im}\left(2D\frac{\delta {h}}{\delta D} +2i\hbar\frac{\delta {h}}{\delta \boldsymbol{\mu}}\cdot(\Omega^*\nabla\Omega) + i\hbar D\text{div}\left(\frac{\delta {h}}{\delta 
    \boldsymbol{\mu}}\right)\right)\\
    &= 2\hbar\frac{\delta {h}}{\delta \boldsymbol{\mu}}\cdot \text{Im}(i\Omega^*\nabla\Omega) 
    - \hbar D \text{div}\left(\frac{\delta {h}}{\delta 
    \boldsymbol{\mu}}\right)\\
    &= -2\hbar\frac{\delta {h}}{\delta \boldsymbol{\mu}}\cdot \text{Re}(\Omega^*\nabla\Omega) 
    - \hbar D \text{div}\left(\frac{\delta {h}}{\delta 
    \boldsymbol{\mu}}\right)\\
        &= -\hbar\nabla D\cdot\frac{\delta {h}}{\delta \boldsymbol{\mu}} 
    - \hbar D \text{div}\left(\frac{\delta {h}}{\delta 
    \boldsymbol{\mu}}\right)\\
    &= -\hbar\text{div}\left(D\frac{\delta {h}}{\delta 
    \boldsymbol{\mu}}\right)\,,
\end{align*}
thus giving the new relation
\begin{align}
 -\hbar\text{div}\left(D\frac{\delta {h}}{\delta 
    \boldsymbol{\mu}}\right)=\text{Im}\left\langle \psi
 \bigg| \frac{\delta h}{\delta \psi}\right\rangle\label{EFcondition2}\,.
\end{align}
Next, we recall that, in the nuclear hydrodynamic frame, the exact factorization Hamiltonian is given by \eqref{EFHydroHamiltonian}
\begin{align*}
  h(\boldsymbol{\mu},D,\psi)= \int 
  \frac{1}{2M}\frac{|\boldsymbol{\mu}+D\boldsymbol{\cal A}|^2}{D}+\frac{\hbar^2}{8M}\frac{(\nabla 
  D)^2}{D}\,\text{d}^3 r + F(D,\psi)\,,
\end{align*}
where
\begin{align*}
  \boldsymbol{\cal A}&:=\braket{\psi|-i\hbar\nabla\psi}\,,\\
  F(D,\psi) &:= \int D\epsilon(\psi,\nabla\psi)\,\text{d}^3r\,,
\end{align*}
from which we can compute the variational derivatives
\begin{align}
   \frac{\delta h}{\delta \boldsymbol{\mu}} &=   \frac{\boldsymbol{\mu}}{MD}+ 
   \frac{\boldsymbol{\cal A}}{M}\,,\\
   \frac{\delta h}{\delta D} &= \frac{|\boldsymbol{\mu}|^2}{2MD^2} + \frac{\boldsymbol{\cal A}^2}{2M} 
   + V_Q + \epsilon\,,\\
  \frac{\delta h}{\delta \psi} &= 
  -\frac{2i\hbar}{M}(\boldsymbol{\mu}+D\boldsymbol{\cal A})\cdot\nabla\psi - \frac{i\hbar}{M}\text{div}(\boldsymbol{\mu}+D\boldsymbol{\cal A})\psi 
  +   \frac{\delta F}{\delta \psi}\,.
\end{align}
Using these we can see how \eqref{EFcondition2} is specialised by checking each 
side individually. Firstly on the left we have,
\begin{align*}
 -\hbar\text{div}\left(D\frac{\delta {h}}{\delta 
    \boldsymbol{\mu}}\right)
     = -\frac{\hbar}{M}\text{div}\left({\boldsymbol{\mu}}+ 
  D {\boldsymbol{\cal A}}\right)\,,
\end{align*}
whilst on the right, 
\begin{align*}
\text{Im}\left\langle \psi
 \bigg| \frac{\delta h}{\delta \psi}\right\rangle
 &= \text{Im}\left\langle \psi
 \bigg|  -\frac{2i\hbar}{M}(\boldsymbol{\mu}+D\boldsymbol{\cal A})\cdot\nabla\psi - \frac{i\hbar}{M}\text{div}(\boldsymbol{\mu}+D\boldsymbol{\cal  A})\psi 
  +   \frac{\delta F}{\delta \psi}\right\rangle\\
  &= \frac{2}{M}(\boldsymbol{\mu}+D\boldsymbol{\cal  A})\cdot \underbrace{\text{Im}\braket{\psi|-i\hbar\nabla\psi}}_{=\text{Im}\boldsymbol{\cal  A}=0} 
  + 
  \frac{\hbar}{M}\text{div}(\boldsymbol{\mu}+D\boldsymbol{\cal  A})\underbrace{\text{Im}\braket{\psi|-i\psi}}_{=-1}+\text{Im}\left\langle \psi
 \bigg| \frac{\delta F}{\delta \psi}\right\rangle\\
 &= -\frac{\hbar}{M}\text{div}\left({\boldsymbol{\mu}}
  +D {\boldsymbol{\cal  A}}\right)+ \text{Im}\left\langle \psi
 \bigg| \frac{\delta F}{\delta \psi}\right\rangle\,.
\end{align*}
Thus, by combining both sides, we obtain the relation \eqref{EFEPcondition}
\begin{align*}
  \text{Im}\left\langle \psi
 \bigg| \frac{\delta F}{\delta \psi}\right\rangle=0\,,
\end{align*}
as required.

\chapter{Geometry of principal fibre bundles}\label{App:Principalbundles}

This appendix is devoted to providing a short exposition on the geometry of principal fibre bundles with particular emphasis on the role of connections and holonomy which play a key role in this thesis.
 For further details, the reader is referred to the standard reference \cite{KobayashiNomizu1963}, with specific applications to the geometric phase and mechanical systems given in \cite{ChruscinskiJamiolkowski2012,BohmEtAl2003, BohmBoyaKendrick1991, WilczekShapere1989, WayThesis2008}. 
To begin, we define the following

\begin{definition}[Principal bundle]
A {\bf principal bundle}, denoted $(P,M,\pi, G)$, is a fibre bundle $\pi:P\rightarrow M$, whose fibres are all isomorphic to the structure group, $G$. As for all bundles it consists of a {\rm base space}, $M$, 
total space $P$ and a projection $\pi:P\rightarrow M$, now such that at each $x\in M$ the preimage $\pi^{-1}(x)\cong 
G$. We denote the right action of the group by $\Phi:G\times P \rightarrow P$ 
which for a given group element maps $p\mapsto \Phi_g(p):=p\cdot g$, for all $p \in P$. 
\end{definition}

Having set the stage by defining a principal bundle, one must consider the tangent spaces and how to compare tangent vectors at different base points. Naturally due to the way a principal bundle is constructed, given a point $p \in P$ there exists a subspace $V_pP$ of the tangent space $T_pP$ 
consisting of all tangent vectors lying on the fibre $G$, called the {\it vertical subspace}. More explicitly we have that 
\begin{align}
  V_pP:= \text{ker}(\pi_*)\,, \qquad \text{where}\quad \pi_*:T_pP\to 
  T_{\pi(p)}M\,,
\end{align}
 is the push-forward of the projection map $\pi$. Clearly we have that $V_pP \cong T_p G \cong 
\mathfrak{g}$, where $\mathfrak{g}$ denotes the Lie algebra of $G$. 
One can then construct the tangent space at $p$ as the pointwise 
direct sum of vector subspaces $T_pP=V_pP \oplus H_pP$, in which $H_pP$ is 
called the {\it horizontal subspace}. However, in general there is no unique method to specify these horizontal subspaces, instead amounting to a choice of {\it connection}. 
A {\it connection} is a smoothly varying assigment of these horizontal subspaces 
such that at each $p\in P$, $H_{\Phi_g(p)}P = (\Phi_g)_*H_p P$ allowing the comparison of tangent spaces between fibres. For the purposes 
of this thesis, we focus on the associated object which determines the horizontal 
subspaces algebraically. To do so, we briefly consider exactly how elements of $\mathfrak{g}$ 
and $V_pP$ are related. Consider a curve $\Gamma(t)$ in the fibre such that $\Gamma(0)=p$ 
and $\text{d}\Gamma/\text{dt}|_{t=0}=v_p \in V_pP$. 
Then, thanks to the transitivity of the right action of $G$ on $P$, we can write $\Gamma(t)=p\cdot g(t)$ 
so that $g(0)=e$ ($e$ the identity element of $G$) and then we have that $v_p=p\,\text{d}g/\text{d}t|_{t=0}:=p\,\xi$ where $\xi \in 
\mathfrak{g}$. Notice, how this is also the infinitesimal generator $\xi_P(p)$ according to Definition \ref{Def:InfGen} and hence we have that all tangent vectors associated to {infinitesimal} generators must be vertical.
 Accordingly for each vertical vector $v_p$ we can associate a Lie 
algebra element $\xi$ corresponding to $\xi_P(p)$. Thus, we now define:

\begin{definition}[Connection 1-form]
  A connection on a principal fibre bundle $(P,M,\pi, G)$ can be specified by 
  $\mathfrak{g}$-valued 1-form $\omega_p:T_pP\rightarrow\mathfrak{g}$, the {\bf connection 1-form}, such that
  \begin{enumerate}[(i)]
  \item    $ \qquad\omega_p(\xi_P(p)) = \xi\,,$
  \item     $\qquad\omega_{\Phi_g(p)}\Big((\Phi_g)_*|_pv_p\Big) = g^{-1}\cdot \xi \cdot g\,.$
\end{enumerate}
  \end{definition}
Now that we have defined a connection, given a curve $c_M(t):[0,1]\rightarrow M$ in the base space, we can uniquely define a curve $c_P(t)$ in the total space such that $\pi(c_P(t))=c_M(t)$ for all $t$ and the tangent vectors to $c_P$ are purely horizontal and project down to tangent vectors in $M$ under $\pi_*$. Such a curve is called a {\it horizontal lift} of $c_M$. At this point, we consider a closed loop $c_M$ in the base starting and ending at $x\in M$. The horizontal lift of this loop $c_P$ begins at $p \in P$ but may not form a closed loop, instead ending at another point on the original fibre related to $p$ by $p\cdot g$.  
This difference $g$ between the start and end point of the curve $c_P$ is an element 
of a subgroup of $G$ defined as follows

\begin{definition}[Holonomy group]\label{def:Holonomy} 
  Consider a principal fibre bundle $(P,M,\pi, G)$ with connection $\omega$. Defining an equivalence relation $p\sim q$ if $p$ and $q$ can be connected by a piecewise smooth horizontal path in $P$, then 
  the {\bf holonomy group} at $p\in P$ is given by
  \begin{align}
    {\rm Hol}_p(\omega):= \{ g\in G:p \sim p\cdot g\}\,.
  \end{align}
  We refer to the group element $g$ as the {\bf holonomy} of the curve $c_P$ 
  associated with the connection $\omega$.
\end{definition}
In practice, it is useful to define the {\it local connection} on a subset of the base 
$U \subset M$ as the $\mathfrak{g}$-valued 1-form $A:T_xU\rightarrow \mathfrak{g} $ 
obtained via a local section $s:U\rightarrow P$ as the pull-back $s^*\omega$ of the globally 
defined connection $\omega$. Then, considering the case when the loop in the base $c_M$ is contained fully in $U$, it can be shown that the holonomy of the 
curve $c_P$ can in fact be given by
\begin{align}
g= \mathcal{P} \,\text{exp} \oint_{c_M}A\,,\label{HolonomyDEF}
\end{align}
where $\mathcal{P}$ denotes the path ordering operator, which is trivial if the group $G$ is Abelian. 

%
%
%
%

\chapter{Calculations for the Schr{\"o}dinger equation reconstruction}\label{App:Reconstruct}
This appendix presents the explicit calculations that reconstruct the Schr{\"o}dinger 
equation from the QHD equations in Sections \ref{Subsec:HolonomicSchr} and \ref{Sec:Osmotic}. 
In both cases, we begin by expanding the following
\begin{align}
  i\hbar\partial_t\psi &= i\hbar(\partial_t R\,\theta + R\,\partial_t 
  \theta)\,.
\end{align} 
Then, we find $\partial_t R$ from equation \eqref{transport} as follows
\begin{align}
 \partial_t R &= - \nabla R \cdot \frac{\bar\bnu}{m} - 
    \frac{R}{2m}\text{div}(\bar\bnu)\,.
\end{align}
Next, using equation \eqref{NuEvo} and \eqref{xiHamiltonJacobi1} we can also 
compute $\partial_t\theta$ and obtain
\begin{align}
\partial_t\theta &= -\frac{i}{\hbar}\left(\frac{|\bar\bnu|^2}{2m}+V +V_Q 
\right)\theta\,.
\end{align}
Hence, at this stage the Schr{\"o}dinger 
equation reads
\begin{align}
i\hbar\partial_t\psi &= \left[-\frac{i\hbar}{m}\left(\frac{\nabla R}{R}\cdot\bar\bnu\right)-\frac{i\hbar}{2m}\text{div}(\bar\bnu)+\frac{|\bar\bnu|^2}{2m} +V_Q\right]\psi 
+ V\psi\,.
\end{align}
This equation holds true for both Sections \ref{Subsec:PhaseConnection} and \ref{Sec:Osmotic}, 
simply by using the appropriate form of $V_Q$ as required. From here we split 
our calculations into two sections. Firstly, for that corrsponding to Section \ref{Subsec:HolonomicSchr} 
and then for Section \ref{Sec:Osmotic}.
\paragraph{Section \ref{Subsec:HolonomicSchr} calculations}
Clearly, at this point we must manipulate the kinetic term to get back to 
$\psi$. To do so, we recall the following relations
\begin{align}
  \frac{\nabla R}{R} = \frac{\text{Re}(\psi^*\nabla\psi)}{|\psi|^2}\,, 
  \qquad
    \bar\bnu =  \frac{\hbar\text{Im}(\psi^*\nabla\psi)}{|\psi|^2} -\hbar
{\bLambda}\,, \qquad
  V_Q=  - \frac{\hbar^2}{2m}\left(\frac{|\nabla R|^2}{R^2} + \text{div}\left(\frac{\nabla 
  R}{R}\right)\right)\,,
\end{align}
and compute term-by-term. Firstly, 
\begin{align*}
  -\frac{i\hbar}{m}\left(\frac{\nabla R}{R}\cdot\bar\bnu\right)&= 
  -\frac{i\hbar^2}{m}\frac{\text{Re}(\psi^*\nabla\psi)\cdot\text{Im}(\psi^*\nabla\psi)}{|\psi|^2|\psi|^2}+ 
  \frac{i\hbar^2}{m}\frac{\text{Re}(\psi^*\nabla\psi)}{|\psi|^2}\cdot{\bLambda}\,.
\end{align*}
Secondly,
\begin{align*}
  -\frac{i\hbar}{2m}\text{div}(\bar\bnu)&= -\frac{i\hbar^2}{2m}\left(\nabla((\psi^*\psi)^{-1})\cdot\text{Im}(\psi^*\nabla\psi)+ 
\frac{\cancelto{0}{\text{Im}(\nabla\psi^*\cdot\nabla\psi)}}{|\psi|^2}+\frac{\text{Im}(\psi^*\Delta\psi)}{|\psi|^2}\right)\\
&= 
-\frac{i\hbar^2}{2m}\left(-(\psi^*\psi)^{-2}(\nabla\psi^*\psi+\psi^*\nabla\psi)\cdot\text{Im}(\psi^*\nabla\psi)+\frac{\text{Im}(\psi^*\Delta\psi)}{|\psi|^2}\right)\\
&= -\frac{i\hbar^2}{2m}\frac{\text{Im}(\psi^*\Delta\psi)}{|\psi|^2} + 
\frac{i\hbar^2}{m}\frac{\text{Re}(\psi^*\nabla\psi)\cdot\text{Im}(\psi^*\nabla\psi)}{|\psi|^2|\psi|^2}\,.
\end{align*}
where in the second line we have used that ${\bLambda} = 
-\nabla\times\bbeta$ so that its gradient vanishes. Thirdly,
\begin{align*}
  \frac{|\bar\bnu|^2}{2m} 
&= \frac{\hbar^2}{2m}\frac{\text{Im}(\psi^*\nabla\psi)\cdot\text{Im}(\psi^*\nabla\psi)}{|\psi|^2|\psi|^2} 
- 
\frac{\hbar^2}{m}\frac{\text{Im}(\psi^*\nabla\psi)}{|\psi|^2}\cdot{\bLambda}+\frac{\hbar^2}{2m}|{\bLambda}|^2\,.
\end{align*}
Finally,
\begin{align*}
  V_Q 
    &= -\frac{\hbar^2}{2m}\left(\frac{\text{Re}(\psi^*\nabla\psi)\cdot\text{Re}(\psi^*\nabla\psi)}{|\psi|^2|\psi|^2} +\frac{\text{Re}(\nabla\psi^*\cdot\nabla\psi)}{|\psi|^2}+\frac{\text{Re}(\psi^*\Delta\psi)}{|\psi|^2}+ 
  \nabla((\psi^*\psi)^{-1})\cdot\text{Re}(\psi^*\nabla\psi)\right)\\
    &= -\frac{\hbar^2}{2m}\left(\frac{\text{Re}(\psi^*\nabla\psi)\cdot\text{Re}(\psi^*\nabla\psi)}{|\psi|^2|\psi|^2} +\frac{|\nabla\psi|^2}{|\psi|^2}+\frac{\text{Re}(\psi^*\Delta\psi)}{|\psi|^2} 
  -2\frac{\text{Re}(\psi^*\nabla\psi)\cdot\text{Re}(\psi^*\nabla\psi)}{|\psi|^2|\psi|^2}\right)\\
   &= -\frac{\hbar^2}{2m}\left(-\frac{\text{Re}(\psi^*\nabla\psi)\cdot\text{Re}(\psi^*\nabla\psi)}{|\psi|^2|\psi|^2} +\frac{|\nabla\psi|^2}{|\psi|^2}+\frac{\text{Re}(\psi^*\Delta\psi)}{|\psi|^2} 
   \right)\,.
\end{align*}
So that all together the kinetic term reads
\begin{align*}
  -\frac{i\hbar}{m}\left(\frac{\nabla R}{R}\cdot\bar\bnu\right)-\frac{i\hbar}{2m}\text{div}(\bar\bnu)+\frac{|\bar\bnu|^2}{2m} 
  +V_Q&= -\frac{\hbar^2}{2m}\left(\frac{\text{Re}(\psi^*\Delta\psi)}{|\psi|^2}+ \frac{i\text{Im}(\psi^*\Delta\psi)}{|\psi|^2}\right) 
  \\
  &+ \frac{i\hbar^2}{m}\left(\frac{\text{Re}(\psi^*\nabla\psi)}{|\psi|^2}+ 
  \frac{i\text{Im}(\psi^*\nabla\psi)}{|\psi|^2}\right)\cdot{\bLambda}+\frac{\hbar^2}{2m}|{\bLambda}|^2\\
  &+\frac{\hbar^2}{2m}\frac{\text{Re}(\psi^*\nabla\psi)\cdot\text{Re}(\psi^*\nabla\psi)}{|\psi|^2|\psi|^2}
\\
&\qquad+\frac{\hbar^2}{2m}\frac{\text{Im}(\psi^*\nabla\psi)\cdot\text{Im}(\psi^*\nabla\psi)}{|\psi|^2|\psi|^2}- \frac{\hbar^2}{2m}\frac{|\nabla\psi|^2}{|\psi|^2}\,,
\end{align*}
at which point we rewrite the following terms
\begin{align*}
  \frac{\hbar^2}{2m}\frac{\text{Re}(\psi^*\nabla\psi)\cdot\text{Re}(\psi^*\nabla\psi)}{|\psi|^2|\psi|^2}
+\frac{\hbar^2}{2m}\frac{\text{Im}(\psi^*\nabla\psi)\cdot\text{Im}(\psi^*\nabla\psi)}{|\psi|^2|\psi|^2}  
= \frac{\hbar^2}{2m}\frac{|\psi^*\nabla\psi|^2}{|\psi|^2|\psi|^2} =  \frac{\hbar^2}{2m}\frac{|\nabla\psi|^2}{|\psi|^2} 
\,,
\end{align*}
so that after the subsequent cancellations one is left with
\begin{align}
    -\frac{i\hbar}{m}\left(\frac{\nabla R}{R}\cdot\bar\bnu\right)-\frac{i\hbar}{2m}\text{div}(\bar\bnu)+\frac{|\bar\bnu|^2}{2m} 
  +V_Q&=-\frac{\hbar^2}{2m}\frac{\Delta\psi}{\psi} +
  \frac{i\hbar^2}{m}\frac{\nabla\psi}{\psi}\cdot{\bLambda} + 
  \frac{\hbar^2}{2m}|{\bLambda} |^2\,.
\end{align}
Then multiplying by $\psi$ and factorising returns the desired result.
\paragraph{Section \ref{Sec:Osmotic} calculations} 
Having introduced the connection correspondong to the amplitude, the calculations simply significantly. It is simple to check that 
using the Helmholtz decomposition for $\bupsilon$ \eqref{john} the kinetic term 
reads
\begin{align}
  \frac{\hbar^2}{2m}\left(-|\bupsilon+{\bnu}|^2 + \text{div}(\bupsilon+{\bnu}) + 
  (\bupsilon+{\bnu})\cdot\nabla\times\bkappa\right)\,.
\end{align}
Then, using that
\begin{align}
 \bupsilon +\bnu &= -\frac{\nabla\psi}{\psi} + 
 \frac{1}{2} \nabla\times\bkappa+ i\bLambda\,,
\end{align}
we evaluate term-by-term. Firstly,
\begin{align*}
   |\bupsilon +\bnu|^2 &= \frac{|\nabla\psi|^2}{\psi^2} + 
   \frac{1}{4}|\nabla\times\bkappa|^2 - |\bLambda|^2 - 
   \frac{\nabla\psi}{\psi}\cdot\nabla\times\bkappa - 2  
   \frac{\nabla\psi}{\psi}\cdot i\bLambda + 
   i\bLambda\cdot\nabla\times\bkappa\,.
\end{align*}
Secondly, 
\begin{align*}
  \text{div}(\bupsilon +\bnu) &= -\frac{\Delta\psi}{\psi}+ \frac{|\nabla\psi|^2}{\psi^2} 
  \,,
\end{align*}
where the term $\text{div}(\bLambda)$ vanishes as $\bLambda = 
-\nabla\times\bbeta$. Finally,
\begin{align}
  (\bupsilon +\bnu)\cdot\nabla\times\bkappa = -
  \frac{\nabla\psi}{\psi}\cdot\nabla\times\bkappa + 
  \frac{1}{2}|\nabla\times\bkappa|^2 +i\bLambda\cdot\nabla\times\bkappa\,.
\end{align}
which after combing together results in equation \eqref{kinetictermsimplified} 
as required. From there, it is simple to verify the final form Schr{\"o}dinger equation.

\chapter{Rasetti-Regge gauge condition}\label{App:RR}
In this appendix we provide some of the supporting calculations for Section 
\ref{Sec:Vortices}. Firstly, we prove explicitly that the so-called gauge 
invariance condition \cite{RasettiRegge1975,Holm03,HolmStechmann2004}
\begin{align}
\bR_{\sigma}\cdot\frac{\delta h}{\delta 
\bR}=0\label{vortexHamiltoniancondition}\,,
\end{align}
is valid for any Hamiltonian that can be expressed in terms of the vorticity 
$h=h(\bomega)$.

We begin by using the chain rule to express 
\begin{align}
  \frac{\delta h}{\delta R_i} &= \int\frac{\delta \omega_j}{\delta R_i}\frac{\delta h}{\delta 
  \omega_j}\,\text{d}^3x\,.
\end{align}
Then, using that $\bomega(\bx)=\int 
\bR_{\sigma}\,\delta(\bx-\bR)\,\text{d}\sigma$, we compute the following (noting that we adopt the notation $\partial_\sigma\bR$ rather than $\bR_\sigma$ when working with indices)
\begin{align*}
\delta \omega_j &= \int \partial_{\sigma} (\delta R_j )\delta(\bx-\bR) + (\partial_{\sigma}  R_j) 
\delta [\delta(\bx-\bR)]\,\text{d}\sigma\\
&=  \int -\delta R_j \,\partial_{\sigma} \delta(\bx-\bR) + (\partial_{\sigma}  R_j) 
\delta [\delta(\bx-\bR)]\,\text{d}\sigma\\
&= \int -\delta R_j (\nabla_{\bR}\delta(\bx-\bR) \cdot \bR_{\sigma}) + \partial_{\sigma}  R_j 
(\nabla_{\bR}\delta(\bx-\bR) \cdot \delta\bR)\,\text{d}\sigma\\
&= \int \delta R_j (\nabla\delta \cdot \bR_{\sigma}) - \partial_{\sigma}  R_j 
(\nabla\delta\cdot \delta\bR)\,\text{d}\sigma\\
&= \int \Big((\nabla\delta \cdot \bR_{\sigma})\delta_{ij} - \partial_{\sigma}R_j\partial_i\delta\Big) \delta R_i 
\,\text{d}\sigma\,,
\end{align*}
so that 
\begin{align}
  \frac{\delta \omega_j}{\delta R_i} &= (\nabla\delta \cdot \bR_{\sigma})\delta_{ij} - 
  \partial_{\sigma}R_j\,\partial_i\delta\,.
\end{align}
Then we compute explicitly
\begin{align*}
    \frac{\delta h}{\delta R_i} &= \int\frac{\delta \omega_j}{\delta R_i}\frac{\delta h}{\delta 
  \omega_j}\,\text{d}^3x\\
  &= \int \Big((\nabla\delta \cdot \bR_{\sigma})\delta_{ij} - 
  \partial_{\sigma}R_j\partial_i\delta\Big)\frac{\delta h}{\delta 
  \omega_j}\,\text{d}^3x\\
    &= \int \partial_j\delta \partial_{\sigma}R_j\frac{\delta h}{\delta 
  \omega_i} + 
  \partial_{\sigma}R_j\partial_i\frac{\delta h}{\delta 
  \omega_j}\delta(\bx-\bR) \,\text{d}^3x\\
      &= \int -\delta(\bx-\bR) \partial_{\sigma}R_j\partial_j\frac{\delta h}{\delta 
  \omega_i} + 
  \partial_{\sigma}R_j\partial_i\frac{\delta h}{\delta 
  \omega_j}\delta(\bx-\bR) \,\text{d}^3x\\
  &= \left[\partial_i\frac{\delta h}{\delta 
  \omega_j}-\partial_j\frac{\delta h}{\delta 
  \omega_i}\right]_{\bx=\bR}\partial_{\sigma}R_j\,,
\end{align*}
which can be shown using vector calculus manipulation to be equivalent to the 
vector equation
\begin{align}
  \frac{\delta h}{\delta \bR} &= \bR_{\sigma}\times\left[\nabla\times\frac{\delta h}{\delta 
  \bomega}\right]_{\bx=\bR}\,.\label{James}
\end{align}
Hence, it follows that the condition \eqref{vortexHamiltoniancondition} is 
immediately satisfied. \\
\newline
In the last part of this appendix we simply elaborate on how \eqref{James} is rewritten in terms of $\bLambda$ to obtain the first equality of \eqref{DerivativeLambda}. 
To do so, we compute the following 
\begin{align*}
  \delta h(\bomega) &= \int \frac{\delta h}{\delta \bomega}\cdot \delta \bomega 
  \,\text{d}^3x\\
  &=  -\frac{\hbar}{m}\int \frac{\delta h}{\delta \bomega}\cdot 
  \nabla\times\delta\bLambda
  \,\text{d}^3x\\
    &=  -\frac{\hbar}{m}\int  \nabla\times\frac{\delta h}{\delta \bomega}\cdot\delta\bLambda
  \,\text{d}^3x\,,
\end{align*}
having used that $\bomega = -\hbar m^{-1}\nabla\times\bLambda$. Then it follows that 
as $\delta h(\bomega)=\delta h (\bLambda)$ we must have that
\begin{align}
  \frac{\delta h}{\delta \bLambda}  &= -\frac{\hbar}{m}  \nabla\times\frac{\delta h}{\delta 
  \bomega}\,,
\end{align}
which allows us to write the derivative $\delta h/\delta \bR$ in terms of $\bLambda$ 
as 
\begin{align}
    \frac{\delta h}{\delta \bR} &= -\frac{m}{\hbar}\bR_{\sigma}\times \frac{\delta h}{\delta 
  \bLambda}\Bigg|_{\bx=\bR}\label{conditionv2}\,.
\end{align}

\chapter{Spacetime formalism for non-Abelian quantum connections}\label{App:Spacetime}
 This appendix reports on some of the preliminary findings that follow from the 
 discussion in Remark \ref{Rem:Spacetime}, considering the generator $\xi=\partial_t U\,U^{-1}$ and gauge connection $\bgamma = -\nabla U\,U^{-1}+U\bgamma_0 U^{-1}$ as a unified object on four-dimensional 
 spacetime. Before employing a spacetime description, we first consider some further relations that can be derived from the dynamical equation for the 
 connection.
 
\begin{proposition}[Gauge connection evolution and interesting expectation relations]\label{Prop:ExpRelations}
  Following Theorem \ref{Theorem:GammaGenEP} we have that the time evolution of the gauge connection is given by the covariant 
  derivative of the generator so that 
  \begin{align}
    \partial_t\gamma_j =   -\partial_j^{\gamma}\xi = -\partial_j\xi -[\gamma_j, \xi]\,.\label{QuantumConnectionEvo}
  \end{align}
  Upon computing expectation values, this equation leads us to the following interesting 
  relations
\begin{align}
  \braket{\partial_t\gamma_j} = -\partial_j\braket{\xi}\,,\qquad
  \braket{\partial_j\xi} = - \partial_t\braket{\gamma_j}\label{interesting}\,.
\end{align}
\end{proposition}
 \paragraph{Proof:} We prove the first relation in \eqref{interesting} as follows
 \begin{align*}
  \braket{  \partial_t\gamma_j } &= \braket{- \partial_j\xi + [\xi,\gamma_j]}\\
  &= -\partial_j\braket{\xi} + \braket{\partial_j\psi|\xi|\psi} + \braket{\psi|\xi|\partial_j\psi} 
  + \braket{[\xi,\gamma_j]}\\
  &= -\partial_j\braket{\xi} + \braket{\psi|\gamma_j\xi|\psi} - \braket{\psi|\xi\gamma_j|\psi} 
  + \braket{[\xi,\gamma_j]}\\
  &= -\partial_j\braket{\xi}\,,
\end{align*}
noting that an anlogous computation proves the second relation in 
\eqref{interesting}.
 \hfill$\square$\\ 
 
  Notice, how \eqref{generalunitarygenerator} implies that $\braket{\xi}=\braket{\psi|\partial_t{\psi}}$ which, as we have seen, is set to 
0 in the Weyl gauge, in turn resulting in $\braket{\partial_t\gamma_j}=0$.  
 
In addition the evolution equation for the connection allows insightful new 
expressions for the time evolution of the Berry connection and curvature:
\begin{proposition}[Time evolution of the Berry connection and curvature]\label{Prop:BerryEvos}
  \hfill \\
  The time evolution of the $\mathfrak{u}(\mathscr{H})$ connection $\gamma$ 
  given by \eqref{QuantumConnectionEvo} results in the following evolution 
  equations for the Berry connection and Berry curvature
  \begin{align}
    {\cal A}_j = -\partial_j\braket{i\hbar\xi} - i\hbar\braket{[\xi, 
    \gamma_j]}\,,\\
    {\cal B}_{ij} = -i\hbar\Big(\braket{[\partial_i\xi, \gamma_j]}-\braket{[\partial_j\xi, 
    \gamma_i]}\Big)\,.
  \end{align}
By comparison with the analogous equation in classical electromagnetism for the magnetic 
vector potential $\partial_t\bA = -\nabla\Phi -\bE$, we identify the Berry electric field 
\begin{align}
{  \cal E}_j := i\hbar\braket{[\xi, \gamma_j]}\,.
\end{align}
The above evolution of the Berry curvature then takes the form $\partial_t\boldsymbol{\cal B}=-\nabla\times\boldsymbol{\cal 
E}$, again analogous to the corresponding equation for the magnetic field in 
classical electromagnetism.
\end{proposition}
 \paragraph{Proof:} Both results follow by direct computation using the 
 evolution equation \eqref{QuantumConnectionEvo}. The evolution of the Berry 
 connection is computed as
 \begin{align*}
   \partial_t{\cal A}_j &= \partial_t\braket{\psi|i\hbar\gamma_j|\psi}\\
   &= \braket{\partial_t\psi|i\hbar\gamma_j|\psi}+ 
   \braket{\psi|i\hbar\partial_t\gamma_j|\psi}+ 
   \braket{\psi|i\hbar\gamma_j|\partial_t\psi}\\
   &= \braket{\xi\psi|i\hbar\gamma_j\psi}-i\hbar\partial_j\braket{\xi} 
+    \braket{\psi|i\hbar\gamma_j|\xi\psi}\\
   &= -i\hbar\braket{\xi\gamma_j} -i\hbar\partial_j\braket{\xi} 
+    i\hbar\braket{\gamma_j\xi}\\
&= -\partial_j\braket{i\hbar\xi} - i\hbar\braket{[\xi, 
    \gamma_j]}\,,
 \end{align*}
 as required, where in the third line we have made use of the first relation in \eqref{interesting}.
 An analogous type of computation proves the result for the Berry 
 curvature.
 \hfill$\square$\\
 
At this point, following the equivalent covariant formulation of classical electromagnetism, Propositions \ref{Prop:ExpRelations} and \ref{Prop:BerryEvos} 
beg us to follow an analogous procedure and incorporate the spatial and temporal 
components into a spacetime 4-vector, in which the temporal component is the 0th index. 
To do so, we introduce the greek index $\mu \in \{0,1,2,3\}$ and define
\begin{align*}
  \partial_{\mu} &= (\partial_0, \partial_j) := (-\partial_t, \partial_j)\,,
  \end{align*}
where we have chosen to use the $(-,+,+,+)$ sign convention for the 
metric on Minkowski space $\mathbb{R}^{1,3}$. 

As a first port-of-call, we begin by considering the $\mathcal{U}(1)$ theory associated to the Berry connection, simply compressing standard results analogous to electromagnetic theory into their spacetime counterparts.
  Firstly, the Berry connection and 
  corresponding scalar potential, can now be combined into a single object
  \begin{align}
  \begin{rcases*}
    \Phi = \braket{\psi|i\hbar\partial_t\psi}\\
    {\cal A}_j = \braket{\psi|-i\hbar\partial_j\psi}
  \end{rcases*}\longrightarrow \quad
  {\cal A}_{\mu} = 
  \braket{\psi|-i\hbar\partial_{\mu}\psi}\label{Berry4potential}\,,
\end{align}
  which we will call the {\it Berry 4-connection}, ${\cal A}_{\mu}$, in which ${\cal A}_0:=\Phi$ is the scalar field. 
The Berry 4-connection can be understood as a $\mathfrak{u}(1)$-connection 
  on Minkowski space so that ${\cal A}_{\mu}\in 
  \Lambda^{1}(\mathbb{R}^{1,3},\mathfrak{u}(1))$.
  
  Similarly, the Berry curvature, corresponding to the Maxwell magnetic field, and the analogous 
  electric field can be combined as
  \begin{align}
  \begin{rcases*}
    {\cal E}_j =-\partial_t{\cal A}_j - \partial_j\Phi\\
    {\cal B}_{ij} = \partial_i{\cal A}_j - \partial_j{\cal A}_i
  \end{rcases*}\longrightarrow \quad
  {\cal F}_{\mu\nu} = \partial_{\mu}{\cal A}_{\nu} - 
  \partial_{\nu}{\cal A}_{\mu}\label{BerryFieldStrength}\,,
\end{align}
 into the {\it Berry field strength tensor}, ${\cal F}_{\mu\nu}$, in which 
${\cal F}_{0j}={\cal E}_j$, ${\cal F}_{ij}={\cal B}_{ij}$ and ${\cal F}_{\mu\mu}=0$. This definition can also be 
understood as the curvature of the Berry 4-connection ${\cal A}_{\mu}$ so that ${\cal F}=\de{\cal A}\in \Lambda^2(\mathbb{R}^{1,3},\mathfrak{u}(1))$.

\begin{remark}[Bianchi identity and Maxwell's equations]
  \hfill \\
  Following the introduction of the Berry field strength, we remark that the Bianchi identity $\de {\cal F}=0$ contains two relations, each a counterpart of Maxwell's equations of classical electromagnetism. 
  Expressed in components this reads
  \begin{align*}
      \partial_{\mu}{\cal F}_{\nu\rho}+  \partial_{\nu}{\cal F}_{\rho\mu}+  
  \partial_{\rho}{\cal F}_{\mu\nu}=0\,,
  \end{align*}
so that upon setting $\mu = 0$, $\nu = i$, $\rho = j$ we recover the Maxwell-Faraday equation, whilst $\mu = i$, $\nu = j$, $\rho = k$ results in Gauss' Law for the Berry 
  curvature, written explicitly as
\begin{align*}
   \partial_t\boldsymbol{\cal B} = -\nabla\times\boldsymbol{\cal E}\,,\qquad
      \nabla\cdot\boldsymbol{\cal B}=0\,,
\end{align*} 
respectively. We note that these equations hold trivially when expressed in terms of $\psi$ and its 
derivatives.
\end{remark}

Having compressed well-known results regarding the $\mathcal{U}(1)$ gauge theory 
associated to the Berry connection into a spacetime formulation, we now look to 
perform the analogous procedure for the new $\mathcal{U}(\mathscr{H})$ theory 
from Section \ref{Sec:NonAbelianQuantumSystems}.

As we have seen, the assumption of unitary evolution for $\psi$ allowed us to 
introduce the generator $\xi \in \mathcal{F}(\mathbb{R}^3, \mathfrak{u}(\mathscr{H}))$ and components of the connection $\gamma_j \in  \mathcal{F}(\mathbb{R}^3, \mathfrak{u}(\mathscr{H}))$ for the temporal and spatial derivatives 
of $\psi$ respectively. We can combine these into a single object:
  \begin{align}
    \begin{rcases*}
\partial_t\psi = \xi\psi \\
\partial_j\psi = -\gamma_j\psi
\end{rcases*} \longrightarrow \quad \partial_{\mu}\psi = -\gamma_{\mu}\psi\,,\label{4DGammaConnection}
  \end{align}
in which $\gamma_0 := \xi$. This can be understood as defining a $\mathfrak{u}(\mathscr{H})$-connection on Minkowski space $\gamma_{\mu}\in\Lambda^1(\mathbb{R}^{1,3}, 
\mathfrak{u}(\mathscr{H}))$. 
The natural next step is to consider generalising the curvature of $\gamma$ to its four-dimensional 
counterpart
\begin{align}
  \Omega_{\mu\nu}:= \partial_{\mu}\gamma_{\nu}-\partial_{\nu}\gamma_{\mu}+[\gamma_{\mu},\gamma_{\nu}] 
  \in \Lambda^2(\mathbb{R}^{1,3},\mathfrak{u}(\mathscr{H}))\,.\label{4DGammaCurvature}
\end{align}
At this point, we consider the new additional components that this curvature possesses. As all diagonal terms are identically zero,
one sees that the only new component is given by
\begin{align*}
 \begin{split} 
 \Omega_{0j} &= -\partial_t\gamma_j - \partial_j\xi + [\xi,\gamma_j]\\
  &= -(-\partial_j\xi + [\xi,\gamma_j])- \partial_j\xi + 
[\xi,\gamma_j] = 0\,,
\end{split}
\end{align*}
and thus $\Omega_{\mu\nu}$ contains no additional information than its purely spatial counterpart. Equivalently in physics terminology we can say that we do not have a Yang-Mills electric field. 

In writing the non-Abelian connection on spacetime we can 
write the generalisation of \eqref{QuantumConnectionEvo} as well as combine the interesting relations from Proposition \ref{Prop:ExpRelations} as 
follows
\begin{corollary}[Generalised evolution equation and interesting relations]
  \hfill \\
  Following the introduction of the spacetime 
  $\mathfrak{u}(\mathscr{H})$-connection in equation \eqref{4DGammaConnection}, the 
generalised spacetime version of the evolution equation \eqref{QuantumConnectionEvo} 
reads
\begin{align}
  \partial_{\mu}\gamma_{\nu}-\partial_{\nu}\gamma_{\mu} = 
  [\gamma_{\nu},\gamma_{\mu}]\,,\label{4Dzerocurvature}
\end{align}
which we recognise as a zero curvature relation for the 4D curvature \eqref{4DGammaCurvature} 
$\Omega_{\mu\nu}=0$. In addition the interesting relations between expectation 
values are combined into the single expression
  \begin{align}
 \braket{\partial_{\mu}\gamma_{\nu}} = 
 \partial_{\nu}\braket{\gamma_{\mu}}\label{interestinggeneralized}\,.
\end{align}
\end{corollary}
We conclude the material in this appendix by considering the relation between 
the non-Abelian $\mathfrak{u}(\mathscr{H})$ and Abelian $\mathfrak{u}(1)$ spacetime 
theories.
\begin{proposition}[Relation between the non-Abelian and Abelian gauge theories]
  \hfill \\
Following the earlier result \eqref{BerryConnectionIsTheExpectation} in 
  which the Berry connection $\boldsymbol{\cal A}$ is expressed as the 
  expectation value of the $\mathfrak{u}(\mathscr{H})$ connection $\gamma$, we 
  have that the spacetime generalisation
  \begin{align}
      {\cal A}_{\mu} = i\hbar\braket{\gamma_{\mu}}\label{Berry4DExpectation}\,.
  \end{align} 
  Similarly, the Berry field strength tensor can be expressed in terms of the 4D 
  curvature \eqref{4DGammaCurvature} as
  \begin{align}
     {\cal F}_{\mu\nu} = i\hbar \braket{[\gamma_{\mu},\gamma_{\nu}]}\,.
  \end{align}
\end{proposition}
\paragraph{Proof:} Whilst the first equation for the Berry 4-connection follows 
by earlier results, the equation for the Berry field strength can be derived as 
follows
\begin{align*}
  {\cal F}_{\mu\nu} &:= \partial_{\mu}{\cal A}_{\nu}-\partial_{\nu}{\cal 
  A}_{\mu}\\
  &= 
  \partial_{\mu}\braket{i\hbar\gamma_{\nu}}-\partial_{\nu}\braket{i\hbar\gamma_{\mu}}\\
  &= i\hbar \braket{\partial_{\nu}\gamma_{\mu}-\partial_{\mu}\gamma_{\nu}}\\
  &= i\hbar\braket{[\gamma_{\mu},\gamma_{\nu}]}\,.
\end{align*}
Here we have subsequently used equation \eqref{Berry4DExpectation} as well as the interesting 
relation \eqref{interestinggeneralized} and the zero curvature relation 
\eqref{4Dzerocurvature}.
\hfill$\square$\\

As discussed in greater detail in Section \ref{Sec:Perspectives}, this 
preliminary investigation opens many further questions which warrant further research.

\renewcommand{\bibname}{References}

\newpage

\pagestyle{fancy}
\fancyhead[RO,LE]{}
\fancyhead[CO,CE]{}
\fancyhead[LO,RE]{}
\fancyfoot[CO,CE]{\thepage}

\subsection*{List of publications}
Foskett, M. S., Holm, D. D., \& Tronci, C. \href{https://link.springer.com/article/10.1007/s10440-019-00257-1}{{Geometry of nonadiabatic quantum hydrodynamics.}} 
\\ {\it Acta Applicandae Mathematicae}, 162(1), 63-103, (2019). 
\\
\newline
Foskett, M. S., \& Tronci, C. \href{https://arxiv.org/pdf/2003.08664.pdf}{Holonomy and vortex structures in quantum hydrodynamics.} 
\\ {arXiv preprint 
arXiv:2003.08664}, {\it Submitted to Math. Sci. Res. Inst. Publ., Cambridge University Press}, 
(2020).

\vspace{1cm}

\subsection*{List of invited talks}
\href{https://sites.google.com/site/geometricmechanics/agm-meetings-2017/second-agm-meeting}{Second Applied Geometric Mechanics meeting  2017-2018 on Geometric Quantum Dynamics, Imperial College, London, 
UK.}
\\ Invited talk: {\it Wavefunction factorisations and gauge structure dynamics in quantum 
chemistry}, 7th December 2017.\\
\newline
\href{https://sites.google.com/site/geometricmechanics/agm-meetings-2019/first-agm-meeting-2019}{First Applied Geometric Mechanics Meeting 2019 on Geometric Quantum Mechanics, University of Surrey, Guildford, UK.}
\\ Invited talk: {\it Geometry of nonadiabatic quantum hydrodynamics}, 25th April 
2019.

\vspace{1cm}

\subsection*{List of poster presentations}
\href{https://memento.epfl.ch/event/nonadiabatic-molecular-dynamics-in-three-different/}{CECAM School: Nonadiabatic Molecular Dynamics in Three Different Flavors, EPFL, Lausanne, 
Switzerland.}
\\ Poster: {\it A geometric mechanics approach to exact factorisation}, 26th February 2018 - 2nd March 2018.

\end{document}